\documentclass[
12pt,
a4paper,
bibtotoc, 
liststotoc, 
idxtotoc, 
onehalfspacing 
]
{icldt}

\usepackage[hyperref 
            ]{ihformat}

\renewcommand{\CVSrevision}{\version$Id: thesis.tex,v 1.43 2009/11/26 16:29:09 ith Exp $}

\begin{document}

%

%
\college{Queen Mary, }
\department{Mathematical Sciences}
\title{Constraining Inflationary Scenarios with Braneworld Models and Second Order
Cosmological Perturbations}
\author{Ian Huston}
\hypersetup{pdftitle={Constraining Inflationary Scenarios with Braneworld Models and
Second Order Cosmological Perturbations},pdfauthor={Ian Huston}}

\declaration{%
I hereby certify that this thesis, which is approximately 45,000 words in length,
has been written by me; that it is the record of the work carried out by me at the
Astronomy Unit, Queen Mary, University of London, and that it has not been submitted
in any previous application for a higher degree.
\\
Chapter~\ref{ch:dbi} primarily contains work done with James E.
Lidsey and published in the Journal of Cosmology and Astroparticle Physics (JCAP)
\cite{lidseyhuston}. Section~\ref{sec:twlargen-multi} also contains work published
in \Rref{lidseyhuston}.
The majority of Chapter~\ref{ch:multibrane} is based on work completed in
collaboration with James E. Lidsey, Steven Thomas and John Ward and published in JCAP
\cite{hltw}. 
Chapters~\ref{ch:perts}, \ref{ch:numericalsystem} and \ref{ch:results} contain work
done with Karim Malik, and large parts of these chapters were published in JCAP
\cite{hustonmalik}.
\\
I have made a major contribution to all the original research presented in this
thesis.
}

\maketitle

%

%
%
\chapter*{Abstract}
\label{ch:abstract}
\addcontentsline{toc}{chapter}{Abstract}
\section*{}
\singlespacing
Inflationary cosmology is the leading explanation of the very early universe. 
Many different models of inflation have been constructed which fit current observational data.
In this work theoretical and numerical methods for constraining the parameter space of a wide class
of such models are described.

First, string-theoretic models with large non-Gaussian signatures are investigated.
An upper bound is placed on the amplitude of primordial gravitational waves produced by ultra-violet
Dirac-Born-Infeld inflation. In all but the most finely tuned cases, this bound is
incompatible with
a lower bound derived for inflationary models which exhibit a red spectrum and detectable
non-Gaussianity. 

By analysing general non-canonical actions, a class of models is found which can
evade the upper bound when the phase speed of perturbations is small. The multi-coincident brane
scenario with a finite number of branes is one such model. 
For models with a potentially observable gravitational wave spectrum the number of coincident branes
is shown to take only small values. 

The second method of constraining inflationary models is the numerical calculation
of second order perturbations for a general class of single field models.
The Klein-Gordon equation at second order, written in terms
of scalar field variations only, is numerically solved. 
The slow roll version of the second order source term is used and the method is
shown to be extendable to the full equation.
This procedure allows the evolution of second order
perturbations in general and the calculation of the non-Gaussianity parameter in
cases
where there is no analytical solution available.


%
%
%
\chapter*{Acknowledgements}
\label{ch:acknowledgements}
\addcontentsline{toc}{chapter}{Acknowledgements}

First, I would like to thank James Lidsey, Karim Malik and Reza Tavakol for all their help and
guidance. Thank you to the past and present members of Room 301, and all my other colleagues in
QMUL who make working there such a pleasant experience.

I am very grateful to Cathy, Sara, Susan, Lisa, Kevin, Fiona and Andrew for putting up with me over
the past few years and making the move to London so enjoyable. And thank you to all my friends for
not letting me talk about work when around them.

I am indebted to my parents, Brian and Phil, my brother and sister, Paul and Anne, 
and my grandparents and extended family
for all their support and encouragement. 

Finally, I could not have survived the last few years and especially the last few months without 
Gr\'{a}inne beside me. Thank you for everything.

\vfill

This work was funded by a combination of a PPARC/STFC Studentship and a Queen Mary Studentship and
by the EPSTAR consortium (Queen Mary).


\tableofcontents
\listoffigures
\listoftables

\onehalfspacing
%

\chapter{Introduction}
\label{ch:shortintro}
In the past cosmology was a speculative science. The scarcity of observational
data meant that many conflicting theories for the evolution of the universe were
entertained, with nothing but personal opinion to differentiate between them.
The explosion in the quantity and quality of observational data in recent years
has led to a much more competitive marketplace of ideas about the physical
beginning of the universe.

The Big Bang scenario has emerged as a cohesive framework for the evolution of
the universe from very early times. The observation of the Cosmic
Microwave Background (CMB) provided much supporting evidence for this scenario
\cite{book:kolbturner}. This
relic radiation, emitted 300,000 years after the Big Bang, continues to
be our primary source of information about the early universe.

The inflationary scenario is an attempt to solve problems with the standard Big
Bang picture and provide an origin for the fluctuations in energy that seeded
the growth of structure in the universe
\cite{Starobinsky:1980te,Guth:1980zm,Albrecht:1982wi,Linde:1981mu,
Starobinsky:1982ee}. These fluctuations link the classical scales of
relativistic gravity with the quantum scales of Planck level physics. There are
many possible realisations of inflation and there has been an explosion in the
number of theoretical models which agree with current observational limits
(for reviews see, for example, Refs.~\cite{book:liddle, Alabidi:2008ej,
Baumann2009}).

Ground and space-based observations have significantly challenged theoretical
cosmological models with a wealth of new data. The Wilkinson Microwave
Anisotropy Probe (WMAP) mission \cite{Komatsu:2008hk}, in conjunction with supernova
surveys and other evidence, have shown that the fluctuations in
the temperature of the CMB are $10^{5}$ times smaller than the background value
and that the magnitude of the fluctuations is roughly independent of the
angular scales at which they are measured. This is in agreement with the
predictions of inflationary models and has led to other scenarios being ruled
out. An upper bound has been placed
on the amplitude of gravitational wave perturbations and bounds have also been
placed on the deviation of the fluctuations from a purely random Gaussian
distribution.

Constraining the parameter space of inflationary models is an important step
towards limiting the number of observationally viable models, and ultimately
towards identifying one such model as the best candidate to describe the
physics of the early
universe.

The goal of this thesis is to constrain inflationary models in two very different
ways: by deriving analytic limits on their parameter spaces, and by demonstrating
a numerical calculation which will allow the investigation of higher order
perturbations. Both these methods have the potential to limit the
parameter space of the models investigated and possibly to rule them out.

In Chapter~\ref{ch:introduction} the foundations are laid for these investigations.
The geometry and physics of the Friedmann-Robertson-Walker universe are presented and
inflationary cosmology is introduced to alleviate problems with the standard Big
Bang scenario. Slow roll conditions are then defined to ensure an adequate duration
of inflation. Despite its elegance, this homogeneous cosmology does not provide an
adequate description of our universe. To understand the inhomogeneities that
are present in reality, first order
cosmological perturbation theory is employed. Models with non-canonical
actions can also be considered. The relationships between observable quantities and
the model parameters are altered in this case, meaning these models could be
distinguished from those with canonical actions. The departure of primordial
perturbations from a Gaussian random distribution could also reveal significant
information about the underlying physics at work.

In Part~\ref{part:dbi} of this thesis, analytical bounds are placed on a class of
non-canonical inflationary models. These models illustrate the dynamics of
extended objects called branes in superstring theory and are considered to be some of
the most promising candidates for achieving inflation using string theory.

Chapter~\ref{ch:dbi-intro} outlines the Dirac-Born-Infeld (DBI) scenario in
terms of the string theoretic background and how it applies in four dimensions
as a realisation of inflation. The six extra dimensions required by string theory
play an integral role in this scenario. These are compactified into a complex
manifold whose geometry allows extended regions called throats to exist. DBI
inflation consists of a brane moving in one of the throats. The inflaton field is
the radial distance of the brane from the tip of the throat. Translating the
higher-dimensional motion into four dimensions introduces a non-canonical term into
the
effective action. The real nature of the action then enforces an upper bound on the
kinetic energy of the inflaton, allowing a sufficiently long period of inflation. The
total inflaton field variation is directly linked to the
amplitude of tensor modes which can be produced.

In Chapter~\ref{ch:dbi} the repercussions of this relationship between the change in
the field value and the tensor mode amplitude are explored further. In the DBI
scenario,
Baumann \& McAllister \cite{bmpaper} placed a conservative
upper bound on the total production of tensor modes during inflation, by assuming the
brane does not propagate further than the length of the throat. By considering only
the
period of observable inflation, which takes place over a much smaller region of the
throat, we have derived a new bound which is considerably stronger. In the
generic
case, the ratio of the amplitudes of the tensor and scalar perturbations must be less
than
$10^{-7}$. This is below  even the most optimistic forecasts for
the sensitivity of future observational experiments. 

If attention is limited to brane motion down the throat, another complementary bound
on the tensor modes can be derived, which depends on the non-Gaussianity of the
scalar modes produced
during inflation. The DBI scenario is inherently non-Gaussian in nature, but, even
assuming the largest levels allowed by observations, the tensor-scalar ratio must
exceed $0.005$. These two bounds are clearly incompatible in the generic
case and only a very fine-tuned selection of model parameters allows the standard DBI
scenario to survive. By taking a more phenomenological approach and allowing the
other parameters to vary,
conditions are found under which the bounds can be relaxed.

A more general class of models which evade the upper bound are identified in
Chapter~\ref{ch:multibrane}. The DBI scenario is characterised by a simple
algebraic relation, in which the sound speed of fluctuations is inversely
proportional to the contribution to the non-Gaussianity. By allowing the
proportionality constant to vary, a new family of actions is derived for which the
bound on the tensor-scalar ratio can be relaxed. 

Instead of considering a single brane moving in the throat, a more natural scenario
might involve multiple branes. These could be created from the energy released by a
brane/anti-brane annihilation and could move up the throat away from the tip.
In \Rref{thomasward}, Thomas \& Ward described the case when these branes are
coincident. When a large number of branes coincide, the resultant action is similar
to the single brane action and is restricted by the bounds on the tensor-scalar
ratio. For a small, finite number of branes, however, the action is non-Abelian in
nature and is one of the family of ``bound-relaxing'' actions described above.
Nevertheless, this model is still constrained by observations and, if a detectable
tensor signal is required, only two or three coincident branes
are allowed. This limit on the number of branes is strongly dependent on the
non-Gaussianity and a tightening of the observational bounds could rule out the
possibility of an observable tensor signal from this model.

In Part~\ref{part:numerical}, the focus of the thesis moves from analytical to
numerical
techniques. Second order cosmological perturbations are numerically calculated for
single field canonical inflationary models.

In Chapter~\ref{ch:perts}, the system of equations for the numerical calculation is
developed. In order to understand non-linear perturbative effects, it is necessary to
examine models using perturbation theory beyond first order. The gauge
transformation for second order perturbations is outlined and the effect on
scalar quantities is considered in the uniform curvature gauge. In
\Rref{Malik:2006ir} the Klein-Gordon equation for second order perturbations was
written in terms of the field perturbations alone. This forms the basis of the
numerical calculation once it is transformed into Fourier space. As the original
equation involves terms quadratic in the first order perturbations, the Fourier
transformed equation contains convolutions of these perturbations. As a first step
towards demonstrating the calculation for the full equation, the slow roll version of
the source term is considered in the second order equation. The second order
perturbations can be linked to observable quantities including the curvature
perturbation and the non-Gaussianity parameter.

The Klein-Gordon equations are the central governing equations of the 
calculation described in Chapter~\ref{ch:numericalsystem}. They must first be
rewritten in a form more suitable for numerical work. This involves changing the
time coordinate to the number of elapsed e-foldings and writing the convolution
terms in spherical polar coordinates. Four different 
potentials will be investigated, each of which has a single field which is slowly rolling. The
parameters for these models are set by
comparing
the calculated power spectrum of first order scalar perturbations with the latest
WMAP data. The initial conditions for the background field and perturbations must
also be specified. The second order perturbations are initially set to zero, to
highlight the creation of second order effects. As this is a novel procedure, a
thorough description of the implementation of the calculation is given. Where
an analytic solution for the convolution terms is possible, this is compared
with the calculated value. Numerical parameters are set by minimising the relative
error in the calculation of one of the terms.

The results of the numerical calculation are presented in Chapter~\ref{ch:results}.
Three different ranges of the discretised momenta are considered and general
results presented for the quadratic potential. 
As expected for a single field, slow
roll model, the second order perturbations are highly suppressed compared to the
first order ones. 
The source term of the second order perturbation equation is similar in form to the
power spectrum of the first order perturbations. It decreases rapidly until
horizon
crossing after which a more steady amplitude is maintained.
The results for all four potentials are also compared. Differences are apparent in
the behaviour of the models after horizon crossing.
This calculation represents only the first step towards a full numerical integration
of the second order Klein-Gordon equation. The next stages towards this goal are
outlined. The second order equation for single field models without the slow roll
assumption is written in the correct form for numerical use and the second order
equations for the two field case are presented in vector form.

In Chapter~\ref{ch:conclusions} the results of the thesis are discussed and some
final conclusions are presented.

\subsection*{Conventions}
\label{sec:conventions}
Throughout this thesis units are chosen such that $\Mpl \equiv (8\pi G )^{-1/2}=
2.4 \times 10^{18}\, {\rm GeV}$ defines the reduced Planck mass and $c=\hbar =1$. 

An overdot ($\dot{~}$) is used for differentiation with respect to proper
time $t$ and a prime ($'$) for differentiation with respect to conformal time
$\eta$. From Chapter~\ref{ch:numericalsystem} onwards,  the dagger symbol
($^\dagger$) denotes differentiation with respect to the
number of e-foldings $\N$.
A subscripted comma denotes partial differentiation by the symbol it
precedes, \eg $f_{,\varphi} = \dfrac{\partial f}{\partial \varphi}$.

The (+++) convention in the notation of Misner \etal \cite{book:kip} is
used throughout.

%
%

\chapter{Inflationary Cosmology}
\label{ch:introduction}

In this chapter the foundations of inflationary cosmology are described. 
In Section~\ref{sec:frw-intro} the
physics of
an isotropic and homogeneous universe is reviewed. The inflationary
scenario is introduced in Section~\ref{sec:inflation-intro}. 
First order cosmological perturbation theory is presented in
Section~\ref{sec:perts-intro} and inflationary models with non-canonical actions are
described in Section~\ref{sec:noncanoninfl}.
The current observational limits on inflationary models are outlined in
Section~\ref{sec:obs-intro} and departures from
Gaussian statistics are parametrised in Section~\ref{sec:fnl-intro}.

\section{The Friedmann-Robertson-Walker Universe}
\label{sec:frw-intro}
The cosmological principle is central to the Friedmann-Robertson-Walker
(FRW\footnotemark) Universe. 
\footnotetext{Lema\^{i}tre is sometimes also included in this group to give
FLRW.}
According to this postulate, there is no privileged
place in the universe and no privileged direction in which to make observations.
These assertions
are formalised by assuming that the universe is homogeneous and isotropic at
every point. This clearly conflicts with the highly inhomogeneous nature of matter
on planetary and solar system scales, but is assumed to hold as larger and larger
scales are considered.
Surveys of the observable universe indicate that this assumption is valid up to
the largest
scales observed \cite{Colless:2001gk, York:2000gk}. 
Historically, homogeneity and isotropy were assumed primarily for simplicity.
Many alternative approaches can be taken. Violating these assumptions can
be done, for example, by specifying a preferred direction or supposing that the
universe is formed by a series of voids connected by filaments. Although many
of these approaches have been disregarded due to lack of evidence, some are still
allowed by observations
\cite{GarciaBellido:2008nz,Alexander:2007xx,Alnes:2005rw,Hunt:2008wp}.

This section outlines the dynamics of the standard Big Bang scenario.
By assuming homogeneity and isotropy, the equations of motion of a
fluid-filled universe can be derived. What follows here is a
standard exposition of well-known physics and has been the subject of
numerous reviews including Refs.~\cite{book:kolbturner, book:kip, book:liddle}.

By imposing both homogeneity and isotropy on a general 4-dimensional metric, the
line element $\d s^2$ of the FRW universe with coordinates $(t,r,\theta,\omega)$
is obtained:
\begin{equation}
 \label{eq:frwmetric-intro}
\d s^2 = -\d t^2 + a^2(t)
  \left( 
    \frac{\d r^2}{1-Kr^2} + r^2\left(\d\theta^2 + \sin^2(\theta)
\d\omega^2\right)
  \right)\,,
\end{equation}
where $K=+1, 0$ or $-1$ depending on whether the universe is closed, flat
or open respectively. The time-like coordinate in the metric is $t$, known as
proper time. 
The spatial part of the FRW metric is multiplied by the scale factor $a(t)$.
This characterises the size of space-like hypersurfaces at
different times. In an expanding universe, $a$ grows with
increasing $t$ and $\dot{a}>0$.
The definition of the Hubble parameter, $H$, captures this expansion:
\begin{equation}
 \label{eq:Hdefn-intro}
  H = \frac{\dot{a}}{a} \,.
\end{equation}

The Einstein equations can be derived by the variational principle from the
action $S$, where $S\equiv S_\mathrm{EH} + S_\mathrm{M}$. This is the sum
of the
Einstein-Hilbert ($S_\mathrm{EH}$) and matter ($S_\mathrm{M}$) actions which are
defined as 
\begin{align}
\label{eq:EHeqn-intro}
 S_\mathrm{EH} &= \frac{1}{16\pi G}\int \d^4 x \sqrt{|g|} \left(R +
2\Lambda_c\right) \,,\\
\label{eq:matteraction-intro}
 S_\mathrm{M} &= \int \d^4 x \sqrt{|g|}
\mathcal{L}_\mathrm{M} \,.
\end{align}
Here $g$ is the determinant of the metric $g_{\mu\nu}$, $R$ is the Ricci
scalar, $G$ is Newton's gravitational constant, $\Lambda_c$ is a cosmological
constant term 
and $\mathcal{L}_\mathrm{M}$ is the sum of the Lagrangian densities
for all the matter fields.
Changing either the matter or gravity actions will affect the
resultant physics. In this work we focus our attention only on the matter
Lagrangian and will use the standard Einstein-Hilbert action throughout.
We can now write down the Einstein equations for a general matter Lagrangian:
\begin{equation}
\label{eq:einstein-intro}
 R_{\mu\nu} - \frac{1}{2}R g_{\mu\nu} = 8\pi G T_{\mu\nu} + \Lambda_c
g_{\mu\nu}\,,
\end{equation}
where $T_{\mu\nu}$ is the stress energy tensor obtained by the variation of the
matter Lagrangian. 
In the definitions above we have included a
cosmological constant term, $\Lambda_c$, for completeness. In the early
universe this term is sub-dominant and will be negligible until much later
\cite{book:liddle}. From now on we will
disregard the contribution of such a term in the early universe.

We concentrate now on the case of a universe filled with a perfect
fluid. Suppose $u^\mu$ is the 4-velocity of this fluid with $u^\mu
u_\mu=-1$. The stress-energy tensor of the fluid is
\begin{equation}
 \label{eq:fluidstress-intro}
  T^\mu_{~\nu} = (E + P)u^\mu u_\nu + P\delta^\mu_\nu \,,
\end{equation}
where $E$ is the matter energy density and $P$ is the isotropic pressure.
The trace of $T$ is given by
\begin{equation}
 \label{eq:Ttrace-intro}
  T^\mu_{~\mu} = -E + 3P\,.
\end{equation}
The Einstein equations and the stress-energy tensor of the
perfect fluid can now be used to derive the equations of motion of the fluid.
From the metric in \eq{eq:frwmetric-intro}, the $00$ and $ij$ components of the
Ricci tensor can be found:
\begin{align}
\label{eq:Ricci00-intro}
 R_{00} &= -3 \frac{\ddot{a}}{a} \,,\\
\label{eq:Ricciij-intro}
 R_{ij} &= \gamma_{ij} \left[ 2\dot{a}^2 +
  a \ddot{a} + 2K \right] \,,
\end{align}
where $\gamma_{ij}$ is the time independent spatial part of the metric in \eq{eq:frwmetric-intro}.
The Friedmann equations are then determined from the Einstein equations
\eqref{eq:einstein-intro}.
The $00$ equation gives
\begin{equation}
 \label{eq:Friedmann1-intro}
 H^2 = \left(\frac{\dot{a}}{a}\right)^2 = \frac{8\pi G}{3} E - \frac{K}{a^2}\,,
\end{equation}
while the trace of the Einstein equations gives the Raychaudhuri or
acceleration equation
\begin{equation}
 \label{eq:Friedmann2-intro}
 \frac{\ddot{a}}{a}  = -\frac{4\pi G}{3}(E + 3P)\,.
\end{equation}
By combining these two equations we can determine a continuity equation for the
energy density:
\begin{equation}
 \label{eq:continuity-intro}
 \dot{E} + 3H(E+P) = 0 \,.
\end{equation}

The last three equations, \eqref{eq:Friedmann1-intro},
\eqref{eq:Friedmann2-intro}
and \eqref{eq:continuity-intro}, will determine the evolution of the
perfect fluid. Two important solutions of these equations are
the radiation and
matter dominated universes. In the standard Big Bang scenario the universe is
dominated by radiation to a good approximation until matter becomes dominant
at later times \cite{book:kolbturner}. These different components
change the rate of expansion
of the universe. For relativistic radiation $P_\mathrm{rad}=E_\mathrm{rad}/3$
and integrating the continuity equation \eqref{eq:continuity-intro} gives
$E_\mathrm{rad}\propto a^{-4}$. Matter conversely is taken to be dust-like with
zero
pressure and so $E_\mathrm{matter}\propto a^{-3}$. The dependence of $a$ on $t$
can then be
found from \eq{eq:Friedmann1-intro}, giving $a\propto t^{1/2}$ and $a\propto
t^{2/3}$ for the radiation and matter eras respectively.

Instead of using proper time as above we could bring
the scale factor outside the whole metric and use conformal time $\eta$
defined by
\begin{equation}
\label{eq:etatime-intro}
 \eta = \int\frac{\d t}{a}\,.
\end{equation}
The metric written in conformal time is then 
\begin{equation}
 \label{eq:frwconformal-intro}
\d s^2 = a^2(\eta)
  \left( -\d\eta^2 +
    \frac{\d r^2}{1-Kr^2} + r^2\left(\d\theta^2 + \sin^2(\theta)
\d\omega^2\right)
  \right)\,,
\end{equation}

As all the coordinates in the line element are now scaled by $a(\eta)$, we
have defined a coordinate grid which does not change as the universe expands.
These ``comoving'' coordinates allow distances to be compared at different
eras with ease. A comoving distance $x$ can be translated into a physical
distance $d$ by
\begin{equation}
 \label{eq:comovingdefn-intro}
 d = ax \,.
\end{equation}
The physical distance changes as the universe expands but the comoving distance
will remain fixed. 

One particularly important distance is the maximum distance light could have
propagated from some initial time $t_i$ to a later time $t$. From
\eq{eq:etalog-intro}, this is simply the conformal time integrated from
the initial time and is called the comoving or particle horizon.
If the initial time is restricted to
being at some finite time in the past, as in the Big Bang scenario, then the
particle horizon will be finite. Two points which are further apart than
this finite distance could never have been in causal contact. This
is the origin of one of the major problems with the standard Big Bang scenario
and will be discussed in the next section.
Rewriting the comoving horizon as
\begin{equation}
\label{eq:etalog-intro}
 \eta = \int_{a_i}^a \frac{\d a'}{a'} \frac{1}{a' H(a')} \,,
\end{equation}
shows that it is also the logarithmic integral of the comoving Hubble
radius $1/aH$. This distance is how far particles can travel in one
``e-folding'', the time for $a$ to expand by one exponential factor. 
The number of e-foldings between two measurements of the scale factor, $a_i$
and $a_f$, is given by
\begin{equation}
\label{eq:nefolddefn-intro}
 \N = \ln \frac{a_f}{a_i}\,.
\end{equation}

Particles
that are separated by more than the Hubble radius cannot be in causal
contact now. Particles separated by more than the
comoving horizon, however, could never have been in causal contact. In addition
to the Hubble parameter $H$, it will be useful
to define the parameter $\H = aH = a'/a$. The comoving Hubble radius is then
$1/\H$.

\section{Inflation}
\label{sec:inflation-intro}
In this section we introduce the inflationary scenario. First we briefly
describe how it solves two major problems with the standard Big Bang
picture: the flatness problem and the horizon problem \cite{book:liddle}. We go on to
describe canonical slow roll inflation, the generation of
perturbations from quantum fluctuations and inflation from non-canonical
actions.

\subsection{Problems with the Big Bang Scenario}
Although remarkably successful in describing the evolution of the universe from
very early in its history, the standard Big Bang scenario suffers from a number
of serious problems. Two of the main problems are described in this section.

\subsubsection{Flatness Problem} 
\label{sec:flatprob}
The Friedmann equation
\eqref{eq:Friedmann1-intro} can be re-written as
\begin{equation}
\label{eq:omegadefn-intro}
\Omega(t) - 1 = \frac{K}{(aH)^2} = \frac{K}{\dot{a}^2} \,,
\end{equation}
where $\Omega(t)=E(t)/E_\mathrm{crit}$ and the critical density
$E_\mathrm{crit}= 3H^2/8\pi G$. 
If $\ddot{a}>0$ then $\Omega$ approaches the critical value $\Omega=1$ over time,
whereas if
$\ddot{a}<0$ it diverges from this value. The flat universe, $K=0$,
is an unstable fixed point in the parameter space.
Current observations confirm that $\Omega=1$ within about 2\%, at a 95\% confidence
level \cite{Komatsu:2008hk}.
During the radiation and matter dominated eras $aH$ is decreasing with time,
so that $\Omega$ diverges away from 1. If the measured value is now very
close to 1 then in the past it must have been even closer. 
The fine-tuning in the initial conditions required for this proximity to $\Omega=1$
is known as the flatness problem.

\subsubsection{Horizon Problem} \label{sec:horizprob}
The particle horizon, also known
as the comoving horizon, defines the maximum separation between two points that
have been in causal contact sometime in the past. During the radiation and
matter eras, this comoving horizon increases monotonically and so length scales
which are now entering the horizon would have been far outside it in
the past. 
The CMB as observed by the WMAP satellite is extremely smooth on scales that
would have been far outside the horizon at the time of last scattering
\cite{Komatsu:2008hk}. These regions of space have very similar energies and
yet according to the Big Bang scenario they could never have been in causal contact.

\subsection{Inflation and Canonical Slow Roll}
\label{sec:slowroll-intro}

Inflation is a period of accelerated expansion in the size of the universe
which took place just after the Big Bang
\cite{Starobinsky:1980te,Guth:1980zm,Albrecht:1982wi,Linde:1981mu,
Starobinsky:1982ee}. During this
expansion phase  the comoving Hubble radius $(aH)^{-1}$ decreases and the isotropic
pressure of the universe is negative \cite{book:liddle, Baumann2009}:
\begin{equation}
\label{eq:infldefn-intro}
 \frac{\d }{\d t}\left( \frac{1}{aH}\right) <0 
\quad \Rightarrow  \quad
\ddot{a}>0
\quad \Rightarrow \quad
E + 3P < 0 \,.
\end{equation}
We can define a new parameter 
\begin{equation}
\label{eq:epsilonHdefn-intro}
\varepsilon_H = -\frac{\dot{H}}{H^2}\,,
\end{equation}
and then rewrite
the Raychaudhuri equation \eqref{eq:Friedmann2-intro} as
\begin{equation}
 \label{eq:Friedeps-intro}
 \frac{\ddot{a}}{a} = H^2 (1-\varepsilon_H)\,.
\end{equation}
This parametrisation illustrates that inflation only occurs when $\varepsilon_H<1$.
In this subsection we describe briefly how inflation solves the problems
outlined above and outline the inflationary dynamics of single scalar field
models. 

\begin{figure}[htbp]
\centering
 \includegraphics[width=\textwidth]{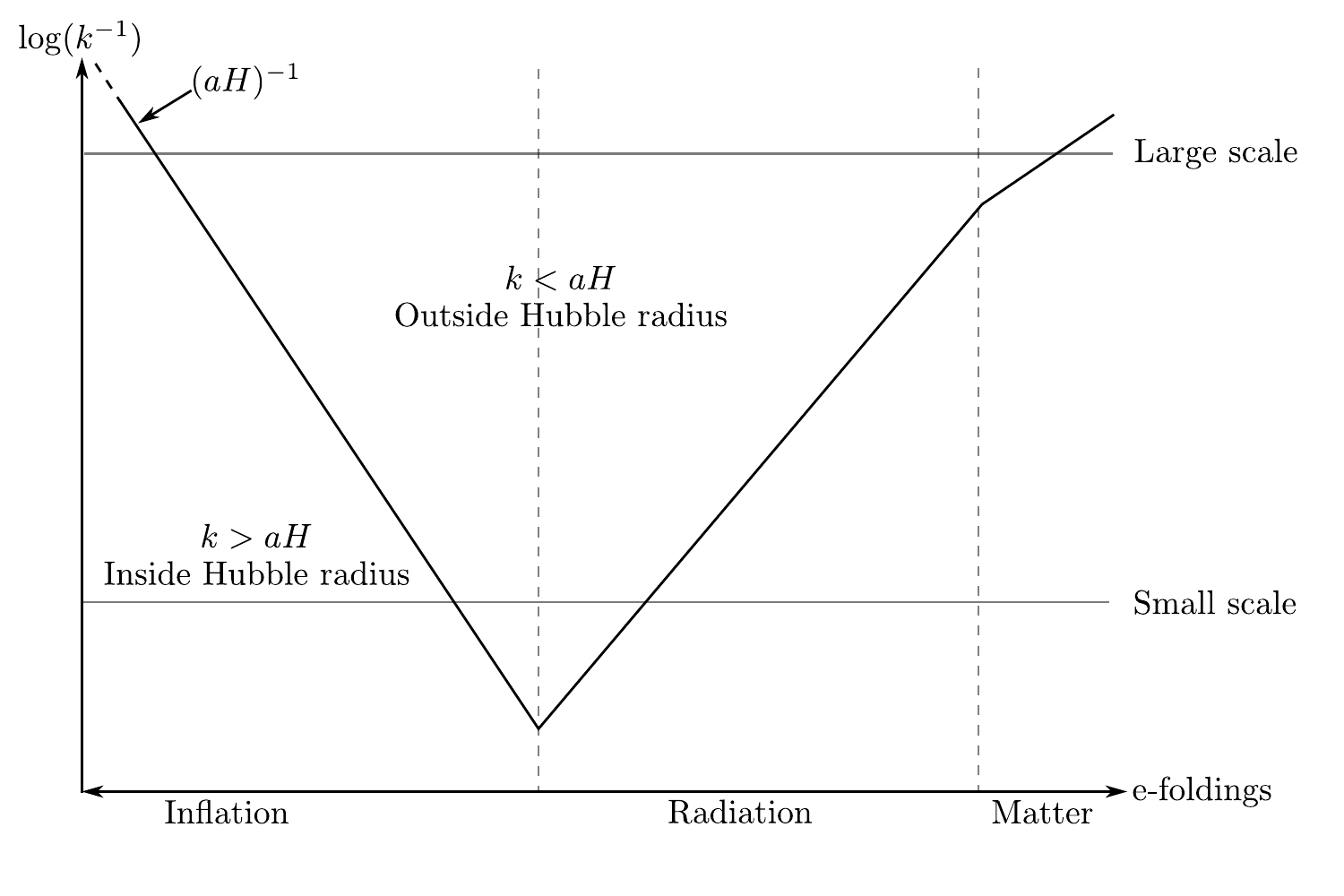}
 \caption[Comoving Scales and the Hubble Radius]{Comoving scales that have recently
entered the horizon would
previously have been inside the horizon, if the inflationary period extended far
enough into the past.}
 \label{fig:comovingscales-intro}
\end{figure}

Both the horizon and flatness problems described above are statements about our
reluctance to impose fine-tuned initial conditions. Inflation removes the need
to fix these conditions at the start of the Big Bang.
A period of decreasing Hubble radius before the radiation period could explain the
homogeneity of temperatures in the CMB at large scales.
Comoving scales that entered the horizon recently, such as
those we observe in the CMB, would have been within the horizon previously. 
During this period, the energy density could reach an equilibrium
value.
Figure~\ref{fig:comovingscales-intro} shows how, by extending
the era of inflation far enough into the past, any comoving length could
previously have been inside the horizon.
Observations require that inflation lasted at least long enough that all the
scales we measure today were previously inside the horizon. 

Now consider the time derivative of $|\Omega -1|$ as defined in
\eq{eq:omegadefn-intro}:
\begin{equation}
 \label{eq:omegaderiv-intro}
 \frac{\d}{\d t}\left(|\Omega -1 |\right) = 3\frac{\d}{\d t}
\left(\frac{1}{aH}\right)\,.
\end{equation}
If the universe is not flat to begin with, a period of inflation of sufficient
duration will push it
towards $\Omega=1$, solving the flatness problem. Instead of an unstable point
in the parameter space, $\Omega=1$ is an attractor during the inflationary
phase. 

To solve the horizon and flatness problems the duration of the inflationary era
must be
sufficiently long. Approximately $50$--$70$ e-foldings is considered standard
\cite{book:liddle}. Inflation
can last longer than this but only the last 50--70 e-foldings will be important
for the length scales of our observable universe.

During inflation the universe is filled by material exhibiting negative isotropic
pressure.
Therefore, whatever drives inflation cannot be matter or radiation in their usual
forms. The simplest proposal is to fill the universe with a single scalar field
$\varphi$. The canonical action for this field is
\begin{equation}
\label{eq:phiaction-intro}
 \mathcal{L}_\mathrm{M} \equiv P(\varphi, X) = X -V(\varphi) \,,
\end{equation}
where $X=-\frac{1}{2}g_{\mu\nu}\partial^\mu\varphi \partial^\nu\varphi$ denotes
the kinetic energy of $\varphi$, $V(\vp)$ is the potential and $P(\varphi, X)$
is called the kinetic
function. In Section~\ref{sec:noncanoninfl} we will consider other
choices for $P$. 

The equation of motion for $\varphi$, for the canonical action $P=X-V$, is
\begin{equation}
 \label{eq:phieom1-intro}
 \ddot{\vp} + 3H\dot{\vp} -\nabla^2 \vp + \frac{\delta V}{\delta \vp} = 0\,.
\end{equation}
If we now restrict ourselves to considering the homogeneous part of the field,
$\vp=\vp(t)$, the $\nabla^2\vp$ term disappears and the functional derivative
of $V$ becomes a standard derivative $V_{,\vp}$. 
With these choices we have the following relations for the matter
energy-density and isotropic pressure:
\begin{align}
\label{eq:EandP-intro}
 E &= \frac{1}{2}\dot{\varphi}^2 + V(\varphi) \,,\\
\label{eq:Pdefn-intro}
 P &= \frac{1}{2}\dot{\varphi}^2 - V(\varphi) \,.
\end{align}
Under these conditions the kinetic function $P(\varphi, X)$ can be identified as
the isotropic pressure. The dynamics of the field are governed by the potential
$V(\varphi)$. Inflation requires $P<-E/3$, so from Eqs.~\eqref{eq:EandP-intro} and
\eqref{eq:Pdefn-intro} inflation
can also be thought of as a period when the potential energy dominates over
the kinetic energy.

Inflation needs to last long enough to solve the problems described above. A generic
potential is not likely to satisfy these requirements without fine-tuning. One
approach is to enforce conditions on the potential under which the inflationary
period is necessarily long. 
We have seen that for
inflation to occur the potential needs to dominate over the kinetic energy. From
\eq{eq:EandP-intro}, this occurs in the limit $P\rightarrow -E$ or equivalently
$\varepsilon_H \ll 1$. However, for this to remain the case for a sufficiently long
period the second derivative of $\vp$ must be small. If we define
another parameter 
\begin{equation}
 \label{eq:etaHdefn-intro}
 \eta_H \equiv -\frac{\d \ln \dot{\vp}}{\d \ln a} =
-\frac{\ddot{\vp}}{H\dot{\vp}}
 =\varepsilon_H -\frac{\dot{\varepsilon_H}}{2H\varepsilon_H}\,,
\end{equation}
then taking $|\eta_H|\ll 1$ ensures that $\dot{\vp}$ and $\varepsilon_H$
change slowly. This allows an inflationary phase of sufficient duration to occur.

The approximations $\varepsilon_H\ll 1$ and $|\eta|\ll 1$ are known as the
slow roll conditions because they force the inflaton field $\vp$ to roll down
the potential $V$ slowly. The parameters $\varepsilon_H$ and $\eta_H$ are the
slow roll parameters. Setting $\ddot{\vp}$ to be small is equivalent to
making the friction $H\dot{\vp}$ in \eq{eq:phieom1-intro} dominant. With
these approximations the
equations of motion for a slowly rolling field become
\begin{align}
 \dot{\vp} &\simeq - \frac{V_{,\vp}}{3H}\,, \\
 H^2 &\simeq \frac{8\pi G}{3}V(\vp)\,.
\end{align}

\section{Perturbations}
\label{sec:perts-intro}

We considered a homogeneous scalar field in the analysis of
Section~\ref{sec:inflation-intro}. Such a field, however, will
lead only to a homogeneous universe later. How does the myriad structure
that we see around us form? From stars to galaxies to clusters, the
gravitational force has concentrated energy density over the history of the
universe, but some initial fluctuation must have been present to begin this
process. One of the main achievements of inflation is to provide a physical
origin for
such initial fluctuations. In Chapter~\ref{ch:perts} we will formally develop
cosmological perturbation theory up to second order. In this section we 
review first order perturbation theory and introduce the observable quantities
important for inflation.

Suppose that a full inhomogeneous scalar field
$\varphi$ is split into a homogeneous background field $\vp_0$, as described
above, and an inhomogeneous perturbation $\dvp{}(\eta,x^i)$ for $i=1,2,3$. For the
following
analysis to be applicable the perturbation must be much smaller than the background
field value. From the amplitude of perturbations in the CMB this approximation
can be seen to be valid \cite{Komatsu:2008hk}. For single field models no mixing
of adiabatic and non-adiabatic modes occurs \cite{Weinberg200804}. Therefore, throughout this
thesis we will only consider adiabatic perturbations and ignore any isocurvature mode present. 

If we suppose that $\epsilon$ is a small quantity then the split in $\vp$
can be written as \cite{Malik:2008im}
\begin{equation}
\label{eq:pertsplit-intro}
\vp(\eta, x^i) = \vp_0(\eta) + \epsilon\dvp{}(\eta, x^i) \,.
\end{equation}
 The perturbation $\dvp{}(\eta, x^i)$ can be further expanded in
powers of $\epsilon$.  We will follow the
custom of not explicitly writing $\epsilon$, instead relying on the order of
the perturbation, denoted by a subscript, to keep track.
If we expand in a Taylor series then up to second order (\ie including terms
up to $\epsilon^2$) we have:
\begin{equation}
 \vp(\eta, x^i) = \vp_0(\eta) + \dvp1(\eta, x^i) + \frac{1}{2}\dvp2(\eta, x^i)
\,.
\end{equation}
There is some freedom in how the split of the perturbations into different
orders is made. We will suppose that the first order perturbation $\dvp1$
contains only linear contributions and the higher order terms contain non-linear
terms.

Instead of working in coordinate space, we can also consider
the perturbation in Fourier space using the definition
\begin{equation}
\label{eq:fourierdefn-intro}
 \dvp{}(\eta, x^i) = \frac{1}{(2 \pi)^3} \int \d^3k \dvp{}(\kvi) \exp (i k_i
x^i)
\,,
\end{equation}
where $k^i$ are the components of the comoving wavenumber vector $\bf{k}$. The
amplitude of this
vector $k=|\bf{k}|$ identifies whether a particular mode is inside or outside
the comoving horizon. Wavemodes inside the comoving horizon are identified by
$k>aH$, while $k<aH$ for those outside the horizon.

We must also consider perturbations in the metric tensor $g_{\mu\nu}$. If the
background metric is the FRW one described in Section~\ref{sec:frw-intro} then
the metric can be written with perturbations, up to first order, as follows:
\begin{align}
 \label{eq:pertmetric-intro}
 g_{00} &= -a^2 (1 + 2\phi_1) \,, \nonumber\\
 g_{0i} &= a^2 B_{1i} \,, \nonumber\\
 g_{ij} &= a^2\left(\delta_{ij} + 2C_{1ij} \right) \,.
\end{align}
The $0$-$i$ and $i$-$j$ perturbations can be decomposed into scalar, vector and
tensor parts \cite{Malik:2008im}:
\begin{align}
\label{eq:svt-intro}
  B_{1i} &= B_{1,i} - S_{1i} \,, \nonumber\\
  C_{1ij} &= -\psi_1\delta_{ij} + E_{1,ij} + F_{1(i,j)} + \frac{1}{2}h_{1ij}\,,
\end{align}
where $F_{1(i,j)} = \frac{1}{2}(F_{1i,j} + F_{1j,i})$.
The vectors $S_{1i}$ and $F_{1i}$ are divergence free and the tensor part
$h_{1ij}$
is divergence free and traceless:
\begin{equation}
 S^k_{1~,k} = 0\,, \quad F^k_{1~,k}=0\,; \qquad h^{ik}_{1~~,k} = 0\,,
  \quad h^k_{1~k}= 0\,.
\end{equation}
In the previous equations $\phi$ is the lapse function, $\psi$ is the curvature
perturbation, $B_1$ and $E_1$ are the scalar part of the shear, $S_{1i}$, and
$F_{1i}$
are the vector parts of the shear, and $h_{1ij}$ is the tensor perturbation
describing gravitational waves.

Splitting an inhomogeneous spacetime into background and perturbation is not a
covariant operation. This leads to an ambiguity in the choice of coordinates
which must be rectified by choosing a gauge. Gauge transformations relate
physical results in one gauge to those in another. To choose a gauge one must
specify how spacetime is foliated, \iec a slicing, and how coordinates in one
spatial hypersurface are related to those in another, \iec a threading
\cite{Malik:2008im}. We will
employ the uniform curvature gauge in which spatial hypersurfaces are flat. This
is also known as the flat gauge.

The gauge transformation vector at first order,  $\xi_1^\mu$, can be split into
scalar
and vector parts
\begin{equation}
\label{eq:xidefn-intro}
 \xi_1^\mu = (\alpha_1, \beta_{1,}^{~i} + \gamma_1^i)\,,
\end{equation}
where the vector part obeys $\gamma_{1~,k}^{~k}=0$. A scalar quantity such as
the inflaton perturbation will transform as \cite{Malik:2008im, Malik:2008yp}
\begin{equation}
 \label{eq:dphitransform-intro}
 \wt{\dvp1} = \dvp1 + \vp_0' \alpha_1\,,
\end{equation}
where a tilde ($\wt{~}$) denotes a transformed quantity. For the metric
perturbations the
transformations for the scalars are
\begin{align}
 \label{transphi1}
\widetilde {\phi_1} &= \phi_1 +\H\alpha_1+\alpha_1'\,,\\
\label{transpsi1}
\widetilde \psi_1 &= \psi_1-\H\alpha_1 \,,\\
\label{transB1}
\widetilde B_1 &= B_1-\alpha_1+\beta_1'\,,\\
\label{transE1}
\widetilde E_1 &= E_1+\beta_1\,,
\end{align}
and for the vectors
\begin{align}
 \label{transS1}
\widetilde {S_{1}^{~i}} &= S_{1}^{~i}-{\gamma_1^i}'\,, \\
\label{transF1}
\widetilde {F_{1}^{~i}} &= F_{1}^{~i}+\gamma_1^i\,. 
\end{align}
The tensor perturbation $h_{1ij}$ does not change under transformation at first
order, but does at subsequent orders.
The flat gauge, which we will use, is the one in which spatial hypersurfaces
are not perturbed by scalar or vector perturbations, so $\wt{\psi_1} = \wt{E_1}
=0$ and $\wt{F_{1i}} = \textbf{0}$.  This is equivalent to the transformation
\begin{equation}
 \alpha_1 = \frac{\psi_1}{\H}\,,\quad \beta_1=-E_1 \,, \quad \gamma^i_1 =
-F^i_1\,.
\end{equation}

A gauge invariant inflaton perturbation variable is the Sasaki-Mukhanov
variable \cite{Mukhanov:1990me, Mukhanov:1988jd,
Sasaki:1986hm}:
\begin{equation}
\label{eq:flatdvp1-intro}
 \wt{\dvp1} \equiv \dvp1 + \vp_0'\frac{\psi_1}{\H}\,,
\end{equation}
In the flat gauge this is just $\dvp1$. We will work in flat
gauge from now on and so will drop the tildes on quantities in that gauge.

Another very important gauge invariant quantity is the comoving curvature
perturbation $\R$. At first order in the flat gauge $\R$ is related to the
inflaton perturbation by \cite{Malik:2008im}
\begin{equation}
\label{eq:flatRdefn-intro}
 \R = \frac{\H}{\vp_0'}\dvp1\,.
\end{equation}
We are interested in the power spectrum of the curvature perturbation as this
is directly related to the temperature fluctuations that we can observe in the
CMB. 

The action \eqref{eq:phiaction-intro},
including perturbations of $\vp$ and $g_{\mu\nu}$ up to first order, is
varied to get the equation of motion of $\dvp1$. 
In the flat gauge the equation can be rewritten in terms of the inflaton field
values only by eliminating the metric perturbations using
\eq{eq:flatdvp1-intro}. 
In Fourier space and in terms of the conformal time $\eta$, the closed form of
the first order perturbation equation of motion is \cite{Malik:2008im}
\begin{multline}
\label{eq:fokg-intro}
\dvp1''(\kvi) + 2\H \dvp1'(\kvi) + k^2\dvp1(\kvi) \\
+ a^2 \left[\Upp +
\frac{8\pi G}{\H}\left(2\vp_{0}' \Uphi + (\vp_{0}')^2\frac{8\pi G
}{\H}\U\right)\right]\dvp1(\kvi) = 0 \,,
\end{multline}
where $\U$ is the background value of the potential $V(\vp)$.
Substituting $u=a\dvp1$ gives the Mukhanov equation \cite{Mukhanov:1990me}
\begin{equation}
 \label{eq:ueq-intro}
 u''(\kvi) + \left[ k^2 -\frac{z''}{z}\right]u(\kvi) = 0\,,
\end{equation}
where $z= a\vp_0'/\H$.

\subsection{Quantum Perturbations}
So far we have considered classical perturbations. However, the generation of
fluctuations is a quantum effect and we need to consider the
perturbations as quantum operators in some vacuum. 

In Minkowski space the quantisation of $u(\kvi)$ is straightforward.
The perturbation modes can be written in terms of quantum operators as
\begin{equation}
 u(\kvi)\rightarrow\hat{u}(\kvi) = 
  w(\kvi) \hat{a}(\kvi) + w^\star(-\kvi)\hat{a}^\dagger(-\kvi) \,.
\end{equation}
The mode function $w(\kvi)$ obeys the same equation of motion as $u(\kvi)$:
\begin{equation}
\label{eq:weqn-intro}
 w''(\kvi) + \left[ k^2 -\frac{z''}{z}\right]w(\kvi) = 0\,.
\end{equation}
The operators $\hat{a}^\dagger$ and $\hat{a}$ are the usual creation and
annihilation operators. They act on quantum states by adding or
removing particles. The zero particle vacuum state, $|0\rangle$, is such that
\begin{equation}
 \hat{a}^\dagger|0\rangle = |1\rangle\,,\quad \hat{a}|0\rangle = 0\,.
\end{equation}
In Minkowski space these operators have the usual commutation relations
\begin{equation}
 [\hat{a}(\kb), \hat{a}^\dagger(\kb')] = (2\pi)^3 \delta(\kb -\kb') \\
\end{equation}
and
\begin{equation}
[\hat{a}^\dagger(\kb), \hat{a}^\dagger(\kb')] = [\hat{a}(\kb), \hat{a}(\kb')] = 0\,.
\end{equation}
The $w$ modes are normalised by the condition \cite{Mukhanov:2005sc}
\begin{equation}
\label{eq:quantcondition-intro}
 w^\star(\kvi)w'(\kvi) - {w^{\star}}'(\kvi)w(\kvi) = i\,.
\end{equation}

In the expanding FRW background the choice of vacuum is not straightforward.
Suppose one observer selects a zero particle state as the vacuum. Another
observer accelerating with respect to the first will see particles being
created in this ``vacuum'' state due to the Unruh effect
\cite{Kinney2009, Unruh1976a}. In selecting the vacuum we must choose one of the
many equivalent options.
To do this we consider the far past where $\eta\rightarrow -\infty$. The wavelengths of all the
modes are then much smaller than the Hubble radius and curvature scale. The modes are
therefore assumed to evolve in flat space. This suggests the Minkowski vacuum as the
most natural vacuum state to select and this choice of vacuum at
early times is known as the Bunch-Davies vacuum.
In the limit $\eta\rightarrow -\infty$ (or equivalently $k/aH\rightarrow \infty$),
the mode equation
\eqref{eq:weqn-intro} becomes
\begin{equation}
  w''(\kvi) + k^2 w(\kvi) = 0\,,
\end{equation}
which has the plane wave solution
\begin{equation}
\label{eq:subhsoln-intro}
 w(\kvi) = \frac{1}{\sqrt{2k}} e^{-ik\eta}\,.
\end{equation}
This is the initial condition for modes which are well inside the horizon.

Now consider the de Sitter limit in which $\varepsilon_H\rightarrow 0$ and $H$
is constant. We have $z''/z = a''/a = 2/\eta^2$ so the mode equation is \cite{Baumann2009}
\begin{equation}
  w''(\kvi) + \left[ k^2 -\frac{2}{\eta^2} \right]w(\kvi) = 0\,.
\end{equation}
A full general solution for $w$ is
\begin{equation}
 w(\kvi) = A\frac{e^{-ik\eta}}{\sqrt{2k}}\left(1 -\frac{i}{k\eta}\right)
	  +B\frac{e^{+ik\eta}}{\sqrt{2k}}\left(1 +\frac{i}{k\eta}\right)\,.
\end{equation}
Taking the condition \eqref{eq:quantcondition-intro} along with the solution
for subhorizon modes in \eq{eq:subhsoln-intro} we find that $A=1$ and $B=0$.
Thus the full solution in de Sitter space is \cite{book:liddle}
\begin{equation}
\label{eq:wfinal-intro}
 w(\kvi) = \frac{e^{-ik\eta}}{\sqrt{2k}}\left(1 -\frac{i}{k\eta}\right)\,.
\end{equation}
Inflation in spacetimes that are close to de Sitter will contain perturbations
with a spectrum defined by \eq{eq:wfinal-intro}. The slow roll approximation is
enough to ensure that inflation occurs in a quasi-de Sitter spacetime.
However, the initial conditions for Fourier modes in \eq{eq:subhsoln-intro}
apply to non slow roll models so long as they are applied well before horizon
crossing.

\subsection{Power Spectra and Spectral Indices}
The power spectrum of the inflaton perturbation $\dvp1 = u/a$ can now be
defined as
\begin{equation}
  \langle \dvp1(\textbf{k}_1) \dvp1(\textbf{k}_2) \rangle 
   \equiv (2\pi)^3 \delta(\textbf{k}_1 + \textbf{k}_2) P_{\dvp{}}^2 (k_1)
   = (2\pi)^3 \delta(\textbf{k}_1 + \textbf{k}_2) \frac{|w(\kb_1)|^2}{a^2}\,,
\end{equation}
where $\langle \ldots \rangle$ denotes the ensemble average. 
If taken over a large enough volume, the ensemble average and spatial average
are equivalent \cite{book:lyth}.
The power spectrum
$P_{\dvp{}}^2$ depends only on the magnitude of the wavenumber
vector, $k=|\bf{k}|$, but has dimensions of $k^{-3}$. A dimensionless power
spectrum
can be defined as
\begin{equation}
 \label{eq:curlPrdefn-intro}
 \mathcal{P}_{\dvp{}}^2= \Delta^2_{\dvp{}} \equiv \frac{k^3}{2\pi^2}
P_{\dvp{}}^2(k)\,.
\end{equation}

In a similar way we can define the power spectrum of the comoving curvature
perturbation $\R = H\dvp1/\dot{\vp}_0$:
\begin{equation}
 \label{eq:Prdefn-intro}
 \langle \R(\textbf{k}_1) \R(\textbf{k}_2) \rangle 
   = (2\pi)^3 \delta(\textbf{k}_1 + \textbf{k}_2) P_\R^2 (k_1)\,,
\end{equation}
and the dimensionless power spectrum
\begin{equation}
 \Pr= \Delta^2_{\dvp{}} \equiv \frac{k^3}{2\pi^2}
P_{\R}^2(k)\,.
\end{equation}

A slow roll inflation model in a quasi-de Sitter spacetime will have the
Fourier mode solution given in \eq{eq:wfinal-intro}. 
After horizon crossing, when $k\ll aH$, this gives $|w|^2 = 1/(2k^3 \eta^2)$ so 
\begin{equation}
\label{eq:pphi-intro}
 \mathcal{P}_{\dvp{}}^2(k) = \left(\frac{H}{2\pi}\right)^2 \,,
\end{equation}
for the scalar perturbation spectrum and
\begin{equation}
 \Pr(k) = \left(\frac{H}{\dot{\vp_0}}\right)^2 \left(\frac{H}{2\pi}\right)^2\,,
\end{equation}
for the comoving curvature perturbation spectrum. Models that are not
slowly rolling usually require their more complicated mode equations to be
numerically solved. 

We have discussed in depth the scalar perturbations but tensor perturbations
can also be produced. The tensor perturbation $h_{ij}$
has two polarisations, $h_s$ for $s=+, \times$. The amplitude of
each can be thought of as a separate scalar field. The analysis for each field
is similar to that above with the substitution $h_s= 2\dvp1/\Mpl$. After horizon
crossing in a quasi-de Sitter space the spectrum for each polarisation is
\begin{equation}
 \mathcal{P}_h^2(k) = \frac{4}{\Mpl^2} \left(\frac{H}{2\pi}\right)^2\,,
\end{equation}
and the overall tensor perturbation spectrum is
\begin{equation}
 \label{eq:Ptdefn-intro}
\Pt(k) = \frac{2}{\Mpl^2} \frac{H^2}{\pi^2}\,.
\end{equation}
The ratio of the tensor to curvature perturbations (tensor-scalar ratio) $r$ is
defined as 
\begin{equation}
 r = \frac{\Pt}{\Pr}\,,
\end{equation}
where $r$ is usually quoted at a particular $k$ but could in principle depend
on $k$. The tensor-scalar ratio can also be written in terms of $\varepsilon_H$:
\begin{equation}
\label{eq:rslowroll-intro}
 r = 16 \varepsilon_H\,.
\end{equation}
As $\varepsilon_H\ll1$ for slow roll models of inflation the amplitude of tensor
perturbations that these models produce is much smaller than the amplitude of
curvature perturbations.

If the curvature perturbation power spectrum, $\Pr(k)$, is independent of
wavenumber $k$, it is said to be scale invariant. The spectral index $n_s$ is a
measure of the deviation from scale invariance:
\begin{equation}
\label{eq:nsdefn-intro}
 n_s -1 = \frac{\d \ln (\Pr(k))}{\d \ln k}\,,
\end{equation}
where $n_s=1$ denotes a scale invariant spectrum. The spectral index of
the tensor power spectrum can be similarly defined:
\begin{equation}
\label{eq:ntdefn-intro}
 n_T = \frac{\d \ln(\Pt(k))}{\d \ln k}\,,
\end{equation}
although this definition means that the spectrum is scale invariant if
$n_T=0$.
The spectral indices and indeed the spectra themselves are usually calculated
at an arbitrary pivot scale. The WMAP results for $\Pr$ and $\Pt$ outlined in
Section~\ref{sec:obs-intro} are quoted at the scale $k=0.002\Mpc^{-1}$.

If there is a non-trivial dependence of $\Pr$ or $\Pt$ on $k$ then higher order
derivatives can be taken to give the running of the quantities. The runnings
of the spectral indices are
\begin{equation}
\label{eq:runningsdefn-intro}
 \alpha_s = \frac{\d \ln n_s}{\d \ln k}\,, \quad 
  \alpha_T = \frac{\d \ln n_T}{\d \ln k}\,.
\end{equation}

In the slow roll approximation $n_s$ and $n_T$ can be written in terms of the slow
roll parameters $\epsilon_H$ and $\eta_H$, evaluated at $k=aH$ using $\d \ln(aH)
\simeq H\d t$:
\begin{align}
 n_s -1 &= -4\epsilon_H + 2\eta_H\,, \\
 n_T &= -2\epsilon_H \label{eq:ntslowroll-intro}\,.
\end{align}
Combining \eq{eq:ntslowroll-intro} and \eq{eq:rslowroll-intro} gives a
powerful consistency condition for slow roll inflation:
\begin{equation}
 \label{eq:consistency-intro}
 r = -8 n_T \,.
\end{equation}
For the slow roll approximation to be valid for single field canonical inflation
models, \eq{eq:consistency-intro} must hold. Current observations are not accurate
enough to test this condition but it is hoped that this will be possible in the
future.

\section{Current Observations}
\label{sec:obs-intro}
There have been rapid improvements in the quantity and quality of cosmological data
sources in the last twenty years. From the launch of the COBE satellite in
1989 \cite{Bennett1994, Bennett1996c}, through the currently
ongoing WMAP mission \cite{spergel, Komatsu:2008hk}, to the recent launch of the
Planck satellite \cite{planck}, space based observations have been at the
forefront of the effort to collect data. Complementing these have been ground
and balloon based missions including CBI
\cite{Mason2003b, Sievers2003, Sievers2007}, VSA \cite{Dickinson2004}, ACBAR
\cite{Kuo2004, Kuo2007} and BOOMERANG \cite{Ruhl2003, Montroy2006,
Piacentini2006}.

Major recent data releases have provided significant confirmation of the FRW
model of the universe. The Hubble parameter today has been measured as $H_0 =
72\pm8\,\mathrm{km}/\mathrm{s}/\Mpc$ by the Hubble Key Project
\cite{Freedman2001}. The WMAP 5-Year data release (WMAP5) \cite{Komatsu:2008hk}
quotes their results combined with
data from Baryon
Acoustic Oscillations in galaxy distributions (BAO) \cite{Percival2007}
and supernova surveys (SN) by the Hubble Space Telescope and others \cite{Riess2004,
Riess2007, Astier2006, Wood-Vasey2007}. 
This combined data constrains the
universe to within two percent of the flat $\Omega =1, K=0$ case outlined in
Section~\ref{sec:frw-intro}. 

The amplitude of the scalar curvature perturbations $\Pr$ was first measured
accurately by the COBE satellite \cite{Bennett1994, Bennett1996c}. The WMAP5
normalisation is taken at a different scale to the COBE result, measuring
\begin{equation}
 \label{eq:wmapnorm-intro}
 \Pr(\kwmap) = 2.457 \e{-9}\,,
\end{equation}
where the pivot scale $\kwmap = 0.002\Mpc^{-1} \simeq 5.25\e{-60}\Mpl$. The
spectral index of scalar perturbations for models with tensor-scalar ratio
$r\ne0$ is given by the combined WMAP5+BAO+SN measurement as
\begin{equation}
 \label{eq:wmapns-intro}
 n_s = 0.968 \pm 0.015\,.
\end{equation}

The detection of B-mode polarisation would provide definitive proof of the existence of primordial
gravitational modes and much observational effort is being expended in the attempt to achieve such
a detection \cite{Seljak:1996gy, Baumann:2008aq,Chiang:2009xs,Piacentini2006,Sievers2007,vpj}.
The observational bound on $r$ from WMAP5 using only the B-mode power spectrum
is weak with $r< 4.7$ at the 95\% confidence level, when $n_s$ is fixed at the best fit value.
Including other polarisation data from the E-mode and TE power spectra reduces this bound to
$r<1.6$, again with $n_s$ fixed. A stronger bound has been obtained with the B-mode
power spectrum by the BICEP experiment, giving $r<0.73$ \cite{Chiang:2009xs}. The
strongest bound to date on the tensor to scalar ratio is given when the temperature power spectrum
data is also included in the WMAP analysis. For the pure WMAP5 data without any restriction on
$n_s$ but with no spectral running the bound is $r<0.43$. When BAO and SN data is combined with the
WMAP5 data the bound on the tensor to scalar ratio becomes
\begin{equation}
\label{eq:rbound-intro}
 r < 0.20\,,
\end{equation}
at the 95\% confidence level.

\section{Non-Canonical Inflation} 
\label{sec:noncanoninfl}

In the previous section we considered the dynamics of a scalar field with a
canonical action $P(\vp, X) = X - V(\vp)$, where $X \equiv -\frac{1}{2}g^{\mu
\nu}\partial_\mu \varphi \partial_\nu \varphi$ is the kinetic energy. 
In this
section we will generalise
that analysis to include non-canonical actions. Non-canonical scalar
field actions appear frequently in string theory derived inflationary models.
In Chapters~\ref{ch:dbi-intro}, \ref{ch:dbi} and \ref{ch:multibrane}
there are explicit examples of non-canonical scenarios.

We will consider an action of the same form as before
\begin{equation}
\label{eq:DBIaction-dbiintro2}
S=\int  \d^4x \sqrt{|g|} \left[ \frac{\Mpl^2}{2} R 
+ P (\varphi , X) \right] \,,
\end{equation}
with minimal coupling to the gravitational sector. Varying this action gives the
stress-energy tensor in
\eq{eq:fluidstress-intro} where $u_\mu = \partial_\mu\vp/\sqrt{2X}$. 
The energy density $E$ is defined as
\begin{equation}
 E = 2X\PX - P\,,
\end{equation}
and for a homogeneous scalar field the kinetic term $P(\vp, X)$ is the
isotropic pressure. 
It proves convenient to define two parameters in terms of the 
kinetic function $P$ and its derivatives \cite{lidser1,lidser3}: 
\begin{align}
\label{eq:defcs-dbiintro}
 \cs^2 &\equiv \frac{\PX}{E_{,X}} =  \frac{\PX}{\PX + 2X P_{,XX}} \,,
\\
\label{eq:deflambda-dbiintro}
\Lambda &\equiv  \frac{X^2 P_{,XX} +
\frac{2}{3}X^3 P_{,XXX}}{X P_{,X} +
2X^2 P_{,XX}}\,.
\end{align}
The first parameter, $\cs$, is called the sound speed of the fluctuations
in the inflaton field. This can be significantly less than unity for non-canonical
actions, 
in contrast to slow roll inflation driven by a canonical 
field such that $\cs = P_{,X} =1$.
Christopherson \& Malik showed in \Rref{Christopherson:2008ry} that $\cs$
is in fact the phase speed of the fluctuations and not the sound speed
which is defined as $\dot{P}/\dot{E}$. However, in common with the rest of the
literature, we will continue to use $\cs$ as defined in \eq{eq:defcs-dbiintro}. 

The generation of quantum perturbations in the non-canonical case is similar to
the canonical one, but now includes contributions from $\cs$. Letting
$u=a\dvp1$, the Mukhanov equation for the
Fourier modes, \eq{eq:ueq-intro}, becomes
\cite{gm, Mukhanov:2005sc}:
\begin{equation}
 u''(\kvi) + \left[\cs^2 k^2 - \frac{z''}{z}\right] u(\kvi) = 0\,,
\end{equation}
where $z$ has been redefined as
\begin{equation}
 z = \frac{a\sqrt{E+P}}{\cs H} = \frac{a\sqrt{2X\PX}}{\cs H} \,.
\end{equation}
We quantise the $u$ modes using the Bunch-Davies vacuum as above and work with the
amplitude $w(\kvi)$. Instead of considering whether modes are
inside the comoving horizon, it is now important to distinguish between modes
inside and outside the sound horizon, defined by $k\cs = aH$. Far inside the sound
horizon, where $k\cs \gg aH$, the mode solution takes a similar asymptotic form to
\eq{eq:subhsoln-intro}:
\begin{equation}
 w(\kvi) = \frac{1}{\sqrt{2k\cs}} e^{-ik\cs \eta}\,.
\end{equation}

Following the same analysis as above, the amplitude of the curvature 
perturbations 
generated during inflation can be found and is given by
\cite{gm}
\begin{equation} 
\label{eq:Ps-dbiintro}
 \Pr = \frac{H^4}{8\pi^2 X}\frac{1}{\cs \PX} \,.
\end{equation}
This expression is only valid after exit from the sound horizon. In contrast
the tensor perturbations are not affected by the change in the action. The
expression for the power spectrum $\Pt$ in \eq{eq:Ptdefn-intro} is still valid.
This should be evaluated after the modes have exited the normal horizon,
\iec when $k < aH$. 
The consistency relation \eqref{eq:consistency-intro} is now defined as
\cite{gm} 
\begin{equation}
\label{eq:rdefn-dbiintro}
  r = 16\cs \varepsilon_H = -8\cs n_T \,.
\end{equation}
Hence, a sound speed different to unity leads to a violation of the 
standard inflationary consistency equation, which might be 
detectable in the foreseeable future \cite{lidser1,lidser2}.

\section{Non-Gaussianity}
\label{sec:fnl-intro}
The initial fluctuations described above have Gaussian statistics, with no
correlations between modes
on different scales. All the information about the perturbations can be obtained from the
two-point function or power spectrum as defined in \eq{eq:Prdefn-intro}. For a
Gaussian random 
field all higher point functions are either zero or combinations of the two-point function. In
particular the three-point function of $\R$, $\langle
\R(x_1)\R(x_2)\R(x_3)\rangle$, will be zero for purely Gaussian $\R$. We can
write the Fourier
transform of the three point function in terms of the bispectrum $B$
\cite{Bartolo:2004if}:
\begin{equation}
 \langle \R(\kb_1)\R(\kb_2)\R(\kb_2)\rangle = (2\pi)^3 \delta^3(\kb_1 + \kb_2 + \kb_3) B(k_1, k_2,
k_3)\,,
\end{equation}
where translation invariance imposes the conservation of the $\kb$ vectors and
the bispectrum
depends only on the magnitude of each wavenumber. Any deviation from Gaussianity
will result in a
non-zero bispectrum value. 
Because of the delta-function, the wavevectors form triangles in Fourier
space and $B$ is a function of only two variables. The bispectrum generated by
inflationary models
takes two main triangular shapes, squeezed and equilateral \cite{Babich:2004gb}.
 
The first parametrisation of non-Gaussianity was defined in
real space in terms of the Gaussian part of the perturbation. As the
non-linearity is localised in real space the parameter is
known as the local non-Gaussianity $\fnlloc$: 
\begin{equation} 
\label{eq:fnllocdefn-intro}
{\cal{R}} = {\cal{R}}_\mathrm{G} + \frac{3}{5} \fnlloc  (
{\cal{R}}_\mathrm{G}^2 -\langle {\cal{R}}_\mathrm{G}^2 \rangle )\,.
\end{equation}
Here the 
quadratic component represents a convolution and 
$\R_\mathrm{G}$ denotes the Gaussian contribution
\cite{Maldacena:2002vr}.
We use the WMAP sign convention for $\fnl$ throughout. 
This is the opposite of the Maldacena convention:
$\fnl^\mathrm{WMAP}=-\fnl^\mathrm{Maldacena}$. One consequence of this choice of sign is that
positive $\fnl$ implies a decrease in temperature in the CMB compared to the Gaussian case. This
can be seen by noting that at linear order the temperature anisotropy in the CMB can be related to
the curvature perturbation by $\mathcal{R}_\mathrm{G}\simeq -5 \Delta T/T$.

The local non-Gaussian parameter $\fnlloc$ can be related to the bispectrum by:
\begin{equation}
 B(k_1, k_2, k_3) = \frac{6}{5}\fnlloc \left[P_\R^2(k_1)P_\R^2(k_2) +
P_\R^2(k_2)P_\R^2(k_3) + P_\R^2(k_3)P_\R^2(k_1)\right]\,.
\end{equation}
If $P_\R^2(k)$ is approximately scale invariant, $P_\R^2(k) = c k^{-3}$, then
the bispectrum becomes \cite{Baumann2009}
\begin{equation}
 B(k_1, k_2, k_3) = \frac{6}{5}\fnlloc c^2\left[\frac{1}{k_1^3 k_2^3} +
\frac{1}{k_2^3 k_3^3} + \frac{1}{k_3^3 k_1^3}\right]\,.
\end{equation}
This expression is maximised if one of the $k_i$ is much smaller than the other two.
Momentum
conservation then requires that they are approximately equal. This
configuration is a squeezed triangle in momentum space where, for example, $k_3 \ll
k_1,k_2$. In
single field inflation $\fnlloc$ is proportional to the slow roll parameters
and therefore expected to be small. Non-linear contributions from the coupling
of the gravitational potential to the curvature perturbation are expected to
produce $\fnlloc$ of order one which would be much larger than the
$O(\varepsilon_H)$ contributions from single field, slow roll inflation
\cite{Bartolo:2004if, Komatsu:2008hk}. Any detection of $\fnlloc$ at greater
than $O(1)$ would present a challenge to such single field slow roll models. 
The current bounds on the non-Gaussianity parameter are not strong but have
been steadily tightening. The WMAP5 bound on the local form of $\fnl$ is
\begin{equation}
\label{eq:fnlloc-wmap5-intro}
 -9<\fnlloc<111\,.
\end{equation}
This observational limit still includes $\fnlloc=0$ at the 95\% confidence level.

The other important case is   
where the three momenta have equal magnitude, which corresponds to the
equilateral triangle limit. Non-Gaussianity of this shape is chiefly produced
by models with non-canonical kinetic terms as defined in
Section~\ref{sec:noncanoninfl}. The equilateral non-Gaussianity parameter
$\fnleq$ can be evaluated in terms of the sound speed $\cs$ and the
$\Lambda$ parameter defined in \eq{eq:deflambda-dbiintro}.
The leading-order contribution to the
non-linearity 
parameter is given by \cite{chenetal,lidser3}
\begin{equation} 
\label{eq:fnldefn-dbiintro}
 \fnleq = -\frac{35}{108}\left(\frac{1}{\cs^2} -1 \right) +
\frac{5}{81}\left( \frac{1}{\cs^2} -1 -2\Lambda \right) \,.
\end{equation}
Data from WMAP3 imposed the bound $|\fnleq| < 300$ on this parameter
\cite{spergel}. The more recent WMAP5 data set
improves on this bound somewhat \cite{Komatsu:2008hk}, and
also indicates that it is distinctly asymmetric. At the $95 \%$ confidence
level, the current bound on the 
equilateral triangle is 
\begin{equation}
\label{eq:fnleq-wmap5-intro}
 -151<\fnleq<253\,.
\end{equation}

The main difference between the local and equilateral types of non-Gaussianity are the eras
and methods of production. Local
non-Gaussianity parametrises non-linear correlations which are local in real space. Non-linear
processes taking place outside the horizon are the cause of these correlations. This is   
Production of this type of non-Gaussianity occurs irrespective of whether the perturbations
are Gaussian when they cross the horizon. 
For single field models the magnitude
of $\fnlloc$ is proportional to the deviation of the scalar curvature power spectrum from scale
invariance and is therefore expected to be small. On the other hand, models with multiple fields 
can produce a large amount of local non-gaussianity by the evolution of a non-inflaton field 
outside the horizon and the subsequent transfer of fluctuations in this field into curvature
perturbations. A detection of non-negligible $\fnlloc$ would therefore be a very strong indication
that multiple degrees of freedom are present in the early universe.

In contrast, equilateral type non-Gaussianity is peaked when the momenta of the three modes are very
similar and is generated by higher order derivative terms. Both the time and space derivatives
become negligible once the modes have left the horizon and therefore any contribution to the
bispectrum peaked in the equilateral shape takes place when the modes are inside the
horizon. The extra derivative terms required are found generally in non-canonical models
which were discussed in Section~\ref{sec:noncanoninfl}. In this case the amplitude of $\fnleq$ is
proportional to the inverse of the sound speed squared and can be large.

In the case of single field DBI inflation, discussed in Part~\ref{part:dbi} of this thesis, the
non-canonical action in \eq{eq:Pdefn-dbiintro}
contains a non-linear function of $\partial_\mu \vp$ in the square-root term. These higher
derivative
terms are related to the magnitude of the equilateral type through \eq{eq:fnldefn-dbiintro}. In the
relativistic limit in which the sound speed is small, $\fnleq$ can become arbitrarily large. Indeed
the current observational limit on $\fnleq$ restricts the degree to which the relativistic limit
can be reached and tighter bounds on $\fnleq$ could make such a limit inconsistent.

In summary there are two main types of non-Gaussianity, which are produced in very different
fashions\footnote{Not all non-linear processes fit into these two categories and other types have
been proposed including one ``orthogonal'' to the equilateral type \cite{Senatore:2009gt}.}. Local
non-Gaussianity is produced outside the horizon and is comprised of correlations which are local in
real space. Equilateral non-Gaussianity is produced by higher derivative terms when similar modes
are inside the horizon. It is generated by models which have non-canonical actions.

\section{Discussion}
\label{sec:disc-intro}

In this chapter the physics of the FRW universe has been described. Inflation has been introduced
to solve problems with the standard Big Bang scenario. To solve these problems the
inflationary
period must be of sufficient duration. This can be ensured by using models which comply with
certain slow roll conditions. 

To explain inhomogeneities in the early universe, cosmological perturbation theory
was presented up
to first order. The power spectrum of scalar perturbations, $\Pr$, the spectral index of this
spectrum, $n_s$, and the ratio of tensor-scalar perturbations, $r$, are the main
observable
quantities against which models can be tested. Slow roll models must also satisfy a
consistency
relation between the tilt of the tensor spectrum and $r$. Current observations favour an almost
scale invariant red spectrum ($n_s<1$) with a low level of tensor signal. The accuracy of the
current data is not yet good enough to meaningfully evaluate the slow roll consistency relation. 

As well as the standard models, one can also construct inflationary models in which the action
takes a non-canonical form. In these models the sound speed of scalar fluctuations, $\cs$, plays a
pivotal role. The predictions for scalar perturbations are altered by a factor of $\cs$, as is the
slow roll consistency relation. Non-canonical models also often exhibit strong non-linear effects
which can be parametrised using the non-Gaussianity parameter $\fnl$. Canonical single field slow
roll models do not predict large amounts of non-Gaussianity. 

In this thesis, inflation is taken to be the mechanism by which inhomogeneities in matter are seeded
and the horizon and flatness problems of the Big Bang are solved. However, the inflationary
paradigm is not without its own challenges.

Chief amongst these is the lack of a unique underlying theory. Many high energy theories have been
shown to produce an inflationary phase. Often, however, these require a great deal of fine-tuning
in order to produce a sufficient number of e-foldings of inflation. Lack of knowledge about the
governing physics at high energy scales hampers our understanding of the cause of inflation and
undermines any analysis of the generic nature of the initial conditions required.

The overall duration of inflation is also unknown. Observations only require that currently
observable scales were previously inside the horizon. Thus the onset of inflation is not
constrained and could occur far in the past. However, allowing such a long inflationary period
typically increases the fine-tuning necessary and can lead to other issues. 

There are further problems with the inflationary paradigm, including the lack of an explanation for
how energy in the inflaton field is transferred to the other constituent parts of the universe, and
indeed the fact that no scalar field has yet been directly observed.
We will continue to employ the inflationary paradigm in this thesis but it is important to
acknowledge that some challenges remain to be overcome.

This chapter laid the foundations for the two main parts of this work in which first
analytic
and then numerical techniques are used to constrain inflationary models. In the next chapter the
DBI brane inflation scenario is presented.


\part{DBI inflation}
\label{part:dbi}
%
%

\chapter{Introduction to Dirac-Born-Infeld Inflation}
\label{ch:dbi-intro}

\section{Introduction}
\label{sec:dbi-intro}

The inflationary scenario provides the 
theoretical framework for the early history 
of the universe. It has now been successfully tested by observations, 
including the five year data from WMAP
\cite{Komatsu:2008hk}. Despite this success, however, the high energy 
physics that resulted in a phase of accelerated expansion is still 
not well understood. String and M-theory attempt to unify the fundamental
interactions 
including gravity. 
The early universe provides a unique window into high energy physics at scales currently
unreachable by particle accelerators.
It is therefore important to 
develop inflationary models within string theory and to confront them with 
cosmological observations.

One class of string theory models that has received 
considerable attention is D-brane inflation
\cite{brane1,brane2,brane3,brane4,brane5,
brane6,brane7,brane8,brane9,brane10,brane11,brane12,brane13,
brane14,brane15,brane16,brane17,Brodie:2003qv,Vikman:2006hk, 
Mukhanov:2005bu,Kallosh:2007wm,brane18,
brane19,brane20,brane21}. 
(For some recent reviews, see
\cite{tyereview,cline,McAllister:2007bg,Lorenz:2007ze,
Bean:2007eh,bean}). 
The Dirac-Born-Infeld (DBI) scenario 
of the compactified type IIB theory is a well-motivated model \cite{brane6,brane11}, 
in which inflation is driven by one or more ${\rm D}$-branes 
propagating in a warped ``throat'' background. 
In the simplest version of the scenario, 
the inflaton parametrises the radial 
position in the throat of a single ${\rm D3}$-brane. 
The brane dynamics are determined by the DBI action in such a 
way that the inflaton's kinetic energy is bounded from above by the warped 
brane tension. The regime where this bound is nearly saturated is 
known as the ``relativistic'' limit.

In Part~\ref{part:dbi} of this thesis we will 
explore the observational consequences of DBI inflation. 
In general, primordial gravitational wave fluctuations
and non-Gaussian statistics in the curvature perturbation provide 
two powerful discriminants of inflationary scenarios. 
The nature of the DBI action is such that the sound 
speed of fluctuations in the inflaton can be much less than the speed of 
light. This induces a large and potentially detectable non-Gaussian 
signal in the density perturbations \cite{brane6,brane11,lidser3,chenetal}.

In this chapter we introduce string theory, warped compactifications and DBI inflation.
In Chapter \ref{ch:dbi} we will derive upper and lower 
limits on the amplitude of the tensor perturbations.  
We will explore how these bounds may be relaxed in Chapter \ref{ch:multibrane} and discuss
multi-brane 
scenarios which permit observable tensor signals. 

\section{String Theory and Extra Dimensions}
\label{sec:extradims}
The desire to unify seemingly disparate theories has been a driving force in
theoretical physics for more than a hundred years. This effort has produced 
the Standard Model (SM) of particle physics which unifies three of the four
fundamental forces in a robust theoretical framework. Since the realisation of
the SM, a clear goal of theoretical physics has been the unification of the
fourth force---gravity---into this framework. String theory is one of the leading
contenders for achieving this unification. 
In this section we will introduce
some of the string theory concepts that will be required later to understand DBI
inflation.
Many review articles and text books have been written about string theory and
its application to cosmology and a short list of recent works includes
Refs.~\cite{cline, Johnson2000, Baumann:2009ni,Kallosh:2007wm,
Linde:2005dd,McAllister:2007bg}.

In string theory there are two main
types of strings, referred to as closed and open. These are distinguished by the fact
that closed
strings form a continuous loop while open strings have two unconnected ends. 
There are several different string theories which are linked in pairs by a process
called
duality. Physical descriptions in one theory can be translated into a
dual description in the other. The dual version often exhibits properties that
are useful for solving problems in the original setting.
We will work in the framework of the Type IIB theory since this has proven
to be the most
fruitful for generating models of cosmological inflation \cite{cline,
Linde:2005dd}.

\subsection{Extra Dimensions}
String theory predicts that the one time-like and three spatial dimensions that
constitute the observable universe do not represent the complete spacetime manifold.
Instead,
our universe is a 10 or
11 dimensional spacetime and physical theories in 3+1 dimensions must
therefore be able to explain why the 
other 6 or 7 dimensions are unobservable. One of the challenges of string
theory is how 
to ``hide'' these extra dimensions in such a way as to 
recover the standard four-dimensional cosmology at low energies.

The early work of Kaluza and Klein (KK) in formulating higher dimensional 
theories laid the groundwork for the current treatment of extra dimensions in
string theory \cite{Kaluza1921, Klein1926}. By
compactifying an extra dimension onto a circle of finite radius an infinite
tower of extra fields are introduced into the lower dimensional theory. The
mass of these fields is inversely proportional to the size of the extra
dimension. The appearance of these
unobserved massive fields is avoided by taking the radius to be extremely small.
This leaves a
massless degree of freedom which must be accounted for in the four-dimensional
effective action.

A similar procedure is undertaken when compactifying string theory from a
10 or 11 dimensional description down to four dimensions (for reviews see
Refs.~\cite{douglas,grana}).
In ten dimensional type IIB theory the six extra dimensions are
compactified into a Ricci flat Calabi-Yau (CY) manifold which can be described
by three complex coordinates \cite{Yau1977}. 
Because any Ricci flat metric can be
rescaled onto another Ricci flat metric, there is no unique solution for the
metric on the CY manifold. Instead a family of solutions exists with many free
parameters. These parameters remain after the compactification, in
analogy to the size of the extra dimension in KK compactification, and can depend
on position in the four-dimensional spacetime. They appear as fields
in the four-dimensional theory and are known as moduli. 
These fields are not subject to any symmetry and so their individual values
at different spacetime points can affect the physics at those points.

\subsection{T-duality}
\label{sec:tduality-dbiintro}
In string theory an extra space time symmetry is present which relates physical properties in
theories with large compactification radius with those in theories with small radius. 
Suppose we have a string theory compactified on a circle of radius $L$. The ``T-duality''
transformation which relates two physical theories with this one compactified dimension is
\begin{equation}
\label{eq:tdualtransform-dbiintro}
 L \rightarrow \wt{L} = \frac{\alpha'}{L}\,.
\end{equation}
Now consider what effect this transformation will have on the momentum of a closed string. Instead
of being a continuum, the momentum takes discrete values
$j/L$ for $j \in \mathbb{Z}$. This is a KK tower of massive states. As we
complete a circuit around the compact dimension, the value of the coordinate
function embedding the string in the background will increase by $2\pi w
L$ for $w \in \mathbb{Z}$. This $w$, called the winding number of the string,
can only be non-zero for closed strings, which can be wrapped around the periodic
dimension.

The total mass of the string contains terms with both the KK tower of states
and the new tower of winding states:
\begin{equation}
\label{eq:closedmass-dbiintro}
 M^2 = \frac{j^2}{L^2} + \frac{w^2 L^2}{\alpha'^2} + \cdots\,,
\end{equation}
where the string parameter $\alpha'$ is related to the string mass scale by
$\alpha'=1/\ms^2$.
If $L$ is taken to infinity, the $w\ne 0$ states become infinitely massive and
only the $w=0$ state is left with a continuum of momentum values. Thus, the
uncompactified result is recovered. However, if $L\rightarrow0$, the $j\ne 0$
states are now infinitely massive as in the standard KK picture. Unlike the
standard case, there is now a continuum of winding states with $w\ne 0$,
again giving an
uncompactified dimension. This major departure from the standard
compactification result is a purely stringy effect. 

The formula for the mass spectrum, \eq{eq:closedmass-dbiintro}, is invariant
when $j$ and $w$ are exchanged given the transformation in \eq{eq:tdualtransform-dbiintro}.
Writing the equations of motion in terms of $\wt{L}$, having interchanged $j$ and
$w$, gives a new theory which is compactified on a circle of radius $\wt{L}$. This is known as the
T-dual theory \cite{Sakai1986,Kikkawa1984b}. The two
theories are physically identical since T-duality is an exact symmetry
of string theory for closed strings. The T-duality applies to all physics in the theory and in
particular also affects open string modes. These behave in a different way under T-duality to closed
strings as will be described below.

\subsection{D-Branes}
\label{sec:dbranes-dbiintro}
The dynamics of extended objects known as branes are particularly important for
building inflationary models.
As mentioned in Section~\ref{sec:tduality-dbiintro}, string
theories are linked by T-duality. Fundamental parameters such as the size of the
extra-dimensions, the string coupling and the
coordinate solutions of the strings are related by such a symmetry.

We introduced T-duality by explaining
its effects on closed strings. But what happens to the open strings in a T-dualised theory? Open
strings, as their name suggests, have two open ends and
consequently cannot have a conserved winding number such as $w$. Suppose once more
that one
of the $D$ dimensions is compactified. As $L\rightarrow0$, the non-zero momentum
states become infinitely massive, but in contrast to the closed case there is now
no continuum of winding states. Thus, the open string now lives in $D-1$ dimensions
similar to the result of standard KK compactification \cite{Johnson2000}.
The endpoints of the open strings 
then observe Dirichlet boundary conditions, taking fixed values in the compactified
direction.
There are still closed strings in this theory, however, and these continue to
move in the full $D$ dimensions after being T-dualised.  

The result is similar if more than
one coordinate is made periodic.
If $D-p-1$ spatial dimensions are compactified, for some $p$, then the ends of
the open
strings can still move freely in the other $p$ spatial dimensions on a $p+1$
dimensional hypersurface. This hypersurface is called a Dirichlet brane or
D$p$-brane. The closed string modes move in the full $D$ dimensions.
In Type IIB theory with supersymmetric strings, an extra condition implies that
only D$p$-branes with $p=1,3,5,7,9$ are stable\footnotemark.
\footnotetext{There is also a $p=-1$ D-instanton in which the time direction
along with all spatial directions is subject to Dirichlet boundary conditions
\cite{Green1992,Green1988,Green1977}.}

D$p$-branes can be considered as dynamical objects in their own right with a
tension given by \cite{Johnson2000}\footnote{$T_p$ here is $\tau_p$ in
\Rref{Johnson2000}.}
\begin{equation}
\label{eq:branetensiondefn-dbiintro}
 T_p = \frac{\ms^{p+1}}{(2\pi)^p \gs}\,,
\end{equation}
where $\gs$ is the string coupling and $\ms$ is the string mass scale.
Their dynamics is governed by the action 
\begin{equation}
\label{eq:gendbiaction-dbiintro}
 S_\mathrm{DBI} = -T_p \int \d^{p+1}\xi \sqrt{-\hat{g}}\,,
\end{equation}
where $\hat{g}_{ab}$ is the induced metric on the brane with internal
coordinates $\xi^a$, for $a=0,\ldots,p$, given by \cite{Johnson2000}
\begin{equation}
 g_{\mu\nu} = \hat{g}_{a b} \frac{\partial \xi^a}{\partial x^\mu} \frac{\partial \xi ^b}{\partial x^\nu}\,.
\end{equation}
\eq{eq:gendbiaction-dbiintro} is the general form of the DBI
action which will be used later.

In the simplest versions of slow roll inflation, only a single scalar field
with a
sufficiently flat potential is required to satisfy the slow roll conditions outlined
in
Section \ref{sec:slowroll-intro}. Since D-branes are charged (with Ramond-Ramond
charge), a D-brane and an anti-D-brane
($\overline{\mathrm{D}}$) separated by some distance will be attracted to each other.
The separation distance
can be identified as a scalar degree of freedom and under appropriate conditions
could play the role of the inflaton
field \cite{brane1,brane2,brane3,brane7,Brodie:2003qv,brane9}. 

As described above, compactifying dimensions
introduces scalar fields known as moduli. These fields must be
accounted for in the dynamics unless some way can be found to stabilise them by
fixing their masses to be large.
Initial efforts to induce inflation using D-branes ignored the issue of moduli
stabilisation. Instead, it was
assumed that whatever stabilisation mechanism was used would have no discernible
effect on the inflationary physics. Kachru \etal \cite{brane4} recognised that
in fact stabilisation will be important and must be taken into account.

\subsection{Warped Throats}
\label{sec:warpedthroats-dbiintro}
The moduli must
be stabilised so that they do not appear in the effective action as
massless fields. This can be achieved by switching on background fluxes in the
compactified space. These fluxes are analogous to magnetic
fields in the higher dimensional space. By Gauss' theorem  the compact space
will now have a quantised non-zero total charge.
In the presence of fluxes, a general form for the ten dimensional metric is
\cite{Baumann:2009ni}:
\begin{equation}
\label{eq:warpedmetric-dbiintro}
 \d s^2 = e^{2A(y)}\eta_{\mu\nu} \d x^\mu \d x^\nu + e^{-2A(y)} g_{mn}\d y^m
\d y^n\,,
\end{equation}
where the function $A(y)$ varies across the compact dimensions $y^m$.
Compactifications in which $A$ varies significantly with $y$ are called warped
compactifications and $e^{A(y)}$ is referred to as the warp factor. These warped
compactifications are qualitatively similar to
the Randall-Sundrum scenario \cite{Randall1999,Brummer2006}.

Flux compactification of type IIB string theory to four dimensions 
results in such a warped geometry, where the six-dimensional CY  
manifold contains one or more throats \cite{douglas,gkp,grana}. 
The metric inside a throat takes the same form as in
\eq{eq:warpedmetric-dbiintro}:
\begin{equation}
\label{eq:conemetric-dbiintro}
\d s_{10}^2= h^2 ( \rho) \d s_4^2 + h^{-2} (\rho ) 
\left( \d\rho^2 +\rho^2 \d s_{X_5}^2 \right) \,,
\end{equation}
where the warp factor $h (\rho)$ is a function of the 
radial coordinate $\rho$ along the throat and $X_5$
is a Sasaki-Einstein five-manifold.

In many cases, the ten-dimensional metric (\ref{eq:conemetric-dbiintro}) can be 
approximated locally by the geometry $AdS_5 \times X_5$, where the 
warp factor is given by $h=\rho /L$ and 
the radius of curvature of the $AdS_5$ space is defined by
\begin{equation}
L^4 \equiv \frac{4\pi^4 \gs N}{{\rm Vol} (X_5) \ms^4} \,,
\end{equation}
such that $\mathrm{Vol}(X_5)$ is the dimensionless volume of 
$X_5$ with unit radius and $N$ is the ${\rm D3}$  charge of the throat.

In the Klebanov-Strassler (KS) background \cite{ks}, the throat 
is a warped deformed conifold and 
corresponds to a cone over the manifold 
$X_5 = T^{1,1}= {\rm SU(2)} \otimes {\rm SU(2)}/{\rm U(1)}$
in the UV limit ($\rho \rightarrow \infty$). This has   
a volume $\mathrm{Vol} (T^{1,1}) = 16\pi^3/27$ and  topology
$S^2\times S^3$, where the $S^2$ is fibred over the $S^3$.

\begin{figure}[htbp]
\centering
\includegraphics{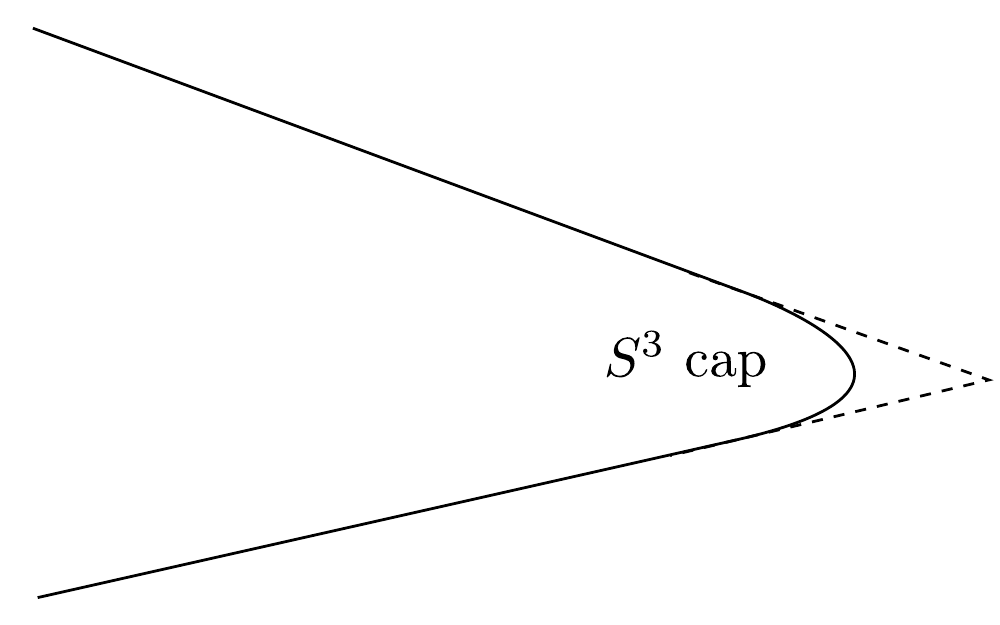}   
\caption[Deformed Conifold]{A conifold can be deformed to remove the singularity at
the tip.}
\label{fig:conifold-dbiintro}
\end{figure}

There are two 3-cycles in the warped throat. The first is the $S^3$ subspace and
is known as the A-cycle. The second, called the B-cycle, is the $S^2$ times a
circle extended in the direction of the throat radius. The three-form fluxes
$F_3$ and $H_3$, aligned with these cycles, are turned on to make the warped
throat a solution of the Einstein equations \cite{cline}. The cycles are threaded with
quantised units of flux $M$ and $K$ given by: 
\begin{align}
\label{eq:fluxdefn-dbiintro}
 \frac{1}{2\pi\alpha'}\int_A F_3 &= M \,, \\
 \frac{1}{2\pi\alpha'}\int_B H_3 &= -K\,, 
\end{align}
where $M,K\in \mathbb{Z}$.
The D$3$ charge of the throat, $N$, is related to the quantised fluxes by
$N=MK$.
The wrapping of the fluxes along the cycles of the conifold smooths out the
conical singularity at the tip of the throat with an $S^3$ cap 
\cite{ks,kt}, as shown in Figure~\ref{fig:conifold-dbiintro}, and the warp factor
asymptotes to 
a constant value in this region.

In this section we have summarised the concepts that will be
required to discuss DBI inflation. The compactified warped throat described
here will provide the setting for this string theoretic realisation of
inflation. 
In the next
section we connect the geometry and physics of the string compactification with
inflationary cosmology and establish the observational parameters that will directly
enable concrete constraints to be formulated. 

%
%
\section{DBI Inflation} 
\label{sec:dbiinflation}
The DBI scenario is based on the compactification of type IIB string theory on a 
Calabi-Yau (CY) three-fold, where the form-field fluxes generate locally
warped regions known as throats, as described above.  The propagation of a 
${\rm D3}$-brane in such a region can drive inflation, where the inflaton 
field is identified with the radial position of the brane 
along the throat. Inflation can occur whether the brane moves towards or away from the tip of the
throat.
Since the radial distance is an open string mode, the field 
equation for the inflaton is determined by a DBI action.

\begin{figure}[htbp]
 \centering
 \includegraphics[width=\textwidth]{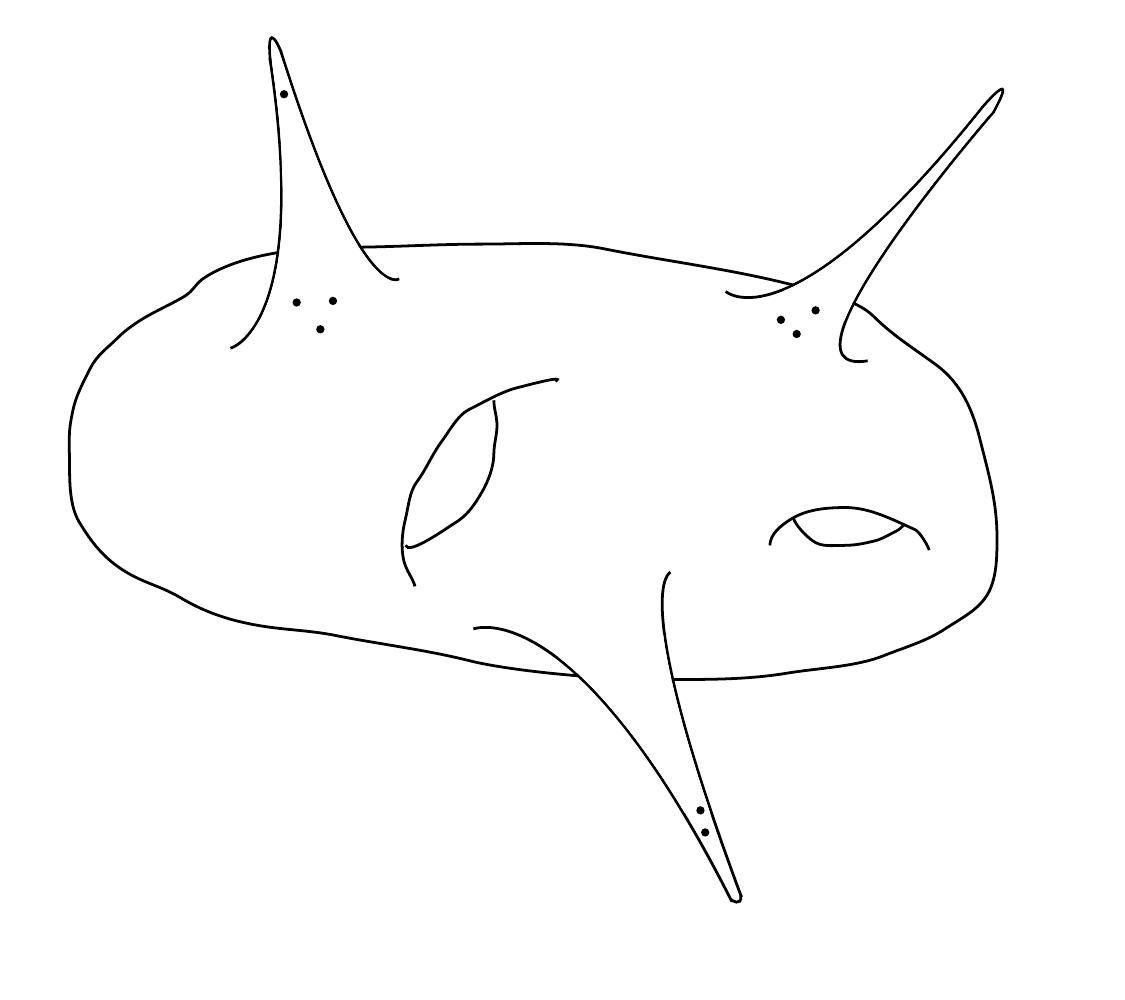}
 \caption[Calabi-Yau Manifold]{A representation of the Calabi-Yau manifold in the 6
compactified
dimensions. Throats are connected to the main bulk. D3-branes appear as dots.}
 \label{fig:braneworld}
\end{figure}

In general, the low-energy world-volume dynamics
of a probe ${\rm D3}$-brane in a warped throat is determined 
by an effective, four-dimensional DBI action, as described in Section
\ref{sec:warpedthroats-dbiintro}.
The inflaton field is related to the radial 
position of the brane by 
$\varphi \equiv \sqrt{T_3} \rho$, where $T_3$ 
is the brane tension defined in \eq{eq:branetensiondefn-dbiintro}. The action is
then given by \cite{brane6}
\begin{align}
\label{eq:DBIaction-dbiintro}
S &=\int  \d^4x \sqrt{|g|} \left[ \frac{\Mpl^2}{2} R 
+ P (\varphi , X) \right] \,,\\
\label{eq:Pdefn-dbiintro}
P( \varphi ,X) &= - T (\varphi)  \sqrt{1 - 2T^{-1} (\varphi ) X}
+T (\varphi)  -V(\varphi)  \,,
\end{align}
where $R$ is the Ricci curvature scalar and $T(\varphi ) = T_3 h^4 (\varphi )$
defines the warped brane tension. As in Section \ref{sec:noncanoninfl} we refer
to $P(\varphi , X)$ as the kinetic function for the inflaton, 
$X \equiv - \frac{1}{2} g^{\mu\nu} \nabla_{\mu} \varphi \nabla_{\nu} \varphi$
is the kinetic energy of the inflaton and $V(\varphi )$ denotes 
the field's interaction 
potential. 
Typically in warped compactifications of 
IIB supergravity, this potential is determined by the 
relevant fluxes and brane interaction terms. 
We will ignore the precise origin
and form of this potential, but simply
note that it is highly sensitive to the string theoretic construction. For the
purpose of this thesis we will simply treat it 
as an arbitrary function of the inflaton field.
(See, for example, Ref. \cite{brane5} for a discussion 
on the precise form that the inflaton potential may take.)

We consider a spatially flat and isotropic cosmology 
sourced by a homogeneous scalar field. 
In this case, the Friedmann equations for a monotonically 
varying inflaton can be expressed in the form \cite{brane6} 
\begin{align}
\label{eq:Friedmann-dbiintro}
3 \Mpl^2 H^2(\varphi ) &= V(\varphi ) -T(\varphi ) 
\left[ 1- \sqrt{1+4\Mpl^4T^{-1} H_{,\vp}^2} \right] \,,\\
\label{eq:phidot-useful}
\dot{\varphi} &= - \frac{2\Mpl^2H_{,\vp}}{\sqrt{1+4\Mpl^4 T^{-1} H_{,\vp}^2}}
\,.
\end{align}

In Section~\ref{sec:noncanoninfl} we introduced the speed of sound of
inflaton fluctuations. For the kinetic function in
\eq{eq:Pdefn-dbiintro}, we find from
\eq{eq:defcs-dbiintro} that
\begin{equation}
\label{eq:csdefn-dbiintro}
\cs = \frac{1}{P_{,X}} = \sqrt{1 -2T^{-1}X}  \,.
\end{equation}
The condition that the sound speed be real 
imposes an upper bound on the kinetic energy 
of the inflaton, $\dot{\varphi}^2 < T(\varphi)$, which 
is independent of the steepness of the potential.
The motion of the brane is said to be relativistic when this bound is 
close to saturation. We will assume throughout Part~\ref{part:dbi} of this thesis that motion takes
place in the relativistic limit in which $\cs\ll1$. 
 
We now define the epoch that is directly 
accessible to cosmological observations as ``observable inflation''. 
We will assume that this phase 
occurred when the brane was located within a 
throat region and moving towards the tip of the throat. 
We denote the parameter values 
evaluated during observable inflation by a subscript star
$(~_*)$. Observable inflation corresponds to about 4 e-foldings  
of inflationary expansion, $\Delta \N_* \simeq 4$,
and occurred somewhere between 30 to 60 e-foldings before the
end of inflation.

The definitions of the slow roll parameters defined in
Section~\ref{sec:slowroll-intro} 
change when $\cs$ is not equal to unity and we
will include a third parameter, $s$, which quantifies the rate of change of $\cs$.
The inflationary dynamics during this phase can  
be quantified in terms of these three parameters: 
\begin{align}
\label{eq:epsdefn-dbiintro}
\varepsilon_H &\equiv -\frac{\dot{H}}{H^2}
= \frac{XP_{,X}}{\Mpl^2H^2} 
= \frac{2\Mpl^2}{\gamma} \left( \frac{H_{,\vp}}{H} \right)^2 \,,\\
\label{eq:defeta-dbiintro}
\eta_H &\equiv  \frac{2\Mpl^2}{\gamma}\frac{H_{,\vp\vp}}{H} \,,\\
\label{eq:defs-dbiintro}
s &\equiv \frac{\dot{\cs}}{\cs H} 
= \frac{2\Mpl^2}{\gamma} \frac{H_{,\vp}}{H}\frac{\gamma_{,\vp}}{\gamma}  \,,
\end{align}
where $\gamma \equiv 1/\cs$. 
We will assume that the quasi-de Sitter conditions 
$\{ \varepsilon_H, |\eta_H | , |s | \}  \ll 1$ apply during observable inflation. 
In this regime, the amplitudes and spectral indices of the two-point functions 
for the scalar and tensor perturbations are given by \cite{gm}
\begin{align}
\label{eq:spectra-dbiintro}
\Pr &= \frac{H^4}{4\pi^2\dot{\varphi}^2} =\frac{1}{8 \pi^2 \Mpl^2}
\frac{H^2}{\cs \varepsilon_H}\,,
\\
\Pt &= \frac{2}{\pi^2} \frac{H^2}{\Mpl^2} \,,
\\
\label{indices}
1-n_s &= 4 \varepsilon_H -2\eta_H  +2s \,,
\\
 n_t &= -2\varepsilon_H  \,,
\end{align}
respectively. $\Pt$ and $n_t$ are evaluated when $k=aH$ but $\Pr$ and $n_s$ are
evaluated 
when the scale with wavenumber $k$ crosses 
the sound horizon $k \cs = aH$.  

A further important consequence of a small sound speed is that departures  
from purely Gaussian statistics may be large 
\cite{brane6,brane11,lidser3,chenetal}. 
DBI inflation produces non-Gaussianity maximised in the equilateral
configuration and the leading contribution is in the form of
\eq{eq:fnldefn-dbiintro}. 
When $\cs\PX=1$ the second term in 
\eq{eq:fnldefn-dbiintro} is identically zero and $\fnleq$ becomes
\cite{chenetal,lidser2}
\begin{equation}
\label{eq:fnlcs-dbiintro}
\fnleq \simeq -\frac{1}{3} \left( \frac{1}{\cs^2} -1 \right) \,.
\end{equation}
When $\cs\ll1$ a significant level of non-Gaussianity is produced.
For a homogeneous field $2X = \dot{\vp}^2$, so from \eq{eq:csdefn-dbiintro} we
find that
\begin{equation}
\label{eq:phiT-dbiintro}
 \dot{\vp}^2 = T(\vp)(1-\cs^2)\,.
\end{equation}
Eqs. \eqref{eq:spectra-dbiintro},
\eqref{eq:fnlcs-dbiintro} and \eqref{eq:phiT-dbiintro}
may then be combined to provide a relation for the warped brane tension: 
\begin{equation}
\label{eq:obswarp-dbiintro}
\frac{T (\varphi)}{\Mpl^4}  = 
\frac{\pi^2}{16} r^2\Pr \left( 1-\frac{1}{3\fnleq} \right) \,.
\end{equation}
%

\section{The Lyth Bound}
\label{sec:lyth-dbiintro}
In the next two chapters we will use a powerful result due to Lyth \cite{lyth}. This
links the change in value of the inflaton field during inflation to the production of tensor
modes. This relation was originally derived for canonical actions but can be straightforwardly
extended to the case of non-canonical actions such as the DBI action.

Eqs. (\ref{eq:Ps-dbiintro}) and (\ref{eq:Ptdefn-intro}) imply 
that the variation of the inflaton field during inflation  
is related to the tensor-scalar ratio by \cite{lyth,bmpaper}
\begin{equation}
\label{eq:genlyth-dbiintro}
\frac{1}{\Mpl^2}
\left( \frac{\d \varphi}{\d \N} \right)^2 = \frac{r}{8 \cs P_{,X}}
\,,
\end{equation}
where $\N$ denotes the number of e-foldings as defined in
\eq{eq:nefolddefn-intro}. 
The total variation in the inflaton field between the epoch of observable 
inflation and the end of inflation is then given by
\begin{equation}
\label{eq:totalfield-dbiintro}
\frac{\Delta \varphi_{\rm inf}}{\Mpl} = 
\left( \frac{r}{8 \cs P_{,X}} \right)_*^{1/2} \Neff \,,
\end{equation}
where
\begin{equation}
\label{eq:Neff-dbiintro}
\Neff \equiv \left( \frac{\cs P_{,X}}{r}\right)_*^{1/2}
\int_0^{\N_{\rm end}}  
\left( \frac{r}{\cs P_{,X}} \right)^{1/2} \d \N \,.
\end{equation}
If $r/(\cs P_{,X})$ varies 
sufficiently slowly during observable inflation, 
the corresponding change in the value of the inflaton  
field is given approximately by \cite{lyth,bmpaper}
\begin{equation}
\label{eq:approxlyth-dbiintro}
\left( \frac{\Delta \varphi}{\Mpl} \right)_*^2 \simeq 
\frac{(\Delta \N_*)^2}{8} \left( \frac{r}{\cs P_{,X}} \right)_* \,.
\end{equation}
This equality links the total variation of the inflaton during observable inflation
with the tensor-scalar ratio, \iec the amplitude of gravitational waves produced
during that period. In Chapter~\ref{ch:dbi} we will show how the dynamics of the
DBI scenario allow an upper limit to be imposed on $r$ using this relation.

In deriving \eq{eq:approxlyth-dbiintro} we have assumed that $r/\cs \PX$ varies slowly during
observable inflation. For the DBI case, $\cs \PX = 1$ and the change in $r$ can be related to the
change in $\epsilon_H$ and $\cs$ through \eq{eq:rdefn-dbiintro}. As we have taken $\epsilon_H,
|\eta_H|,|s|\ll 1$ the tensor-scalar ratio will indeed vary slowly over the observable epoch.
For more general models where $\cs \PX \ne 1$ we have that
\begin{equation}
 \frac{\d }{\d \N}\left[ \frac{r}{\cs \PX}\right] = 16\frac{\epsilon_H}{\PX}\left( 2\epsilon_H
-2\eta_H\right)\,.
\end{equation}
Therefore $r/\cs\PX$ varies slowly as long as $\PX$ is not too small, \iec close to
$\mathcal{O}(\epsilon_H^2)$. This will not be the case in the models studied in Chapters
\ref{ch:dbi} and \ref{ch:multibrane}.

\section{Discussion}
\label{sec:summary-dbiintro}
In this chapter we have introduced the Dirac-Born-Infeld inflationary scenario.
Many attempts have been made to provide an inflationary expansion phase in the
early universe using string theory. In compactifying from ten dimensions down to
four,
complicated geometries and additional fluxes must be used to stabilise the
remaining moduli fields. 

The DBI scenario 
is a particular example of the non-canonical inflationary para\-digm described in
Section~\ref{sec:noncanoninfl}. 
The radial position of a D3-brane in a warped throat is identified as the inflaton field. While the
brane propagates up or down the throat, the kinetic energy of the inflaton is bounded
above by requiring
the sound speed of fluctuations to be real. This bound holds no matter how steep the potential of
the field. The relativistic limit takes the bound to be close to saturation and the sound speed to
be small.
In the case of DBI inflation the speed of sound
parameter
takes the simple form $\cs = 1/\PX$. The previously derived
results for $\Pr$ and $n_s$, as well as the redefined slow roll parameters 
\eqref{eq:epsdefn-dbiintro}--\eqref{eq:defs-dbiintro} can then be expressed in terms
of this parameter. 

Significant non-Gaussianity in the density perturbations spectrum can be generated
due to the small sound speed of the inflaton fluctuations.
This non-Gaussianity can be related to the brane tension and tensor-scalar ratio
through \eq{eq:obswarp-dbiintro}. The tensor-scalar ratio can also by related to the variation in
the inflaton field by the Lyth bound \eqref{eq:genlyth-dbiintro}. This relation can be refined by
focusing only on the period of observable inflation. 
In the next chapter we will derive two
competing bounds on $r$ which will strongly constrain the parameter space for
DBI models.

%
%

\chapter{Observational Bounds on DBI Inflation}
\label{ch:dbi}
%
%
%
\section{Introduction}
\label{sec:intro-dbi}

In this chapter two bounds on the amplitude of primordial gravitational waves will be derived, which
severely challenge the standard
DBI inflationary scenario. By considering the field range of observable
inflation inside a warped throat, the tensor-scalar ratio $r$ will be
constrained to be less than $10^{-7}$. In contrast a lower bound of $r\gtrsim
0.005$ will be derived when the power spectrum of scalar perturbations has a red
spectral index. These clearly incompatible bounds can be relaxed by using a more
general form of the DBI action.

The gravitational wave background generated in DBI 
inflation was initially investigated by Baumann \& McAllister (BM) 
\cite{bmpaper}. By exploiting a relationship due originally 
to Lyth \cite{lyth}, these authors derived a field-theoretic upper limit 
to the tensor amplitude and concluded that 
rather stringent conditions would need to be satisfied for these 
perturbations to be detectable.      
Moreover, the special case of 
DBI inflation driven by a quadratic potential is incompatible with the WMAP3 
data when this constraint is imposed \cite{bean}.

Our aim in this chapter is to derive observational constraints on DBI inflation
that are 
insensitive to the details of the throat geometry and the inflaton potential. 
In general, there are two realisations of the scenario, 
which are referred to as the ultra-violet (UV) and infra-red (IR) 
versions. These are characterised respectively by whether the brane is 
moving towards or away from the tip of the throat. 
We focus initially on the UV scenario 
and derive an upper bound on 
the gravitational wave amplitude in terms of observable 
parameters. This limit arises by considering 
the variation of the inflaton field during the era when 
observable scales cross the Hubble radius, and 
we find in general that the tensor-scalar ratio must satisfy 
$r \lesssim 10^{-7}$. This 
is below the projected sensitivity of future CMB
polarisation 
experiments \cite{Baumann:2008aq,vpj}. 

On the other hand, the WMAP5 data 
favours a red perturbation spectrum, with 
$n_s<1$, when  
the scalar spectral index is effectively constant \cite{Komatsu:2008hk}. 
For models which generate such a spectrum, 
we identify a corresponding lower limit on the 
tensor modes such that $r \gtrsim 0.1 (1-n_s)$. 
This is incompatible with the upper bound 
on $r$ when $1-n_s \simeq 0.03$, as inferred
by the observations. 

Therefore a reconciliation between theory and observation 
requires either a relaxation of the upper limit on $r$ or a blue 
spectral index $(n_s >1)$. The DBI scenario would need 
to be generalised in a suitable way for the upper bound on $r$
to be weakened. Necessary conditions are identified that a 
generalised action must satisfy for the BM constraint and our newly derived
bound to be relaxed. 
Such conditions are shown in Chapter~\ref{ch:multibrane} to be
realised in a recently proposed IR version of DBI inflation driven
by multiple coincident branes \cite{thomasward}. 

%
%
\section{An Upper Bound on the Primordial Gravitational Waves}
\label{sec:upper-dbi}
%

\begin{figure}[htbp]
 \centering
 \includegraphics[width=\textwidth]{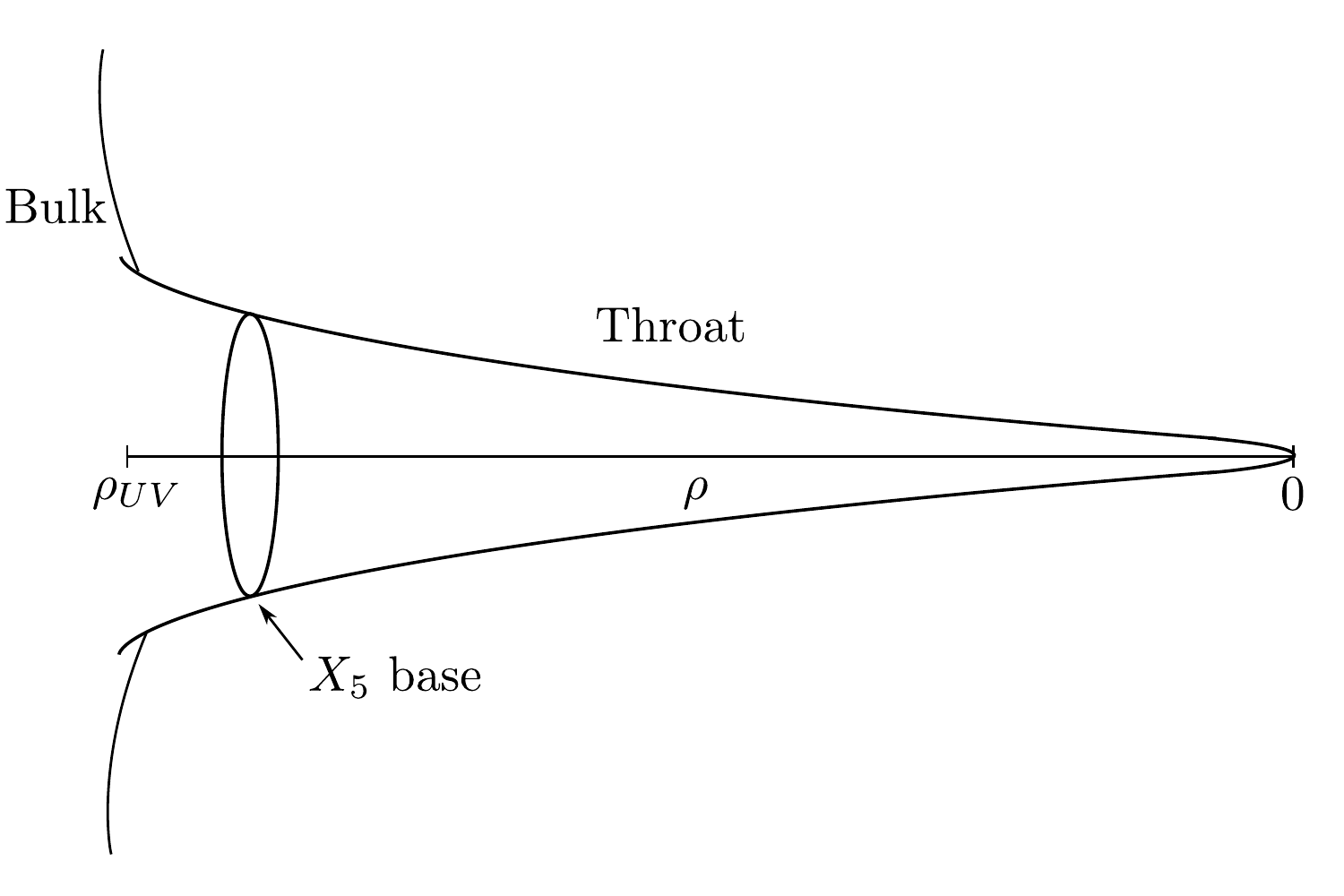}
 \caption[Warped Throat Geometry]{Geometry of the warped throat. The radial
coordinate $\rho$ is measured from the tip of the throat to $\rho_{UV}$ at the join
with the bulk manifold.}
 \label{fig:throat-geom}
\end{figure}

In \Rref{bmpaper} Baumann \& McAllister
derived a field-theoretic upper bound on the tensor-scalar ratio. They achieved
this by noting that the four-dimensional Planck mass is related 
to the volume of the compactified CY manifold, $V_6$, such that 
$\Mpl^2=V_6 \kappa_{10}^{-2}$, where $\kappa_{10}^2 \equiv 
\frac{1}{2} (2\pi )^7 \gs^2 \ms^{-8} = \pi /T_3^{2}$ for a 
${\rm D3}$-brane\footnote{We parametrise the Planck scale 
in terms of the ${\rm D3}$-brane tension out of convenience, 
and note that there is no physical relationship between the two.}.
In general, the compactified volume 
is comprised of bulk and throat contributions, 
$V_6 = V_{6\,{\rm bulk}}+V_{6\,{\rm throat}}$. The latter is 
given by
\begin{equation}
\label{eq:throatvolume}
V_{6\,\mathrm{throat}} = \mathrm{Vol}(X_5)  
\int_0^{\rho_{UV}} \d\rho \frac{\rho^5}{h^4(\rho )} \,,
\end{equation}
where $\rho_{UV}$ denotes the radial coordinate at 
the edge of the throat (defined as the region 
where $h (\rho_{UV})$ is of order unity). The geometry of the throat is shown
in Figure~\ref{fig:throat-geom}.

If one assumes that the bulk volume is 
non-negligible relative to 
that of the throat ($V_{6\,\mathrm{throat}} < V_{6}$), 
it follows that $\Mpl^2> V_{6\,\mathrm{throat}}\kappa_{10}^{-2}$. 
For a warped $AdS_5 \times X_5$ geometry, this leads to an 
upper limit on the total variation of the inflaton field in 
the throat region in terms of the D3 charge:
\begin{equation}
\label{eq:BMbound-dbi}
\frac{\varphi_{UV}}{\Mpl}   < \frac{2}{\sqrt{N}} \,.
\end{equation}

Condition (\ref{eq:BMbound-dbi}) may be converted into a 
corresponding limit on the tensor-scalar ratio by noting from 
the definition (\ref{eq:epsdefn-dbiintro})
that $\dot{\varphi}^2 /\Mpl^2 = 2\varepsilon_H H^2/P_{,X}$.
This implies that the variation of the inflaton field 
is given by the Lyth bound \eqref{eq:genlyth-dbiintro} 
\cite{lyth,bmpaper}:
\begin{equation}
\label{eq:rtheory}
\frac{1}{\Mpl^2} \left( \frac{\d\varphi}{\d \N} \right)^2 =
\frac{r}{8} \,,
\end{equation}
where $\N$ is the number of e-foldings as defined in \eq{eq:nefolddefn-intro}. 
Since $\vp_*$, the field value during observable inflation,  is
less than $\varphi_{UV}$,
this results in an upper bound on the observable tensor-scalar ratio
\cite{bmpaper}: 
\begin{equation}
\label{eq:BMboundr}
r_*  < \frac{32}{N (\Neff)^2} \,.
\end{equation}
The effective number of e-foldings, $\Neff$, defined in
\eq{eq:Neff-dbiintro}, 
is a model-dependent parameter that quantifies 
how $r$ varies during the final stages of inflation. Since $\cs\PX = 1$ in the
standard DBI model, 
it follows that $\Neff = \N_\mathrm{end}$ if $r$ is constant during inflation,
where $\N_\mathrm{end}$ is the total number of e-foldings
from the epoch of observable inflation until inflation ends.

Typically, one expects $30 \lesssim {\Neff} \lesssim 60$, 
although smaller values may be possible if the slow roll conditions are 
violated after observable scales have crossed the horizon. 
Furthermore, $N \gg 1$ is necessary 
for backreaction effects to be negligible \cite{bmpaper}. 
Hence, the constraint \eqref{eq:BMboundr} 
imposes a strong restriction on DBI inflationary models. 
On the other hand, the numerical value 
of $\Neff$ is uncertain.  
Our aim here is to focus on the range of values covered by the 
inflaton field during the observable stages of inflation. 
This will result in a constraint on the tensor modes that 
can be expressed in terms of observable parameters.

To proceed, we denote the change in the value of the inflaton field over 
observable scales by 
$\Delta  \varphi _{*} = \sqrt{T_3} \Delta \rho_{*}$. 
Since the brane moves towards the tip of the throat in 
UV DBI inflation, it follows that $\rho_{*} > \rho_{end} >0$, which 
implies that  
\begin{equation}
\label{eq:importantbound}
\rho_{*} > |\Delta \rho _{*}| \,.
\end{equation}
This change in the inflaton value will correspond 
to a fraction of the throat volume, 
$| \Delta V _{6\,*}|  < {V_{6\,\mathrm{throat}}} \lesssim {V_6} $,
where equality in the second limit arises if
the bulk volume is negligible. Hence, 
$| \Delta \varphi_* |$ is bounded such that  
\begin{equation}
\label{eq:halfwayconstraint}
\left( \frac{\Delta \varphi}{\Mpl} \right)^2_{*} < 
\frac{T_3 \kappa_{10}^2 (\Delta \rho_{*})^2}{|\Delta V_{6\,*}|} \,.
\end{equation}

The observations of the CMB 
that directly constrain the primordial tensor perturbations only 
cover multipole values in the range $2 \le l \lesssim 100$. 
This is equivalent to ${\Delta \N_{*}} \simeq {4}$ 
e-foldings of inflationary expansion and, in general,   
corresponds to a narrow range of inflaton values. 
To a first approximation, therefore, the fraction of the throat volume 
(\ref{eq:throatvolume}) that is accessible to cosmological 
observation can be estimated to be 
\begin{equation}
\label{eq:trapezium}
| \Delta V_{6\,*} | \simeq \mathrm{Vol}(X_5) 
\frac{|\Delta \rho_*| \rho^5_{*}}{h^{4}_{*}} \,.
\end{equation}
Combining the inequality (\ref{eq:importantbound}) with \eq{eq:trapezium} 
then implies that 
\begin{equation}
\label{eq:trapeziumlimit}
|\Delta V _{6\,*}| > \mathrm{Vol}(X_5) 
\frac{(\Delta \rho_* )^6}{h^{4}_*}  \,.
\end{equation}
Substituting the condition \eqref{eq:trapeziumlimit} into the bound
\eqref{eq:halfwayconstraint} gives
\begin{equation}
\label{eq:hbound6}
\left( \frac{\Delta \varphi}{\Mpl} \right)_*^6 < \frac{\pi T_3}{\Vol} 
\left( \frac{h_*}{\Mpl} \right)^4 \,,
\end{equation}   
and using $T(\vp) = T_3 h^4$ and 
\eq{eq:obswarp-dbiintro} yields the upper limit
\begin{equation}
\label{eq:boundpower6}
\left( \frac{\Delta \varphi}{\Mpl} \right)^6_{*} 
< \frac{\pi^3}{16\mathrm{Vol}(X_5)} r^2 \Pr 
\left( 1- \frac{1}{3\fnleq} \right)  \,.
\end{equation}
Hence, employing the Lyth bound \eqref{eq:rtheory} in the form
$(\Delta \varphi_* / \Mpl )^2 \simeq 
r (\Delta \N_*)^2 /8$  
results in a very general upper limit on the tensor-scalar ratio: 
\begin{equation}
\label{eq:generalbound}
r_{*} < \frac{32 \pi^3}{(\Delta \N_*)^6 \mathrm{Vol}(X_5)} 
\Pr \left( 1- \frac{1}{3\fnleq} \right) \,.
\end{equation}

Condition (\ref{eq:generalbound}) is 
only weakly dependent on the level of non-Gaussianity 
when $-\fnleq > 5$ and we may therefore neglect the 
factor involving this parameter. 
Substituting the WMAP5 normalisation 
$\Pr \simeq 2.5 \times 10^{-9}$ then implies that
\begin{equation}
\label{eq:upperbound}
r_{*} < \frac{2.5\times 10^{-6}}{( \Delta \N_*)^6 \mathrm{Vol}(X_5)} \,.
\end{equation}
Furthermore, the most optimistic 
estimate for the minimum number of e-foldings that could be 
probed by observation is $\Delta \N_{*} \simeq 1$, whereas
a generic compactification arises when 
the volume of the Einstein five-manifold is $\mathrm{Vol}(X_5) 
\simeq {\cal{O}} (\pi^3)$ \cite{ks}. This yields a model-independent 
upper bound on the tensor-scalar ratio for standard UV DBI inflation:
\begin{equation}
\label{eq:standardbound}
r_* < 10^{-7} \,.
\end{equation}
The bound \eqref{eq:standardbound} is the main result of this section.
This value of $r$ is significantly below the sensitivity 
of future CMB polarisation experiments, which will measure 
${r} \gtrsim 10^{-4}$ \cite{Baumann:2008aq,vpj}. 
If CMB  
observations are able to span the full range of e-foldings such that
$\Delta \N_* \simeq 4$, this constraint is strengthened to 
${r_*} \lesssim {2 \times 10^{-11}}$.

Before concluding this section, we should explicitly outline all the assumptions that have lead to
\eq{eq:standardbound}. First, we are considering the relativistic limit where $\cs\ll1$. We are
also restricting ourselves to considering the UV scenario where a brane moves towards the tip of
the throat. This ensures that \eq{eq:importantbound} is satisfied. For the Lyth bound to take the
form in \eq{eq:approxlyth-dbiintro}, we have assumed that $r$ varies slowly during the observable
period of inflation. This is justified as the change in $r$ can be written in terms of the
quasi deSitter parameters $\epsilon_H, \eta_H$ and $s$ and we have assumed their magnitudes are
much less than unity.

The estimate (\ref{eq:trapezium}) was derived under the assumption  
that the integrand in \eq{eq:throatvolume}  
is constant. This inevitably introduces errors into the bound
(\ref{eq:generalbound}). However, the two limiting cases of interest 
in KS-type geometries arise 
when the warp factor scales either as $h \propto \rho$
or as $h \simeq {\rm constant}$ \cite{ks,kt}. In both cases
the integral (\ref{eq:throatvolume}) can be performed analytically. 
Indeed, if we specify $h \propto \rho^{\alpha}$ for some constant $\alpha$,  
evaluate the integral from $\rho_{*}$ 
to $\rho_{*}+\Delta \rho_{*}$, and expand to second-order in a 
Taylor series, we find that  
\begin{equation}
\label{eq:limits}
\Delta V_{6\,*} \simeq \mathrm{Vol}(X_5) \frac{\rho^5_{*}}{h^{4} 
(\rho_{*} )}(\Delta \rho_*) 
\left[ 1 +\frac{(5-4 \alpha )}{2} 
\frac{(\Delta \rho_*)}{\rho_{*}} \right]  \,.
\end{equation}
This implies that the error in \eq{eq:trapezium} 
is no greater than 
about $3 (\Delta \rho_* / \rho_*)$ if  
$0 \le \alpha \le 1$. More generally, it follows that a similar
error will arise for {\em any} warp factor 
$h \propto \rho^{\alpha (\rho )}$, where the function 
$\alpha (\rho)$ satisfies $0 \le \alpha (\rho ) \le 1$ 
over observable scales. 
We conclude, therefore, that \eq{eq:trapezium} 
provides a sufficiently good estimate of the volume element  
for a generic warp factor\footnote{As we shall see in the following section, 
even an order of magnitude error will make little 
difference to our final conclusions.}.

In order to neglect the $\fnleq$ term in \eq{eq:boundpower6} we have assumed that $-\fnleq>5$. As
$\cs$ has been taken to be small this is expected to be the case. The volume of the Sasaki-Einstein
manifold $X_5$ is taken to be $\mathcal{O}(\pi^3)$ in keeping with the values for known solutions.
The WMAP5 normalisation of the scalar perturbation power spectrum has also been used. Finally, in
going from \eq{eq:upperbound} to the final numerical figure in \eq{eq:standardbound} the most
``optimistic'' value, $\Delta \N_*\simeq 1$, has been chosen as this leads to the least
restrictive bound on $r$.  As described above a more realistic value of $4$ would severely
constrain $r$ due to the strong dependence of \eq{eq:upperbound} on $\Delta\N_*$.
%
%
%
%
\section{A Lower Bound on the Primordial Gravitational Waves} 
\label{sec:lower-dbi}

The analysis of the previous section 
indicates that standard versions of UV DBI inflation generate a 
tensor spectrum that is unobservably 
small. Therefore, $r=0$ can be assumed as a prior when discussing the WMAP5
data.
However, in this case the data 
disfavours a scale-invariant density spectrum at close to the $3 \sigma$ level
($2.78\sigma$)
when the running in the spectral index, $\alpha_s \equiv \d n_s/\d\ln k$, 
is negligible \cite{Komatsu:2008hk}.  
Furthermore, a blue spectral index 
is only marginally consistent with the data when $r\ne 0$ and $\alpha_s=0$. 
(The inferred upper limit is $n_s < 1.018$.)
Although the results from WMAP5 do allow for a blue spectrum if there is 
significant negative running in the spectral index, we will 
focus in this section 
on models that generate a red spectral index $n_s<1$, since these are preferred by the current
data.

In general, the spectral index may be related to the tensor-scalar ratio. 
After differentiating \eq{eq:csdefn-dbiintro} 
with respect to coordinate time, and employing Eqs. (\ref{eq:phidot-useful}) 
and (\ref{indices}), we find that\footnotemark
\begin{equation}
\label{eq:nsconstraint}
1-n_s = 4 \varepsilon_H +\frac{2s}{1-\gamma^2} \mp 
\frac{2\Mpl^2}{\gamma} \frac{T_{,\vp}|H_{,\vp}|}{TH}  \,,
\end{equation}
where the minus (plus) sign corresponds to 
a brane moving down (up) the warped throat.
\footnotetext{The relationship between $\dot{\vp}$ and $H_{,\vp}$ is defined in
\eq{eq:phidot-useful}. When $\dot{\vp}<0$ and the brane moves down the throat
(UV case) $H_{,\vp}>0$. Alternatively in the IR case when $\dot{\vp}>0$ we have
$H_{,\vp}<0$. In order to remove the ambiguity we rewrite \eq{eq:phidot-useful}
using $-H_{,\vp} = \mp |H_{,\vp}|$ where the minus (plus) sign corresponds to
the UV (IR) case.}
The second term in \eq{eq:nsconstraint}
can be converted into observable parameters
by defining the `tilt' of the non-linearity parameter  \cite{brane14}: 
\begin{equation}
\label{eq:nnl-defn-dbi}
\nnleq \equiv \frac{\d \ln |\fnleq|}{\d\ln k}\,. 
\end{equation}
This implies that $s=  3 \fnleq \nnleq /[2(1-3\fnleq)]$ and     
substitution of Eqs.~(\ref{indices})--(\ref{eq:fnlcs-dbiintro}) 
into \eq{eq:nsconstraint} then yields
\begin{equation}
\label{eq:obscon1}
1-n_s = \frac{r}{4} \sqrt{1-3\fnleq} + \frac{\nnleq}{1-3\fnleq}
\mp \sqrt{\frac{r}{8}} \left( \frac{T_{,\vp}}{T} \Mpl \right)_*  \,.
\end{equation}

In \Rref{lidser2}, brane inflation near the tip of a KS-type 
throat was considered, where the warped brane tension asymptotes to a 
constant value. In this regime, \eq{eq:obscon1} reduces to 
the condition $r \simeq 2.3 (1-n_s)/\sqrt{-\fnleq}$ when $|\fnleq|$ is 
sufficiently large to be detectable by Planck, 
\iec $|\fnleq| > 5$. 
It then follows from the  
WMAP5 best-fit value $n_s \simeq 0.968$ and lower limit 
$\fnleq > -151$ \cite{Komatsu:2008hk} that the gravitational wave amplitude 
is bounded both from above and below such that $0.001 \lesssim r 
\lesssim 0.01$. These bounds follow from 
current WMAP5 limits on the spectral index and the 
non-linearity parameter, but do not take into account the 
field-theoretic upper bound that must be imposed 
on the variation of the inflaton field during inflation.

More generally, in UV DBI inflation where the brane moves towards the 
tip of the throat, it is reasonable to assume 
that the warp factor decreases monotonically 
with the radial coordinate over the observable range of inflaton values, 
\iec $dh/d \rho \ge 0$. This condition is satisfied for  
$AdS_5 \times X_5$ compactifications and KS-type solutions. 
Consequently, the third term in 
\eq{eq:obscon1} will be semi-negative definite, 
which implies that 
\begin{equation}
\label{eq:halfbound}
\frac{r}{4} \sqrt{1-3\fnleq} + \frac{\nnleq}{1-3\fnleq} 
> 1-n_s \,.
\end{equation}

Condition (\ref{eq:halfbound}) is a consistency relation on UV DBI 
inflation in terms of observable parameters and it 
may be combined with the upper bound 
(\ref{eq:generalbound}) to confront the scenario with observations.
Firstly, let us assume that the tensor-scalar ratio is negligible. 
The WMAP5 data implies that $1-n_s > 0.026$ at $1\sigma$, and this is only 
compatible with condition (\ref{eq:halfbound}) if 
\begin{equation}
\label{eq:nnlbound}
\nnleq \simeq -2s > -3 (1-n_s ) \fnleq > -0.078 \fnleq \,.
\end{equation}
However, when $-\fnleq \gg 1$, this would violate the slow roll conditions
that must be satisfied for a consistent 
derivation of the perturbation spectra 
(\ref{eq:spectra-dbiintro}). For example, the conservative 
bound $|s| < 0.1$ with $1-n_s \simeq 0.05$ is violated if  
$-\fnleq > {\cal{O}} (5)$. 

In view of this, let us consider the case where the tensor 
perturbations are non-negligible. 
The magnitude of the second term in condition (\ref{eq:halfbound}) 
is suppressed by a factor of $(-\fnleq)^{3/2} \gg 1$ 
relative to the first. This is expected to 
be a significant effect in DBI inflation. 
Consequently, 
by saturating the WMAP5 limit $\fnleq > -151$, we arrive at 
a lower bound on the tensor-scalar ratio which applies   
to any model for which the ratio $|\nnleq/\fnleq|$ is 
negligible:
\begin{equation}
\label{eq:lowerbound}
r_* >  \frac{4(1-n_s)}{\sqrt{-3\fnleq}} > \frac{1-n_s}{6} \,.
\end{equation}
This second bound requires $r > 0.005$ for the WMAP5 best-fit value 
$1-n_s \simeq 0.032$, which is incompatible with the upper limit 
(\ref{eq:standardbound}).

In general, therefore, it is difficult to simultaneously satisfy 
the bounds on $r$ with the WMAP5 data
in standard UV DBI inflation. There is a 
small observational window where a blue spectrum is consistent 
with the data, in which case the lower limit 
(\ref{eq:lowerbound}) does not apply. 
However, if the tensor modes are negligible,
as implied by the inequality (\ref{eq:standardbound}), the 
data strongly favours a red spectral index with $n_s < 0.974$,
and this violates the condition (\ref{eq:lowerbound}). A significant 
detection of a red spectral index requires either a 
violation of the slow roll conditions or a sufficiently 
small value for the volume of $X_5$. 
In particular, combining the limits
(\ref{eq:upperbound}) and (\ref{eq:lowerbound}) results in the condition 
\begin{equation}
\label{eq:upperboundvol}
\mathrm{Vol}(X_5) < \frac{2 \times 10^{-5}}{(1-n_s) 
(\Delta \N_{*} )^6}  \,,
\end{equation}
and we find that $\mathrm{Vol}(X_5) \lesssim 10^{-7}$ 
is required for typical values $1-n_s \simeq 0.05$ and 
$\Delta \N_* \simeq 4$. 
This 
is comparable to the limit on the volume derived for the special case of a 
quadratic inflaton potential \cite{bmpaper}.  

As noted in Refs.~\cite{bmpaper} and \cite{bean}, condition 
\eqref{eq:upperboundvol} may be achieved 
if $X_5$ corresponds to a $Y^{p,q}$ space. Previously only two five-dimensional Sasaki-Einstein
metrics were explicitly known, $S^5$ and $T^{1,1}$ on $S^2\times S^3$. The $Y^{p,q}$ metrics
described in \Rref{gauntlett} are a countably infinite number of Sasaki-Einstein metrics on
$S^2\times S^3$. The metrics are parametrised by the two topological numbers $p$ and $q$, which are
coprime when the $Y^{p,q}$ is topologically $S^2\times S^3$. The volume of one of these manifolds
is proportional to $1/p$. Hence by setting $q=1$ and letting $p$ become large, this volume can be
made arbitrarily small \cite{gauntlett}. On the other hand, the largest volume occurs for $p=2$,
$q=1$ giving $\mathrm{Vol}(Y^{2,1})\simeq 0.29\pi^3$. 
Small volumes could also be realised 
by orbifolding the $S^2$ symmetry of a KS-type throat.

On the other hand, 
the upper limit (\ref{eq:upperbound}) on the gravitational waves 
follows as a consequence of assuming 
the constraint (\ref{eq:importantbound}). This  
could be violated in IR versions of the scenario, where
observable scales crossed the Hubble radius when the 
brane was near the tip of the throat and $\varphi \ll \Mpl$
\cite{brane12,brane14}. 
Nonetheless, we emphasise that the upper bound (\ref{eq:upperbound})
on the tensor modes 
will also apply to any IR DBI model for which 
$|\Delta \varphi_* | < \varphi_*$.  
In view of the above discussion, 
we will proceed in the following section
to discuss a framework for generalising the DBI scenario so 
that the constraints on the tensor modes can be satisfied. 
%
%
%
%
\section{Relaxing the Upper Bounds}
\label{sec:relaxing-dbi}
%
In this section, we take a phenomenological 
approach and consider the following kinetic function which has a more general form than the DBI one
but still contains a square root term:
\begin{equation}
\label{eq:genaction-dbi}
P= -f_A (\varphi ) \sqrt{1-f_B (\varphi ) X} -f_C (\varphi) \,,
\end{equation}
where $f_i (\varphi )$ are unspecified functions of the inflaton 
field. We will assume 
implicitly that these functions have a suitable form for 
generating a successful phase of inflation.
A direct comparison with \eq{eq:Pdefn-dbiintro} 
indicates that the standard DBI action can be recovered by setting $f_A f_B =2$. This implies
that $\cs P_{,X} =1$ and greatly simplifies the form of \eq{eq:rtheory}. 
Another important property in the DBI case is that the warp factor uniquely determines 
the kinetic structure of the action, i.e., $h^4 \propto f_A \propto f_B^{-1}$.  
In view of this, it is interesting to consider whether
the gravitational wave constraints could be weakened by relaxing one 
or both of these conditions.

We can differentiate $P(X, \vp)$ in \eq{eq:genaction-dbi} to find:
\begin{align}
 \PX &= \frac{f_A f_B}{2\sqrt{1-f_B X}} \,, \\
 P_{,XX} &= \frac{f_A f_B}{2}\frac{f_B}{2(1-f_B X)^{\frac{3}{2}}} \,.
\end{align}
The sound speed of fluctuations in 
the inflaton, defined in \eq{eq:defcs-dbiintro}, is then given by
\begin{equation}
\label{eq:cs-dbi}
\cs = \sqrt{1-f_B X} = \frac{f_Af_B}{2} \frac{1}{P_{,X}}  \,,
\end{equation}
and the scalar power spectrum \eqref{eq:Ps-dbiintro} by
\begin{equation}
\label{eq:spectrum-dbi}
\Pr = \frac{1}{2\pi^2}\frac{H^4}{f_Af_B\dot{\varphi}^2}  \,.
\end{equation}
However, the consistency equation (\ref{eq:rdefn-dbiintro}) and 
non-Gaussianity constraint (\ref{eq:fnlcs-dbiintro}) remain unaltered 
for this more general class 
of models \cite{lidser2}. It 
follows, therefore, 
that the CMB normalisation condition (\ref{eq:obswarp-dbiintro}):
\begin{equation}
\label{eq:obswarp2-dbi}
\frac{T (\varphi)}{\Mpl^4}  = 
\frac{\pi^2}{16} r^2\Pr \left( 1-\frac{1}{3\fnleq} \right) \,,
\end{equation}
generalises to a constraint on the value of $f_A (\varphi_*)$:
\begin{equation}
\label{eq:observef1}
\left( \frac{f_A}{\Mpl^4} \right)_{*} \simeq \frac{\pi^2}{16} r^2\Pr
\left( 1- \frac{1}{3\fnleq} \right)  \,.
\end{equation}
Finally, the expression for the scalar spectral index
follows by generalising the derivation of \eq{eq:obscon1} given by
\begin{equation}
\label{eq:obscon1-repeat}
1-n_s = \frac{r}{4} \sqrt{1-3\fnleq} + \frac{\nnleq}{1-3\fnleq}
\mp \sqrt{\frac{r}{8}} \left( \frac{T_{,\vp}}{T} \Mpl \right)_*  \,.
\end{equation}
It 
is straightforward to show that for the more general kinetic function this expression becomes
\begin{equation}
\label{eq:nsconstraintW}
1-n_s = \frac{r}{4} \sqrt{1-3\fnleq}
 +\frac{\nnleq}{1-3\fnleq} \mp \sqrt{\frac{r}{4f_A f_B}} \left( 
\frac{f_{A,\vp}}{f_A} \Mpl \right)_*  \,.
\end{equation}

\subsection{A More General BM Bound}

The BM bound \eqref{eq:BMboundr} restricts the maximal 
variation of the scalar field $\varphi$ in the full throat region for DBI inflation. 
This is determined by expression \eqref{eq:BMbound-dbi} 
for generic warped geometries that are asymptotically 
$AdS_5 \times X_5$ away from the tip of the throat. However, in Section~\ref{sec:lyth-dbiintro} the
Lyth bound was also defined for general
non-canonical actions. For the more general kinetic function
\eqref{eq:genaction-dbi}, the BM bound becomes
\begin{equation}
\label{eq:bmboundgen-dbi}
 r < \frac{32}{N(\Neff)^2}\cs\PX = \frac{16}{N(\Neff)^2}f_A f_B \,.
\end{equation}
To use this bound we must be able to calculate $\Neff$ over the full range of e-foldings of
inflation. This requires knowledge of the behaviour of $f_A$ and $f_B$ over that range. 

A more cautious approach would be to restrict our attention to 
the observable stage of inflation.
Assuming that the variation
of $f_Af_B = 2 \cs\PX$ is negligible during that epoch, we can use
\eq{eq:approxlyth-dbiintro} which states that 
\begin{equation}
\label{eq:genphivary1}
\left( \frac{\Delta \varphi}{\Mpl} \right)^2_{*} \simeq 
\frac{(\Delta \N_{*} )^2}{8} \left(\frac{r}{\cs\PX}\right)_*  
= \frac{(\Delta \N_{*} )^2}{4} \left(\frac{r}{f_A f_B}\right)_*  \,.
\end{equation}
In addition, 
if observable scales leave the horizon 
while the brane is inside the throat, the change in the field value 
must satisfy $| \Delta \varphi_*|<\varphi_{UV}$. It follows from \eqs{eq:bmboundgen-dbi} and
\eqref{eq:genphivary1}, therefore, that 
\begin{equation}
\label{eq:genBMbound}
r_*< \frac{32}{N (\Delta \N_{*})^2} (\cs\PX)_* = \frac{16}{N (\Delta \N_{*})^2}
(f_Af_B)_* \,.
\end{equation}

Condition 
(\ref{eq:genBMbound}) will be referred to as the 
generalised BM bound. 
We have been
conservative by restricting our discussion to the 
observable phase of inflation. A stronger condition is obtained by using
\eq{eq:bmboundgen-dbi}, which is equivalent to substituting $\Delta \N_* \rightarrow 
\Neff$. If 
$f_Af_B$ remains nearly constant over the last $\N$ 
e-foldings of inflation, 
then $\Neff$ may be as large as $60$ and the right hand side of \eq{eq:genBMbound} will be reduced
by a factor of $225$. 
Thus, the generalised bound 
(\ref{eq:genBMbound}) should be regarded as a necessary 
(but not sufficient) condition to be satisfied by the tensor modes.

Given expressions (\ref{eq:observef1}) and (\ref{eq:genBMbound})
we can either constrain $r$ using a specified value for $f_B$, or find a necessary
condition on the value of $f_B(\varphi_*)$
for the generalised BM bound to be satisfied using $r$ and $\Pr$:
\begin{equation}
\label{eq:f2bound}
\frac{f_B (\varphi_*)\Mpl^4}{N} > \frac{(\Delta \N_*)^2}{\pi^2} 
\frac{1}{r\Pr} \,.
\end{equation}

In UV models, identical arguments that led to 
the lower limit (\ref{eq:lowerbound}) on the tensor-scalar ratio
will also apply in this more general context if, as expected, $f_A (\varphi) $ 
is a monotonically increasing function. 

A necessary condition for
the lower and upper limits
(\ref{eq:lowerbound}) and (\ref{eq:genBMbound}) to be compatible, therefore, is  
that  
\begin{equation}
\label{eq:evade}
f_Af_B > \frac{N (\Delta \N_*)^2 (1-n_s)}{4\sqrt{-3\fnleq}} \,.
\end{equation}

In IR scenarios, however, the positive sign will apply in the 
last term of the right-hand side of \eq{eq:nsconstraintW}.
Hence, assuming $f_{A,\vp} >0$ and neglecting the term proportional to 
$\nnleq/\fnleq$ yields another {\em upper} limit on the tensor-scalar ratio:
\begin{equation}
\label{eq:IRupper}
r_* < \frac{4(1-n_s)}{\sqrt{-3\fnleq}}  \,.
\end{equation}
Combining conditions (\ref{eq:f2bound}) and (\ref{eq:IRupper}) 
therefore leads to a constraint on $f_B (\varphi_*) $
for the generalised BM bound to be satisfied in IR inflation: 
\begin{equation}
\label{eq:f2IRlower}
\frac{f_B\Mpl^4}{N} > \frac{(\Delta \N_*)^2}{4\pi^2}
\frac{\sqrt{-3\fnleq}}{(1-n_s)\Pr}  \,.
\end{equation}

To summarise, for the more general kinetic function in \eq{eq:genaction-dbi}, the parameters
$f_A$ and $f_B$ must satisfy \eqs{eq:observef1} and \eqref{eq:f2bound}, where we have restricted our
interest to the era of observable inflation. For UV models the lower and upper bounds on the
tensor-scalar ratio will be compatible if \eq{eq:evade} is satisfied. For IR models two upper
bounds on $r$ have been found, which when combined constrain $f_B$ as in \eq{eq:f2IRlower}. This
constraint will prove useful in Section~\ref{sec:ir-largen-bound-multi}.

\subsection{The New Upper Bound for General Models}

The newly derived upper bound on $r$ for DBI models, \eq{eq:generalbound}, arises 
because the warp factor in standard DBI models 
completely specifies the kinetic 
energy of the inflaton field. Deriving a corresponding bound for 
the generalised model (\ref{eq:genaction-dbi}) would be more involved, 
since the CMB normalisation (\ref{eq:observef1}) only 
directly constrains the function 
$f_A (\varphi )$ and this may not necessarily depend on the warp factor. 
Instead, the constraint \eqref{eq:hbound6}, given by
\begin{equation}
\label{eq:hbound6-repeat-dbi}
\left( \frac{\Delta \varphi}{\Mpl} \right)_*^6 < \frac{\pi T_3}{\Vol} 
\left( \frac{h_*}{\Mpl} \right)^4 \,,
\end{equation} 
can be 
combined with \eq{eq:genphivary1} to derive a limit 
on the tensor-scalar ratio in terms of the warp factor and $P(X,\vp)$. We find
that 
\begin{equation}
\label{eq:LHbound}
r_* < \frac{8}{(\Delta \N)_*^2} \left(\frac{\pi T_3}{\Vol} \right)^{1/3} 
\left( \frac{h_*}{\Mpl} \right)^{4/3} \left( \cs P_{,X} \right)_* \,.
\end{equation}
This bound is valid for any $P(X,\vp)$ in the warped throat including the
generalised DBI function given in \eq{eq:genaction-dbi}.
For a specific model where the warp factor and 
the functions $f_i (\varphi )$ are determined by particle 
physics considerations,   
condition \eqref{eq:LHbound} 
may be interpreted as a bound that relates 
the tensor modes directly to the value of the inflaton field during observable 
inflation. This constraint provides a consistency 
check that any given model must satisfy 
irrespective of the form of the inflaton potential. 

It is worthwhile to compare \eq{eq:LHbound} with the BM bound for the full evolution given in
\eq{eq:bmboundgen-dbi}. To evaluate the BM bound requires knowledge of $f_A$ and $f_B$ over the
whole of the inflationary era. 
In contrast, using \eq{eq:LHbound} only requires values during observable inflation. However
$\cs\PX=f_A f_B/2$ must be assumed to be slowly varying for \eq{eq:LHbound} to
be valid. As discussed in Section~\ref{sec:lyth-dbiintro} this is a reasonable assumption for
models in which $\PX$ is larger than $\mathcal{O}(\epsilon_H^2)$ during observable inflation. 

As the two bounds provide upper limits on $r$ their relative strength can be compared. The bound
\eqref{eq:LHbound} is stronger than the full throat BM bound if
\begin{equation}
\label{eq:LHstronger}
h_*^{4/3}N < 20 \left( \Vol \gs \right)^{1/3}  
\left( \frac{\ms}{\Mpl} \right)^{-4/3} 
\frac{( \Delta \N )_*^2}{\Neff^2} \,.
\end{equation}
For typical field-theoretic values $\Vol \simeq \mathcal{O}(\pi^3)$, $\ms \simeq
0.1 \Mpl$ 
and  $\gs \simeq 10^{-2}$, this implies that
\begin{equation}
\label{eq:LHstronger1}
h_*^{4/3} N < 300 \frac{(\Delta \N)^2_*}{\Neff^2} \,.
\end{equation}
If the more conservative approach outlined above is taken, the BM bound for observable
inflation, \eq{eq:genBMbound}, is weaker than \eq{eq:LHbound} when
\begin{equation}
\label{eq:LHstronger-weakBM}
h_*^{4/3} N < 300 \,.
\end{equation}

To summarise this section, new bounds have been derived which generalise those described in
Section~\ref{sec:upper-dbi} to the case of the kinetic function in \eq{eq:genaction-dbi}.
The expressions \eqref{eq:bmboundgen-dbi}, \eqref{eq:genBMbound} and \eqref{eq:LHbound} imply that
the 
bounds on $r$
could be relaxed if $2\cs\PX = f_Af_B \gg 1$ on observable scales. 
It is therefore important to develop string-inspired models 
where this condition arises naturally. We will explore this possibility in the
next chapter.

%
%
%
\section{Review of Other DBI Based Models} 

\label{sec:others-dbi}
\begin{figure}[htbp]
 \centering
 \includegraphics[width=0.8\textwidth]{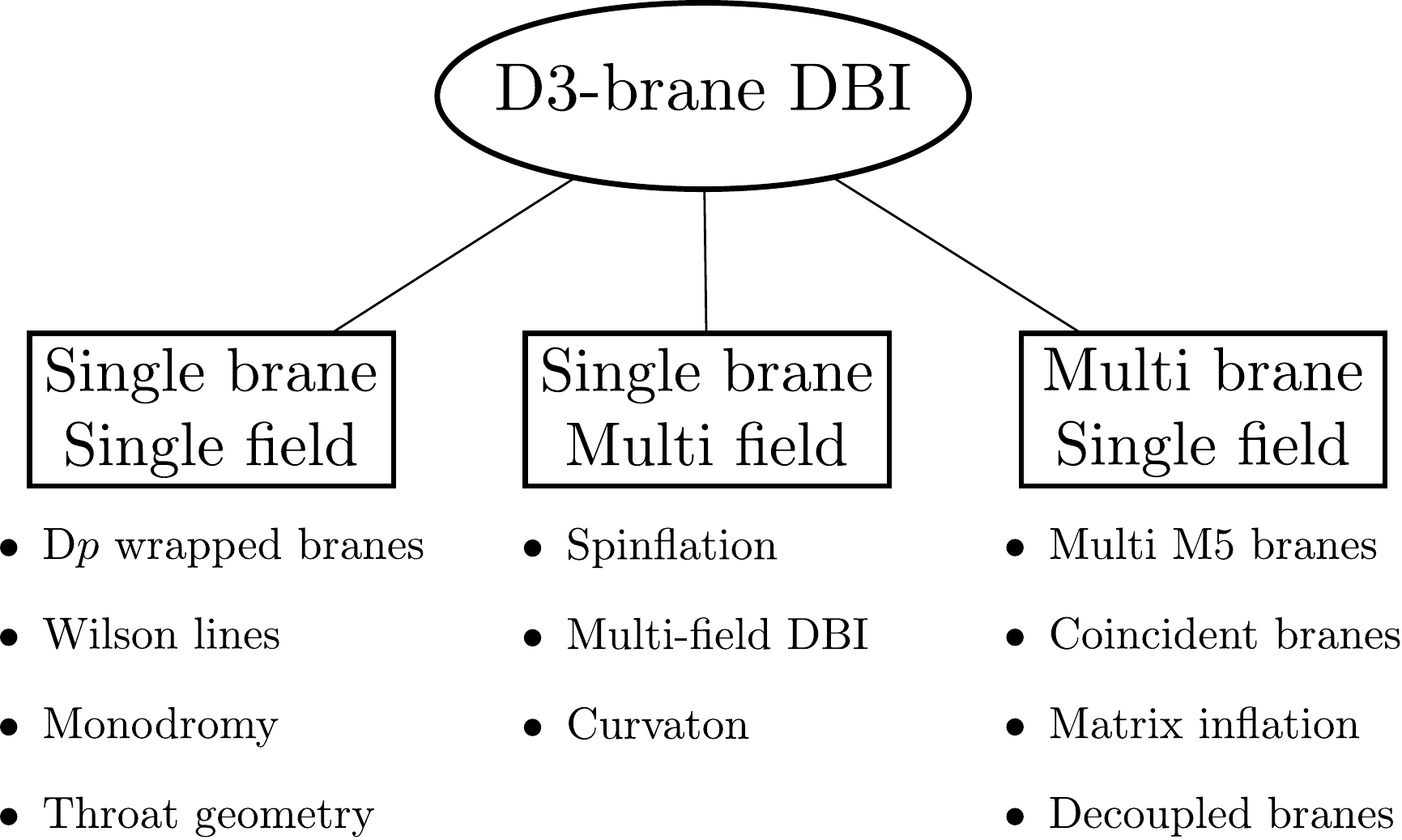}
 \caption[Recent DBI Inspired Models]{A schematic of recent DBI inspired models}
 \label{fig:review-dbi}
\end{figure}

We have found that the standard DBI model appears to be in conflict with
observations. Many attempts have since been made to modify the original scenario in
order to evade the bounds derived above. These new models can be classified according
to
whether they involve single or multiple fields and single or multiple branes.
Figure~\ref{fig:review-dbi} lists some of the models in
each category.

The most straightforward extensions of the DBI model are single field, single brane
models. 
These have a single degree of freedom, as in the D3-brane model, but
they rely on other physical mechanisms to ease the bounds on $r$.
A natural extension to the single ${\rm D3}$-brane model 
is to consider a ${\rm D}p$-brane wrapped around a $(p-3)$-cycle of the 
internal space.
This leads to a change in the relationship
between $\rho$ and $\vp$ from that defined in Section~\ref{sec:dbiinflation}
\cite{Kobayashi:2007hm, Becker:2007ui, Ward:2007gs}.  
For example, Becker \etal \cite{Becker:2007ui} 
have proposed a model in which inflation is driven by a wrapped ${\rm D5}$-brane. 
In this case, the range of allowed values for the inflaton 
becomes independent of the throat charge, $N$, which weakens the upper bound on 
the tensor-scalar ratio to $r \lesssim 0.04$. (Strictly speaking there is a weak
dependence on the charge since $\Delta \varphi \sim N^{-1/4}$.)
However, in arriving at this bound, it was assumed that 
backreaction effects of any fluxes in the throat were 
negligible. 
Kobayashi \etal \cite{Kobayashi:2007hm} 
considered both ${\rm D5}$ and
${\rm D7}$ wrapped brane models, but concluded that the former case
required an excessively large background charge in 
order to relax the bounds on $r$. 
This requirement is highly constraining, but is
still not as restrictive as the value of the charge required by the single brane scenario,
which effectively rules this model out. 
Thus, a wrapped brane configuration is preferable to
the single D3-brane model, but the parameter space of the former is still severely
limited by the WMAP5 observations \cite{Alabidi:2008ej}.

Another interesting proposal is warped Wilson line DBI. In this scenario, moduli
fields associated with Wilson
lines play the role of the inflaton \cite{Avgoustidis:2008zu}. This scenario is
T-dual to the
standard DBI model with non-parallel branes. In general, the model describes the
physics of a single brane with multiple position fields and multiple Wilson line
fields.
In \Rref{Avgoustidis:2008zu}, observational predictions were derived for
the case when the brane position is fixed and only one Wilson line degree of freedom
is used. This implementation is therefore a single brane, single field model.
By following the method outlined in Section~\ref{sec:upper-dbi} for this single field model,
a lower bound on $r$ was derived, instead of the upper bound \eqref{eq:upperbound}
\cite{Avgoustidis:2008zu}. The lower bound \eqref{eq:lowerbound} remains valid for
this scenario. 
There are, therefore, two lower bounds on $r$ and the inconsistency of the
standard DBI model is not replicated.

Changing the physical setting can also allow larger field ranges, which in turn can relax the
bounds on $r$. One such example is the case of a D4-brane in compactified manifolds
containing monodromies. 
The large field variations in this single brane, single field model lead to possibly
observable
tensor modes
\cite{Silverstein:2008sg}. Although formulated in Type IIA string theory, the
monodromy scenario has a simple inflationary interpretation as a large field, slow roll
model with a potential $V(\vp)\propto\vp^{2/3}$. 

The tensor-scalar ratio and other observable quantities are
significantly altered if the throat geometry is not of the $AdS_5$ type, even in the case of the
standard D3-brane model \cite{Gmeiner:2007uw}. In
\Rref{Butti:2004pk}, a one parameter family of solutions was found, which interpolates
between the
Klebanov-Strassler (KS) \cite{ks} and Maldacena-Nu\~{n}ez
\cite{Maldacena:2000yy} throats. As the throat geometry moves away from KS, more
non-Gaussianity is produced whereas the tensor-scalar ratio is reduced.
The choice of throat geometry, therefore, could affect the bounds on $r$
and must be considered when models are compared.

The next class of models that can be investigated are the single brane, multi-field
configurations. 
The warped throat is six-dimensional, so it is natural to consider cases where
the D3-brane is not restricted to a radial trajectory. This was investigated in
Refs.~\cite{spinflation} and \cite{Huang:2007hh}. Increasing the degrees of
freedom in this way introduces the possibility of entropy mode production. There are also
changes in the predictions for the amount and type of non-Gaussianity produced
and the constraints on $r$ can be eased \cite{Arroja:2008yy, Langlois:2009ej,
Langlois:2008qf, Langlois:2008wt, Mizuno:2009mv, Mizuno:2009cv, RenauxPetel:2009sj}.
The bounds on $r$ could also be affected if a non-negligible part of the
curvature perturbation was produced by a curvaton field \cite{Lyth:2001nq}.
Curvaton fields arise generically in scenarios containing warped throats,
particularly for propagation near the tip \cite{Li:2008fm, Kobayashi:2009cm}. 

Investigations have also been made into models with multiple branes, each of which
has a single dynamical field. In Refs.~\cite{Cai:2008if} and \cite{Cai:2009hw}, no
interactions between branes were considered, and the branes could conceivably propagate in different
throats. The action for $n$ decoupled branes in the
relativistic limit is the sum of $n$ copies of the DBI action
\eqref{eq:DBIaction-dbiintro}. The power spectrum of curvature perturbations is
enhanced by a factor of $n^{3/2}$ with respect to the single brane case. Consequently, the value
of the tensor-scalar ratio will be reduced.

We have not yet addressed models with multiple branes but only one effective
degree of freedom. Multiple M5-branes in M-theory act with an effective
single degree of freedom, but the Lyth bound is now significantly weakened. Large
field ranges and an observable tensor signal are therefore possible \cite{Krause:2007jr}. 
Another proposal is that of $n$ D3-branes which are coincident and propagating in a
warped throat \cite{thomasward, hltw, Ward:2007gs, Berndsen:2009ww}. The
non-Abelian nature of the interactions between the branes differentiates this model
from other multi-brane models and the model is also known as ``Matrix Inflation''.
In Chapter~\ref{ch:multibrane} we
investigate this model in the relativistic limit
for both large and small $n$, and show how the constraints derived in this
chapter can be applied.

%
%
%
\section{Discussion}
\label{sec:conclusion-dbi}

In this chapter, we have derived an upper limit on
the amplitude of the primordial gravitational wave spectrum
generated during UV DBI inflation. We considered   
the maximal inflaton field variation   
that can occur during the observable stages of inflation and assumed  
only that the brane was propagating inside the throat during that epoch. 
The bound (\ref{eq:upperbound}) is valid for an arbitrary inflaton potential and 
warp factor (modulo some weak caveats) and can be expressed 
entirely in terms of observable parameters, once the volume of 
the five-dimensional sub-manifold of the throat has been specified. 
The inferred upper limit on $r$ is surprisingly strong. 
We find that the standard UV  
scenario predicts tensor perturbations that are undetectably small, 
at a level ${r_*} \lesssim {10^{-7}}$. 

The current WMAP5 data 
favours models that generate a red spectral index, $n_s<1$,
when both the gravitational waves and running in the scalar 
spectral index are negligible. For UV versions of the scenario, 
we have identified a corresponding 
lower limit on $r$ which applies in this region of 
parameter space, $r_* \gtrsim 0.1 (1-n_s)$. It is clear that 
the standard scenario 
cannot satisfy both the upper and lower bounds 
on the tensor modes for the observationally favoured value 
$1-n_s \simeq 0.03$.

The generality of our 
analysis implies that modifying either the inflaton potential 
or the form of the warp factor is unlikely to resolve this discrepancy. 
On the other hand, there are a number of possible ways of reconciling  
theory with observation. In general, 
either the upper or lower limit on $r$ needs to be relaxed. 
Weakening the latter would require a violation of the slow roll 
conditions or a blue spectral index. 
A value of $n_s >1$ is compatible with WMAP5 if the running of the 
spectral index 
is sufficiently negative, but is only marginally
consistent if just the tensor modes are non-negligible.  The 
upper limit on $r$ can be weakened by reducing 
the volume of $X_5$ or 
by generalising the DBI action. Furthermore, it need not necessarily 
apply in IR versions of the scenario, although the BM bound will still hold
in such cases. 
We considered a generalised version of the 
DBI action and identified a necessary condition on the form of such  
an action for the BM bound to be relaxed.

In conclusion therefore, we have shown that primordial gravitational wave constraints 
combined with cosmological observations of the density perturbation
spectrum act as a powerful discriminant of DBI inflationary models. 
They also serve as an important observational guide for identifying viable 
generalisations of the scenario. In Chapter~\ref{ch:multibrane} we will explore
one particular generalisation, the multi-coincident brane scenario
introduced in \Rref{thomasward}.

%
%

\chapter{Multi-Coincident Brane Inflation}
\label{ch:multibrane}

\section{Introduction}
\label{sec:intro-multi}
We have seen in Chapter~\ref{ch:dbi} that the standard DBI inflationary model
is severely constrained by current observational data. The amplitude of tensor
perturbations is bounded from above by $r\le 10^7$. When the brane is moving
towards the tip of the throat, a complementary lower bound
on $r$ can be derived such that the two bounds are incompatible using
current observational data. In this chapter we will explore
how to evade, reconcile and weaken these bounds by considering a more general
class of models that exhibit properties similar to the standard DBI scenario.

In Section \ref{sec:relaxing-multi} we consider the special algebraic
properties of the DBI action.
We identify a general class of non-canonical inflationary models
where the leading-order contribution to the non-Gaussianity of the 
curvature perturbation is determined 
entirely by the speed of sound of the inflaton fluctuations. 
In these models, the bounds on $r$ can be relaxed 
if significant non-Gaussianities are generated.

As reviewed in Section~\ref{sec:others-dbi}, many alternative ways to relax these
bounds have been proposed, including
models based upon multiple fields, the addition of
angular momentum as another degree of freedom and using
different throat geometries. However, in most cases
the extra degrees of freedom introduced in these models do not solve the problem
\cite{Alabidi:2008ej}. The 
bounds are relaxed only by a small fraction, and therefore these models should still be 
regarded as unsatisfactory since they require an extreme amount of fine tuning in order
to satisfy the observational constraints.

One alternative possibility is to consider 
multiple brane configurations\footnote{In certain limits this approach 
is actually dual to considering wrapped branes \cite{Ward:2007gs}.}. 
One scenario in which multiple branes are expected is after brane flux annihilation, in which
branes travelling down the throat annihilate with the trapped flux, creating new branes
\cite{thomasward, DeWolfe:2004qx, Kachru:2002gs}. These are then attracted by other branes and
fluxes in other throats and propagate toward the bulk.
In \Rref{thomasward} Thomas \& Ward argue that it is unlikely that only a single brane is left
after the flux annihilation process, due to the large amount of fine tuning necessary to achieve
this. Instead it is more likely that a number of branes remain.

In the case where  
$n$ branes are localised initially at equal distances, $l > l_s$, and 
subsequently follow the same trajectory, 
the effective theory is equivalent to that of $n$ copies of the
action for a single brane. A more general initial condition, particularly
for branes created in the IR region of the throat
\cite{brane13, DeWolfe:2004qx, Kachru:2002gs}, is that
the branes should be separated over a range of scales, 
with a subset being coincident and the remainder being widely separated.

In Section~\ref{sec:multibranes-multi} we introduce the multi-brane
model with $n$ coincident branes described in \Rref{thomasward}. We will
consider two limiting cases of this model. The large $n$ case is similar in
form to the original DBI model, and we will show in
Section~\ref{sec:twlargen-multi} that it can be constrained using the
formalism derived in Chapter~\ref{ch:dbi}. In contrast, the effective action in the
relativistic limit for a finite number of
branes is shown to be in the
class of actions for which the bounds on $r$ can be relaxed.
In Section~\ref{sec:finiten-multi} we find that
such models 
can in principle lead to a detectable 
gravitational wave background if 
the number of coincident branes is sufficiently small.

%
%
\section{Relaxing the Upper Bounds on the Tensor-Scalar Ratio}
\label{sec:relaxing-multi}
%
In Chapter~\ref{ch:dbi-intro} we described the standard DBI scenario, in
which the kinetic function $P(\vp, X)$ takes the form given in
\eq{eq:Pdefn-dbiintro}:
\begin{equation}
\label{eq:DBIkinetic}
P (\varphi , X) = -T(\varphi ) \sqrt{1-2T^{-1}(\varphi ) X} + T(\varphi ) - V(\varphi ) \,,
\end{equation}
where $T(\varphi ) = T_3 h^4 (\varphi )$ 
is the warped brane tension and  $V(\varphi )$ is the 
inflaton potential. The standard DBI scenario is
algebraically special, 
in the sense that the kinetic function satisfies the constraints 
\begin{equation}
\label{eq:cspx1}
\cs P_{,X} = 1 , \qquad  \Lambda = \frac{1}{2} \left( 
\frac{1}{\cs^2} -1 \right) \,.
\end{equation}
We saw in Section~\ref{sec:relaxing-dbi} that the 
bounds \eqref{eq:genBMbound} and \eqref{eq:LHbound} 
on the tensor-scalar ratio could in principle be 
significantly relaxed in models where $(\cs P_{,X})_* \gg 1$. 
In view of the second relation in \eq{eq:cspx1}, it is of interest 
to begin by taking a phenomenological approach and to consider the more  
general class of models that satisfy the relation
\begin{equation}
\label{eq:defalpha}
\frac{1}{\cs^2} -1 = \alpha \Lambda \,,
\end{equation}
for some positive constant $\alpha$. 

\subsection{Approximate Solution}
\label{sec:approx-multi}
A large non-Gaussian signature in the curvature perturbation is 
typically generated in models where the sound speed of fluctuations 
is small.
By substituting \eq{eq:defalpha} into the definition of $\fnleq$ in
\eq{eq:fnldefn-dbiintro}, we can see that for this case
\begin{equation}
 \fnleq \propto \frac{1}{\cs^2} \propto \Lambda\,.
\end{equation}
Recall that the definitions of $\cs^2$ and $\Lambda$ in \eqs{eq:defcs-dbiintro} and
\eqref{eq:deflambda-dbiintro} are in terms of $P$ and its derivatives with respect
to $X$. We can require that the magnitude of $\fnleq$ is large by considering
scenarios in the limit where $\cs^2$ is small and $\Lambda$ is large, or
equivalently:
\begin{equation}
\label{eq:Plimits}
X^2 P_{,XXX} \gg XP_{,XX} \gg P_{,X} \,.
\end{equation}
Having taken this limit, the constraint \eqref{eq:defalpha} reduces to the 
third-order, non-linear, partial differential equation
\begin{equation}
\label{eq:pde-multi}
P^2_{,XX} = \frac{\alpha}{6} P_{,X} P_{,XXX} \,.
\end{equation}
Changing the dependent variable to $\Upsilon \equiv P_{,XX}/P_{,X}$ 
reduces \eq{eq:pde-multi} to
\begin{equation}
\label{eq:Qdefn-multi}
\alpha \Upsilon_{,X} = (6-\alpha )\Upsilon^2 \,,
\end{equation}
and it is straightforward to integrate \eq{eq:Qdefn-multi} 
exactly. The remaining integrations can also be performed analytically 
and the general solution to \eq{eq:pde-multi} for $\alpha \ne
6$ is given by\footnote{The special case $\alpha =6$ results in an 
exponential dependence of the kinetic function on $X$. However, we 
do not consider this model further, since it does not lead to a 
weakening of the gravitational wave constraints.}
\begin{equation}
\label{eq:gensoln-multi}
P (\varphi , X) = -f_1 (\varphi ) \left[ 1-f_2 (\varphi ) X 
\right]^l -f_3(\varphi ) \,,
\end{equation}
where $f_i (\varphi )$ are arbitrary functions of the scalar 
field and
\begin{equation}
l \equiv \frac{2(\alpha -3 )}{\alpha -6} \, . 
\end{equation}
The inequalities \eqref{eq:Plimits} 
are satisfied in the relativistic limit, where $X \simeq 1/f_2$, justifying their use.
We consider the inflationary dynamics in the relativistic limit in what follows but for
completeness we show in Appendix~\ref{sec:apx-multi} that \eq{eq:defalpha} can be solved
analytically without this approximation.

\subsection{Consequences}
\label{sec:consequences-multi}

The standard DBI scenario is recovered from \eq{eq:gensoln-multi} for
$l=1/2, \alpha = 2$ (or $s=-1$ in the exact case following a redefinition of
$f_i$). 
More generally, however, \eq{eq:gensoln-multi} implies that  
\begin{align}
\label{eq:consequence-multi}
\cs P_{,X} &\simeq \frac{l f_1 f_2}{\sqrt{2(1-l)}} \left( 
1- f_2X \right)^{(2l-1)/2} \,,
\\
\cs^2 &\simeq \frac{1-f_2X}{2(1-l)} \,,
\end{align}
when $X \simeq 1/f_2$. Self-consistency therefore
requires $l<1$. Moreover
we find from \eq{eq:fnldefn-dbiintro} that
\begin{align}
\label{eq:consequence3-multi}
\fnleq &\simeq \frac{-\beta}{1-f_2X}\,, & \beta &\equiv \frac{5(59-55l)}{486}
\,,
\\
\label{eq:consequence4-multi}
\fnleq &\simeq -\frac{\sigma}{\cs^2}\,, & \sigma &\equiv 
\frac{5}{972} \left( \frac{59-55l}{1-l} \right) \,.
\end{align}
Hence substituting Eqs.~\eqref{eq:consequence-multi} and
\eqref{eq:consequence3-multi} 
into the BM bound \eqref{eq:genBMbound} and the bound \eqref{eq:LHbound}
implies that
\begin{equation}
\label{eq:weakBM-multi}
r_*< \frac{32}{N \Neff^2} \frac{l f_1 f_2}{\sqrt{2(1-l)}}
\left( -\frac{\fnleq}{\beta} \right)^{(1-2l)/2}
\end{equation}
and
\begin{equation}
\label{eq:weakLH-multi}
r_* < \frac{10}{(\Delta \N)_*^2} \left( \frac{T_3}{\Vol} \right)^{1/3} 
\left( \frac{h_*}{\Mpl} \right)^{4/3}
\frac{l f_1 f_2}{\sqrt{2(1-l)}}
\left( -\frac{\fnleq}{\beta} \right)^{(1-2l)/2}
\end{equation}
respectively.

We conclude, therefore, that 
the upper limit on the tensor-scalar ratio could be significantly 
relaxed if $l <1/2$, since the non-linearity parameter is at present only 
weakly constrained at $\fnleq >-151$. Although it is possible 
to phenomenologically construct a model which has a value of $l$ in this 
range, it is clearly preferable to identify  UV complete models
that satisfy this requirement within a string theory context. 
Unfortunately this is quite difficult to achieve since the inflaton 
will either be associated with an open or closed string mode. 
The open strings are governed by relativistic actions of the 
DBI form, whilst closed strings arise from compactification of Einstein gravity
and are typically put into a canonical form.
However, there do exist classes of open string
models which satisfy the above requirement, 
namely those associated with multiple coincident branes.

More specifically, if the branes are
spatially separated, the effective action is algebraically equivalent 
to that of a single brane. It will therefore not satisfy the 
bound on $l$\footnote{In this discussion, we are ignoring 
the non-trivial backreaction of these branes on the background, and therefore 
one should be careful about the range of validity of the effective action.}. 
In the remainder of this chapter we will examine the case of $n$ coincident
branes as described in \Rref{thomasward}. The large $n$ limit of this configuration will also
fall into the class of models with $l=1/2$, which is equivalent to the single
brane case. On the other hand, if it is assumed that 
$n$ is finite, the special properties associated with the 
matrix degrees of freedom become important and this 
results in a kinetic function satisfying $l \le 1/2$.

%
%
%
\section{The Multi-Coincident Brane Model} 
\label{sec:multibranes-multi}

We have seen how the form of the kinetic function
$P$ can significantly change the
strength of the bound \eqref{eq:LHbound} on the tensor-scalar ratio, depending
on its explicit
form. One model in which a suitable form for 
$P$ is realised is the multiple coincident
brane model as outlined by Thomas \& Ward in \Rref{thomasward}.
In this model, the flux annihilation process 
generates $n$ coincident branes that are initially located at the 
bottom of a throat region. The dynamics of this configuration
is determined by the non-Abelian world-volume theory \cite{myers1,myers2}. 
This theory exhibits extra stringy degrees of freedom which arise due to the 
fuzzy nature of the geometry.

In general the open string degrees of freedom for $n$ coincident branes 
combine to fill out representations of $U(n)$, as opposed to $U(1)^n$ 
in the case of separated branes. This introduces a non-Abelian 
structure into the theory. In the single brane case, the fluctuations of the 
brane are characterised by induced scalar fields on the world-volume. 
However, for multiple branes
these scalars must be promoted to matrix representations of some gauge group. 
Typically the transverse space of any given compactification will always admit
an $SO(3)$ isometry. Scalars can therefore be chosen to 
transform under representations of the algebra of $SO(3) \sim SU(2)$ by making 
the identifications
\begin{equation}
\varphi^i = R_m \alpha^i \hspace{1cm} i =1, 2, 3 \,,
\end{equation}
where $R_m$ is some scale with canonical mass dimension, and the $\alpha^i$ are
specified to be the irreducible generators satisfying the commutator
\begin{equation}
[\alpha^i, \alpha^j] = 2i \epsilon_{ijk} \alpha^k \,,
\end{equation}
and the conditions
\begin{equation}
\frac{1}{n} Tr(\alpha^i \alpha^j) = \hat{C} \delta^{ij} = (n^2-1) \delta^{ij}
\,,
\end{equation}
where $\hat{C}$ is the quadratic Casimir of the gauge group.
The irreducibility condition corresponds to the configuration being in the
lowest energy state. It is therefore an additional fine-tuning
of the initial conditions.

The Myers prescription requires a symmetrised trace (denoted $STr$) to 
be made over the gauge group. This implies that 
the symmetric averaging must be taken over all the group dependence 
before taking the trace:
\begin{equation}
 \label{eq:str-defn-multi}
 STr (A_1 \ldots A_s) = \frac{1}{s!} Tr(A_1 \ldots A_s + \text{all
permutations})\,.
\end{equation}
For a large number of branes, $n \gg 1$, the symmetric trace can be
approximated with a trace, 
which results in the usual DBI action multiplied by a potential term (as
described in Refs.~\cite{thomasward, Kachru:2002gs}). 
However for finite $n$, the symmetrisation becomes more important and it
is essential to have some means of performing this operation. A prescription for
treating the symmetric trace at finite $n$ was proposed
in Refs.~\cite{Ramgoolam:2004gw} and \cite{McNamara:2005ry}, 
using highest weight methods and chord diagrams. 
The result is that the $STr$ acts on different spin representations of $SU(2)$ 
in the following manner:
\begin{align}
\label{eq:str-even-multi}
STr (\alpha^i \alpha^i)^q &= 2(2q+1)\sum_{i=1}^{n/2}(2i-1)^{2q} , 
&& n\; \mathrm{ even}\,, 
\\
\label{eq:str-odd-multi}
STr (\alpha^i \alpha^i)^q &= 2(2q+1)\sum_{i=1}^{(n-1)/2} (2i)^{2q} , 
&& n\; \mathrm{ odd}\,.
\end{align}

In order for the solution to converge in this prescription, 
it is also necessary to modify the definition of the radius of the 
$SU(2)$ sphere. In the large $n$ limit, this is given by
\begin{equation}
\label{eq:largen-rho-multi}
\rho^2 = \lambda^2 R_m^2 \frac{1}{n} Tr(\alpha^i \alpha^i) = \lambda^2 R_m^2
\hat{C} \,,
\end{equation}
where $\lambda \equiv 2\pi l_s^2 = 2\pi \ms^{-2}$, 
whereas for finite $n$, it becomes
\begin{equation}
\label{eq:finiten-rho-multi}
\rho^2 = \lambda^2 R_m^2 \mathrm{Lim}_{q \to \infty} \left(\frac{STr (\alpha^i
\alpha^i)^{q+1}}{STr(\alpha^i \alpha^i)^q} \right) 
= \lambda^2 R_m^2 (n-1)^2 \,.
\end{equation}
This converges to the large $n$ result in the appropriate limit.
This point is important, since the warp factor 
of the four-dimensional theory is typically of the form $h= h(\rho)$.
The next two sections will examine this coincident brane model in both the
large and finite $n$ limits. 

\section{Coincident Brane Inflation with a Large Number of Branes}
\label{sec:twlargen-multi}
Taking the limit of a large number of coincident branes significantly
simplifies the non-Abelian action. The symmetrised trace can
now be replaced with a normal trace operator and the expression for $\rho$
takes the form in \eq{eq:largen-rho-multi}. 
 For the case where a fuzzy two-sphere is 
embedded in a three-cycle in the $X_5$ manifold, 
the kinetic structure of the action is given in the large $n$ limit by 
\cite{thomasward}
\begin{equation}
\label{eq:largeP-multi}
P=-nT_3 \left[ h^4(\varphi ) W(\varphi ) 
\sqrt{1-2 T_3^{-1} h^{-4}(\varphi) X}
-h^4(\varphi ) + V (\varphi ) \right] \,,
\end{equation}
where
\begin{equation} 
\label{eq:defW}
W (\varphi ) \equiv \sqrt{1+ C^{-1}h^{-4}(\varphi ) \varphi^4}
\end{equation}
defines the so-called `fuzzy' potential, 
$C = \pi^2 \hat{C}T_3^2/\ms^4$ is a model-dependent constant and 
$\hat{C} \simeq n^2$ in the large $n$ limit. 

The kinetic term \eqref{eq:largeP-multi} is clearly of the same form as the
single-brane DBI term, with $l=1/2$ in the scheme outlined in
Section~\ref{sec:relaxing-multi}. We can therefore apply the analysis of
Section~\ref{sec:relaxing-dbi} to investigate whether the bounds described in
Chapter~\ref{ch:dbi} can be relaxed.
Comparison with \eq{eq:genaction-dbi},
\begin{equation}
\label{eq:genaction-multi}
P= -f_A (\varphi ) \sqrt{1-f_B (\varphi ) X} -f_C (\varphi) \,,
\end{equation}
implies that $f_Af_B =2nW$ and $f_B=2/(T_3h^4)$. Hence, 
the new features of this model relative to the standard single-brane 
scenario are parametrised in terms of the fuzzy potential $W (\varphi )$. 
This configuration is conjectured to be dual to 
a ${\rm D5}$-brane which is wrapped around a two-cycle 
of the throat \cite{dual1,dual2,dual3}. 

\subsection{Bound on \texorpdfstring{$n$}{n} during IR Propagation}
\label{sec:ir-largen-bound-multi}

The regime $W \gg 1$ is of interest for 
relaxing the gravitational wave constraints\footnote{Note that 
the case $n \gg 1$ and
$W \sim 1$ will not significantly relax the BM bound, 
since we require $n \ll N$ for backreaction effects to be negligible.}. 
The generalised BM bound for IR models, with branes propagating towards the bulk, is given by 
\eq{eq:f2IRlower}:
\begin{equation}
\label{eq:f2IRlower-multi}
\frac{f_B\Mpl^4}{N} > \frac{(\Delta \N_*)^2}{4\pi^2}
\frac{\sqrt{-3\fnleq}}{(1-n_s)\Pr}  \,.
\end{equation}
As we know $f_B$, this may be expressed as 
a limit on the value of the warp factor $h(\varphi_*)$ on CMB scales: 
\begin{equation}
\label{eq:warptorelax}
N T_3 \left(\frac{h_*}{\Mpl}\right)^4 < 
\frac{8\pi^2 (1-n_s)\Pr}{\sqrt{-3\fnleq}(\Delta \N_*)^2} \,.
\end{equation}

We now consider whether this limit can be satisfied for reasonable choices 
of parameters when the warped compactification corresponds to 
an $AdS_5$ or KS throat, respectively. Recall that the warp 
factor for the $AdS_5$ throat is given by $h=\varphi/(\sqrt{T_3}L)$.  
Condition (\ref{eq:warptorelax}) therefore reduces to a constraint on the 
value of the inflaton during observable inflation: 
\begin{equation}
\label{eq:phivalue-dbi}
\frac{\varphi_*^4}{\Mpl^4} < 
\frac{8\pi^2 (1-n_s)\Pr}{\sqrt{-3\fnleq} ( \Delta \N_*)^2} 
\frac{T_3 L^4}{N} \,.
\end{equation}
However, non-perturbative string effects are expected to become 
important below a cutoff scale, $\varphi_{\rm cut} = 
h_{\rm cut} \sqrt{T_3} L$, where $h_{\rm cut}$ is the value of the 
warp factor at that scale. For consistency, therefore, one requires 
$\varphi_*>\varphi_{\rm cut}$, so that 
\begin{equation}
 N T_3 \left(\frac{h_\mathrm{cut} }{\Mpl}\right)^4 < 
\frac{8\pi^2 (1-n_s)\Pr}{\sqrt{-3\fnleq} ( \Delta \N_*)^2} \,,
\end{equation}
which implies an upper limit on the 
${\rm D3}$-brane charge: 
\begin{equation}
\label{eq:Nlimit-dbi}
N< \frac{64\pi^5 \gs (1-n_s)\Pr}{\sqrt{-3\fnleq} ( \Delta \N_*)^2}
\left( \frac{\Mpl}{h_{\rm cut} \ms} \right)^4   \,.
\end{equation}
Assuming the typical values $\ms \sim 10^{-2}\Mpl$, 
$\Delta \N_* \simeq 4$ and 
$h_{\rm cut} \sim 10^{-2}$ implies
\begin{equation}
N < 1.76 \times 10^8 (1-n_s)(-\fnleq)^{-1/2} \,, 
\end{equation}
and for $1-n_s <0.05$ and $-\fnleq>5$ the inequality becomes
\begin{equation}
 N < 4\times 10^6\,.
\end{equation}

For an $AdS_5$ throat, the fuzzy potential $W$ is a constant,
and the condition that $W \gg 1$ becomes 
\begin{equation}
\label{eq:Chatlimit}
\hat{C} \ll \frac{4\pi^2\gs N}{\mathrm{Vol}(X_5)} \,.
\end{equation}
Hence, combining inequalities 
(\ref{eq:Nlimit-dbi}) and (\ref{eq:Chatlimit}) implies that
\begin{equation}
\label{eq:nlimit-dbi}
\hat{C} \ll 
\frac{2 (2\pi)^7 (1-n_s)\Pr}{\sqrt{-3\fnleq} ( \Delta \N_*)^2}
\frac{\gs^2}{\mathrm{Vol} (X_5) }
\left( \frac{\Mpl}{h_{\rm cut}\ms} \right)^4  \,,
\end{equation}
and specifying $\gs \sim 10^{-2}$ and 
$\mathrm{Vol}(X_5) \simeq \pi^3$ then yields the limit  
\begin{equation}
\hat{C} \ll 2.25\times 10^{6} (1-n_s)(-\fnleq)^{-1/2} < 5\times 10^4,
\end{equation}
 or equivalently,
\begin{equation}
\label{eq:nbound-multi}
n \ll 225.
\end{equation}
In deriving the action \eqref{eq:largeP-multi} the number of coincident branes was assumed to be
large. However we have now found that for the case of branes propagating towards the bulk, the
number of such branes is bounded from above.
Furthermore, since $f_Af_B \simeq {\rm constant}$ for the $AdS_5$ throat, 
the stronger form of the inequality (\ref{eq:genBMbound}) may be used. The right hand side of
inequality \eqref{eq:nlimit-dbi} would be reduced by a factor of $(   \Neff /\Delta \N_*)^2$ by
substituting 
$\Delta \N_* \rightarrow \Neff$. This ratio 
could be as high as $(60/4)^2 \simeq 200 $, leading to $n$ being less than $15$. In this case
the assumption of large $n$ would clearly be inconsistent and the model would be ruled out. 

\subsection{Bound on D3 Charge at the Tip of the Throat}

Since the branes are initially located at the tip of the 
throat, another case of interest is the IR limit of the KS geometry, where 
the warp factor asymptotes to a constant value 
\cite{gkp}:
\begin{equation}
\label{eq:KStip-dbi}
h_{\rm tip} = \exp \left( - \frac{2\pi K}{3M\gs} \right) \,.
\end{equation}
In this case, the generalised BM bound (\ref{eq:warptorelax}) becomes
\begin{equation}
\label{eq:loglimit-dbi}
\frac{8\pi K}{3M\gs} -\ln N > 4 \ln \left( \frac{\ms}{\gs^{1/4}\Mpl} \right)
-\ln \left( 
\frac{64\pi^5 (1-n_s)\Pr}{\sqrt{-3\fnleq} ( \Delta \N_*)^2}
\right) \,.
\end{equation}
The radius of the three-sphere at the tip of the KS 
throat is of the order $(\gs M)^{1/2}$ in string units 
and this must be large (and at the very least should exceed
unity) for the supergravity approximation to be reliable. 
Substituting this requirement into expression (\ref{eq:loglimit-dbi}) 
results in a necessary (but not sufficient) condition 
on the ${\rm D3}$-brane charge for the 
generalised BM bound to be satisfied:
\begin{equation}
\label{eq:Nlowerlimit-dbi}
\frac{1}{N} \exp \left( \frac{8\pi \gs N}{3}  \right)
> \frac{\sqrt{-3\fnleq} ( \Delta \N_*)^2}{64\pi^5 
(1-n_s)\Pr} \frac{1}{\gs} \left( \frac{\ms}{\Mpl} \right)^4 \,.
\end{equation}

Recall that a necessary condition for the backreaction 
of the branes to be negligible is 
$N \gg n$ and this implies that the 
exponential term in (\ref{eq:Nlowerlimit-dbi}) will dominate unless 
the string coupling constant is extremely small. Hence, for 
the parameter estimations quoted above, we 
deduce the lower limit 
\begin{equation}
\label{eq:Nconstraint-dbi}
N- 12 \ln N > -6.8 +12 \ln \left( \frac{\sqrt{-\fnleq}}{1-n_s} \right) \,,
\end{equation}
which becomes $N \gtrsim 10^2$ for $1-n_s \simeq 0.05$ and $-\fnleq>5$.

In general, however, the $K$ and $M$ units of flux are not independent. 
F-theory compactification on Calabi-Yau four-folds
provides a geometric way of parametrising  
type IIB string compactifications
\cite{witten1,witten2,witten3,sethi,gkp,klemm}. 
Global tadpole cancellation constrains the topology of the four-fold
and this restricts the brane and flux configurations.  
When the KS system is embedded into F-theory, the  
constraint is given by \cite{gkp}
\begin{equation}
\label{eq:Ftheory}
\frac{\chi}{24} = n + MK \,,
\end{equation}
where $\chi$ is the Euler characteristic of the four-fold.  
Hence, $N = MK < \chi /24$ and together with condition 
(\ref{eq:Nconstraint-dbi}), this implies that
\begin{equation}
\label{eq:chilimit}
\chi > 2400 \,,
\end{equation}
for $N > 10^2$.
It is known that the Euler number for four-folds 
corresponding to hypersurfaces in weighted projective spaces
can be as high as $\chi \le 1,820,448$ \cite{klemm},
so there are many compactifications that could 
in principle satisfy the generalised BM bound.
On the other hand, the above limit on the Euler characteristic 
does allow us to gain some insight into the 
topology of the extra dimensions, since compactifications 
which result in a small Euler characteristic would be  
incompatible with the generalised BM bound. 

%
%
\section{Coincident Brane Inflation with a Finite Number of Branes}
\label{sec:finiten-multi}

\subsection{\texorpdfstring{The Finite $n$ Model}{The Finite n Model}}
\label{sec:finitensub-multi}
The coincident brane model outlined in Section~\ref{sec:multibranes-multi}
takes a significantly different form if, instead of assuming a large number of
coincident branes, there are now only a small finite number. The prescription for
the symmetrised trace given in Eqs.~\eqref{eq:str-even-multi} and
\eqref{eq:str-odd-multi} must be used, where $\rho$ is determined from
\eq{eq:finiten-rho-multi}. 
The resulting kinetic function for $n$ coincident branes in 
the finite $n$ limit is therefore given by
\begin{equation}
\label{eq:genP-multi}
P = -T_3 STr \left(h^4(\rho) \sum_{k,p=0}^{\infty}(-\XR)^k Y^p (\alpha^i
\alpha^i)^{k+p}\binom{1/2}{k} \binom{1/2}{p} + V(\rho)-h^4(\rho) \right) \,,
\end{equation}
where 
\begin{equation}
Z \equiv \lambda^2 h^{-4}(\rho), \hspace{0.5cm} Y \equiv 4\lambda^2 R_m^4
h^{-4}(\rho),
\hspace{0.5cm} \binom{1/2}{q}
\equiv \frac{\Gamma(3/2)}{\Gamma(3/2-q)\Gamma(1+q)} \,.
\end{equation}
Note that the second and third terms in \eq{eq:genP-multi} 
are singlets under the $STr$ and therefore contribute terms proportional 
to $n$. The physics of these branes away from the large $n$ limit is particularly interesting as
discussed further
in Refs.~\cite{thomasward} and \cite{Ward:2007gs}.

It was shown in \Rref{hltw} that the functional forms of the kinetic function, $P$, and
corresponding energy density, $E$, for all the solutions with $n>2$
can be derived recursively from the $n=2$ solution. We will use the notation
$P_n$ and $E_n$ to denote the non-singlet sector of the kinetic function and energy density for
the $n$-brane solutions. The full pressure and energy densities are then 
given by $P = P_n - nT_3(V-h^4)$ and $E = E_n + nT_3 (V-h^4)$, respectively. Using
\eq{eq:str-even-multi} and the expressions
\begin{align}
 \sum_{k=0}^\infty A^k \binom{1/2}{k} &= \sqrt{1+A}\,,\\
 \sum_{k=0}^\infty A^k \binom{1/2}{k} 4k &= \frac{2A}{\sqrt{1+A}}\,,
\end{align}
then implies that the terms $P_2$ and $E_2$ can be derived:
\begin{align}
\label{eq:2brane}
P_2 \left[Z,Y\right] &= - 2 T_3 h^4 \left(\frac{(1+2Y -
(2+3Y)\XR)}{\sqrt{1+Y}\sqrt{1-\XR}}\right) \,, \nonumber \\
E_2 \left[Z,Y\right] &= 2 T_3 h^4 \left(\frac{(1+2Y -
Y\XR)}{\sqrt{1+Y}(1-\XR)^{3/2}}\right) \,.
\end{align}
The recursion relation described in \Rref{hltw} can then be written for odd $n$ as
\begin{align}
\label{eq:oddbrane}
P_n^{(O)} &= \left(\sum_{k=1}^{(n-1)/2} P_2 \left[(2k)^2Z, (2k)^2Y\right]
\right)-nT_3(V-h^4) \,, \nonumber \\
E_n^{(O)} &= \left(\sum_{k=1}^{(n-1)/2} E_2 \left[(2k)^2Z, (2k)^2Y \right]
\right)+
nT_3(V-h^4) \,,
\end{align}
and for even $n$ as
\begin{align}
\label{eq:evenbrane}
P_n^{(E)} &= \left(\sum_{k=1}^{n/2} P_2 \left[(2k-1)^2Z, (2k-1)^2Y\right]
\right)-nT_3(V-h^4) \,, \nonumber \\
E_n^{(E)} &= \left(\sum_{k=1}^{n/2} E_2 \left[(2k-1)^2Z, (2k-1)^2Y \right]
\right)+
nT_3(V-h^4) \,.
\end{align}
If we let $\delta_n$ = 1 when $n$ is even and $\delta_n=0$ when $n$ is odd, we can
combine these two expressions \cite{Berndsen:2009ww}:
\begin{align}
\label{eq:combinedPn-multi}
 P_n &= \left(\sum_{k=1}^{(n-1+\delta_n)/2} P_2 \left[(2k-\delta_n)^2Z,
(2k-\delta_n)^2Y\right]
\right)-nT_3(V-h^4) \,, \nonumber \\
E_n &= \left(\sum_{k=1}^{(n-1+\delta_n)/2} E_2 \left[(2k-\delta_n)^2Z,
(2k-\delta_n)^2Y \right] \right)+ nT_3(V-h^4) \,.
\end{align}
In \Rref{Berndsen:2009ww} it was shown that this recursive definition for
$P_n$ reproduces \eq{eq:largeP-multi} in the large $n$ limit.

The backreaction of the multiple branes introduces corrections of the form $n/N$
\cite{hltw}. Ensuring that this ratio is small allows the continued use of the
supergravity analysis. As we will see in the next section it will not be
difficult to constrain this model when $N\gg n$.

%
%
\subsection{Bounds on the Tensor-Scalar Ratio for Finite \texorpdfstring{$n$}{n}} 
\label{sec:multibounds-multi}
In the last section we introduced the multi-coincident brane model in the limit
of a finite number of branes. In this section we will consider this model in
the context of the class of actions derived in Section~\ref{sec:relaxing-multi},
and show that current observational data can strongly constrain the ability of
this model to produce an observable tensor signal.

The last term appearing in the summation of $P_n$ in \eq{eq:combinedPn-multi}
is 
\begin{equation}
P_n^{\mathrm{last}} = P_2\left[(n-1)^2Z, (n-1)^2Y\right]\,.
\end{equation} 
This 
means that for all $n$, this term can be expressed in the form 
\begin{equation}
\label{eq:unifiedgenP-multi}
P_n^\mathrm{last} = -2T_3 \left\{ \frac{h^4 \left[ 1+2(n-1)^2Y
- [ 2+3(n-1)^2Y] (n-1)^2\XR  \right]}{\sqrt{1+(n-1)^2Y}
\sqrt{1-(n-1)^2\XR}} 
 \right\}.
\end{equation}

Inspection of Eqs.~\eqref{eq:2brane}--\eqref{eq:combinedPn-multi} implies that 
the relativistic limit is realised for any finite number of branes when 
$(n-1)^2 \XR \rightarrow 1$. In this case, the dominant contribution 
to the summations appearing in \eq{eq:combinedPn-multi}
will arise from the last term, \eq{eq:unifiedgenP-multi}. In the relativistic limit, 
therefore, the kinetic function appearing in the effective action simplifies to 
\begin{equation}
\label{eq:unifiedP-multi}
P = 2T_3 \left\{ h^4 \sqrt{1+(n-1)^2Y} 
\left( 1- \frac{2X}{T_3h^4} \right)^{-1/2} 
 \right\} - n T_3 \left(V - h^4 \right) \,,
\end{equation}
where 
\begin{equation}
\label{eq:defY-multi}
Y \equiv \frac{4}{(n-1)^4 \lambda^2 T_3^2} \left( \frac{\varphi}{h} \right)^4 \,,
\end{equation}
\begin{equation}
\label{eq:defXR2-multi}
\XR \equiv \frac{2}{(n-1)^2 h^4 T_3}X \,,
\end{equation} 
and we have effectively imposed the relativistic condition 
\begin{equation}
\label{eq:rellimit-multi}
X \simeq \frac{1}{2} T_3 h^4 \,,
\end{equation}
in the numerator of \eq{eq:unifiedgenP-multi}.  
For the $n=2$ and $n=3$ cases, we have verified by direct 
calculation that when one calculates the speed of sound 
(\ref{eq:defcs-dbiintro}) and the non-linearity parameter 
(\ref{eq:fnldefn-dbiintro}) from the general expressions (\ref{eq:2brane}) and (\ref{eq:oddbrane}) 
and then imposes the relativistic
limit (\ref{eq:rellimit-multi}), one arrives at the identical result 
by starting explicitly with \eq{eq:unifiedP-multi}.

At this point we should consider the validity of 
the function in \eq{eq:unifiedP-multi}. 
The recursive relation for $P$ in \eqref{eq:combinedPn-multi} converges to the
expression in \eq{eq:largeP-multi} in the limit of large $n$
\cite{Berndsen:2009ww}. 
There must exist some 
value of $n$, beyond which $P$ appears to resemble \eq{eq:largeP-multi},
rather than the 
approximate form proposed in \eq{eq:unifiedP-multi}. 
For a range of background 
solutions, numerical calculations suggest that the approximation is valid
when $n$ is less than $\mathcal{O}(10)$ \cite{hltw}. 

As there are a large
number of 
parameters in the theory, it is possible to find solutions 
where $n \gg 10$. However, a larger background flux would then be necessary, which
would result 
in a situation where the conformal 
Calabi-Yau condition is no longer valid. In view of this, we focus on
the sector of the theory where $n \le 10$, which implies that the 
backreaction is under control and that the kinetic function
is still of the required form.

\eq{eq:unifiedP-multi} is precisely of the form given by the 
general solution (\ref{eq:gensoln-multi}), where $l=-1/2$ 
and\footnote{This is the
case $\alpha =18/5$ or $s=-1/3$ in the analytic solution
\eqref{eq:thirdint-multi} which after redefinition of the $f_i (\varphi)$
becomes:

\begin{equation}
P = \frac{-f_1\left[8 - 4f_2X^{1/3}
-\left(f_2X^{1/3}\right)^2\right]}{\sqrt{1-f_2X^{1/3}}} -f_3 \, .
\end{equation}
This expression appears in a slightly different 
form to that in (\ref{eq:unifiedgenP-multi}). 
However in deriving (\ref{eq:unifiedgenP-multi}) we assumed the
relativistic limit, which in turn imposes a non-trivial 
relation between $X$ and $\varphi$. Using this, and with a 
suitable redefinition of the functions, we can
transform the above expression into the required form.} 
\begin{equation}
\label{eq:fdefns-multi}
f_1 (\varphi) = -2T_3 h^4 \sqrt{1+(n-1)^2Y} , \qquad 
f_2 (\varphi) = \frac{2}{T_3 h^4} \,.
\end{equation}
We may, therefore, immediately conclude from \eq{eq:consequence4-multi} that
$\fnleq
\simeq -0.3/\cs^2$. Moreover, since $\beta \simeq 0.9$ in this scenario, 
Eqs.~\eqref{eq:consequence-multi} and \eqref{eq:consequence3-multi} reduce to  
\begin{equation}
\label{eq:csPXlimit}
\cs P_{,X} \simeq -1.3 \sqrt{1+(n-1)^2Y} \fnleq \,.
\end{equation}

We first consider the bound in \eq{eq:LHbound}. This applies at least for all
UV scenarios. It follows after substitution of the relativistic limit
(\ref{eq:rellimit-multi}) into the scalar perturbation amplitude,
\eq{eq:Ps-dbiintro},
that 
\begin{equation}
\label{eq:usefulPs-multi}
\Pr \simeq -\frac{1}{50} \frac{H^4}{T_3 h^4\sqrt{1+(n-1)^2Y}}
\frac{1}{\fnleq}\,.
\end{equation}
Substituting the tensor-scalar ratio into  
\eq{eq:usefulPs-multi} then results in a constraint on the magnitude of 
the warp factor during observable inflation:
\begin{equation}
\label{eq:warpfactor-multi}
\frac{h^4_*}{\Mpl^4} \simeq \frac{-1}{2 T_3 \sqrt{1+(n-1)^2Y}} 
\frac{r^2 \Pr}{\fnleq} \,.
\end{equation}
Eqs.~\eqref{eq:csPXlimit} and \eqref{eq:warpfactor-multi} may now be
substituted into 
the bound \eqref{eq:LHbound} to yield
\begin{equation}
\label{eq:actualLH-multi}
r_* < \frac{1100}{(\Delta \N )_*^6} 
\frac{[1+(n-1)^2Y]}{\Vol} \Pr (\fnleq)^2 \,.
\end{equation}

It is clear that the parameter $Y$ 
must be sufficiently large if the tensor perturbations 
are to be non-negligible. For the $AdS_5 \times X_5$ throat, this parameter  
takes the constant value    
\begin{equation}
\label{eq:YAds-multi}
\YAdS \equiv \frac{4\pi^2 \gs N}{(n-1)^4 \Vol} \,.
\end{equation}
As before, we choose natural field-theoretic values for the volume, 
$\Vol \simeq \pi^3$, and the string coupling, 
$\gs \simeq 10^{-2}$, and further assume that 
$(n-1)^2 Y \gg 1$. 
We again assume that the tensor-scalar ratio does not change significantly
over the entire range of scales that are accessible to cosmological
observation, which corresponds to $\Delta \N_* \simeq 4$. 
After substitution of the above values, therefore, 
the bound (\ref{eq:actualLH-multi}) simplifies to
\begin{equation}
\label{eq:AdSupper-multi}
r_* < 2.8 \times 10^{-13} \frac{N}{(n-1)^2} (\fnleq)^2 \,.
\end{equation}

As in Section~\ref{sec:twlargen-multi},
global tadpole cancellation constrains the magnitude of
the background charge $N$ such that $N < \chi /
24$. 
The maximal known value of the Euler number implies the upper limit of 
\begin{equation}
\label{eq:Nlimit-multi} 
N < 75852
\end{equation}
for known solutions, although in principle higher values are possible. 
Imposing the WMAP5 bound $\fnleq>-151$ in \eq{eq:AdSupper-multi}
and noting that $n \ge 2$ for consistency then implies an absolute
upper limit 
on the tensor-scalar ratio:
\begin{equation}
\label{eq:rupper-multi}
r_* < 5 \times 10^{-4} \, .
\end{equation}

This limit is below the sensitivity of the Planck satellite 
$(r \gtrsim 0.02 )$ \cite{planck}. On the other hand, 
the projected sensitivity of future CMB polarisation experiments 
indicates that a background of primordial 
gravitational waves with $r_* \gtrsim 10^{-4}$ 
should be observable \cite{songknox,vpj, Baumann:2008aq}. In view of this, 
it is interesting to consider whether
a detectable gravitational wave background could in principle 
be generated in this class of multi-brane inflationary 
models. We find from \eq{eq:AdSupper-multi} that this would require 
\begin{equation}
\label{eq:nlimit-multi}
n < 1 -5.3 \times 10^{-5} \sqrt{N} \fnleq < 1-0.014 \fnleq \,,
\end{equation}
where the theoretic limit \eq{eq:Nlimit-multi} for 
known compactifications has been imposed in the 
second inequality. We may deduce, therefore, that  
since we require $n \ge 2$ for consistency, a detectable tensor 
signal will require $-\fnleq >70$. This implies that an observation of 
the tensors should also be 
accompanied by a sufficiently large --- and detectable --- non-Gaussianity. 
In other words, this class of models could  
be ruled out if tensors are observed in the absence of any
non-Gaussianity. On the other hand, the current 
limit of  $\fnleq >-151$ implies that $n \le 3$ is required 
for the tensors to be observable. 
Consequently, if tensor perturbations are detected, this would rule 
out all models with $n \ge  4$ or, alternatively, would require presently 
unknown configurations with $N$ exceeding the bound (\ref{eq:Nlimit-multi}).

In the above analysis we assumed that the string coupling 
took the value $\gs \simeq 10^{-2}$. For the $AdS_5 \times X_5$ throat, 
the bound (\ref{eq:actualLH-multi}) depends proportionally on $\gs$ and can 
therefore be weakened by allowing for larger values of the string coupling. 
For example, increasing this parameter by a factor of $4$ 
to $\gs \simeq 0.04$ (so that it is still in the perturbative regime)
relaxes the limit on the number of branes for the tensors to be detectable to 
$n \le 5$. Similarly, considering a smaller value for the 
volume of the Einstein manifold $X_5$ will also weaken the upper limit.

Let us re-iterate that this limit on $n$ is well within the 
regime of validity for the theory, which we have argued is 
self-consistent for $n<10$. Moreover, since the constraint (\ref{eq:nlimit-multi})
arises using the absolute maximal bound on the known 
Euler characteristics, it suggests that in realistic scenarios $n$ will 
always be much smaller than this. Indeed, one could argue that 
only the $n=2$ and $n=3$ theories are
likely to be valid over a large distribution of the flux landscape.

We must also ensure that our approximation $(n-1)^2Y \gg 1$ 
is valid for consistency.  
For the parameter values we have chosen this requires that 
$\gs N \gg  (n-1)^2$ 
and this is satisfied if the condition \eqref{eq:nlimit-multi} 
holds. Note also that we require $N \gg n$ for the supergravity 
approximation to be under control and for backreaction effects to 
be negligible. This is also satisfied when \eq{eq:nlimit-multi}  
holds.

Finally, it should be emphasised that the derivation of the bound in
\eq{eq:LHbound} 
underestimates the Planck mass by assuming that 
the volume of the throat is much smaller 
than the volume of the compactified Calabi-Yau 
three-fold. It is likely, therefore, 
that the actual constraint on $r$ would be much stronger. Consequently, 
although the bound \eqref{eq:nlimit-multi}  
does marginally allow for detectable tensors if $n$ is sufficiently 
small, in practice this constraint would be further tightened by a more 
complete calculation. Nonetheless, our analysis does not necessarily 
rule out these models as viable candidates for inflation. Rather, it  
suggests that it will be difficult to construct a working model 
that results in a detectable tensor signal.

%
%
%
\section{Discussion}
\label{sec:disc-multi}

The relativistic DBI brane scenario represents an attractive, 
string-inspired realisation of the inflationary scenario. In
Chapter~\ref{ch:dbi} we showed that
cosmological data has placed very strong constraints on the simplest 
UV models based on a single ${\rm D3}$-brane. The strength 
of these constraints follows from field-theoretic upper limits 
on the tensor-scalar ratio, $r$, which in turn arise because 
the effective DBI action satisfies special  
algebraic properties. This provides motivation 
for considering generalisations of the scenario, in particular to 
multi-brane configurations.

In this chapter we have identified a phenomenological class of 
effective actions for which the constraints 
on $r$ are relaxed, if significant (and detectable) 
non-Gaussian curvature perturbations are generated during inflation. 
We have provided approximate and exact derivations of this class of models which coincide
in the relativistic limit. It would be interesting to 
investigate whether the effective action (\ref{eq:gensoln-multi}) with values
of $l \ne - 1/2$ arises in string-inspired settings or elsewhere.

In Section~\ref{sec:multibranes-multi} we introduced the coincident $n$-brane
model of Thomas \& Ward \cite{thomasward}. We examined the predictions of this
model in two limits, arbitrarily large $n$ and small finite $n$. The large $n$
model
is similar to the single brane case. Using the results of
Section~\ref{sec:relaxing-dbi}, we showed that it is strongly
constrained by current observations.

The finite $n$ model is of more theoretical interest as it exhibits the
non-Abelian nature of the scenario. In \Rref{hltw} a recursive approach was
derived to calculate the pressure and energy densities for $n>2$ models using
the $n=2$ results. In the relativistic limit, these finite $n$ models are
included in the class of actions derived in Section~\ref{sec:relaxing-multi}
which relax the bounds on $r$. The backreaction of these models can also be
kept well under control.

We proceeded to consider the question of whether the upper limits on 
$r$ could be relaxed to such an extent 
that a background of primordial gravitational waves 
might be detectable in future CMB experiments. The vast majority of 
string-inspired inflationary models that have been proposed to date 
generate an unobservable tensor background. We 
found that a detectable signal is possible, in principle, 
for typical string-theoretic parameter values 
if the number of coincident branes is either $2$~or~$3$. 
This is consistent with known F-theory configurations and 
current WMAP5 limits on the non-Gaussianity. Furthermore, 
we found that the level of non-Gaussianity must satisfy $-\fnleq 
\gtrsim 70$ if such configurations are to generate a detectable tensor 
signal. This is well within the projected sensitivity 
of the Planck satellite \cite{planck}.

Our analysis invoked an $AdS_5 \times X_5$ warped throat geometry. However, we 
made no assumptions regarding the form of the inflaton potential, other 
than imposing the implicit requirement that the universe underwent a phase of 
quasi-exponential expansion. In this sense, therefore, we have 
yet to explicitly establish that these inflationary models will 
be able to generate a measurable tensor signal. 
Nonetheless, since such a detection would provide a unique observational 
window into high energy physics, our results 
provide strong motivation for considering the cosmological consequences 
of these multi-brane configurations when specific choices for the 
inflaton potential are made. In particular, it would be interesting 
to employ the techniques developed in
to identify the regions of parameter space that are consistent 
with current cosmological observations.  

In Part~\ref{part:dbi} of this thesis we have concentrated on using analytic
techniques to constrain string-inspired inflationary models. 
In Part~\ref{part:numerical},
numerical techniques will be developed with the goal of constraining
inflationary models using second-order perturbation theory.


\part{Numerical Simulations of the Evolution of Second Order Perturbations}
\label{part:numerical}
%

\chapter{Cosmological Perturbations}
\label{ch:perts}

\section{Introduction}
\label{sec:intro-numerical}

Cosmological perturbation theory is an essential tool for the analysis
of cosmological models. It will become more so as the quantity and quality
of observational
data continues to improve. With the recent launch of the
Planck satellite, the WMAP mission reaching its eighth
year, and a host of other new experiments, we will have access to more
information about the early universe than ever before
\cite{planck,Komatsu:2008hk}.

To distinguish between theoretical models, 
it is necessary to go beyond the standard statistical analyses that
have been so successful in recent years. As a result, much interest
has been focused on non-Gaussianity as a new tool to classify and
test models of the early universe. Perturbation theory beyond first
order will be required to make the best possible use of 
the data. In Chapter~\ref{ch:introduction} cosmological perturbations at first order
were introduced.  In Part~\ref{part:numerical} of this thesis, we outline an
important
step in the understanding of perturbation theory beyond first order, demonstrating
that second order perturbations are readily amenable to numerical
calculation, even on small and intermediate scales inside the horizon.

Inflationary model building has for the past few years focused on
meeting the requirements of first order perturbation theory, namely
that the power spectra of scalar and tensor perturbations, as defined in
Eqs.~\eqref{eq:Prdefn-intro} and \eqref{eq:Ptdefn-intro}, should match
those observed in the CMB.  Inflationary
models are classified and tested according to their predictions for the scalar power
spectrum, scalar spectral index and tensor-scalar ratio. 
An important observable that arises at second order is the non-Gaussianity
parameter $\fnl$. As described in Section~\ref{sec:fnl-intro}, this parameter is
not yet well constrained by observational data in comparison with the
other quantities. In Part~\ref{part:dbi}, however, it was shown that $\fnl$ can
already be used to rule out models with particularly strong non-Gaussian signatures.

There are two main approaches to studying 
non-Gaussianity and higher order effects.  
The first uses non-linear theory and a gradient expansion in various
forms, either explicitly, \eg
Refs.~\cite{Salopek:1990jq,Rigopoulos:2005xx}, or through the
$\Delta \N$ formalism, \eg
Refs.~\cite{Starobinsky:1982ee,
Starobinsky:1986fxa, Sasaki:1995aw, Sasaki:1998ug,
Lyth:2004gb,Lyth:2005fi,Langlois:2006vv}.

However, a gradient expansion approach is restricted and can only be applied on
scales much larger than the particle horizon.  
The
second approach uses cosmological perturbation theory developed by Bardeen
\cite{Bardeen:1980kt} and extends it to second order,
\eg~Refs.~\cite{Tomita:1967,Mukhanov:1996ak,Bruni:1996im,
  Acquaviva:2002ud,Nakamura:2003wk,Noh:2004bc,
  Bernardeau:2002jy,Maldacena:2002vr,
  Finelli:2003bp,Bartolo:2004if,Enqvist:2004bk,Lyth:2005du,Seery:2005gb,
  Malik:2003mv, Barnaby:2006cq}\footnotemark.
\footnotetext{For an extensive list of references and a recent review on these
issues see Ref.~\cite{Malik:2008im}.}
This approach works on all scales, but can be more complex in comparison to the
$\Delta \N$ formalism. The two methods lead to identical results on large scales
\cite{Malik:2005cy}. We
will follow the Bardeen approach here.

In Section~\ref{sec:fnl-intro} the first order perturbations of the inflaton field
were taken to be purely Gaussian. It is therefore 
necessary to go to second order if we are to understand and estimate
the non-Gaussian contribution of any inflationary model (for a recent
review see Ref.~\cite{Malik:2008im}). Deriving the equations of motion is
not trivial at second order and only recently was the Klein-Gordon
equation for scalar fields derived in closed form, taking into account the
metric backreaction \cite{Malik:2006ir}. This allows for the first time
a direct and complete computation of the second order perturbation, in
contrast with previous attempts which have focused only on certain
terms in the expression, for example \Rref{Finelli:2006wk}.

In this chapter the equations of motion for first and second order field
perturbations are described. These form the basis of the numerical calculation
undertaken in Chapters~\ref{ch:numericalsystem} and \ref{ch:results}.
Chapter~\ref{ch:numericalsystem} describes the numerical implementation of the
calculation, including the initial conditions used and the
computational requirements. We outline the numerical steps
taken in the system and examine the current constraints
on the calculation. The calculation is based on the slow roll version of the second
order equation, but solves the full non-slow roll equations for the
background and first order systems.
We present the results of the calculation
in Chapter~\ref{ch:results}, including a comparison of the second
order scalar field perturbation calculated for specific inflationary potentials. 

The models tested in this calculation are single field models with a canonical
action. Significant second order corrections
are expected only in models with a non-canonical action or multiple fields, or
when slow roll is violated. Numerical simulations will be particularly
useful in analysing models with these characteristics.
Section~\ref{sec:next-res} discusses planned future work to extend the current
numerical system to deal with models beyond the standard single field, slow
roll inflation.

In Section~\ref{sec:perts-num} of this chapter the Klein-Gordon equations
for first and second order perturbations are introduced. These will be the central
governing equations of the numerical calculation. They are
initially written in
terms of the metric perturbations and
then described in closed form, \iec in terms of the field perturbations
alone. In
Section~\ref{sec:slowroll}, the second order equation is written in a slow roll
approximation.
Section~\ref{sec:observable-perts} describes the observable quantities that can be
calculated from second order field perturbations. In Section~\ref{sec:disc-perts} we
discuss the results of this chapter.

%
%
%
%
\section{Perturbation Equations}
\label{sec:perts-num}

In this section we will briefly review the derivation of the first and
second order perturbation equations in the uniform curvature gauge and describe
the slow roll approximation that will be used. There are
many reviews on the subject of cosmological perturbation theory, and
here we will follow Ref.~\cite{Malik:2008im}.  The full closed
Klein-Gordon equation for second order perturbations was recently
derived in Ref.~\cite{Malik:2006ir} and the results of that paper will be
outlined below.

\subsection{Second Order Perturbations}
\label{sec:fosoperts-num}
In Section~\ref{sec:perts-intro} cosmological perturbations of a single scalar field
were introduced at first order. 
The methods adopted in that section can be extended at second
order to
find gauge invariant quantities. Recall that scalar quantities such as the inflaton
field, $\varphi$, can be split into an homogeneous background, $\vp_0$, and
inhomogeneous perturbations. Up to second order $\vp$ becomes
\begin{equation}
 \varphi(\eta, x^\mu) = \vp_0(\eta) + \dvp1(\eta, x^i) + \frac{1}{2}\dvp2(\eta, x^i)
\,.
\end{equation}

The metric tensor $g_{\mu\nu}$ must also be perturbed up to second order. In
\eq{eq:pertmetric-intro} the vector and tensor perturbations were included at first
order. Here we consider only the scalar metric perturbations \cite{Malik:2008im}:
\begin{align}
\label{eq:metric1-num}
g_{00}&= -a^2\left(1+2\phi_1+\phi_2\right) \,, \nonumber\\
g_{0i}&= a^2\left(B_1+\frac{1}{2}B_2\right)_{,i}\,, \nonumber\\
g_{ij}&= a^2\left[\left(1-2\psi_1-\psi_2\right)\delta_{ij}
+2E_{1,ij}+E_{2,ij}\right]\,,
\end{align}
where $\delta_{ij}$ is the flat background metric, $\phi_1$ and $\phi_2$ are the
lapse functions at first and second order, $\psi_1$ and $\psi_2$ are the curvature
perturbations, and $B_1$, $B_2$, $E_1$ and $E_2$ are the
scalar perturbations describing the shear.
In addition to the first order transformation vector \eqref{eq:xidefn-intro}, there
is
now a second order transformation vector
\begin{equation}
 \label{eq:xi2defn-perts}
\xi_2^\mu = (\alpha_2, \beta_{2,}^{~~i}) \,,
\end{equation}
where the spatial vector part of the transformation has been ignored. 

As before, we
can write down how a second order scalar quantity (such as $\dvp2$) will be
transformed \cite{Malik:2005cy}:
\begin{equation}
\label{eq:dvp2transform-perts}
 \wt{\dvp2} = \dvp2 + \vp_0'\alpha_2 + \alpha_1\left(\vp_0'' \alpha_1 + \vp_0'
\alpha_1' + 2\dvp1'\right) + \left(2\dvp1 + \vp_0'\alpha_1\right)_{,i}
\beta_{1,}^{~~i} \,,
\end{equation}
where a tilde ($\wt{~}$) denotes a transformed quantity. 
The metric curvature perturbation transformation at first order is straightforward,
$\wt{\psi_1} = \psi_1 -\H \alpha_1$, but at second order it becomes more complicated
\cite{Malik:2008im}:
\begin{equation}
 \label{eq:psi2transform-perts}
\wt{\psi_2} = \psi_2 -\H\alpha_2 -\frac{1}{4}\mathcal{X}^k_{~k} +
 \frac{1}{4}\nabla^{-2} \mathcal{X}^{ij}_{~~ij}\,,
\end{equation}
where $\mathcal{X}_{ij}$ is given by
\begin{align}
 \label{Xijdef}
\mathcal{X}_{ij} \equiv 
&2\Big[\left(\H^2+\frac{a''}{a}\right)\alpha_1^2
+\H\left(\alpha_1\alpha_1'+\alpha_{1,k}\xi_{1}^{~k}
\right)\Big] \delta_{ij}\nonumber\\
&
+4\Big[\alpha_1\left(C_{1ij}'+2\H C_{1ij}\right)
+C_{1ij,k}\xi_{1}^{~k}+C_{1ik}\xi_{1~~,j}^{~k}
+C_{1kj}\xi_{1~~,i}^{~k}\Big] \nonumber\\
&+2\left(B_{1i}\alpha_{1,j}+B_{1j}\alpha_{1,i}\right)
+4\H\alpha_1\left( \xi_{1i,j}+\xi_{1j,i}\right)
-2\alpha_{1,i}\alpha_{1,j}+2\xi_{1k,i}\xi_{1~~,j}^{~k} \nonumber \\
&+\alpha_1\left( \xi_{1i,j}'+\xi_{1j,i}' \right)
+\left(\xi_{1i,jk}+\xi_{1j,ik}\right)\xi_{1}^{~k}
+\xi_{1i,k}\xi_{1~~,j}^{~k}+\xi_{1j,k}\xi_{1~~,i}^{~k} \nonumber \\
&+\xi_{1i}'\alpha_{1,j}+\xi_{1j}'\alpha_{1,i}
\,,
\end{align}
and $B_{1i}$ and $C_{1ij}$ were defined in \eq{eq:svt-intro}.

We will work in the uniform curvature gauge where spatial 3-hypersurfaces are flat.
This implies that
\begin{equation}
 \label{eq:gauge-num}
\wt\psi_1=\wt\psi_2=\wt E_1=\wt E_2=0 \,.
\end{equation}
These relations can be used at first and then at second order to define gauge
invariant variables. It follows from Section~\ref{sec:perts-intro} that the first
order
transformation variables in the flat gauge satisfy $\alpha_1 = \psi_1/\H$ and
$\beta_1
= -E_1$. At second order, for the transformation of scalar quantities, as in
\eq{eq:dvp2transform-perts}, we require only $\alpha_2$. This is found by using
\eq{eq:psi2transform-perts} to have the form
\begin{equation}
 \label{eq:alpha2-perts}
\alpha_2=\frac{\psi_2}{\H}+\frac{1}{4\H}\left[
\nabla^{-2}\X^{ij}_{~~,ij}-\X^k_{ k}\right]\,,
\end{equation}
where the first order gauge variables have been substituted into $\X_{ij}$.

The Sasaki-Mukhanov variable, \iec the field perturbation on uniform curvature
hypersurfaces \cite{Sasaki:1986hm,Mukhanov:1988jd}, was given at first order in
\eq{eq:flatdvp1-intro} as
\begin{equation}
\label{eq:Q1I-num}
\wt{\dvp1}=\dvp1+\frac{\vp_{0}'}{\H}\psi_1\,.
\end{equation}
At second order the Sasaki-Mukhanov variable becomes more complicated
\cite{Malik:2005cy,Malik:2003mv}:
\begin{align}
\label{eq:Q2I-num}
\wt{\dvp2} = &\dvp2
+\frac{\vp_0'}{\H}\psi_2
+\frac{\vp_0'}{4\H}\left(
\nabla^{-2}\X^{ij}_{~~,ij}-\X^k_{k}\right) \nonumber\\
&+\frac{\psi_1}{\H^2}\Big[\vp_0'' {\psi_1}
+\vp_0'\left(\psi_1'-\frac{\H'}{\H}\psi_1\right)+2\H\delta\vp_1'\Big]
+\left(2\dvp1+\frac{\vp_0'}{\H}\psi_1\right)_{,k}\xi_{1\mathrm{flat}}^k \,,
\end{align}
where $\xi_{1\mathrm{flat}} = -(E_{1,i} +F_{1i})$. From now on we will
drop the tildes and only refer to variables calculated in the flat gauge.

The potential of the scalar field can also be separated into homogeneous and
perturbed sectors:
\begin{align}
 V(\varphi) &= \U + \delta V_{1} + \frac{1}{2}\delta V _{2}\,,\quad \\
 \delta V_{1} &= \Uphi \dvp1 \,,\quad \\
 \delta V_{2} &= \Upp \dvp1^2 + \Uphi\dvp2 \,.
\end{align}

Finally, the Klein-Gordon equations describe the evolution of the scalar field and
are
found by requiring the perturbed energy-momentum tensor $T_{\mu\nu}$ to obey the
energy conservation equation $\nabla_\mu T^{\mu\nu}=0$ (see for example
\Rref{Malik:2005cy}). For the
background field, $\vp_0$, the Klein-Gordon equation is 
\begin{equation}
\label{eq:KGback-num}
\vp_{0}''+2\H\vp_{0}'+a^2 \Uphi = 0\,.   
\end{equation}
The first order equation is
\begin{equation}
\label{eq:KGflatsingle-num}
\dvp1''+2\H\dvp1'+2a^2 \Uphi \phi_1
-\nabla^2\dvp1-\vp_{0}'\nabla^2 B_1
-\vp_{0}'\phi'_1 + a^2 \Upp \dvp1
=0\,,
\end{equation}
and the second order version is given by
\begin{align}
\label{eq:KG2flatsingle-num}
\dvp2'' &+ 2\H\dvp2'-\nabla^2\dvp2+a^2 \Upp \dvp2
+ a^2 \Uppp (\dvp1)^2 +2a^2 \Uphi \phi_2
-\vp_{0}'\left(\nabla^2 B_2+\phi_2'\right)\nonumber\\
&+ 4\vp_{0}' B_{1,k}\phi_{1,}^{~k}
+2\left(2\H\vp_{0}'+a^2 \Uphi\right) B_{1,k}B_{1,}^{~k}
+4\phi_1\left(a^2 \Upp \dvp1-\nabla^2\dvp1\right) \nonumber\\
&+ 4\vp_{0}'\phi_1\phi_1'
-2\dvp1'\left(\nabla^2 B_1+\phi_1'\right)-4{\dvp1'}_{,k}B_{1,}^{~k} \nonumber \\
&= 0\,,
\end{align}
where all the variables are now in the flat gauge.

In order to write the Klein-Gordon equations in closed form, the Einstein field
equations \eqref{eq:einstein-intro} are also required at first and second order.
These are not reproduced here, but are presented for example in Section~II~B of
\Rref{Malik:2006ir}. 

\subsection{Fourier Transform}
\label{sec:fourier-perts}
%
In general, the dynamics of the scalar field becomes clearer in Fourier space.
However, terms at second order of the form
$\left(\dvp1(x^i)\right)^2$ require the use
of the convolution theorem (see for example \Rref{Vretblad:2005}).
Following Refs.~\cite{book:liddle} and \cite{Malik:2006ir} we will write
$\dvp{}(k^i)$
for the Fourier component of $\dvp{}(x^i)$ such that
\begin{equation}
 \dvp{}(\eta, x^i) = \frac{1}{(2 \pi)^3} \int \d^3k \dvp{}(\kvi) \exp (i k_i
x^i)
\,,
\end{equation}
where $\kvi$ is the comoving wavenumber.
In Fourier space, the closed form of the first order Klein-Gordon equation then
transforms into
\begin{multline}
\label{eq:fokg}
 \dvp1(\kvi)'' + 2\H \dvp1(\kvi)' + k^2\dvp1(\kvi) \\
+ a^2 \left[\Upp +
\frac{8\pi G}{\H}\left(2\vp_{0}' \Uphi + (\vp_{0}')^2\frac{8\pi G
}{\H}\U\right)\right]\dvp1(\kvi) = 0 \,.
\end{multline}

The second order equation requires a careful
consideration of terms that are quadratic in the first order perturbation. In
particular, we
require convolutions of the form
\begin{equation}
 f(x^i)g(x^i) \longrightarrow \frac{1}{(2 \pi)^3} \int \d^3q \d^3p\, \delta^3(\kvi
-\pvi -\qvi) f(\pvi)
g(q^i) \,.
\end{equation}
For convenience we will group together those terms with gradients of $\dvp1(x^i)$
and denote them by $F$. 
The full closed form, second order Klein-Gordon
equation in Fourier space is then given by \cite{Malik:2006ir}
\begin{align}
\label{eq:SOKG-real-num}
\dvp2''(\kvi) &+ 2\H \dvp2'(\kvi) + k^2 \dvp2(\kvi) \nonumber \\
&+ a^2\left[\Upp + \frac{8\pi G}{\H}\left(2\vp_{0}'\Uphi
+ (\vp_0')^2\frac{8\pi G}{\H}\U \right) \right]\dvp2(\kvi) \nonumber \\
&+ \frac{1}{(2\pi)^3}\int \d^3q \d^3p\, \delta^3(\kvi -\pvi -\qvi) \biggl\{
\biggr.\nonumber\\
&\quad \frac{16\pi G}{\H} \left[ Q \dvp1'(\pvi) \dvp1(\qvi) + \vp_{0}' a^2\Upp
\dvp1(\pvi)\dvp1(\qvi) \right]  \nonumber \\
&+ \left(\frac{8\pi G}{\H}\right)^2\vp_{0}'\left[2a^2\Uphi\vp_{0}'
\dvp1(\pvi)\dvp1(\qvi) + \vp_{0}'Q\dvp1(\pvi)\dvp1(\qvi) \right] \nonumber  \displaybreak[0]\\
&- 2\left(\frac{4\pi G}{\H}\right)^2\frac{\vp_{0}' Q}{\H} \left[Q\dvp1(\pvi)
\dvp1(\qvi) +
\vp_{0}' \dvp1(\pvi) \dvp1'(\qvi)\right] \nonumber \\
\biggl. &+ \frac{4\pi G}{\H} \vp_{0}' \dvp1'(\pvi) \dvp1'(\qvi) 
 + a^2\left[\Uppp + \frac{8\pi G}{\H}\vp_{0}' \Upp\right] \dvp1(\pvi)
\dvp1(\qvi) \biggr\} \nonumber \\
&+ F(\dvp1(\kvi), \dvp1'(\kvi)) = 0\,,
\end{align}
where we have defined the parameter $Q=a^2 (8\pi G \U \vp_0'/\H + \Uphi)$ for
convenience.
The $F$ term contains gradients of $\dvp1$ in real space and therefore
the convolution integrals include additional factors of $k$ and
$q$. The form of $F$ is given by \cite{Malik:2006ir}
\begin{align}
 \label{eq:Fdvk1-fourier-num}
&F\left(\dvp1(\kvi),\dvp1'(\kvi)\right)
= \frac{1}{(2\pi)^3}\int \d^3p\, \d^3q\,\delta^3(\kvi-\pvi-\qvi) 
\Bigg\{ \nonumber \\
&\quad 2\left(\frac{8\pi G}{\H}\right)\frac{p_kq^k}{q^2}
\delta\vp_{1}'(\pvi)\left(Q\dvp1(\qvi)+\vp_{0}'\dvp1'(\qvi)\right)
+p^2\frac{16\pi G}{\H}\dvp1(\pvi)\vp_{0}'\dvp1(\qvi) \nonumber \\
&\quad 
+\left(\frac{4\pi G}{\H}\right)^2
\frac{\vp_{0}'}{\H}\Bigg[
\left(p_lq^l-\frac{p^iq_jk^jk_i}{k^2}\right) 
\vp_{0}'\delta\vp_{1}(\pvi)\vp_{0}'\delta\vp_{1}(\qvi)
\Bigg]\nonumber \displaybreak[0]\\
&\quad +2\frac{Q}{\H}\left(\frac{4\pi G}{\H}\right)^2 
\frac{p_lq^lp_mq^m+p^2q^2}{k^2q^2}
\Bigg[\vp_{0}'\delta\vp_{1}(\pvi)
\left(Q\dvp1(\qvi)+\vp_{0}'\dvp1'(\qvi)\right)
\Bigg]
\nonumber \displaybreak[0]\\
&\quad +\frac{4\pi G}{\H}
\Bigg[
4Q\frac{q^2+p_lq^l}{k^2}\left(
\dvp1'(\pvi)\dvp1(\qvi)\right)
-\vp_{0}'p_lq^l \delta\vp_{1}(\pvi)\delta\vp_{1}(\qvi)
\Bigg]
\nonumber\\
&\quad +\left(\frac{4\pi G}{\H}\right)^2
\frac{\vp_{0}'}{\H}\Bigg[
\frac{p_lq^lp_mq^m}{p^2q^2}
\left( Q\dvp1(\pvi)+\vp_{0}'\dvp1'(\pvi)\right)
\left(Q\dvp1(\qvi)+\vp_{0}'\dvp1'(\qvi)\right)
\Bigg]\nonumber \displaybreak[0]\\
&\quad +\frac{\vp_{0}'}{\H}
\Bigg[
8\pi G\left(\frac{p_lq^l+p^2}{k^2}q^2\dvp1(\pvi)\dvp1(\qvi)
-\frac{q^2+p_lq^l}{k^2}\dvp1'(\pvi)\dvp1'(\qvi)
\right)
\nonumber\\
&\quad +\left(\frac{4\pi G}{\H}\right)^2
\frac{k^jk_i}{k^2}\Bigg(
2\frac{p^ip_j}{p^2}
\left(Q\dvp1(\pvi)+\vp_{0}'\dvp1'(\pvi)\right)
Q\dvp1(\qvi)
\Bigg)\Bigg]
\Bigg\}\,.
\end{align}

\subsection{Slow Roll Approximation}
\label{sec:slowroll}

In order to establish the viability of a numerical calculation of the evolution of
second order perturbations from the
Klein-Gordon equation, Chapters~\ref{ch:numericalsystem} and \ref{ch:results} will be limited to
the framework of the slow roll approximation. 
This involves taking
\begin{align}
 &\vp_{0}'' + \H \vp_{0}' \simeq 0\,,\quad \\
&\frac{\left(\vp_{0}'\right)^2}{2a^2} \ll \U\,,
\end{align}
such that $Q\simeq0$ and $\H^2 \simeq (8\pi G/3) a^2 \U$. 
In Chapter~\ref{ch:introduction} the slow roll parameter $\varepsilon_H$ was
defined in \eq{eq:epsilonHdefn-intro}. 
In this chapter and the rest of Part~\ref{part:numerical}, a different slow 
roll parameter will be used, denoted by
$\bar{\varepsilon}_H$ and defined in Refs.~\cite{Malik:2006ir} and
\cite{Seery:2005gb}. 
This new parameter is the square-root of $\varepsilon_H$ and is given by
\begin{equation}
\label{eq:bareps-defn-perts}
 \bar{\varepsilon}_H = \sqrt{4\pi G} \frac{\vp_{0}'}{\H} = \sqrt{\varepsilon_H}\,.
\end{equation}
The second slow roll parameter is still $\eta_H = \varepsilon_H -
\varepsilon_H'/2\H \varepsilon_H$. Following \Rref{Malik:2006ir} we
will implement the slow roll approximation by keeping terms up to and including
$\mathcal{O}(\bar{\varepsilon}^2_H)$ and terms which are
$\mathcal{O}(\bar{\varepsilon}_H \eta_H)$. 
Within this approximation the second order equation (\ref{eq:SOKG-real-num})
simplifies dramatically, and with the $F$ term included it reduces to
\begin{align}
 \label{eq:KG2-fourier-sr-num}
&\dvp2''(\kvi)+2\H\dvp2'(\kvi)+k^2\dvp2(\kvi)
+\left(a^2
\Upp-{24 \pi G}(\vp_{0}')^2\right)
\dvp2(\kvi) \nonumber\\
&+ \frac{1}{(2\pi)^3}\int \d^3p\ \d^3q\ \delta^3(\kvi-\pvi-\qvi) \Bigg\{
a^2\left(\Uppp
+ \frac{8\pi G}{\H}\vp_{0}' \Upp\right)
 \dvp1(\pvi)\dvp1(\qvi) \nonumber \\
&\qquad +\frac{16\pi G}{\H}a^2
\vp_{0}'\Upp\dvp1(\pvi)\dvp1(\qvi)\Bigg\}
\nonumber \displaybreak[0] \\
&+ \frac{1}{(2\pi)^3}\frac{8\pi G}{\H}
\int \d^3p\ \d^3q\ \delta^3(\kvi-\pvi-\qvi)  \Bigg\{
\frac{8\pi G}{\H}\frac{p_l q^l}{q^2}\vp_{0}'\dvp1'(\pvi)
\dvp1'(\qvi) \nonumber\\
&\qquad+ 2p^2\vp_{0}' \dvp1(\pvi) \dvp1(\qvi)
+\vp_{0}'
\Bigg(
\left(\frac{p_lq^l+p^2}{k^2}q^2-\frac{p_lq^l}{2}\right)
\dvp1(\pvi)\dvp1(\qvi) \nonumber \\
&\qquad+\left(\frac{1}{2}-\frac{q^2+p_lq^l}{k^2}\right)
\dvp1'(\pvi)\dvp1'(\qvi)\Bigg)
\Bigg\} \nonumber \\
&=0 \,.
\end{align}
The numerical simulation described in Chapter~\ref{ch:numericalsystem} will solve the
slow roll version of the second order equation given above,
\eq{eq:KG2-fourier-sr-num}, together with the complete first
order equation (\ref{eq:fokg}) and background equation
(\ref{eq:KGback-num}). 
%
%
%
%
\section{Observable Quantities}
\label{sec:observable-perts}

Cosmological perturbations at second order are becoming increasingly important now
that
statistical quantities beyond the power spectrum and spectral index are being
investigated. Observations, however, do not tell us anything about the inflaton
field directly. In this section the second order perturbations described above will
be related to observable quantities in order to demonstrate how a numerical
calculation could
be employed in the near future to gain further insight into the nature of the field
that drives inflation.

The temperature fluctuations observed in the CMB can
be directly related to the curvature perturbation $\R$. In
Section~\ref{sec:perts-intro}, $\R$ was defined at first order in terms of $\dvp1$.
When the second order contribution is included the total comoving curvature
perturbation is defined as
\begin{equation}
\label{eq:Rfulldefn-perts}
 \R = \R_1 + \frac{1}{2}\R_2\,.
\end{equation}
The first order term is related to the inflaton perturbation in the flat gauge by
$\R_1
= \H\dvp1/\vp_0'$. The second order part includes terms quadratic in $\dvp1$ and so
in Fourier space requires convolutions. We are interested in the value of $\R$ after
horizon crossing for the calculation of $\Pr$ and a determination of the
non-Gaussianity produced during inflation. This allows us to neglect gradient terms
in real space or terms proportional to $k$ in Fourier space.
In this limit the real space expression for $\R_2$ is \cite{Malik:2005cy}
\begin{equation}
 \label{eq:R2real-perts}
\R_2(\eta, x^i) = \frac{\H}{\vp_0'}\dvp2 - 2\frac{\H}{\left(\vp_0'\right)^2}
\dvp1'\dvp1 + \frac{\dvp1^2}{\left(\vp_0'\right)^2}\left(\H\frac{\vp_0''}{\vp_0} 
 - \H' -2\H^2 \right)\,.
\end{equation}
Using the background evolution equation \eqref{eq:KGback-num} and transforming to
Fourier space implies that \eq{eq:R2real-perts} can be written as
\begin{align}
 \R_2(\eta, \kvi) &= \frac{\H}{\vp_0'}\dvp2(\eta, \kvi) 
  + \frac{1}{(2\pi)^3}\int \d^3q\, \d^3p\, \delta(\kvi-\qvi-\pvi) \Bigg\{
\nonumber\\
 &\qquad -2\frac{\H}{\left(\vp_0'\right)^2} \dvp1(\eta, \pvi)\dvp1'(\eta, \qvi) \nonumber\\
&\qquad -\frac{1}{\left(\vp_0'\right)^2} \left(
  2\H^2 \frac{\vp_0'}{\vp_0} + \frac{a^2 \H}{\vp_0}\Uphi + (8\pi G)a^2 \U
 \right)
\dvp1(\eta,\pvi) \dvp1(\eta, \qvi)\Bigg\}\,.
\end{align}
Once the numerical calculation has been carried out at first and second order as
described in Chapter~\ref{ch:numericalsystem} this quantity can be evaluated after
horizon crossing.

In Chapter~\ref{ch:introduction} the non-Gaussianity parameter $\fnlloc$ was defined
in terms of $\R$ in \eq{eq:fnllocdefn-intro}. Writing \eq{eq:fnllocdefn-intro} in
Fourier space using \eq{eq:Rfulldefn-perts} implies that
\begin{multline}
 \R(\kvi) = \R_1(\kvi)
  + \frac{3}{5}\fnlloc \Bigg( \frac{1}{(2\pi)^3}\int \d q^3\, \R_1(\qvi)
\R_1(\kvi-\qvi) \\ 
  \quad- \left\langle \frac{1}{(2\pi)^3} \int \d q^3\, \R_1(\qvi)
\R_1(\kvi-\qvi) \right\rangle \Bigg) \,,
\end{multline}
where $\langle \rangle$ denotes the expectation value.
A good approximation of the local non-Gaussianity produced is then given by
\begin{multline}
 \label{eq:fnlloc-perts}
\fnlloc = \frac{5}{6} \R_2(\kvi) \Bigg[ \frac{1}{(2\pi)^3}\int \d q^3\, \R_1(\qvi)
\R_1(\kvi-\qvi) \\
  \quad- \left\langle \frac{1}{(2\pi)^3} \int \d q^3\, \R_1(\qvi)
\R_1(\kvi-\qvi) \right\rangle \Bigg]^{-1} \,.
\end{multline}
Calculating $\dvp2$ and $\R_2$ therefore provides direct insight into the behaviour
and
production of the non-Gaussianity parameter $\fnlloc$.

To go beyond the local shape of the non-Gaussianity it is necessary to calculate the full
bispectrum of the perturbations. In practice the bispectrum of the curvature perturbation on
uniform density hypersurfaces, $\zeta$, is used in setting observational limits. At first order this
is simply related to the comoving curvature perturbation by $\zeta_1 = -\R_1$. At second order the
relationship is more complicated. For large scales outside the horizon, $\zeta_2$ can be related to
the field perturbations in real space using \cite{Malik:2005cy}
\begin{equation}
 \zeta_2(x^i) = -\frac{\H}{\vp_0'}\dvp2(x^i) - \left[4 - 3\frac{(\vp_0')^2 - a^2\U}{(\vp_0')^2/2
+a^2\U} \right]\left(\frac{\H}{\vp_0'} \right)^2 \dvp1(x^i)^2\,. 
\end{equation}
In Fourier space this again introduces a convolution integral of the first order perturbations.

The bispectrum of $\zeta$ is given by
\begin{equation}
 \langle \zeta(\kb_1)\zeta(\kb_2)\zeta(\kb_3)\rangle = 
        (2\pi)^3 \delta(\kb_1+\kb_2+\kb_3) B(k_1,k_2,k_3)\,,
\end{equation}
where translation invariance introduces the delta function. The $k$ dependence of the bispectrum 
is usually separated from an overall amplitude factor and considered as a shape function
$F(k_1,k_2,k_3)$. The bispectrum is then of the form \cite{Liguori:2010hx, Babich:2004gb}
\begin{equation}
 \langle \zeta(\kb_1)\zeta(\kb_2)\zeta(\kb_3)\rangle = 
        A (2\pi)^3 \delta(\kb_1 + \kb_2 + \kb_3) F(k_1,k_2,k_3)\,,
\end{equation}
and for a particular shape function $F$ the best estimator for $A$ when the non-Gaussianity is
small is given by \cite{Babich:2004gb}
\begin{equation}
\label{eq:fnlampl-perts}
\hat{A} = \frac{\sum_{\kb_i} \zeta(\kb_1)\zeta(\kb_2)\zeta(\kb_3) F(k_1,k_2,k_3)/ 
                (\sigma^2_{k_1}\sigma^2_{k_2}\sigma^2_{k_3}) }
                {\sum_{\kb_i} F(k_1,k_2,k_3)^2/(\sigma^2_{k_1}\sigma^2_{k_2}\sigma^2_{k_3})} \,,
\end{equation}
where $\sigma_{k_i}$ is the variance of the mode and the sums run over all the triangles in $k$
space subject. If $\kb_3$ is chosen to be the longest side of the triangle then the triangle
inequality enforces
\begin{equation}
 k_3 \le k_1 + k_2\,.
\end{equation}
\eq{eq:fnlampl-perts} provides a blueprint for how to evaluate the bispectrum in terms of a
particular given shape. To compare a primordial bispectrum with the observed temperature bispectrum
from the CMB it is necessary to construct the spherical harmonics of the bispectrum and use
transfer functions to relate the primordial values with the observed values. We have not
carried out this procedure in this thesis. However, one of the goals of our future work is to
undertake such a comparison of the numerically generated bispectrum with the observed quantity. 

As mentioned above, the shape most often used in comparisons with observations is the local shape
given by the
ansatz in \eq{eq:fnllocdefn-intro}. The expression for $F_\mathrm{local}$ is
\cite{Babich:2004gb, Komatsu:2010fb}
\begin{equation}
 F_\mathrm{local}(k_1,k_2,k_3) = 2N\fnlloc \left(\frac{1}{k_1^3 k_2^3}
                                  +\frac{1}{k_2^3 k_3^3} + \frac{1}{k_1^3 k_3^3}\right) \,,
\end{equation}
where the spectrum has been taken to be scale invariant and $N$ is a normalisation factor. 

This is not the only shape
that has been considered and, as we have seen, non-canonical inflationary actions generate a
bispectrum which is peaked
when the magnitudes of the $k$ modes are approximately equal. The form of $F$ for the equilateral
case is \cite{Komatsu:2010fb}
\begin{multline}
 F_\mathrm{eq}(k_1,k_2,k_3) = 6 N \fnleq \left\{ -\frac{1}{k_1^3 k_2^3} -\frac{1}{k_2^3 k_3^3}
                                -\frac{1}{k_3^3 k_1^3}  \right.\\
                                \left.-\frac{2}{(k_1 k_2 k_3)^2}  + \left[ \frac{1}{k_1 k_2^2 k_3^3}
+ \mathrm{5\, perms} \right]
                                \right\} \,.
\end{multline}

A third form has been proposed which is nearly orthogonal to the other two shapes
\cite{Senatore:2009gt}. These shapes all have the property that they are separable functions of
each $k_i$ or can be constructed from these separable functions. This property eases analytic
calculations but clearly does not hold for generic shapes. 
There has been a proposal to define the
shape of the bispectrum in terms of a set of basis vector shapes
\cite{Fergusson:2009nv,Liguori:2010hx}. This would remove the need for only separable shapes to be
considered and would allow for a straightforward analysis of the bispectrum from its primordial
value up to the observed bispectrum in the CMB. 

We have seen that the second order scalar perturbation is not the direct observable quantity of
interest. The bispectrum of the curvature perturbations, which contains a contribution from the
second order non-linear part, can be compared with observations either by use of various
shape functions or through a full analysis with transfer of the primordial values. A future goal of
our work is to compare the bispectrum obtained numerically with that from observations.

%
%
%
%
\section{Discussion}
\label{sec:disc-perts}

In this chapter, the equations of motion for a single scalar
field
up to second order in cosmological perturbations have been introduced. The second
order gauge
transformation has been discussed and the transformation components determined for
the uniform curvature gauge.
In Chapter~\ref{ch:introduction} first order classical perturbations were quantised
in the Minkowski spacetime limit and normalised using the Wronskian condition in
\eq{eq:quantcondition-intro}. This constraint also fixes the quantisation for other
orders of the perturbation, including $\dvp2$ \cite{Seery:2008qj}. 

The perturbation equations are better understood in Fourier space, although the cost
of adopting this approach is the need to employ convolution integrals of the first
order
perturbations. When written in Fourier space, the second order Klein-Gordon equation
can
be described entirely in terms of the field perturbations and background quantities. 

\eq{eq:SOKG-real-num}, first derived in \Rref{Malik:2006ir}, is valid on all scales
inside and outside the horizon. When a particular slow roll approximation is made
this equation simplifies to that found in \eq{eq:KG2-fourier-sr-num}. This slow roll
version of the equation will be the central governing equation of the numerical
calculation described in the next chapter.

Finding numerical solutions of \eq{eq:KG2-fourier-sr-num} is the first step towards
solving the full equation \eqref{eq:SOKG-real-num} for a single field and
ultimately the multi-field equation given in \Rref{Malik:2006ir}. Understanding
cosmological perturbations beyond linear order is critical if higher order statistical
effects are to be accurately calculated. Section~\ref{sec:observable-perts}
outlined the connection between $\dvp2$ and observable quantities such as the
comoving curvature perturbation and the non-Gaussianity of the perturbations. Going
beyond single field, slow roll models, non-linear effects become more important. In
Chapters~\ref{ch:numericalsystem} and \ref{ch:results} the first step is taken
towards calculating higher order perturbations for these models.

%

\chapter{Numerical System and Implementation}
\label{ch:numericalsystem}
%
%
\section{Introduction}
\label{sec:intro-num}

Our goal in Part~\ref{part:numerical} of this thesis is to show that, just as at
first
order, a direct numerical calculation of the second order perturbations of a
scalar field system is achievable. 
In this chapter the implementation of this system is outlined.
The structure of the numerical
system follows the work done at first order by Martin
\& Ringeval \cite{Martin:2006rs, Ringeval:2007am} and previously by
Salopek \etal \cite{Salopek:1988qh}.

The most important difference between an analytic and numerical treatment of the
equations presented in Chapter~\ref{ch:perts} is the requirement to specify a finite
numerical range of a finite number of $k$ modes to be calculated.
The upper cutoff in $k$, which marks the smallest scale considered, is well
motivated by the difficulty in observing primordial perturbations at very small
scales. 
On the other hand, we also need to specify the largest scale (or smallest $k$) that
we
will consider. Analytically, this is often taken to be the size of the universe,
with $k=0$ being the equivalent mode. One immediate problem with this
specification, however, is that
the Bunch-Davies vacuum initial conditions diverge. The standard approach to this
problem is to implement a cutoff
at large scales beyond which the amplitude of perturbations vanishes. This is a
pragmatic approach, but recently there has been some evidence that a sharp
cutoff similar to this could be responsible for the lack of power at large
scales in the WMAP data \cite{Lyth:2007jh, spergel, Sinha:2005mn,Kim:2009pf}.
 
The main concern is that the $k$ range covers most, if not all, of the modes observed
to date in the CMB. The WMAP team rely for their main results in 
\Rref{Komatsu:2008hk}  
 on $\ell$ multipoles in the range $\ell \in [3, 1000]$,
which corresponds approximately\footnotemark  to $k\in \left[0.92 \e{-60}, 3.1 \times
10^{-58}\right]\Mpl = \left[3.5\e{-4}, 0.12\right] \Mpc^{-1}$.
\footnotetext{The approximate conversion for $\ell$ is $\ell\simeq \frac{2k}{H_0}$ and a
Megaparsec
is given in Planck units as $1\Mpc^{-1} \simeq 2.6247\e{-57} \Mpl$.}
We will consider three different ranges of $k$ modes when producing the results in
Chapter~\ref{ch:results}, all of which contain the WMAP pivot scale
$\kwmap=0.002\Mpc^{-1}$. The
choice of $k$ range is flexible with the only numerical constraint being that the
number of modes at second order should be equal to $2^l +1$, where $l$ is a positive
integer. This enables
faster integration using the Romberg method, as explained below.

In Chapter~\ref{ch:perts} the Klein-Gordon equations of motion were described for the
background field, and the first and second order field perturbations. These form the
basis of the numerical calculation in this chapter. In Section~\ref{sec:eqs-num},
these equations are rewritten in a form more amenable to numerical work. The time
variable is changed from conformal time, $\eta$, to the number of e-foldings, $\N$.
The convolution terms which are present in the Fourier space equations are then
written in terms of spherical polar coordinates and split into smaller units.

Four inflationary potentials were chosen in order to test the numerical calculation.
These are defined in
Section~\ref{sec:pots-num} and the steps taken to establish the values of the
required parameters are outlined. The initial conditions used for the first order
perturbations are the Bunch-Davies vacuum conditions, as specified in
Section~\ref{sec:perts-intro}. At second order the perturbations are set to be
identically zero at the beginning of the simulation, as explained in
Section~\ref{sec:initconds-num}. This section also describes the method and timing of
the initialisation of the variables.

In Section~\ref{sec:impl-num} the numerical method is discussed. The calculation can
be split into four stages, each of which is described in depth, along with the
logistical constraints and software environment. The numerical code is tested in
Section~\ref{sec:tests-num} by comparing the computed value with an analytic result
where this is possible. The choice of parameters in the calculation is determined by
the results of this comparison. In Section~\ref{sec:disc-numerical}, the results of
this chapter are summarised and discussed.

\section{Numerical Equations}
\label{sec:eqs-num}

The Klein-Gordon equations in Chapter~\ref{ch:perts} are not appropriate for a
numerical calculation and in this section we rearrange them into a more
suitable form. This involves a change of time coordinate and grouping of terms
into smaller units for calculation.
The second order slow roll equation \eqref{eq:KG2-fourier-sr-num} can be further
simplified by performing the $p$
integral and changing to spherical polar coordinates $q, \theta$ and $\omega$, where
$q=|\textbf{q}|$. The $\d^3q$ integral then becomes
\begin{equation}
 \int \d^3q \longrightarrow \int_{0}^{\infty} q^2 \d q \int_{0}^{\pi}\sin \theta
\d\theta 
   \int_{0}^{2\pi}\d\omega \,.
\end{equation}
For each $k$ mode equation we take the $\theta=0, \omega=0$ axis in the
direction of $\kvi$, so that the angle between $\kvi$ and $\qvi$ is
$\theta$, and the scalar product is $q_i k^i = q k \cos\theta$. 
The argument of
each $\dvp1$ or $\dvp1'$ term depends on $\theta$ through
$|\kvi-\qvi|=\sqrt{k^2 + q^2 -2kq \cos\theta}$ and
so must remain inside the $\theta$ integral. There is no $\omega$ dependence
in $\dvp1$ with this choice of axes, so the last integral is straightforward
to evaluate.

In the slow roll case there are only four different $\theta$ dependent terms,
here labelled $\A$--$\D$:
\begin{align}
\label{eq:AtoD-num}
 \A(\kvi,\qvi) &= \int_0^\pi \sin(\theta) \dvp1(\kvi-\qvi) \d\theta \,,
\nonumber\\
 \B(\kvi,\qvi) &= \int_0^\pi \cos(\theta)\sin(\theta) \dvp1(\kvi-\qvi)
\d\theta \,,\nonumber\\
 \C(\kvi,\qvi) &= \int_0^\pi \sin(\theta) \dvp1'(\kvi-\qvi) \d\theta \,,
\nonumber\\
 \D(\kvi,\qvi) &= \int_0^\pi \cos(\theta) \sin(\theta) \dvp1'(\kvi-\qvi)
\d\theta \,.
\end{align}
When written in terms of the variables defined in \eqs{eq:AtoD-num},
the slow roll equation
\eqref{eq:KG2-fourier-sr-num} becomes:
\begin{equation}
\label{eq:KG2-fourier-sr-aterms}
\dvp2''(\kvi)+2\H\dvp2'(\kvi)+k^2\dvp2(\kvi)
+\left(a^2
\Upp-{24 \pi G}(\vp_{0}')^2\right)
\dvp2(\kvi)
+ S(\kvi) = 0 \,,
\end{equation}
where $S(\kvi)$ is the source term which will be determined before the
second order system is run:
\begin{align}
\label{eq:KG2-src-sr-aterms}
&S(\kvi) = \frac{1}{(2\pi)^2}\int \d q\ \Bigg\{
a^2\Uppp q^2 \dvp1(\qvi) \A(\kvi,\qvi) \nonumber\\
&+ \frac{8\pi G}{\H}\vp_{0}' \Bigg[ 
\left( 3a^2\Upp q^2 + \frac{7}{2}q^4 + 2k^2q^2\right) \A(\kvi,\qvi)
-\left(\frac{9}{2} + \frac{q^2}{k^2}\right)kq^3 \B(\kvi,\qvi)
\Bigg]\dvp1(\qvi) \nonumber\\
&+ \frac{8\pi G}{\H}\vp_{0}' \Bigg[
-\frac{3}{2}q^2 \C(\kvi,\qvi) + \left(2-\frac{q^2}{k^2}\right)kq \D(\kvi,\qvi) 
\Bigg]\dvp1'(\qvi) \Bigg\} \,.
\end{align}
 The full set of equations which must be evolved are
then \eq{eq:KGback-num} for the background, \eq{eq:fokg} for the first
order perturbations and \eqs{eq:KG2-fourier-sr-aterms} and
\eqref{eq:KG2-src-sr-aterms} for the second order and source terms.

A more appropriate time variable for the numerical simulation is the
number of e-foldings \eqref{eq:nefolddefn-intro}. We employ 
\begin{equation}
\label{eq:def-ntime}
\N = \log ( a / a_{\mathrm{init}} )
\end{equation}
as our time variable instead of conformal time. This is measured from
$a_{\mathrm{init}}$, the value of $a$ at the beginning of the
simulation. We will use a dagger\footnotemark\ ($\dN{}$) to denote differentiation
with respect to $\N$.
\footnotetext{This should not be confused
with the use of $^\dagger$ as Hermitian conjugate, which is confined to
Section~\ref{sec:perts-intro}.}
Derivatives with respect to conformal time, $\eta$, and coordinate time, $t$, are
then given by
\begin{align}
 \pd{ }{\eta} &= \frac{\d \N}{\d \eta}\pd{}{\N} = \H \pd{}{\N} \,,\\
 \pd{ }{t} &= \frac{\d \eta}{\d t} \frac{\d \N}{\d \eta}\pd{}{\N} = H
\pd{}{\N}\,,
\end{align}
respectively, where $H = \d \ln a/\d t$ and $\H = aH$.

If $a$ is set to be unity at the present epoch, we can calculate
$a_{\mathrm{init}}$ once the background run is complete and the number of e-foldings
of inflation has been determined.
The value of $a$ at the end of inflation, $a_\mathrm{end}$, is calculated by
connecting it with $a_0$ (see for example
Eq.~(3.19) in \Rref{book:liddle} or Eq.~(7) in \Rref{Peiris:2008be}). 
The relation between them is given by
\begin{equation}
 \label{eq:connection-num}
\frac{a_\mathrm{end}}{a_0} = \frac{a_\mathrm{end}}{a_\mathrm{reh}}
			     \frac{a_\mathrm{reh}}{a_\mathrm{eq}}
			     \frac{a_\mathrm{eq}}{a_\mathrm{0}}\,,
\end{equation}
where $a_\mathrm{reh}$ and $a_\mathrm{eq}$ are the values of $a$ at the end of
reheating and matter-radiation equality. Using the relationship between energy
densities and the scale factor relevant to the matter and radiation dominated
eras, together with the Friedmann
equation \eqref{eq:Friedmann1-intro}, we can write
\begin{equation}
 \label{eq:conn2-num}
\log\left(\frac{a_\mathrm{end}}{a_0}\right) =
 -\frac{2}{3}\log\left(\frac{H_\mathrm{end}}{\Mpl}\right)
 + \frac{1}{6}\log\left(\frac{H_\mathrm{reh}}{\Mpl}\right)
 + \frac{1}{2}\log\left(\frac{H_\mathrm{eq}}{\Mpl}\right)
 + \log\left(\frac{a_\mathrm{eq}}{a_0}\right)\,,
\end{equation}
where $H_\mathrm{end}$, $H_\mathrm{reh}$ and $H_\mathrm{eq}$ are the values of $H$
at the end of inflation, at the end of reheating and at matter-radiation equality
respectively.
The value
of $a_\mathrm{eq}$ is taken to be $4.15\e{-5} (\Omega_m h^2)^{-1}$ and
$H_\mathrm{eq}=4.63\e{-54} \Omega_m^2 h^4 \Mpl$ \cite{Peiris:2008be, book:dodelson}.
The mean value for $\Omega_m h^2$ determined by WMAP5 + BAO + SN measurements is
$\Omega_m h^2 =0.1369$ \cite{Komatsu:2008hk}. With these values and taking
$a_0=1$, the scale factor at the end of inflation is given by
\begin{equation}
\label{eq:aend-num}
a_\mathrm{end} \simeq e^{-72}
\left(\frac{H_\mathrm{end}}{\Mpl}\right)^{-\frac{2}{3}} 
\left(\frac{H_\mathrm{reh}}{\Mpl}\right)^{\frac{1}{6}}\,.
\end{equation}

In Chapter~\ref{ch:results}, it is assumed that
reheating occurs instantaneously at the end of
inflation such that $H_\mathrm{reh} = H_\mathrm{end}$. This gives
$a_\mathrm{end} \simeq 10^{-29}$ and approximately 65
e-foldings from the end of inflation to the present. It also fixes the horizon
crossing
time of the WMAP pivot scale, $\kwmap=0.002\Mpc^{-1}$, to be about 60 e-foldings
before
the end of inflation. Because the Hubble parameter is not kept fixed during the
numerical calculation the
number of e-foldings between horizon crossing and the end of inflation will depend
on the form of the inflationary potential and the evolution of $H$.

The background and first order equations, written in terms of the new time
variable $\N$, are
\begin{equation}
\ddN{\vp_{0}} + \left(3 + \frac{\dN{H}}{H}\right)\dN{\vp_{0}} + \frac{\Uphi}{H^2} = 0
\,,
\label{eq:bgntime}
\end{equation}
and
\begin{multline}
\ddN{\dvp1}(\kvi) + \left(3 + \frac{\dN{H}}{H}\right)\dN{\dvp1}(\kvi) 
 + \Bigg[ \left(\frac{k}{aH}\right)^2 + \frac{\Upp}{H^2} + \frac{8\pi G}{H^2}
 2\dN{\vp_{0}}\Uphi \\
+ \left.\left(\frac{8\pi G}{H}\right)^2
\left(\dN{\vp_{0}}\right)^2\U \right]\dvp1(\kvi) = 0\,, \label{eq:fontime}
\end{multline}
respectively.
The corresponding second order perturbation equation takes the form
\begin{multline}
 \label{eq:KG2-fourier-sr-ntime}
\ddN{\dvp2}(\kvi) + \left(3 + \frac{\dN{H}}{H}\right)
\dN{\dvp2}(\kvi)+ \left(\frac{k}{aH}\right)^2\dvp2(\kvi) \\
+ \left(\frac{\Upp}{H^2}-{24 \pi G}(\dN{\vp_{0}})^2\right)
\dvp2(\kvi) +S(\kvi) = 0 \,,
\end{multline}
with the source term given by
\begin{align}
\label{eq:KG2-source-ntime}
S(\kvi) = \frac{1}{(2\pi)^2}\int \d q\ &\Bigg\{
\frac{\Uppp}{H^2} q^2 \dvp1(\qvi) \A(\kvi,\qvi) \nonumber\\
&+\, \frac{8\pi G}{(aH)^2}\dN{\vp_{0}} \Bigg[ 
\left( 3a^2\Upp q^2 + \frac{7}{2}q^4 + 2k^2q^2\right) \A(\kvi,\qvi) \nonumber\\
&-\left(\frac{9}{2} + \frac{q^2}{k^2}\right)kq^3 \B(\kvi,\qvi)
\Bigg]\dvp1(\qvi) \nonumber\\
&+\, 8\pi G \dN{\vp_{0}} \Bigg[
-\frac{3}{2}q^2 \tilde{\C}(\kvi,\qvi) + \left(2-\frac{q^2}{k^2}\right)kq
\tilde{\D}(\kvi,\qvi) 
\Bigg]\dN{\dvp1}(\qvi) \Bigg\}\,,
\end{align}
where 
\begin{align}
\label{eq:cdtilde-num}
 \tilde{\C}(\kvi,\qvi) &= \frac{1}{aH} \C(\kvi-\qvi) = \int_0^\pi \sin(\theta)
\dN{\dvp1}(\kvi-\qvi) \d\theta \,,\nonumber \\
 \tilde{\D}(\kvi,\qvi) &= \frac{1}{aH} \D(\kvi-\qvi) = \int_0^\pi \cos(\theta)
\sin(\theta) \dN{\dvp1}(\kvi-\qvi)
\d\theta \,.
\end{align}

The arguments of $\dvp1$ and $\dN{\dvp1}$ in the $\A$--$\tilde{\D}$ terms require
special consideration. 
To compute the integrals, $\theta$ is sampled at 
\begin{equation}
\label{eq:nthetadefn}
N_\theta = 2^l + 1
\end{equation}
points in the range
$[0,\pi]$ (for some $l\in \mathbb{Z}^+$ to allow Romberg integration) and the
magnitude of
$\kvi-\qvi$
is
found using
\begin{equation}
 |\kvi -\qvi| = \sqrt{k^2 + q^2 - 2kq\cos(\theta)}\,.
\end{equation}
While $\dvp1(\kvi)=\dvp1(k)$, the value of $|\kvi-\qvi|$ is at most
$2\kmax$, where $k,q \in [\kmin,\kmax]$. This means that to calculate
the source term for the $k$ range described we require that $\dvp1$
and $\dN{\dvp1}$ be known in the range $[0, 2\kmax]$. In
Section \ref{sec:impl-num}, we will 
show that this first order upper bound does not significantly affect
performance. On the other hand, $|\kvi-\qvi|$ can also drop below the
lower cutoff of calculated $k$ modes. As discussed above a sharp cutoff will be
implemented and $\dvp1(k)=0$ used for the values below
$\kmin$. When the spacing of the discrete $k$ values, $\Delta k$,  is
approximately $\kmin$ the cutoff affects only the $k=q$ modes and
is only significant close to $\kmin$. 
Section \ref{sec:tests-num}
describes how the accuracy is affected by changing $\Delta k$ and
other parameters. Without extrapolating outside our computed $k$ range
it appears to be very difficult to avoid taking $\dvp1=0$ for a small number of
$k$ values below $\kmin$.

The value of $|\kvi-\qvi|$ will not in general coincide with the computed $k$
values of $\dvp1$. We use linear interpolation between the neighbouring $k$ values to
estimate $\dvp1$ at these points. We leave to future work the
implementation of a more
accurate and numerically more intensive interpolation scheme.

Throughout the discussion above we have not specified any particular inflationary
potential, $V$. Indeed the numerical code can use any reasonable
single field potential provided that it drives a period of inflationary expansion in
the e-folding range being simulated. In the next section the four potentials which
have been tested are outlined. 

\subsection{Potentials and Parameters}
\label{sec:pots-num}
\begin{table}[htb]
\begin{center}
\begin{tabular}{ccc}
\toprule
Potential & Parameter & Value\\
\midrule
$\msqphisq$ & $m$ & $6.32\e{-6}\Mpl$\\
$\lambdaphifour$ & $\lambda$ & $1.55\e{-13}$\\
$\phitwooverthree$ & $\sigma$ & $3.82\e{-10}\Mpl^{\frac{10}{3}}$\\
$\msqphisqwithV$ & $m_0$ & $1.74\e{-6}\Mpl$\\
\bottomrule
\end{tabular}
\caption[Parameter Values for the Four Potentials]{The parameter values for the four
potentials, chosen so
that $\mathcal{P}^2_{\mathcal{R}_1}(\kwmap)$ is in agreement with the WMAP5 value.}
\label{tab:params-num}
\end{center}
\end{table}

\begin{figure}[htbp]
\centering%
\subfloat[The potentials in terms of $\varphi$.][The potentials in terms of
$\varphi$ over the course of the background evolution.]{
\includegraphics[width=0.8\textwidth]
  {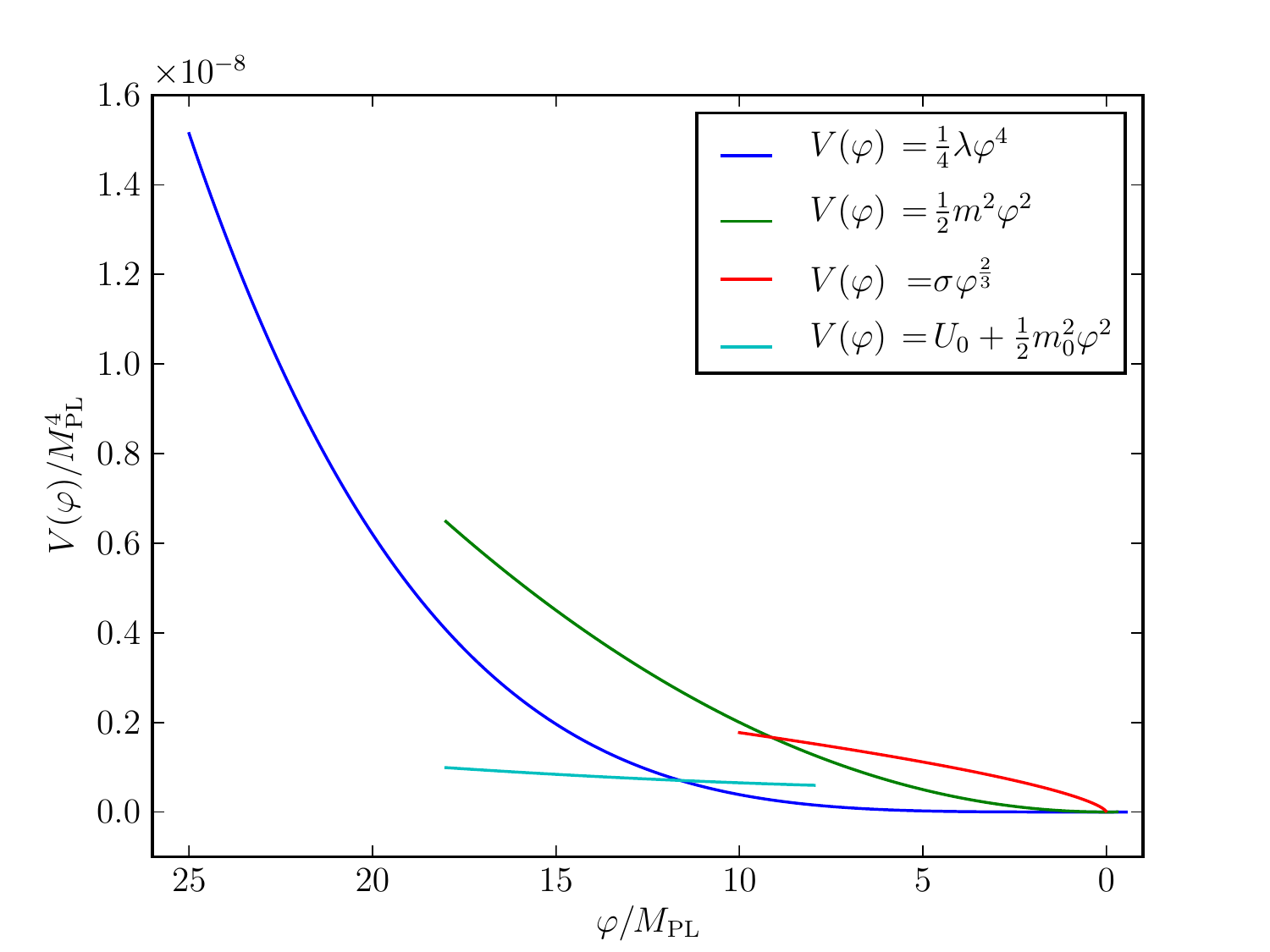}
}\\%
\subfloat[The potentials in terms of $\N$.][The potentials in terms of $\N$ for the
last 70 e-foldings of inflation.]{
 \includegraphics[width=0.8\textwidth]
  {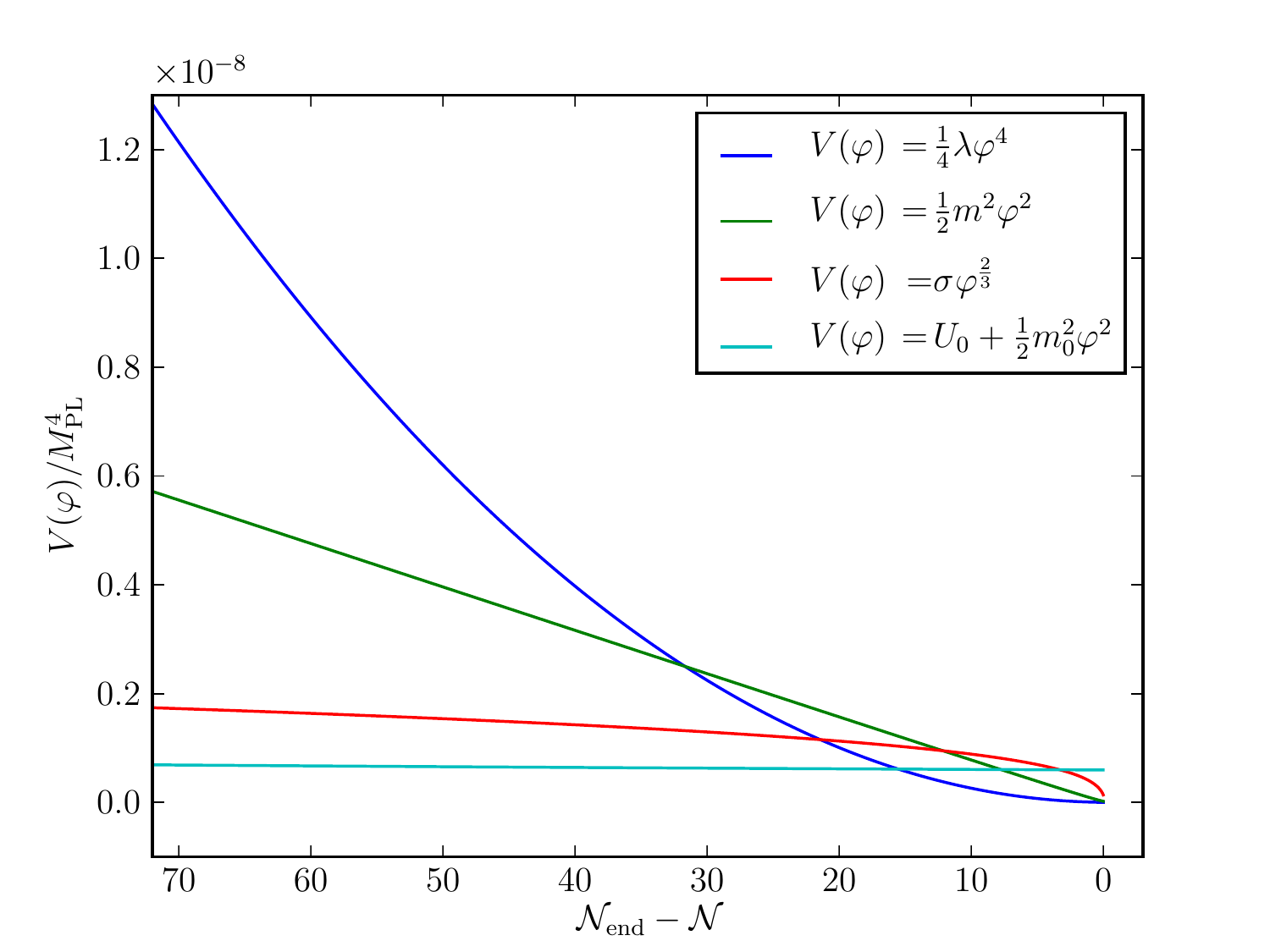}
}
\caption[Comparisons of the Potentials]{Two comparisons of the four potentials.}
\label{fig:cmp-pot-num}
\end{figure}

To demonstrate the numerical calculation four different single field, slow roll
potentials were chosen. These are not intended to represent an exhaustive selection,
but they do exhibit an interesting variety of behaviours. The potentials used are:
\begin{enumerate}
 \item $V(\vp) = \msqphisq$. This is the original chaotic inflation model which is
still in good agreement with the observational data \cite{Alabidi:2008ej}.
 \item $V(\vp) = \lambdaphifour$. Although increasingly in tension with observations
 this is a standard large field model.
 \item $V(\vp) = \phitwooverthree$. This potential with an unusual fractional index
is the effective potential resulting from the monodromy inflation model of D$4$
branes, where observable tensor modes are possible \cite{Silverstein:2008sg,
Alabidi:2008ej}.
 \item $V(\vp) = \msqphisqwithV$. This is a contrived toy model which requires inflation to
be terminated by hand. We will set inflation to end when $\vp \simeq 8$. By taking a value
of $U_0 = 5\e{-10}\Mpl^{4}$ a blue spectrum ($n_s>1$) can then be obtained
\cite{Linde:1993cn,Komatsu:2008hk}.
\end{enumerate}
\begin{figure}[htbp]
\centering%
\subfloat[$V(\vp)=\msqphisq$]{
 \includegraphics[width=0.43\textwidth]{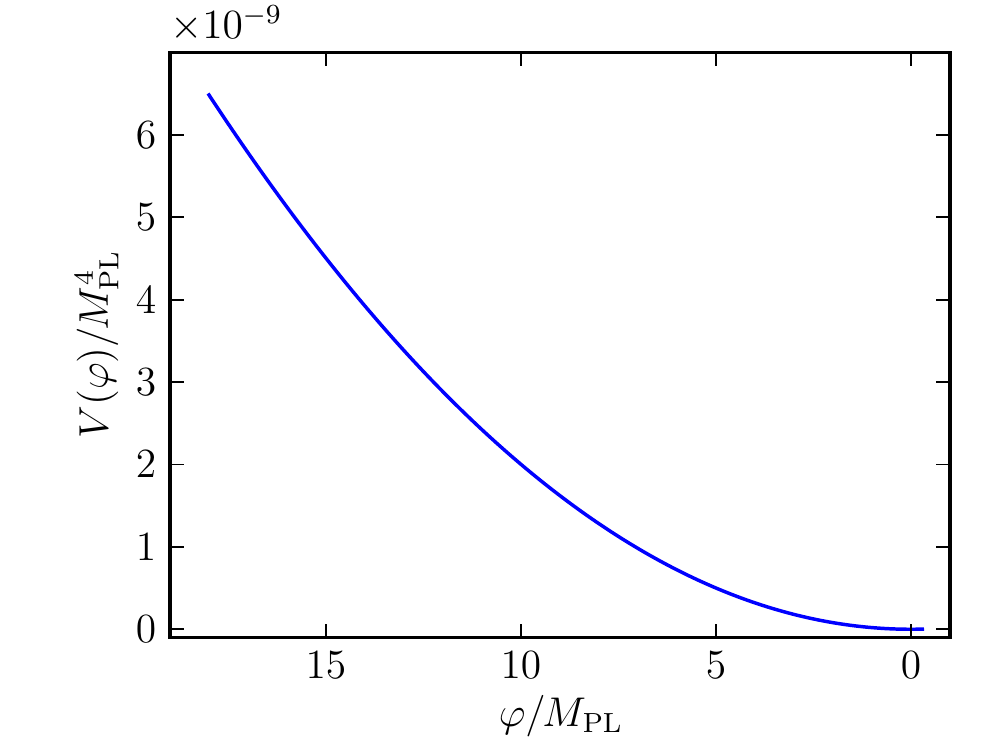}
}\qquad%
\subfloat[$V(\vp)=\lambdaphifour$]{
 \includegraphics[width=0.43\textwidth]{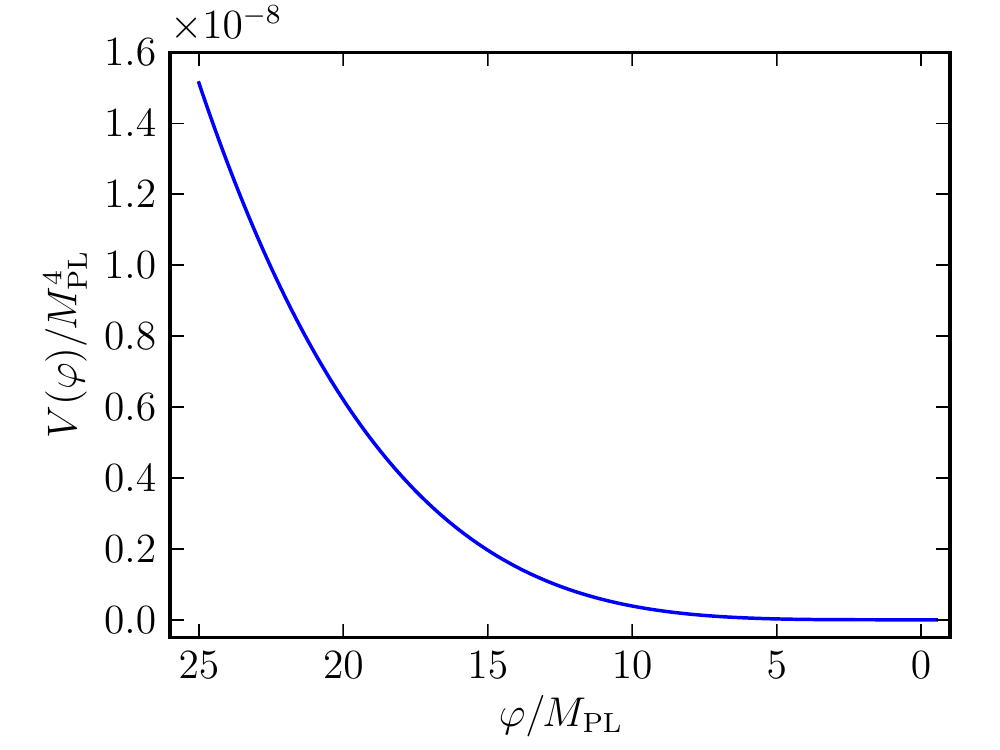}
}\\%
\subfloat[$V(\vp)=\phitwooverthree$]{
 \includegraphics[width=0.43\textwidth]{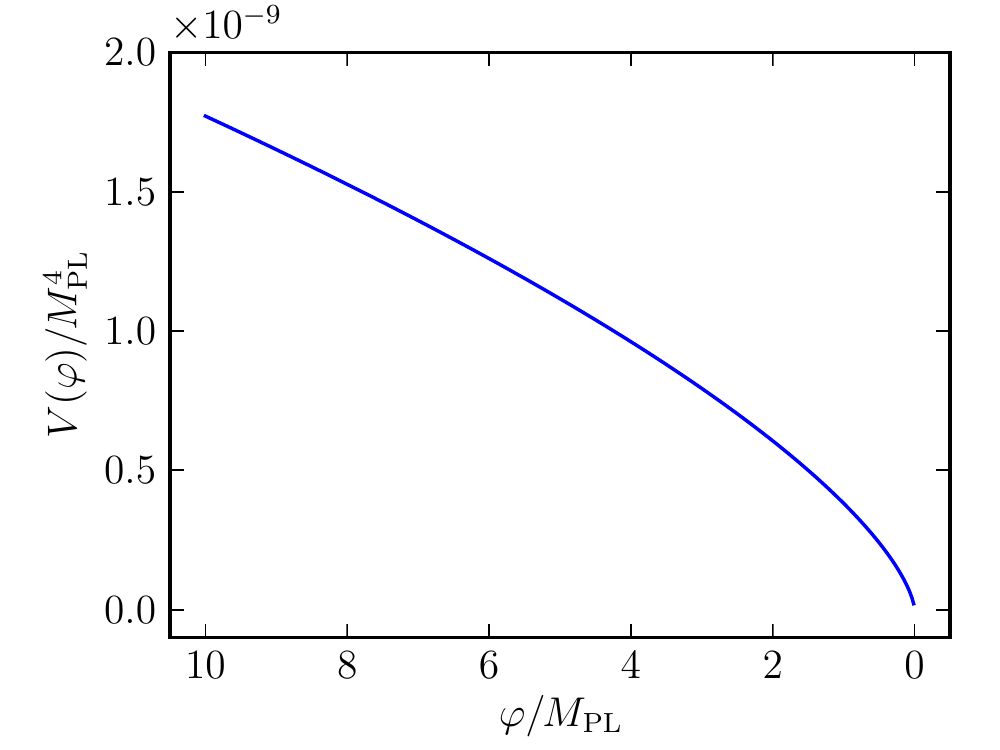}
}\qquad%
\subfloat[$V(\vp)=\msqphisqwithV$]{
 \includegraphics[width=0.43\textwidth]
  {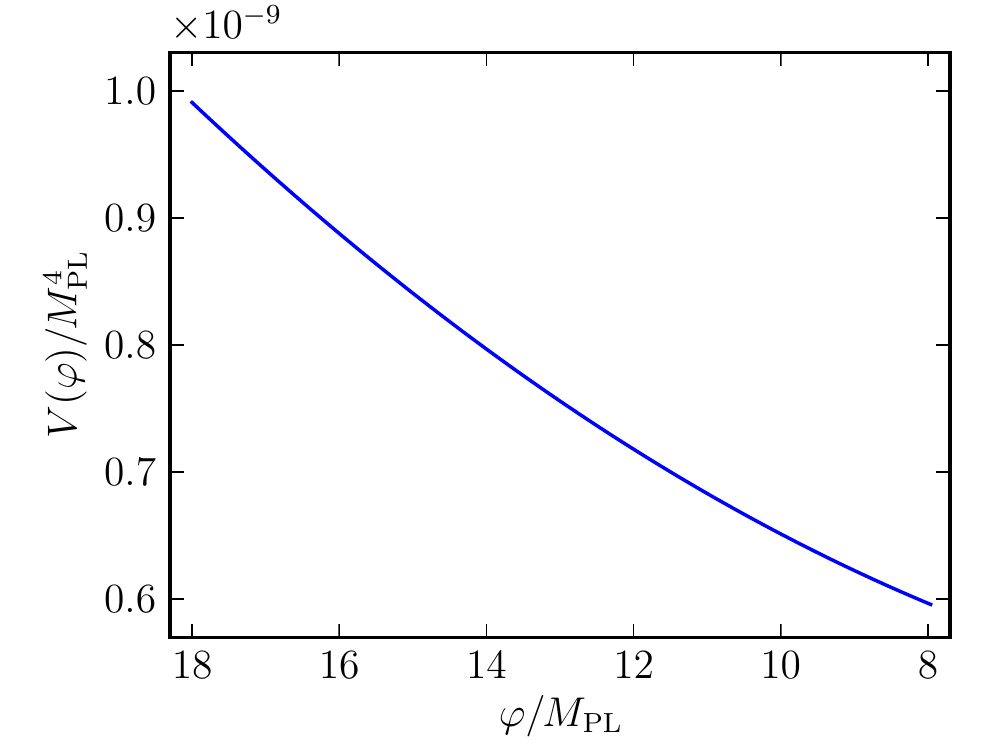}
}
\caption[The Four Potentials]{Plots of the four different potentials investigated.}
\label{fig:potentials-num}
\end{figure}
In Figures~\ref{fig:potentials-num} and \ref{fig:cmp-pot-num} the potentials are
plotted over the course of their evolution. Throughout the rest of this
chapter we will use the quadratic model to
demonstrate the calculation unless otherwise stated. In Chapter~\ref{ch:results} the
results for each potential will be compared.

\begin{figure}[htbp]
\centering%
\subfloat[Fixing $m$ for $V(\vp)=\msqphisq$.]{
 \includegraphics[width=0.43\textwidth]{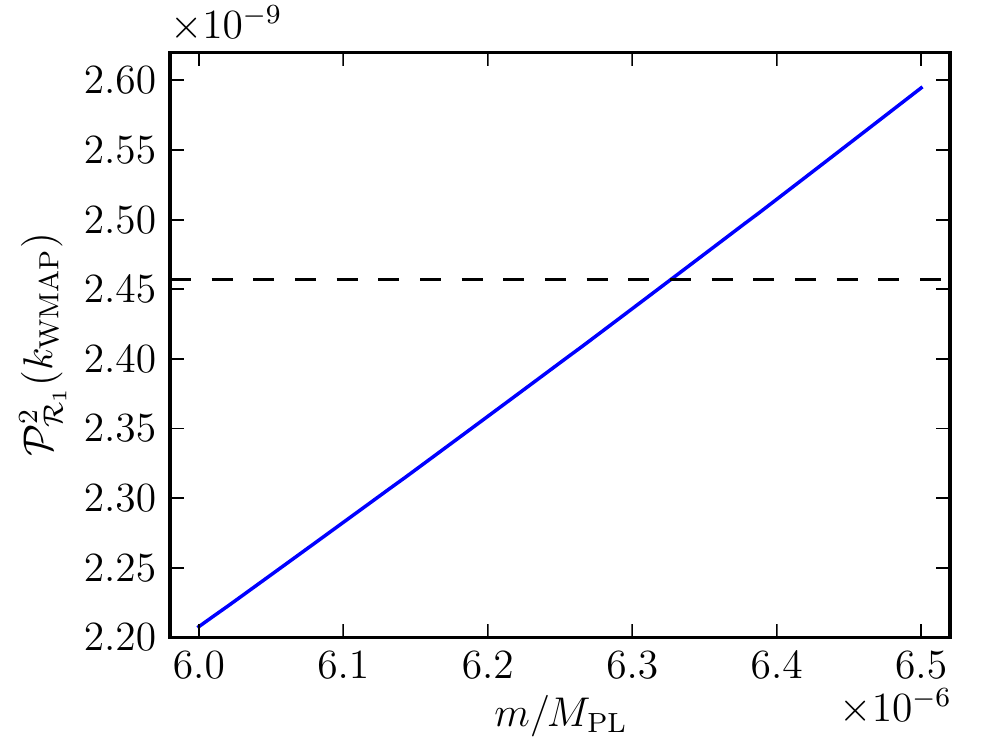}
}\qquad%
\subfloat[Fixing $\lambda$ for $V(\vp)=\lambdaphifour$.]{
 \includegraphics[width=0.43\textwidth]{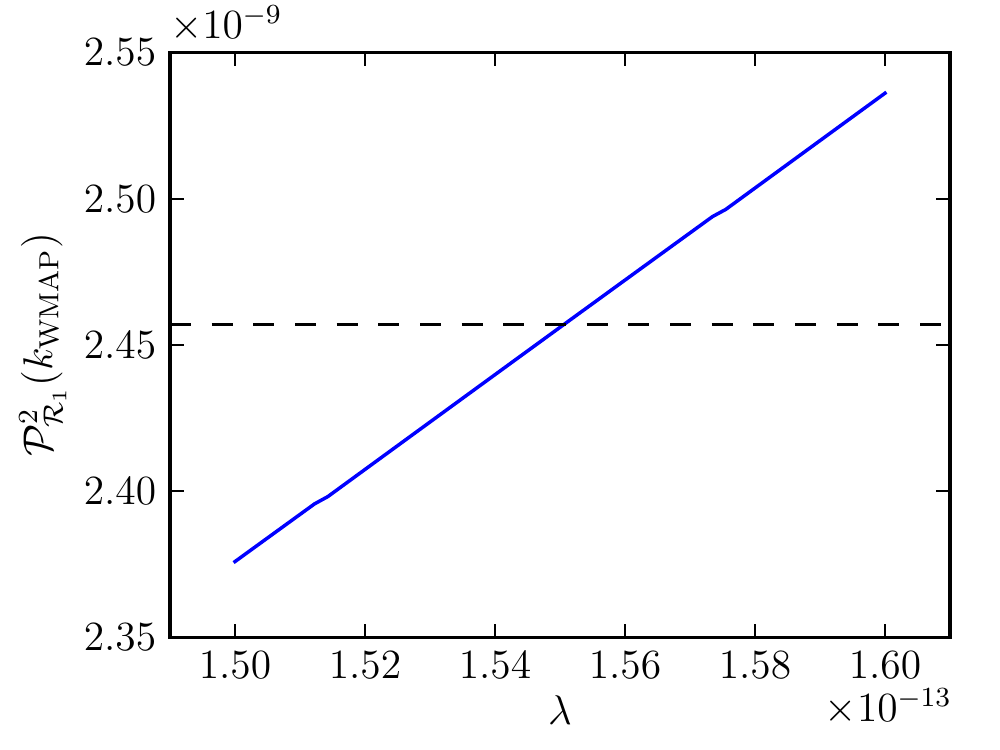}
}\\%
\subfloat[Fixing $\sigma$ for $V(\vp)=\phitwooverthree$.]{
 \includegraphics[width=0.43\textwidth]{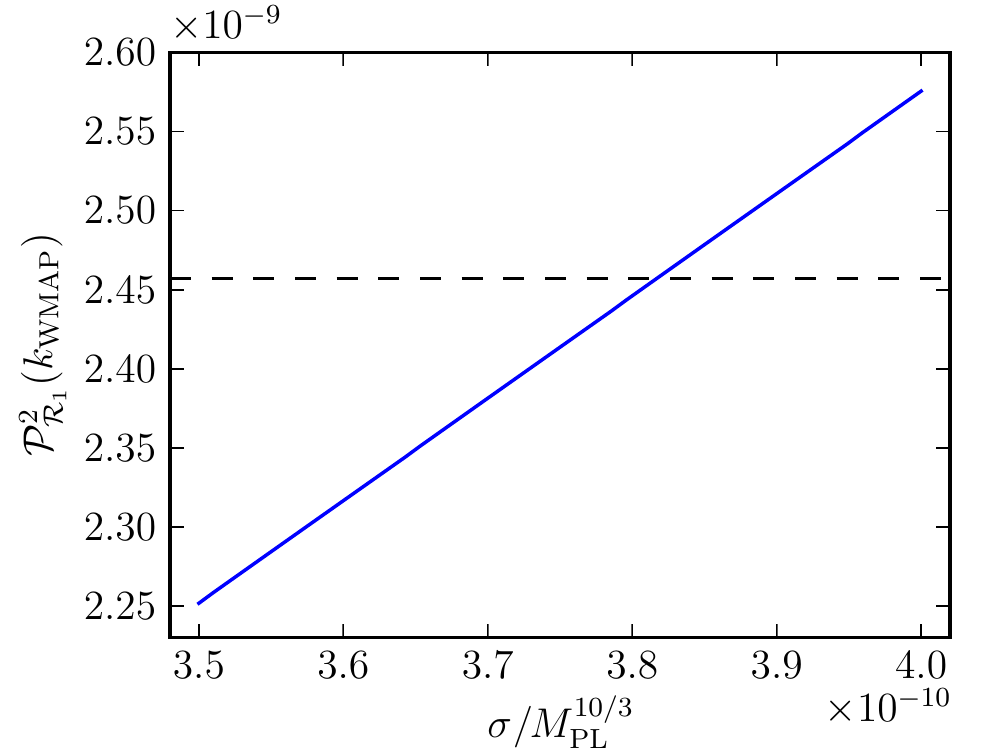}
}\qquad%
\subfloat[Fixing $m$ for $V(\vp)=\msqphisqwithV$.]{
 \label{fig:mv0-params-num}
 \includegraphics[width=0.43\textwidth]
  {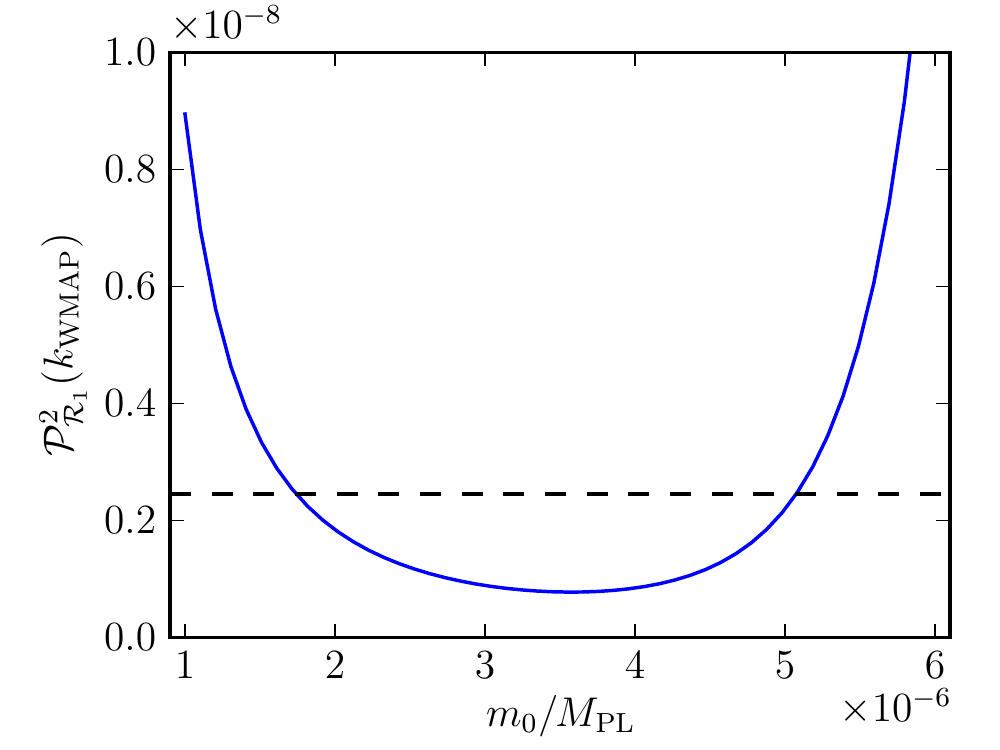}
}
\caption[Parameter Values for the Potentials]{Parameter values for the
different potentials are chosen by requiring consistency with the WMAP5
normalisation of the first order power spectrum.}
\label{fig:params-num}
\end{figure}

For each of the chosen potentials there is one free parameter that needs to be
determined.
We choose the parameters $m$, $\lambda$, $\sigma$ and $m_0$ so that
$\mathcal{P}^2_{\mathcal{R}_1}$ calculated
for each model is in agreement with the
WMAP5 value at the pivot scale
$\kwmap=0.002\Mpc^{-1} \simeq5.25\times10^{- 60} \Mpl$. 
The dependence of $\mathcal{P}^2_{\mathcal{R}_1}(\kwmap)$ on each of the parameters
can be seen in
Figure~\ref{fig:params-num}.
Requiring $\mathcal{P}^2_{\mathcal{R}_1}(\kwmap)=2.457\e{-9}$
gives the values shown in Table~\ref{tab:params-num}. Here we have chosen the lower
of the two possible values of $m_0$ shown in Figure~\ref{fig:mv0-params-num}.

\subsection{Initial Conditions} 
\label{sec:initconds-num}

The background system requires initial conditions for $\vp_{0}$ and
$\dN{\vp_{0}}$. These initial conditions and the range of
e-foldings to be simulated must be selected with the choice of
potential in mind. Not only must the
e-folding range include an inflationary period, but the $k$ modes to
be calculated at first and second order must begin inside the horizon. For example,
the initial value $\vp_0 = 18\Mpl$ for the
$\msqphisq$ model 
gives the background evolution described below and shown in Figure~\ref{fig:eps}. As
the evolution quickly reaches the attractor solution, the choice for $\dN{\vp_0}$
is not particularly important; changing the initial value adds or subtracts a small
number of e-foldings of evolution before the modes are initialised
\cite{Ringeval:2007am, Martin:2006rs}.

The initial conditions are set for each $k$ mode a few e-foldings
before horizon crossing. This follows Salopek
\etal
\cite{Salopek:1988qh} and is justified on the basis that the mode is
sufficiently far inside the
horizon for the Minkowski limit to be taken. This initial time,
$\N_{\mathrm{init}}(k)$, is calculated to be when
\begin{equation}
 \frac{k}{(aH)|_{\mathrm{init}}} = 50 \,.
\end{equation}
The range of e-foldings being used must include the starting point for
all $k$ modes, but the parameter on the right hand side, here chosen to
be 50, can be changed if needed.  We use the small wavelength solution
of the first order equations described in Section~\ref{sec:perts-intro} as the
initial conditions \cite{Salopek:1988qh}, with
\begin{align}
\label{eq:foics}
 \dvp1|_{\mathrm{init}} &= \frac{\sqrt{8\pi G}}{a}
\frac{e^{-i k\eta}}{\sqrt{2k}} \,,\\
 \dN{\dvp1}|_{\mathrm{init}} &= -\frac{\sqrt{8\pi G}}{a}
\frac{e^{-i k\eta}}{\sqrt{2k}} \left(1 + i \frac{k}{a H}\right) \,,
\end{align}
where conformal time $\eta$ can be calculated from 
\begin{equation}
\label{eq:etacalc-num}
\eta=\int \d\N/aH \simeq
-(aH(1-\bar{\varepsilon}^2_H))^{-1}\,,
\end{equation} 
when $\bar{\varepsilon}_H$ changes slowly. For
example $\kwmap$ is initialised
about $65$ e-foldings before the end of inflation and crosses the horizon about $5$ e-foldings
later.
We also use these formulae in the calculation of the source term in \eq{eq:KG2-source-ntime} to
determine the value of $\dvp1$ for a $k$ mode before its numerical evolution has
begun.

We are interested in the production of second order effects by the
evolution of the the Gaussian first order modes and we make no
assumptions about the existence of second order perturbations before
the simulation begins. Therefore, we set the initial condition for each second order
perturbation mode to be $\dvp2=0, \dN{\dvp2}=0$ at
the time when the corresponding first order perturbation is initialised.
One argument in favour of this choice of initial conditions is that far in the past the perturbations
are assumed to be Gaussian and therefore the second (and higher) order perturbations would be identically zero. 

A numerical solution for the second order perturbation equation will contain a homogeneous solution
and a particular solution. 
The homogeneous part of the solution of
the slow roll equation, \eq{eq:KG2-fourier-sr-ntime}, can be calculated analytically 
as done in Appendix~\ref{sec:init-apx}. On their own the initial conditions we have chosen above do not  
remove this homogeneous solution from the result for $\dvp2$ in general. In order to do this, and keep only the 
particular solution to the equation, it is necessary to ensure that the homogeneous solution is the trivial $(0,0)$ 
solution throughout the evolution. We have not attempted to do this in this thesis but it is an important
issue for further study in the future. Approaches to removing the homogeneous part of the solution include
calculating a semi-analytic value for the second order initial conditions which equals the particular solution
 or numerically trying to select the trivial
homogeneous solution by introducing a ramping function to the source term in \eq{eq:KG2-source-ntime}.
In summary, the results quoted in Chapter~\ref{ch:results} for the second order perturbations include both a homogeneous 
and particular solution. Extraction of either of these parts from the full result remains an issue for future study.

\section{Implementation} 
\label{sec:impl-num}

\begin{figure}[htbp]
\centering
 \includegraphics[width=0.8\textwidth]{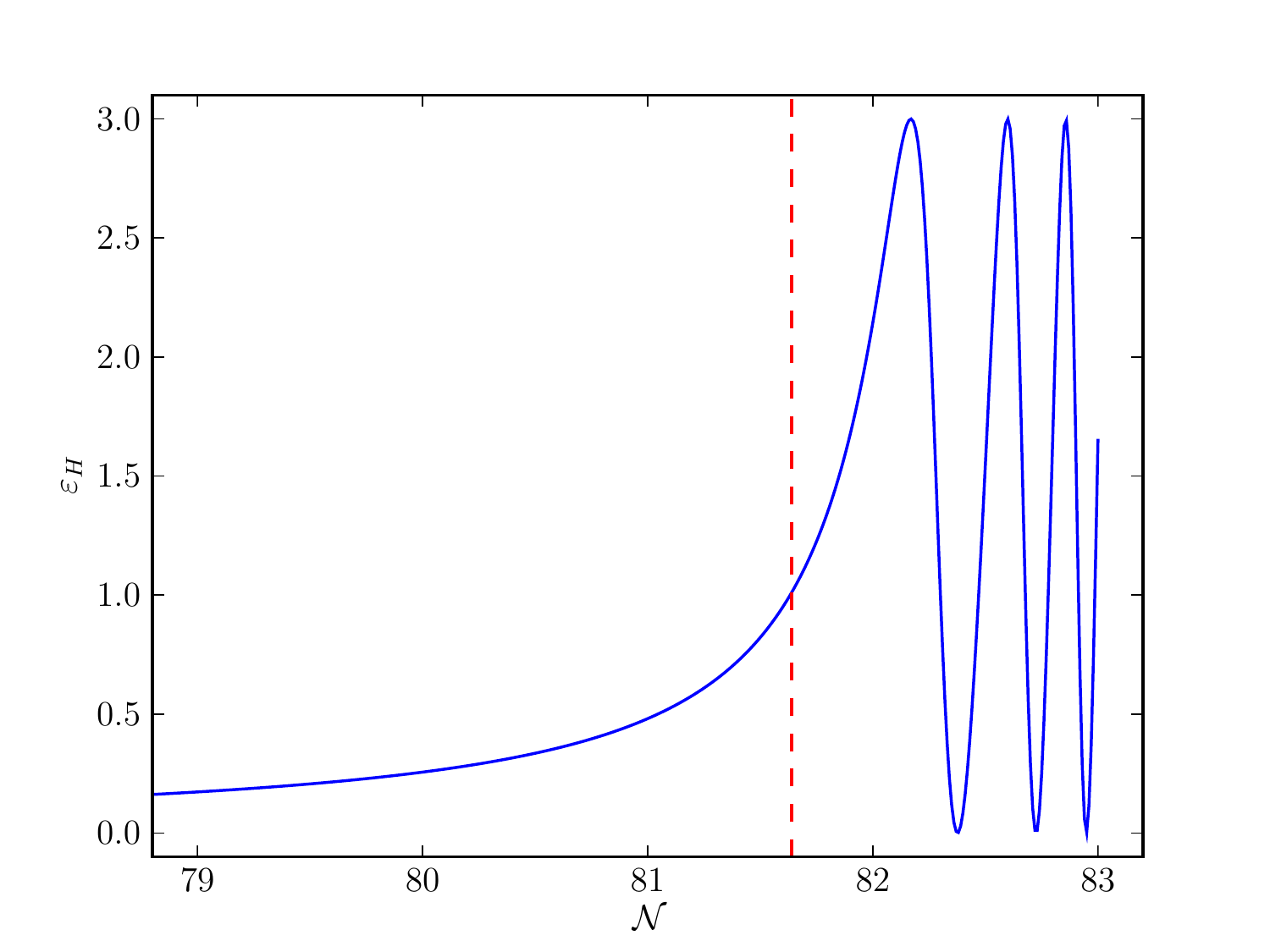}
 \caption[The Value of $\varepsilon_H$ Near the End of Inflation]{The end of
inflation is
determined by calculating when
   $\bar{\varepsilon}^2_H = \varepsilon_H=-\dN{H}/H=1$ (red dashed line). On the
horizontal axis,
   $\N$ is the number of e-foldings from the start of the
   simulation.}
\label{fig:eps}
\end{figure}

The current implementation of the code is mainly in the Python\footnote{Python
website: \url{http://www.python.org}} programming language (with compiled Cython
components) and uses the
Numerical and Scientific Python modules for their strong compiled array support
\cite{scipy}. The core of the model computation is a
Runge-Kutta fourth order method (see, for example, Eq.~(25.5.10) in
\cite{abramowitz+stegun}).  Following
Refs.~\cite{Martin:2006rs} and \cite{Ringeval:2007am}, the numerical calculation
proceeds in four stages. The background equation (\ref{eq:bgntime}),
rewritten as two first order equations, is
evolved from the specified initial state until some end time required
to be after the end of the inflationary regime.  The end of inflation
occurs when $\d^2a/\d t^2$ is no longer positive and the parameter
$\bar{\varepsilon}^2_H = \varepsilon_H= -\dN{H}/H$ becomes greater than or
equal to unity
(see Figure~\ref{fig:eps}). This specifies a new end time for the first
order run, although the simulation can run beyond the strict end of
inflation if required. For the $V(\vp)=\msqphisqwithV$ model, the end of inflation is
set by hand to remove the need for a second, inflation-terminating field. The initial
conditions for the first order system are then set as above.

The system of ordinary differential equations for the first order
perturbations in \eq{eq:fontime} is integrated using a standard fourth order
Runge-Kutta method. A fixed time step method is used in order to
simplify the construction of the  second order source term.
This is also necessary since it is not known a priori
which time steps would be required at
second order if an adaptive time step system were used. The first
order equations are separable in terms of $k$ and so it is
straightforward to run the system in parallel and collate
the results at the end. However, as will be discussed below, the first
order calculation is not computationally expensive in comparison with
the other stages and only takes of the order of a few minutes for around
$8000$ time steps with $\Delta \N=0.01$ and $1025$ $k$ modes.

Once the first order system has been solved, 
the source term for the second order system must be calculated. As the
real space equation for the source involves terms quadratic in the
first order perturbation, it is necessary to perform a convolution in
Fourier space, as shown in \eq{eq:KG2-fourier-sr-aterms}.  We do not transform
back into real space due to the presence of both
gradient operators and their inverses. 
Instead, the slow roll version of
the source term integrand was used, although it is worth remarking that the method
can also be
applied to the full equation. This stage is computationally the most
intensive, and can be run in parallel since the calculation at each time step is
independent of the others. The nature of the convolution integral and the
dependence of the first order perturbation on the absolute value of
its arguments requires that twice as many $k$ modes are calculated at
first order than are desired at second order.  As
the first order calculation is computationally cheaper than the source
term integration, this does not significantly lower the possible
resolution in $k$-space, which is still limited by the source term
computation time.  Once the integrand is determined, it is fed into a
Romberg integration scheme. As for $\theta$,  which was
discretised by $N_\theta$ points in \eq{eq:AtoD-num}, this requires that the
number of $k$ modes is
\begin{equation}
\label{eq:nk-constraint-num}
N_k=2^l + 1\,,
\end{equation}
for some\footnotemark $\,l\in\mathbb{Z}^+$. 
\footnotetext{The number of discretised $k$ modes $N_k$ does not need to be equal to
  $N_\theta$.}
This requirement can be relaxed by opting for a less
accurate and somewhat slower standard quadrature routine.

The second order system is finally run with the source term and other
necessary data being read as required from the memory or disk. The
Runge-Kutta method calculates half time steps for each required point.
For example, if $y(x_n)$ is known and $y(x_{n+1})=y(x_n+h)$ is required
(for step size $h$), the method will calculate the derivatives of $y$
at $y(x_n), y(x_n +h/2)$ and $y(x_n + h)$. As we need to specify the
source term at every calculated time step, the requested time step for
the second order method must be twice that used at first order.  This decreases the
accuracy of the method, but does not require the use
of splines and interpolation techniques to determine background and
first order variables between time steps.

The second order system is similar in run time to the first order 
system. However, the
source integration is more complex and involves the
integration of $N_k^2\times N_\theta$ values at
each time step.
When $N_k=1025$ and $N_\theta=513$, the first order evolution lasts around 200
seconds. The source calculation, on the other hand, takes approximately 200 seconds
for each time step. Each of the four terms $\A-\wt{\D}$ is approximately 16 gigabytes
in size at each time step for these values of $N_k$ and $N_\theta$. However, only the
integrated result is stored for use
in the second order run. This is approximately $16$ kilobytes in size for each time
step. 
Results for each stage are stored in the open HDF5 standard
\cite{pytables, hdf5}, which can
deal efficiently with large
files, is very portable and allows for data analysis independent of the
Python/Numpy programming environment.

The full calculation contains around 8000 time steps, making the source term
calculation approximately 470 hours long. Each time step is independent of the
others, however, so parallelisation of the system is straightforward. The results in
Chapter~\ref{ch:results} were obtained on the Virgo Cluster in the Astronomy Unit at
Queen Mary, University of London. The code was run on ten nodes, each containing
four Opteron cores with a clock speed of $1994$Mhz. With this configuration the run
time of the source term calculation is reduced to under twelve hours.

\section{Code Tests}
\label{sec:tests-num}

The numerical code has been tested in a variety of controlled
circumstances in order to quantify the effects of different parameter options. In
particular, it is important to establish whether the values
specified for the number of discretised $\theta$'s, $N_\theta$, the size
of the
spacing of the discretised $k$ modes, $\Delta k$, and the range of
$k$ values significantly impact on the results. The sections
of the code that solve ODEs are straightforward and follow standard algorithms.

As mentioned above, the WMAP results \cite{Komatsu:2008hk} use
observations in the range $k\in [0.92 \e{-60}, 3.1 \times
  10^{-58}]\Mpl = [3.5\e{-4}, 0.12] \Mpc^{-1}$. We will
consider three different $k$ ranges both in our results and the tests
of the code\footnotemark:
\begin{align}
\label{eq:Krangedefns}
K_1 &= \left[1.9\e{-5}, 0.039\right]\Mpc^{-1}\,, &\Delta k = 3.8\e{-5}\Mpc^{-1}
\,,\nonumber\\
K_2 &= \left[5.71\e{-5}, 0.12\right]\Mpc^{-1}\,, &\Delta k =
1.2\e{-4}\Mpc^{-1}\,,
\nonumber\\ 
K_3 &= \left[9.52\e{-5}, 0.39\right]\Mpc^{-1}\,, &\Delta k = 3.8\e{-4}\Mpc^{-1}
\,.
\end{align}
\footnotetext{The $k$ ranges in $\Mpl$ are:
\begin{align*}
\label{eq:Krangedefns-mpl}
K_1 &= \left[0.5\e{-61}, 1.0245\e{-58}\right]\Mpl\,, &\Delta k = 1\e{-61}\Mpl\,,
\nonumber\\
K_2 &= \left[1.5\times 10^{-61}, 3.0735\e{-58}\right]\Mpl\,, &\Delta k =
3\e{-61}\Mpl\,,
\nonumber\\ 
K_3 &= \left[0.25\e{-60}, 1.02425\e{-57}\right]\Mpl\,, &\Delta k = 1\e{-60}\Mpl
\,.
\end{align*}
}
The first, $K_1$, has a very fine resolution but covers only a small portion of the WMAP range. 
The next, $K_2$, is closest to the WMAP range and still has quite a fine resolution. 
The final
range, $K_3$, has a larger $k$ mode step size, $\Delta k = 1\e{-60}\Mpl =
3.8\e{-4}\Mpc^{-1}$, and
covers a greater range than the others. It extends to much smaller scales than WMAP can observe.

\begin{figure}[htbp]
 \centering
 \includegraphics[width=0.8\textwidth]
    {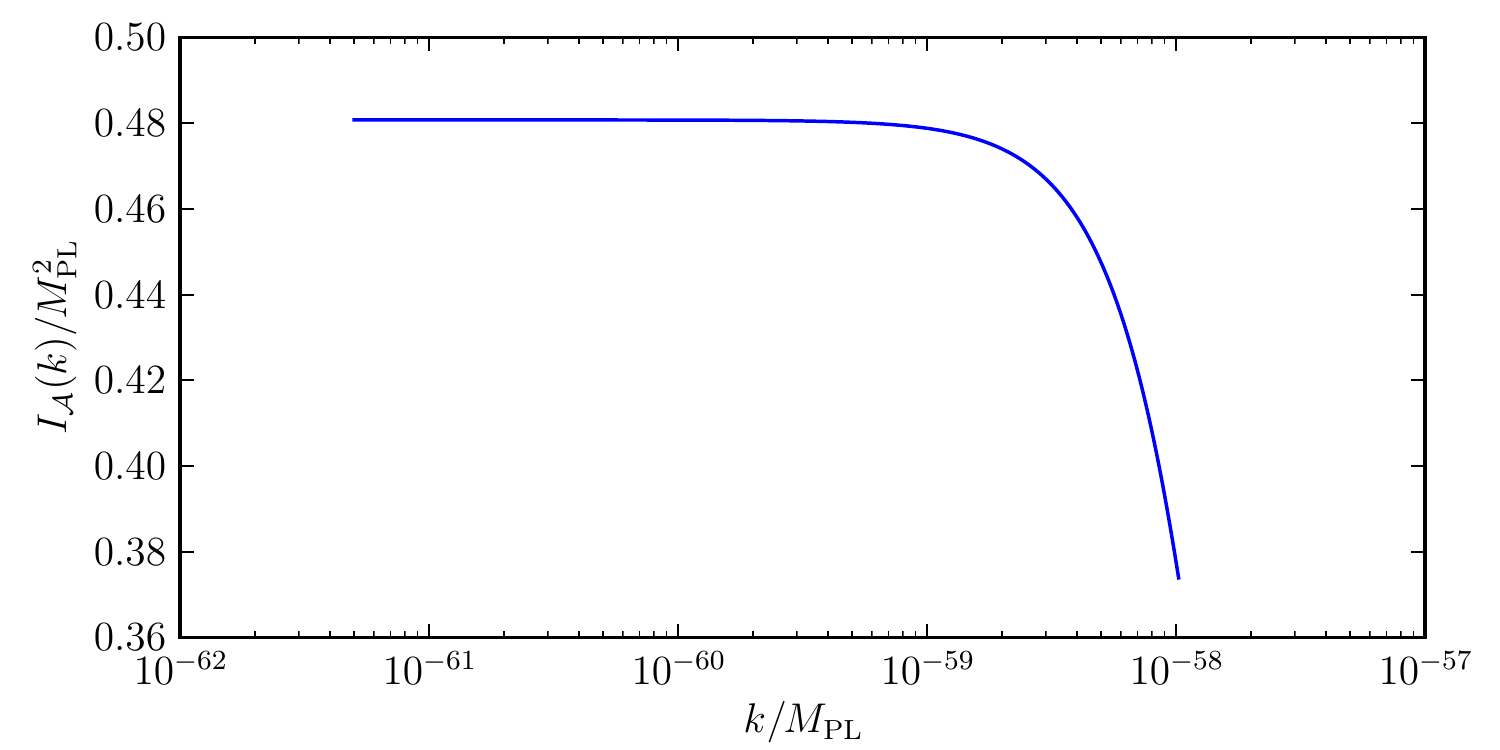}
 \caption[The Analytic Solution of $I_\A$]{The analytic solution of $I_\A$ given in
\eq{eq:err-analytic-num} for
$k\in K_1$. The value of $\alpha$ is set as $2.7\e{57}$.}
 \label{fig:inta-num}
\end{figure}

\begin{figure}[htbp]
\centering
 \subfloat[Relative error for different $N_\theta$][The relative error for different
$N_\theta$, the number of
discretised $\theta$s, keeping the other parameters fixed and using the $K_3$
range. The upper blue line ($N_\theta=129$) and middle green line
 ($N_\theta=257$) have relative errors at least an order of magnitude larger
than the lower red line ($N_\theta=513$).]{
 \includegraphics[width=0.43\textwidth]{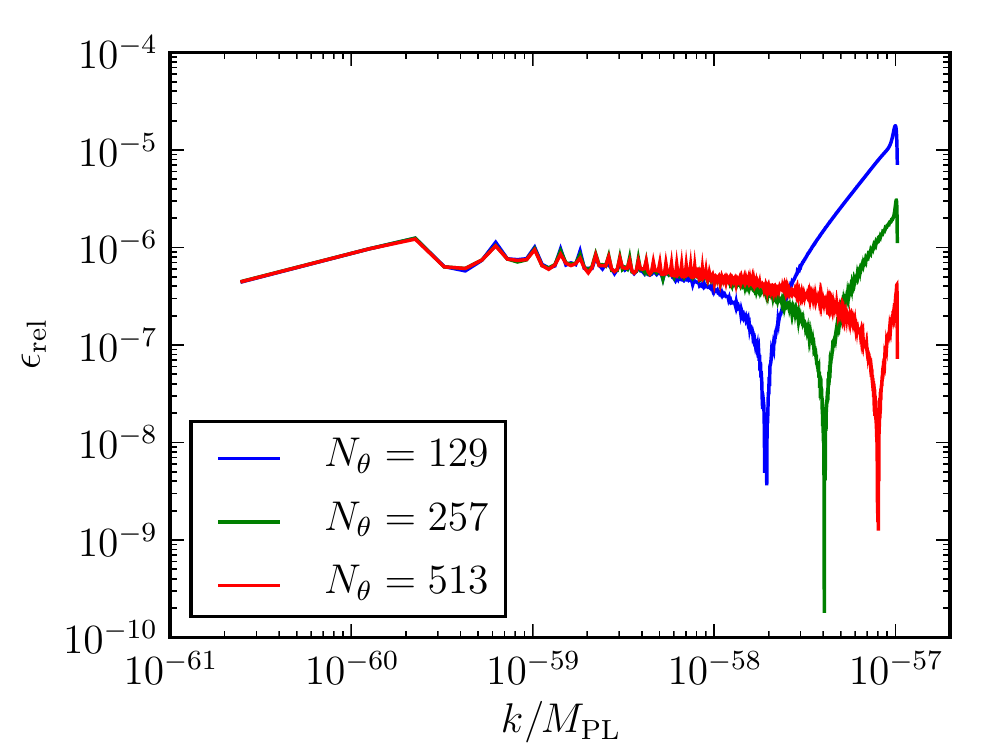}
 \label{fig:err-nthetas}
}\qquad
\subfloat[Relative error for three different $k$ ranges][The relative error for the
three
different $k$ ranges $K_1$, $K_2$,
$K_3$ (starting from the left). The parameter $\Delta k$
is set equal to $1\e{-61}\Mpl, 3\e{-61}\Mpl, 1\e{-60}\Mpl$ respectively.]{
 \includegraphics[width=0.43\textwidth]{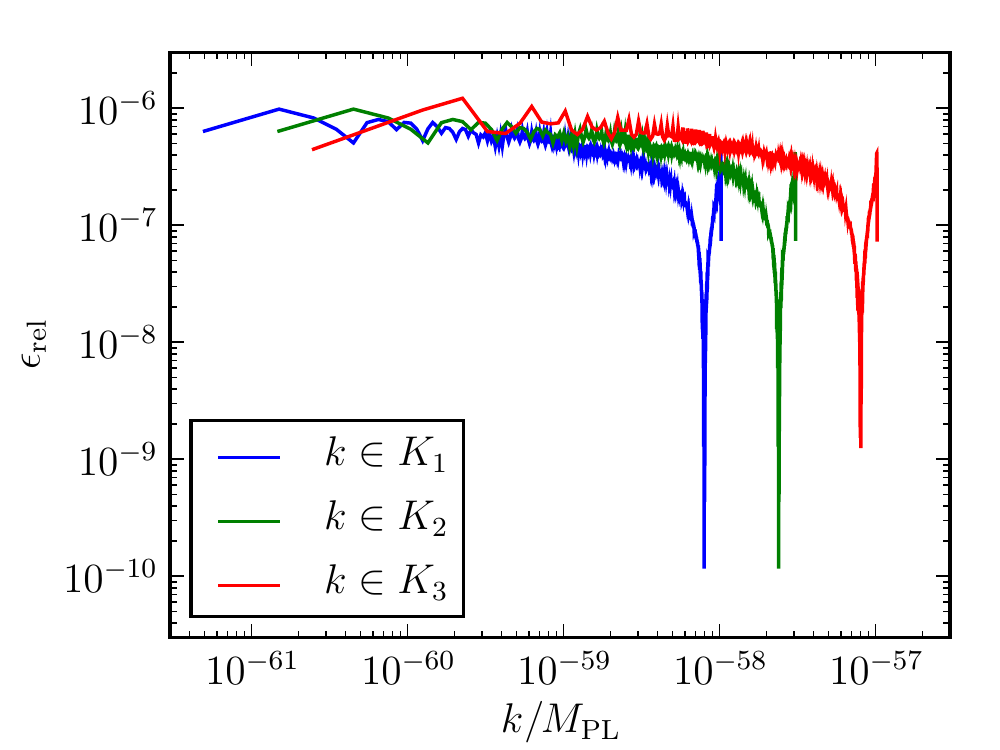}
 \label{fig:err-kranges}
}
\caption[Relative Errors in $I_\A$ for Different $N_\theta$ and $k$
Ranges]{Comparison of relative errors in $I_\A$ for
different $N_\theta$ and $k$ ranges.}
\label{fig:err-comparison}
\end{figure}

The central calculation in the code is of the
convolution of the perturbations for the source term,
\eq{eq:KG2-source-ntime}. The first of the $\theta$ dependent terms in
\eq{eq:AtoD-num}, $\A$,
can be convolved analytically for certain smooth choices of $\dvp1(k)$. 
Taking $\dvp1(k)$ to be similar in form to the initial conditions
(\ref{eq:foics}) gives $\dvp1(k)\propto 1/\sqrt{k}$ with proportionality constant
$\alpha$.
If $I_\A$ denotes the following integral of the $\A$ term:
\begin{equation}
 I_\A (k) = \int \d q^3 \dvp1(\qvi) \dvp1(\kvi-\qvi) 
          = 2\pi \int_{\kmin}^{\kmax} \d q\, q^2 \dvp1(\qvi) \A(\kvi, \qvi)\,,
\end{equation}
then substituting $\dvp1(k) = \alpha/\sqrt{k}$ implies that
\begin{equation}
 I_\A(k) = 2\pi\alpha^2 \int_{\kmin}^{\kmax} \d q\, q^{\frac{3}{2}}
\int_{0}^{\pi} \d\theta\, (k^2 + q^2 -2k q \cos{\theta})^{-1/4} \sin{\theta}\,. 
\end{equation}
This has the analytic solution
\begin{align}
\label{eq:err-analytic-num}
 I_\A(k) = -\frac{\pi\alpha^2}{18 k}\Bigg\{ 
	&3 k^3 \Bigg[ \log\Biggl(\frac{\sqrt{\kmax-k} + \sqrt{\kmax}}{\sqrt{k}}
			    \Biggr)
	 + \log\Biggl( \frac{\sqrt{k+\kmax} +\sqrt{\kmax}}{\sqrt{\kmin+k} +
		      \sqrt{\kmin}}\Biggr) \nonumber \\
	&+\frac{\pi}{2} - \arctan\left( \frac{\sqrt{\kmin}}{\sqrt{k-\kmin}}\right)
	\Bigg] \nonumber\\
        &-\sqrt{\kmax}\Bigg[ \left(3k^2 + 8\kmax^2 \right)\left(\sqrt{k+\kmax} -
	  \sqrt{\kmax-k}\right) \nonumber \\
	&\qquad + 14k\kmax\left(\sqrt{k+\kmax}+\sqrt{\kmax-k}\right)\Bigg]\nonumber\\
	&+\sqrt{\kmin}\Bigg[ \left(3k^2 + 8\kmin^2 \right)\left(\sqrt{k+\kmin} +
	  \sqrt{k-\kmin}\right) \nonumber \\
	&\qquad +14k\kmin \left(\sqrt{k+\kmin} -
         \sqrt{k-\kmin} \right) \Bigg] \Bigg\} \,.
\end{align}
The $k$ dependence of $I_\A$ can be seen in Figure~\ref{fig:inta-num}. 
We have tested our code against this analytic solution for various
combinations of $k$ ranges and $N_\theta$. The relative error
\begin{equation}
 \epsilon_\mathrm{rel} = \frac{|\mathrm{analytic}- \mathrm{calculated}
|}{|\mathrm{analytic}|}
\end{equation}
is small for all the tested cases, but certain combinations of
parameters turn out to be more accurate than others. The relative errors of
all the following results are not affected by the choice of $\alpha$ so
we will keep its numerical value fixed throughout as $2.7\e{57}$.

\begin{figure}[htbp]
 \centering
 \includegraphics[width=0.8\textwidth]{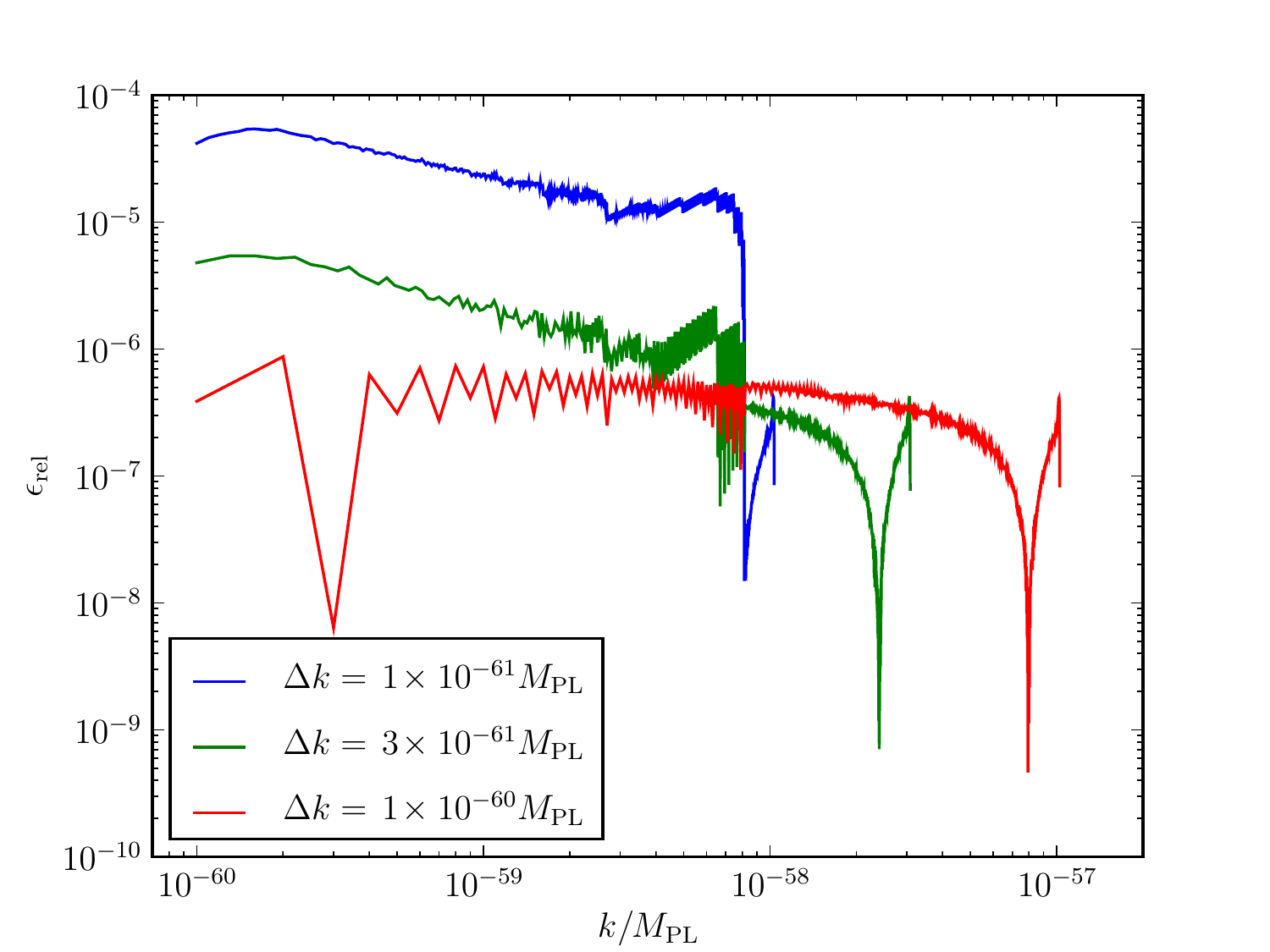}
 \caption[Relative Error in $I_\A$ for Different $\Delta k$]{The relative error in
the
integral $I_\A$ for different values of $\Delta k$.
The other parameters are fixed: $\kmin=1\times10^{-60}\Mpl, N_k = 1025$ and
$N_\theta=513$. 
The value of $\Delta k$ is less than $\kmin$ for the upper 
blue line ($\Delta k=1\e{-61}\Mpl$) and the middle green line ($\Delta
k=3\e{-61}\Mpl$). These have relative errors at least an order of magnitude larger
than the lower red line for which $\Delta k=\kmin=1\e{-60}\Mpl$.}
 \label{fig:err-deltaks}
\end{figure}

We first tested the effect of changing $N_\theta$, the number of
samples of the $\theta$ range $[0,\pi]$.  Figure~\ref{fig:err-nthetas}
plots these results for the $k$ range $K_3$ with $\Delta k =
1\e{-60}\Mpl$. Only three values of $N_\theta$ are shown for clarity. It
can be seen that increasing $N_\theta$ decreases the relative error when the other
parameters are kept constant, as one
might expect.

As mentioned above the choice of $k$ range is especially important as
the convolution of the terms depends strongly on the minimum and
maximum values of this range. Indeed, this is clear from the analytic
solution in \eq{eq:err-analytic-num}. Figure~\ref{fig:err-kranges}
shows the difference in relative error for the three different $k$
ranges described above with 
$\Delta k= 3.8\e{-5}, 1.2\e{-4}$ and $3.8\e{-4}\Mpc^{-1}$
($\Delta k= 1\e{-61}, 3\e{-61},1\e{-60}\Mpl$),
respectively. The accuracy is similar in all three cases.

Another important check is whether the resolution of the $k$ range is
fine enough. Varying $\Delta k$ can not be done in isolation if the
constraint \eqref{eq:nk-constraint-num} for $N_k$ is to
be satisfied. For this test the end of the $k$ range was changed with $\Delta k$
but the other parameters were kept fixed at $\kmin=1\e{-60}\Mpl=3.8\e{-4}\Mpc^{-1},
N_k = 1025$ and $N_\theta=513$. Figure~\ref{fig:err-deltaks} plots
these results again for three indicative values.  For $\Delta
k<\kmin$, there is a marked degradation in
the accuracy of the method for the upper two lines. This is understandable as many
interpolations of multiples of $\Delta k$ below $\kmin$ will be set to
zero. Once $\Delta k$ is greater than $\kmin$, the relative error is
insensitive to further increases in the value of $\Delta k$. (This is not shown in
the figure.)

\begin{figure}[htbp]
 \centering
 \includegraphics[width=0.8\textwidth]{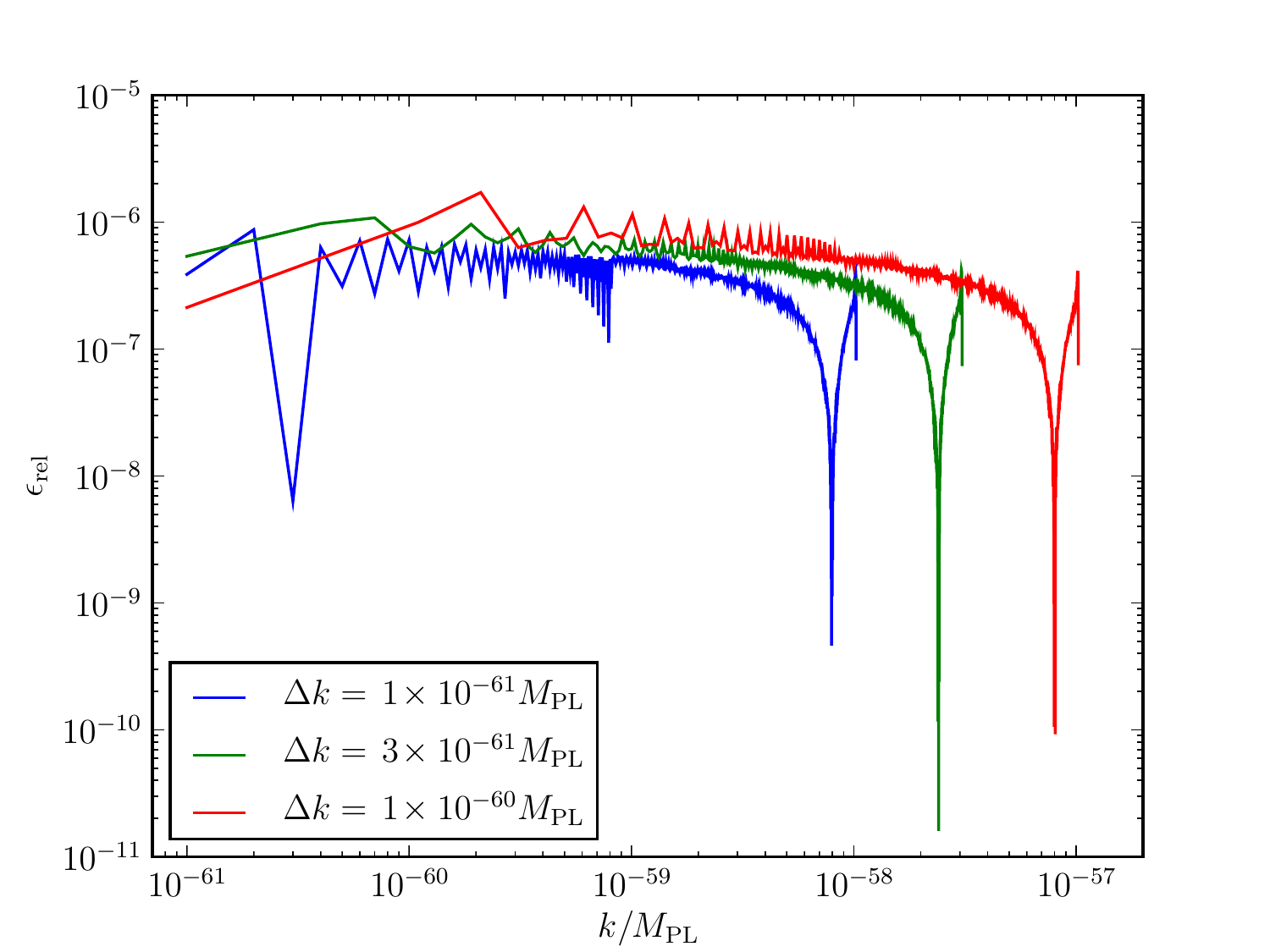}
 \caption[Relative Error in $I_\A$ with fixed $\kmin$]{The 
relative error in the integral $I_\A$ for three different values of $\Delta k$. In
contrast to Figure~\ref{fig:err-deltaks}, $\kmin=1\e{-61}\Mpl=3.8\e{-5}\Mpc^{-1} \le
\Delta k$ for each case.}
 \label{fig:err-deltak-kmin}
\end{figure}
 
\begin{figure}[htbp]
\centering%
\subfloat[The relative error in $I_\B$ for each $k$ range.]{
 \includegraphics[width=0.8\textwidth]{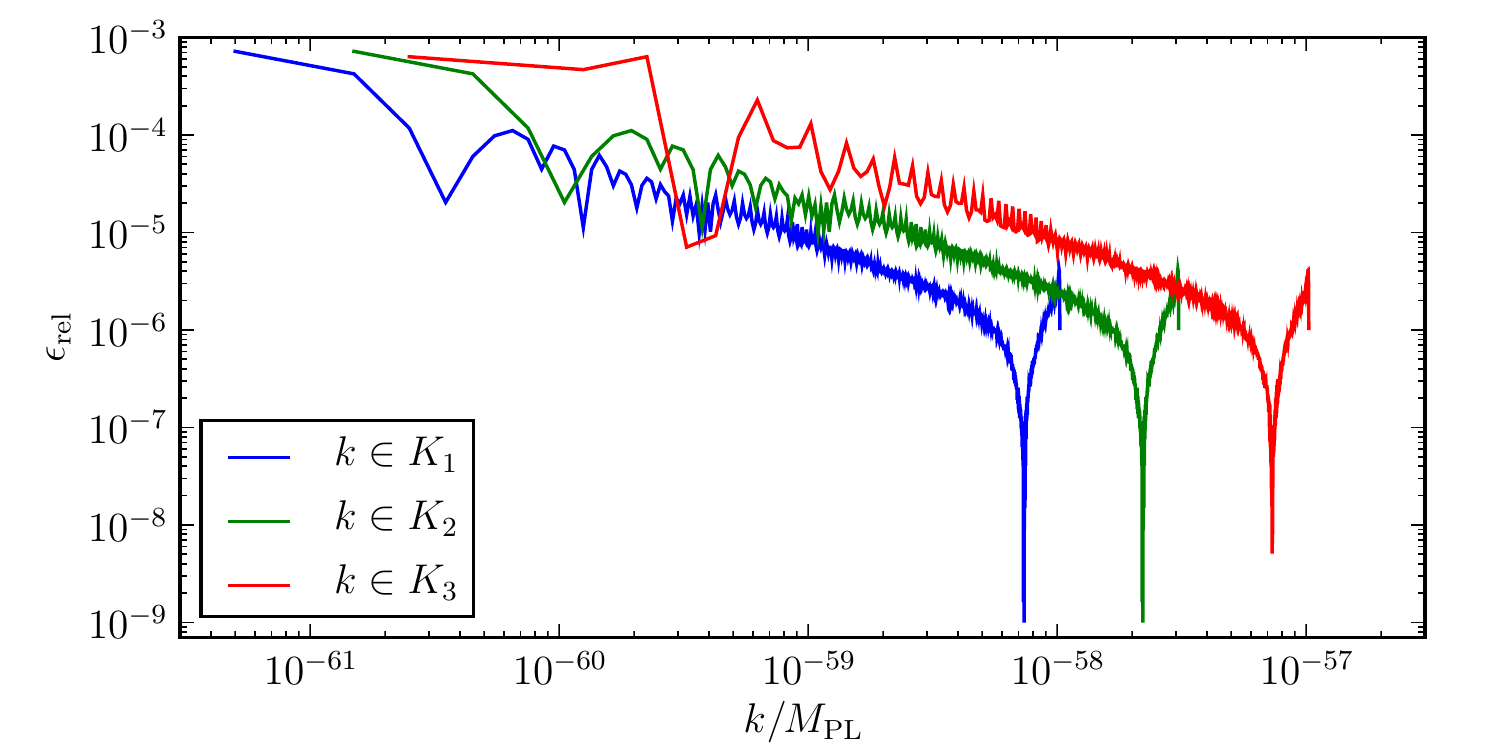}
}\\%
\subfloat[The relative error in $I_{\wt{\C}}$ for each $k$ range.]{
 \includegraphics[width=0.8\textwidth]{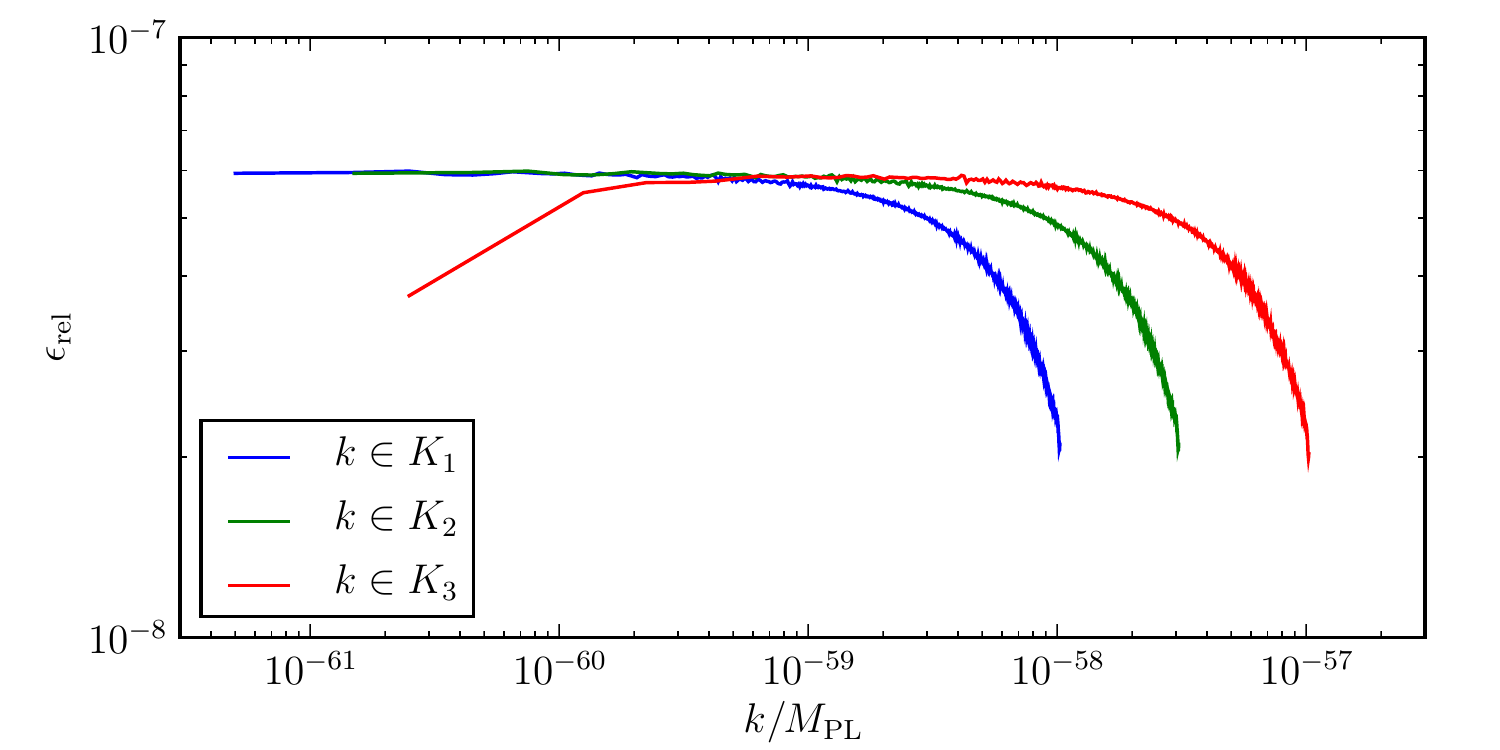}
}\\%
\subfloat[The relative error in $I_{\wt{\D}}$ for each $k$ range.]{
 \includegraphics[width=0.8\textwidth]{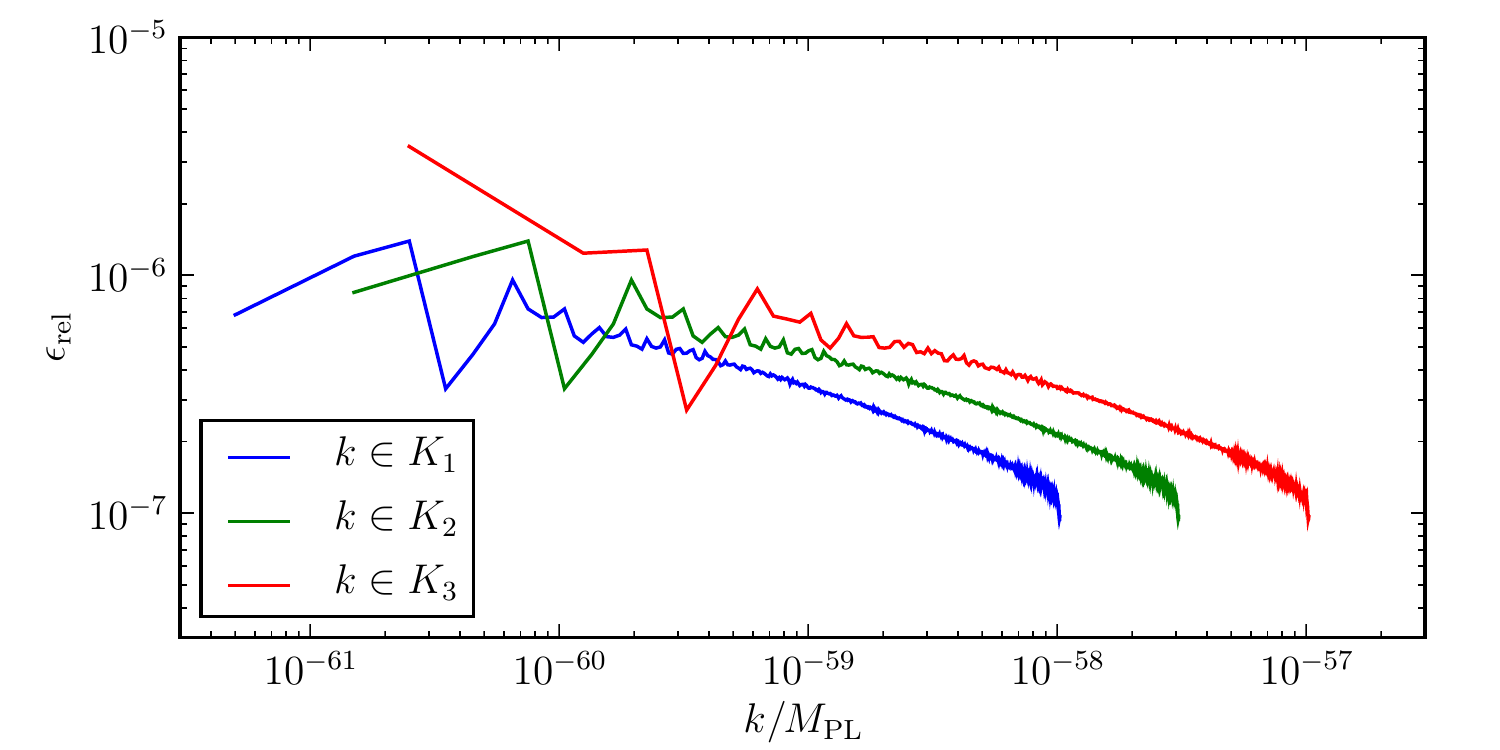}
}%
\caption[Relative Errors in $I_\B$, $I_{\wt{\C}}$ and $I_{\wt{\D}}$]{The relative
errors in $I_\B$,
$I_{\wt{\C}}$ and $I_{\wt{\D}}$ for each of the $k$ ranges with $\beta=10^{-62}$.}
\label{fig:rel-bcd-num}
\end{figure}

The analytic solutions for the $\B$, $\wt{\C}$ and $\wt{\D}$ terms are given in
Appendix~\ref{sec:apx-codetests}. 
The relative errors between the analytic and calculated values for $I_\B$,
$I_{\wt{\C}}$ and $I_{\wt{\D}}$ are shown in Figure~\ref{fig:rel-bcd-num} for the
three final $k$ ranges, with $\beta=10^{-62}$. The errors
for the $I_{\wt{\C}}$ and $I_{\wt{\D}}$ terms are very small, being of the order of
$10^{-8}$ and $10^{-6}$, respectively. The relative error for the $I_\B$ term is
larger, especially for small $k$ values. However, the error is still below $0.08\%$
for each of the $K_1$, $K_2$ and $K_3$ ranges.

It should be noted that these tests only show the relative errors in the
computation of integrals of the four terms in \eq{eq:AtoD-num}. They
do not represent
errors for the full calculation. However, they do show that the accuracy is good
compared with the analytic result. 

%
%
%

\section{Discussion}
\label{sec:disc-numerical}
This chapter has described the implementation of the numerical calculation of second
order cosmological perturbations.
The Klein-Gordon equations in Chapter~\ref{ch:perts} are the central focus of this
computation. In Section~\ref{sec:eqs-num} these equations were rewritten using $\N$
as the time variable, a choice more suitable for numerical work. The convolution
integrals in \eq{eq:KG2-fourier-sr-num} can be expressed in spherical polar
coordinates and split into four sub-terms $\A$--$\wt{\D}$ in \eqs{eq:AtoD-num} and
\eqref{eq:cdtilde-num}. Computing the source term
\eq{eq:KG2-source-ntime}, which is written using $\A$--$\wt{\D}$, is the most complex
and
time consuming part of the calculation.

To demonstrate the numerical code, four different potentials have been chosen and
these were described in Section~\ref{sec:pots-num}. One parameter for each
potential was determined by comparing the calculated $\mathcal{P}^2_{\mathcal{R}_1}$
with the WMAP5 value.

In Section~\ref{sec:initconds-num} the initial conditions for the computed
quantities were explained. Each $k$ mode is initialised well inside the horizon
using the Bunch-Davies vacuum conditions from Section~\ref{sec:perts-intro}. The
second order perturbations are initially set to zero. This choice
concentrates focus
on the generation of second order effects during the observable period of
inflation.

The technical implementation of the code was discussed in
Section~\ref{sec:impl-num}. There are four stages in the procedure. First, the
background fields are evolved over a specified time period. The end time of
inflation is then determined by the condition $\varepsilon_H=1$ and the scale factor
calculated for this time. With this
information the initialisation of the first order modes can be performed. In the
second stage, the first order perturbation equations are solved. These results are
used in the third stage to calculate the source term \eqref{eq:KG2-source-ntime} at
each
time step and for each $k$ value. Finally, the second order perturbation equations
are solved using
the source term results.

To test the source term calculation the numerical results were compared to 
analytic solutions in Section~\ref{sec:tests-num}. Numerical parameters such as
$N_\theta$ and $N_k$ were set by minimising the relative error between the two
approaches.

In Chapter~\ref{ch:perts} the evolution equations for second order perturbations
were introduced. In this chapter the practical implementation of a numerical
calculation of these perturbations was discussed. In Chapter~\ref{ch:results} the
results of this numerical calculation will be examined and the next steps towards an
improved procedure will be described.
%

\chapter{Results and Future Work}
\label{ch:results}

\section{Introduction}
\label{sec:intro-res}

The main result of Part~\ref{part:numerical} of this thesis is the numerical
integration of
the Klein-Gordon equation of motion for second order scalar field
perturbations, \eq{eq:KG2-fourier-sr-num}. The slow
roll approximation of the source term for second order perturbations was employed,
but the complete versions of the evolution equations were used for the
background and first
order perturbations. In this chapter the results of the numerical calculation will be
presented. This represents the
first step towards a full calculation of the Klein-Gordon equation at second order.
In addition to the new results obtained, plans will be described for future
work aimed at improving the numerical system and increasing its range
of applicability. 

As a proof of concept, the numerical system was tested with four different
potentials, $V(\vp)=\msqphisq$, $\lambdaphifour$, $\phitwooverthree$ and
$\msqphisqwithV$, and results computed across three
different $k$ ranges. As expected, the second order perturbation for a single,
slowly rolling inflaton field that we have calculated is extremely
small in comparison with the first order term. However, there are
differences apparent between the potentials, which will be outlined in
Section~\ref{sec:compare-res}.

We have listed the potential parameters $m$, $\lambda$, $\sigma$ and $m_0$
in Table~\ref{tab:params-num}. These were found using the WMAP5 normalisation
at $\kwmap=0.002 \Mpc^{-1} = 5.25 \e{-60}\Mpl$ \cite{Komatsu:2008hk}.
We have also outlined in \eq{eq:Krangedefns} the three $k$ ranges that have been
used:
\begin{align}
\label{eq:Kranges-res}
K_1 &= \left[1.9\e{-5}, 0.039\right]\Mpc^{-1}\,,\quad \Delta k =
3.8\e{-5}\Mpc^{-1} \,,\nonumber\\
K_2 &= \left[5.71\e{-5}, 0.12\right]\Mpc^{-1}\,, \quad \Delta k =
1.2\e{-4}\Mpc^{-1} \,,
\nonumber\\ 
K_3 &= \left[9.52\e{-5}, 0.39\right]\Mpc^{-1}\,, \quad \Delta k =
3.8\e{-4}\Mpc^{-1} \,.
\end{align}
Many of the results will be quoted for $\kwmap$ which lies in all three of these
ranges.

Given that the first order perturbations for the chosen potentials produce an
almost scale invariant power spectrum with no running, it is no surprise that
the results from the three different $k$ ranges are very similar. The second
order source term is somewhat dependent on the lower bound of $k$ (upper bound
on size). This is also to be expected and in the scale invariant case a logarithmic
divergence can
be shown to exist \cite{Lyth:2007jh}. We have implemented an arbitrary sharp
cutoff at $\kmin$, below which 
$\dvp1$ is taken to be zero. 
As mentioned in Chapter~\ref{ch:numericalsystem}, there is some evidence to suggest
that a similar cutoff might be supported by the WMAP data
\cite{Sinha:2005mn,Kim:2009pf}. 

In Section~\ref{sec:results}, the numerical results for the computation described in
Chapter~\ref{ch:numericalsystem} are presented. Comparisons of the results from the
four different test potentials will be made in Section~\ref{sec:compare-res}. Since
this represents the first stage towards a full calculation of the source term, the
next
steps that will be required are outlined in Section~\ref{sec:next-res}. Finally,
in Section~\ref{sec:disc-num} we discuss some of the consequences of our results.
%

\section{Results}
\label{sec:results}
\subsection{Results for \texorpdfstring{$V(\vp)=\msqphisq$}{m-squared Model}}
\label{sec:msqphisq-res}

\begin{figure}[htbp]
 \centering
 \includegraphics[width=0.8\textwidth]{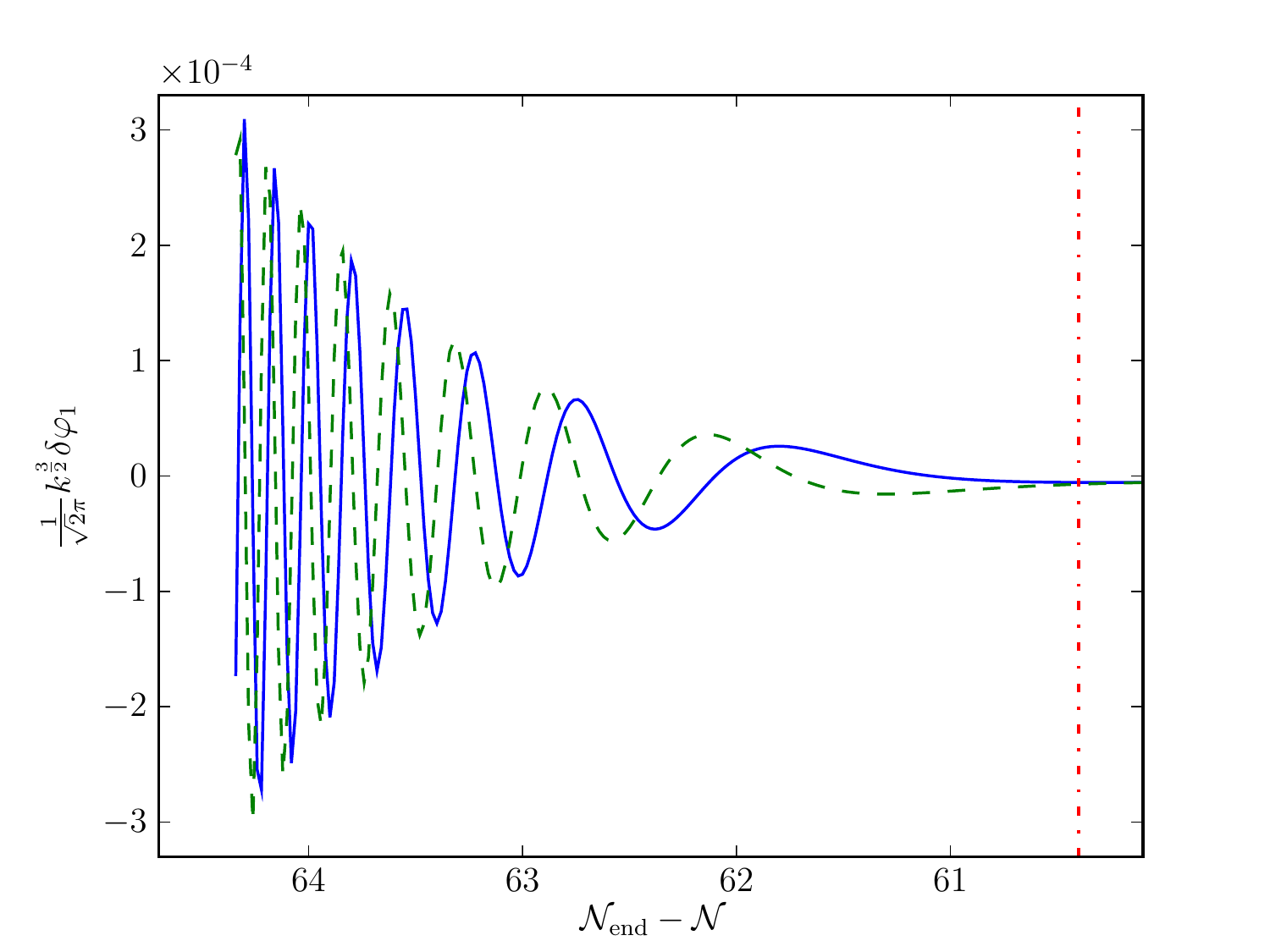}
 \caption[First Order Perturbation]{The first order perturbation $\dvp1$ rescaled by
$k^{3/2}/(\sqrt{2}\pi)$ from the beginning of the simulation until around
horizon crossing (red dot-dashed line). The real (blue) and imaginary (green
dashed) parts of the perturbation are shown for the scale $\kwmap$.}
\label{fig:dp1}
\end{figure}

At first order the solutions obtained for the quadratic potential agree with previous
work in
Refs.~\cite{Salopek:1988qh, Martin:2006rs, Ringeval:2007am}. Oscillations
are damped until horizon crossing (when $k=aH$) after which the
curvature perturbation becomes conserved. Figure~\ref{fig:dp1} shows the evolution of
the real and imaginary parts of the first order perturbation from
when the initial conditions are set, at $k/aH=50$, to just after horizon
crossing. The horizontal axis for most of the following figures parametrises the
number of e-foldings remaining until the end of inflation ($\N_\mathrm{end}-\N$),
instead
of the time variable $\N$ employed in the calculations.

\begin{figure}[htbp]
 \centering
 \includegraphics[width=0.8\textwidth]{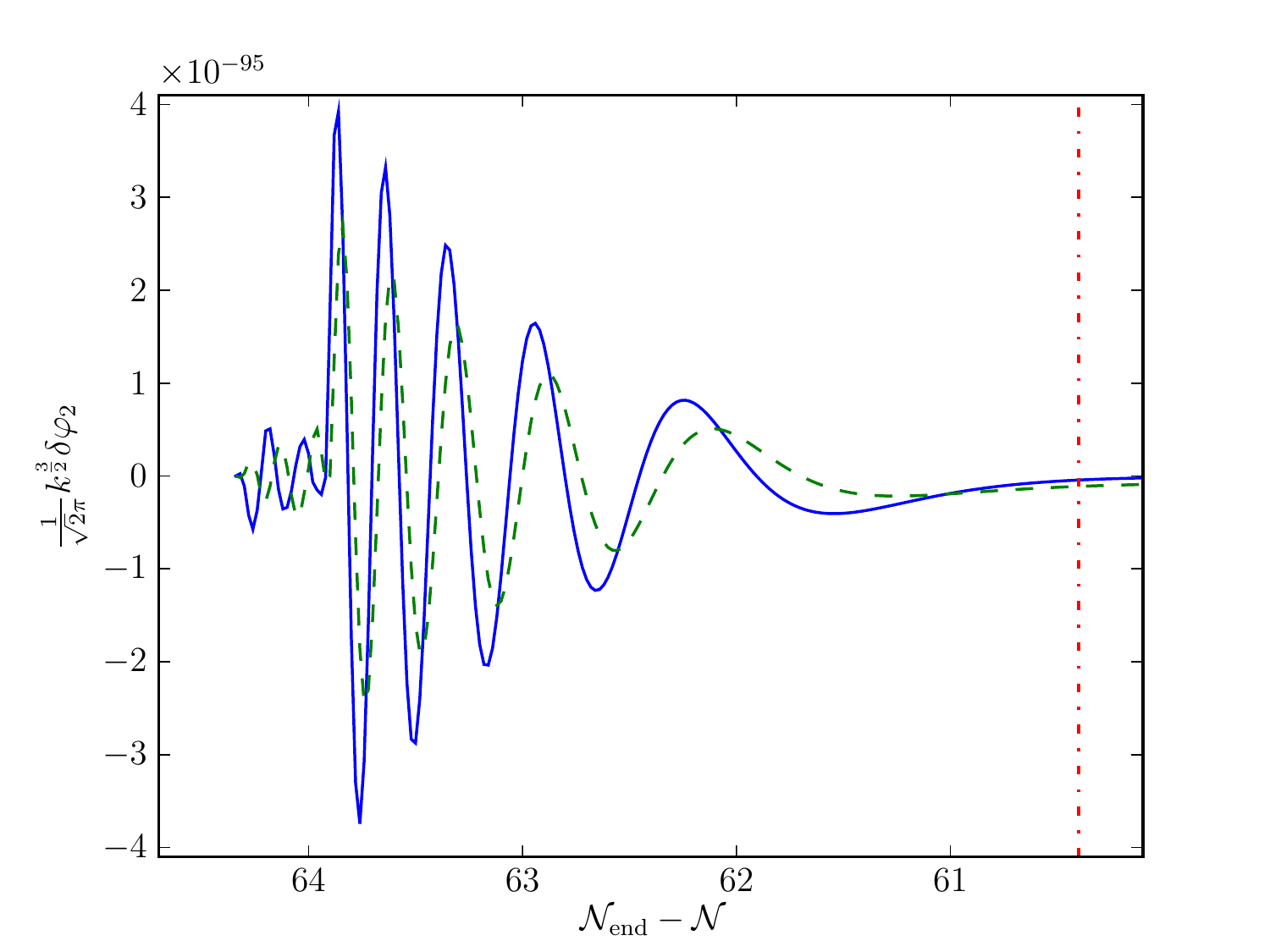}
 \caption[Second Order Perturbation]{The real (blue line) and imaginary (green
dashed) components of the
second order
perturbation $\dvp2(\kwmap)$ from the beginning of the simulation until around
the time
of horizon exit (red dot-dashed line).}
\label{fig:dp2realimag}
\end{figure}

Figure~\ref{fig:dp2realimag} shows the evolution of the second
order perturbations for the scale $\kwmap$. As mentioned above, the
overall amplitude of the second order perturbation is many orders of
magnitude smaller than the corresponding first order one. 
The results given for $\dvp2$ in this
chapter are for the full solution which includes a homogeneous and particular solution as described in
Section~\ref{sec:initconds-num}.
 In Figures~\ref{fig:dp1}
and \ref{fig:dp2realimag} the field values have been rescaled by
$k^{3/2}/(\sqrt{2}\pi)$ to allow for a better appreciation of the
magnitude of the resulting power spectra.
%

\begin{figure}[htbp]
\centering
\subfloat[Absolute magnitude of the source term.]{
 \includegraphics[width=0.8\textwidth]{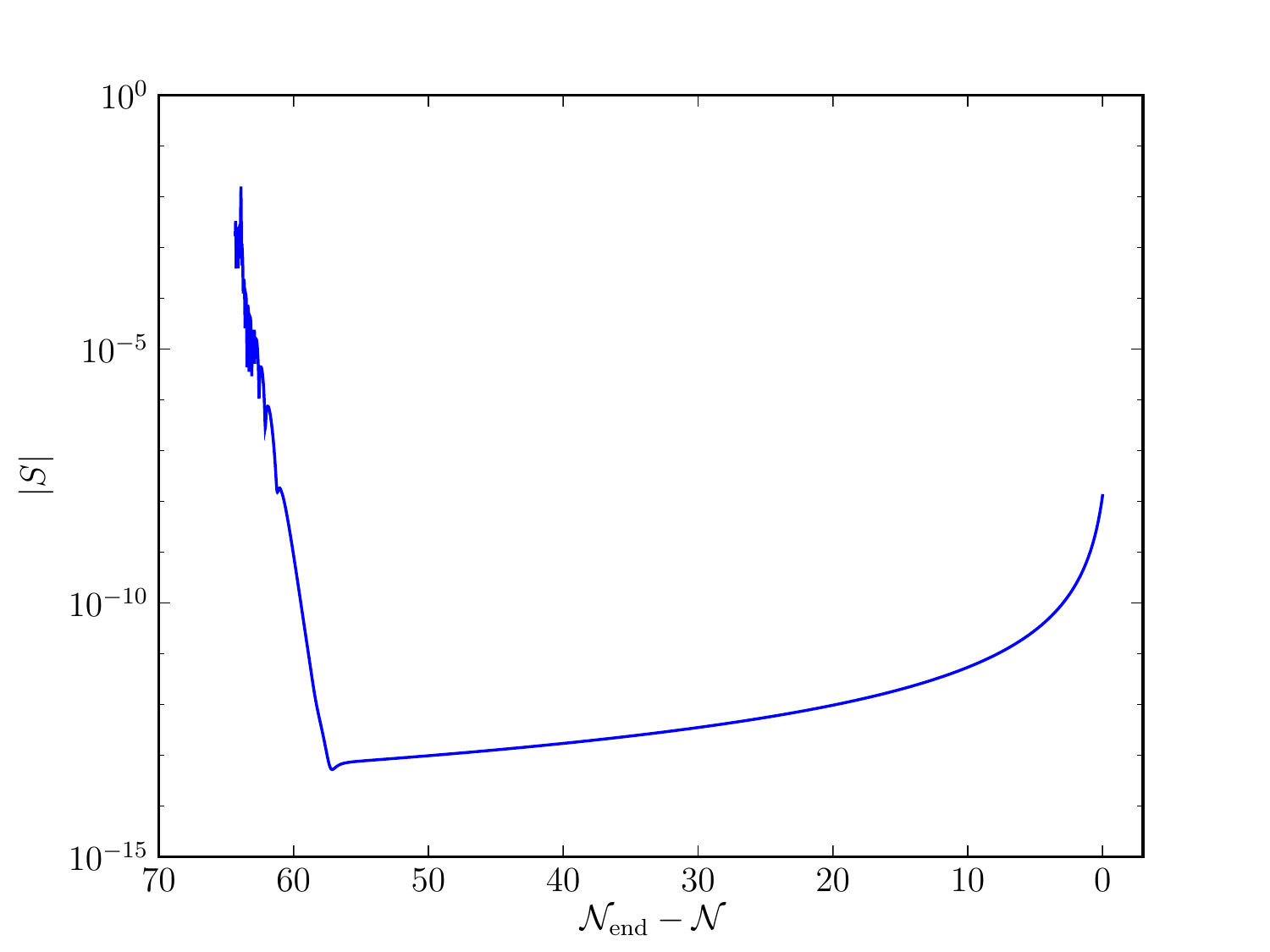}
\label{fig:src-full}
}\\
\subfloat[First Order Power Spectrum of Scalar Perturbations][Power spectrum of
first order scalar
perturbations $\mathcal{P}^2_{\delta\varphi_1} =
\frac{k^3}{2\pi^2}|\delta\varphi_1|^2$.]{
\includegraphics[width=0.8\textwidth]{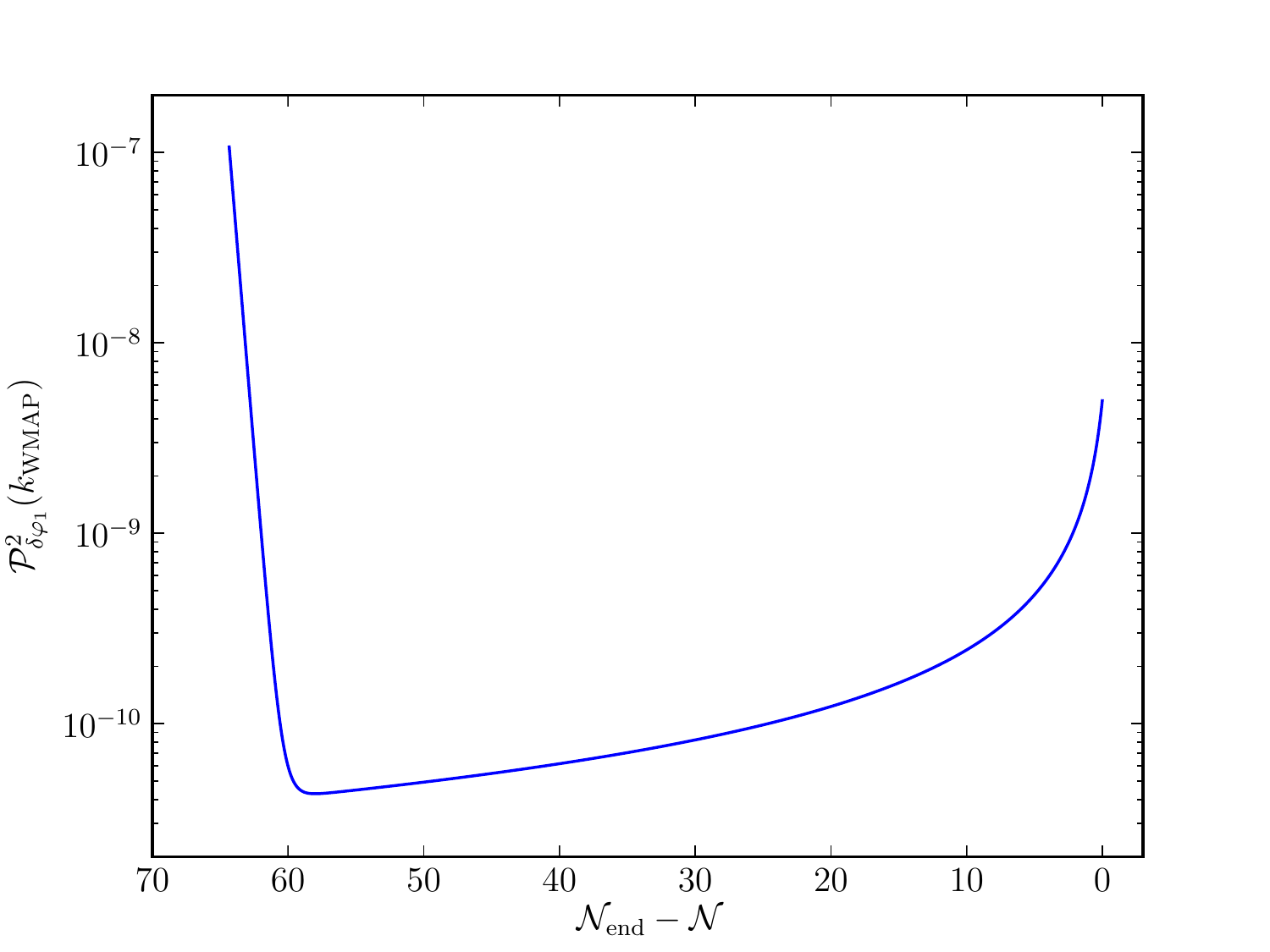}
\label{fig:Pphi-kwmap}
}
\caption[Source Term and First Order Power Spectrum for $\kwmap$]{Source term and 
first order power spectrum for the WMAP pivot scale $\kwmap$.}
\end{figure}

\begin{figure}[htbp]
\centering
 \includegraphics[width=0.8\textwidth]{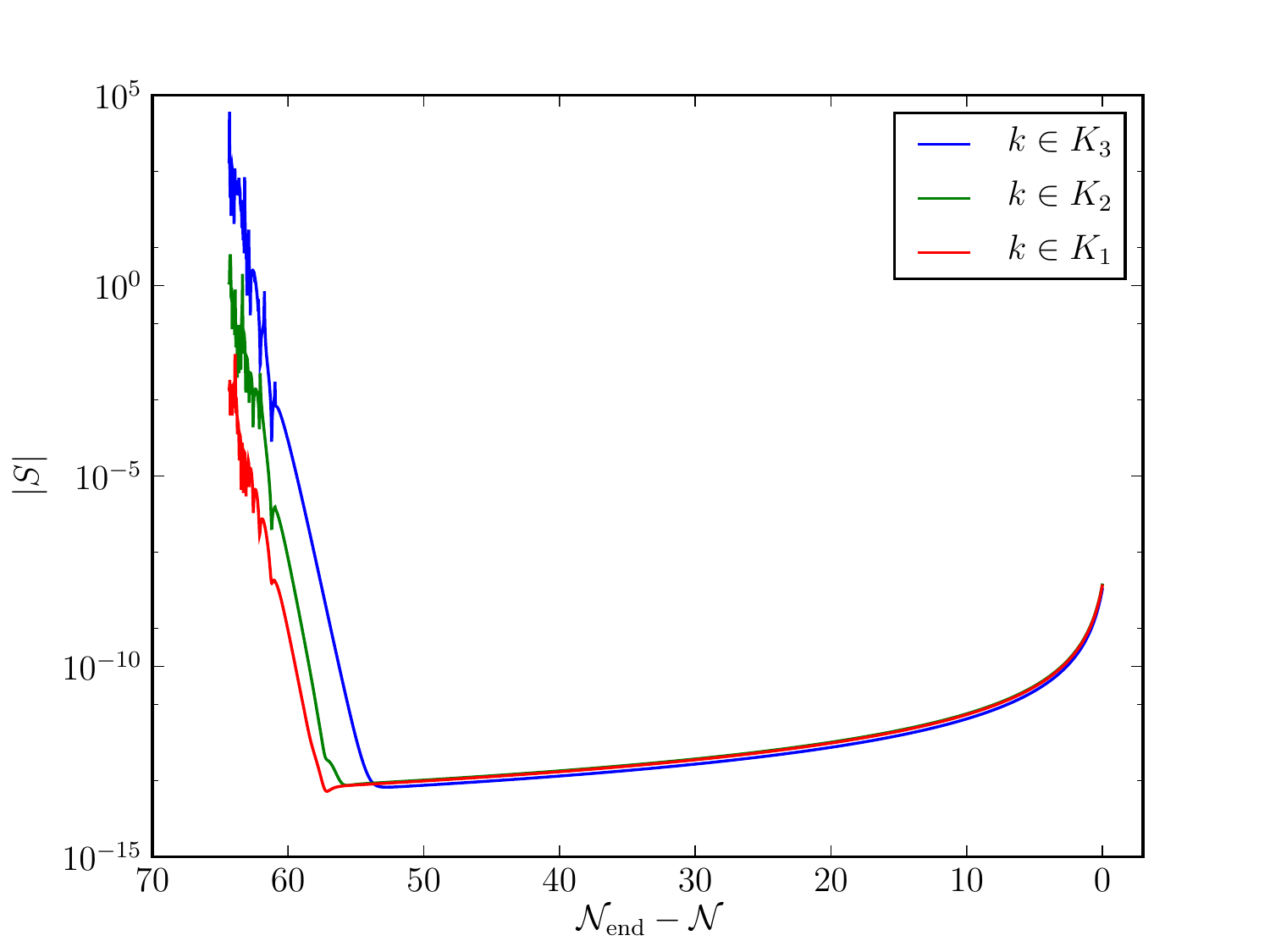}
\caption[Comparison of Source Term for Different $k$ Ranges]{A comparison of the
source
term \eqref{eq:KG2-source-ntime}, for the scale $\kwmap=5.25\e{-60}\Mpl$, over the
three different ranges $K_1$,
$K_2$ and $K_3$, which were specified in \eq{eq:Kranges-res}. Before horizon
crossing there is a significant difference in the amplitude of the source term for
$\kwmap$. After horizon crossing, however, the magnitude of $S$ is independent of
the choice of $K_i$.
} 
\label{fig:src-kwmap-3ranges}
\end{figure}

\begin{figure}[htbp]
\centering
 \includegraphics[width=0.8\textwidth]{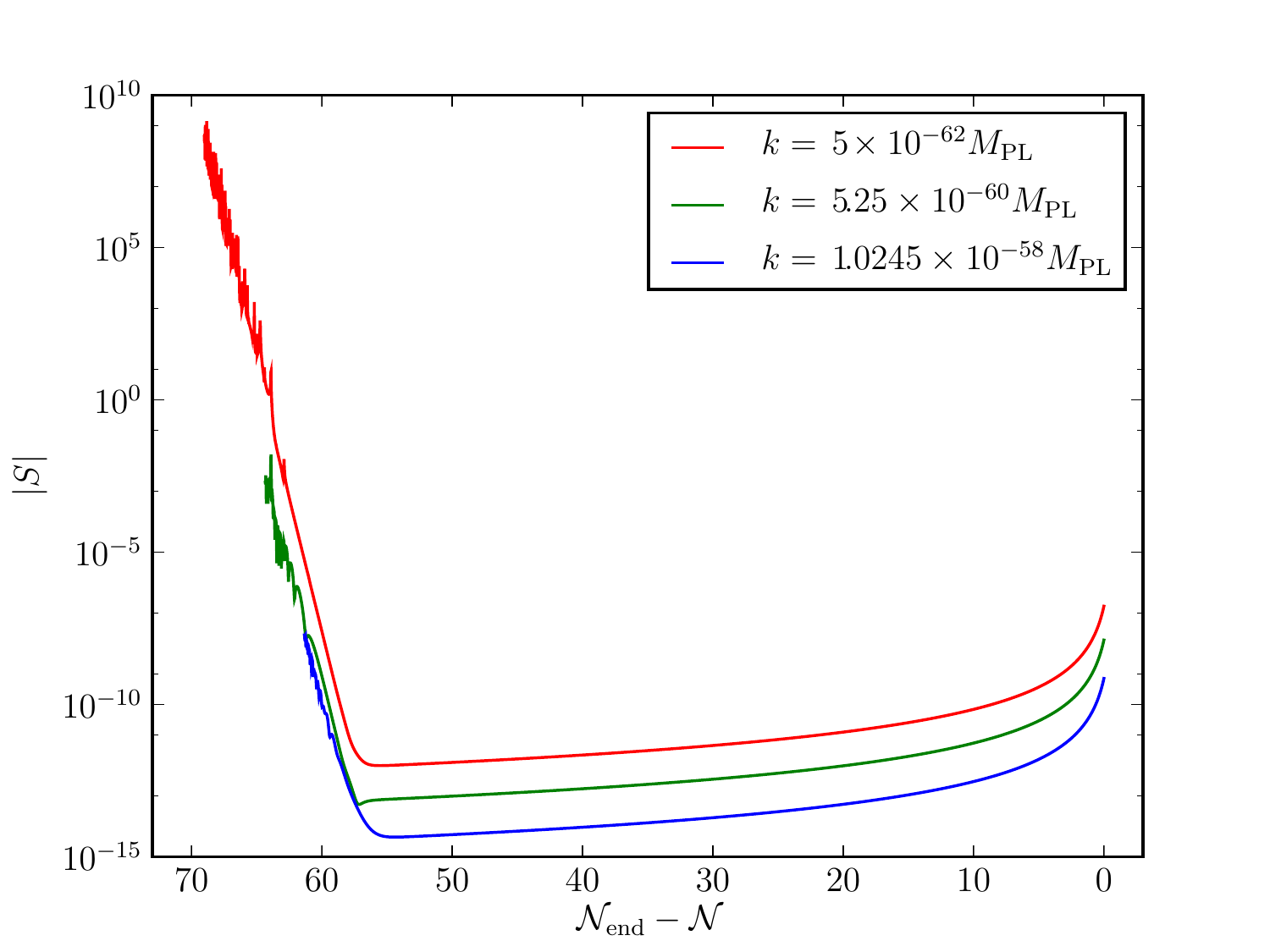}
\caption[Source Term at Three Different Values of $k$]{The source term
\eqref{eq:KG2-source-ntime} for three different $k$
values in the $K_1$ range, including the WMAP
pivot scale, $\kwmap=5.25\e{-60}\Mpl$ (middle green line). As the value of $k$
increases or equivalently the scale decreases, the magnitude of the source term
decreases. The calculation of the source term for each $k$ value starts from the
time step at which the corresponding first order perturbation is initialised, \iec
when $k/aH = 50$.
}
\label{fig:src-3ks}
\end{figure}

\begin{figure}[htbp]
\centering
 \includegraphics[width=0.8\textwidth]{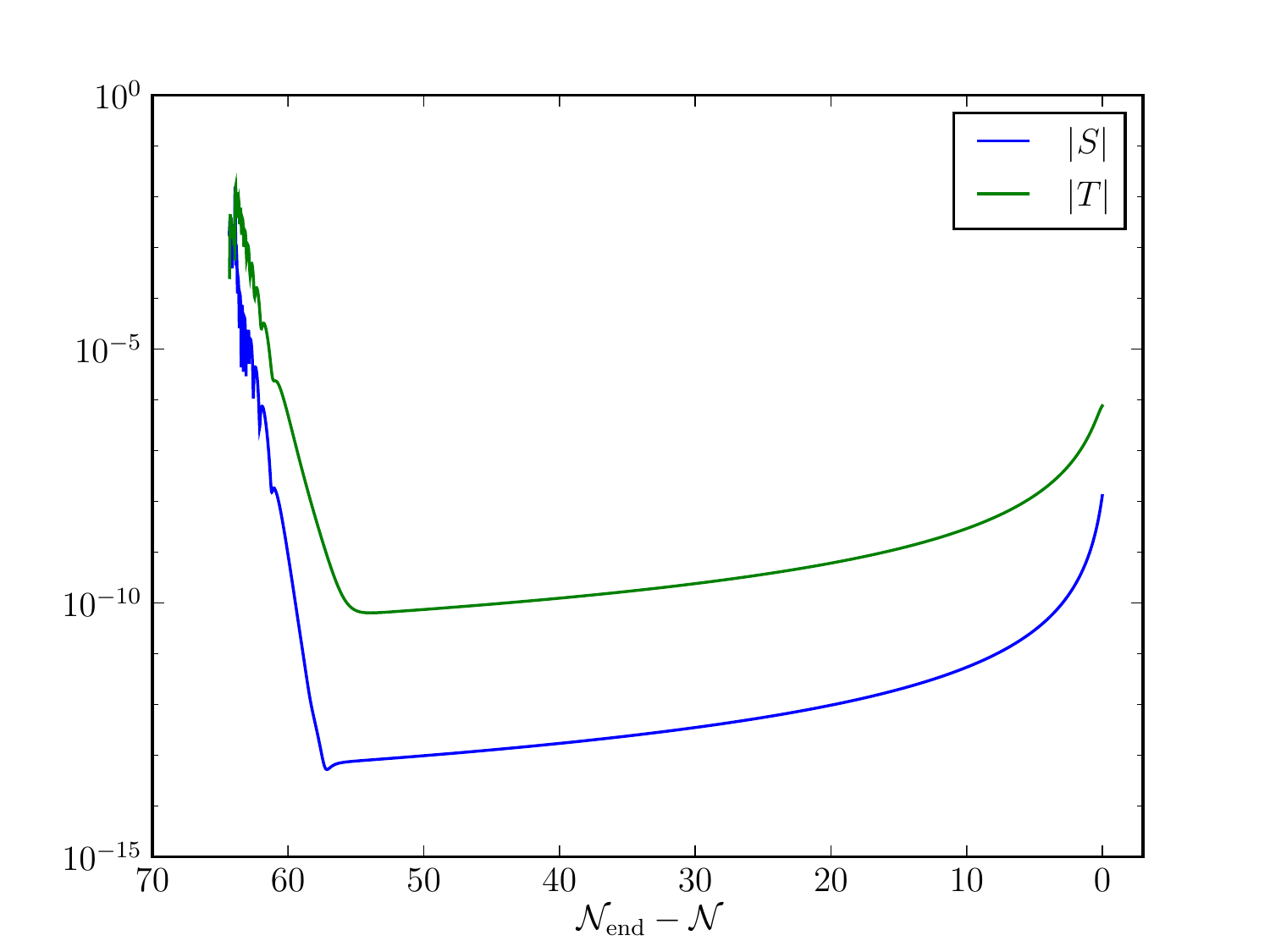}
\caption[Source Term Compared to $T$ Term]{The source term (lower blue line), as
defined in \eq{eq:KG2-source-ntime}, is
compared with the $T$ term
(upper green line), as defined in \eq{eq:Tdefn}, for $\kwmap$. The source term is of
comparable magnitude at the beginning of the simulation.}
 \label{fig:src-vs-others}
\end{figure}

\begin{figure}[htbp]
\centering
 \includegraphics[width=0.8\textwidth]{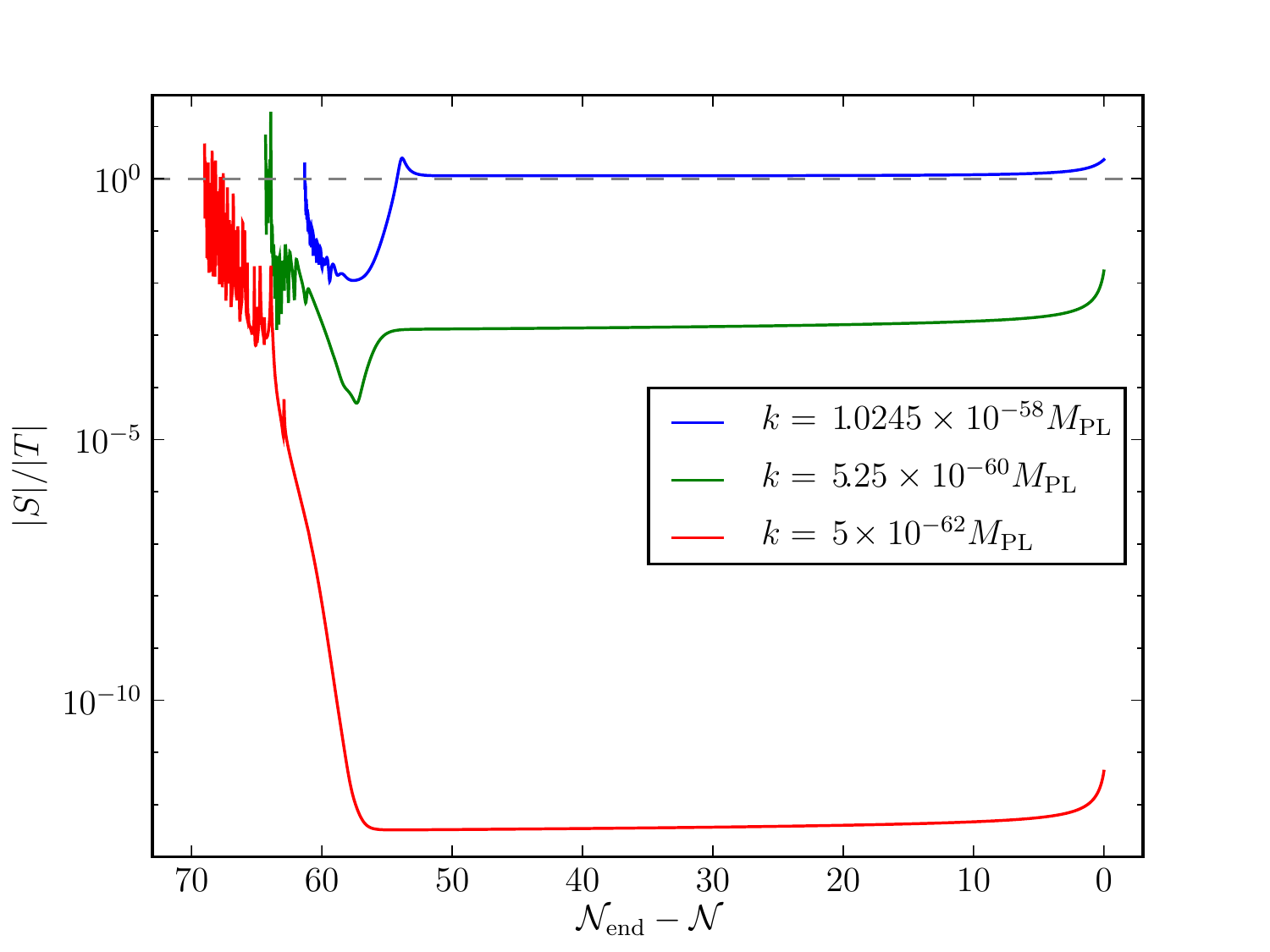}
\caption[Quotient of $S$ and $T$]{The quotient of the $S$ term,
\eq{eq:KG2-source-ntime}, and the $T$ term, \eq{eq:Tdefn}, for three
different $k$ values in the range $K_1$,
including the WMAP pivot scale, $\kwmap=5.25\e{-60}\Mpl$.  
For small values of $k$ the source term is not comparable to the magnitude of $T$
after horizon crossing. However, for larger $k$ values (smaller scales) the two
terms have comparable magnitude.
It is, therefore,
important to calculate the source term over the full range of e-foldings.}
 \label{fig:s-over-t-3ks}
\end{figure}

The source term $S(\kvi)$ is calculated using \eq{eq:KG2-src-sr-aterms} at each time
step using the results of the first order and background simulations. This term
drives the production of second order perturbations as shown in
\eqs{eq:KG2-fourier-sr-num} and
\eqref{eq:KG2-fourier-sr-ntime}. Figure~\ref{fig:src-full} shows the
absolute magnitude of the source term for a single $k$ mode, $\kwmap$,
for all time steps calculated. 
The source term is large at early times, and closely follows the form
of the spectrum of the first order perturbations, as can be seen from
Figure~\ref{fig:Pphi-kwmap}.
Figure~\ref{fig:src-kwmap-3ranges} shows how the source term depends on
the choice of $k$ range.  After horizon crossing, the source term
is independent of the specific choice of $K_i$ ($i=1,2,3$). Before horizon crossing,
however, there is a
strict hierarchy with the smaller $k$ ranges, $K_1$ and $K_2$, leading to
smaller source
contributions.  As discussed in Section \ref{sec:tests-num}, $\Delta k$
should be at least as large as $\kmin$ in order for the error to be reduced to
a minimum. In Figure~\ref{fig:src-3ks} the source term is plotted at three different
values of $k$ for the range $K_1$. As $k$ increases, or equivalently the length scale
decreases, the magnitude of the source term after horizon crossing decreases. 
It is informative to compare the magnitude of the source term with the
other terms in the second order evolution equation
(\ref{eq:KG2-fourier-sr-ntime}). We denote these other terms by $T$:
\begin{equation}
\label{eq:Tdefn}
 T(\kvi) = \left(3 + \frac{\dN{H}}{H}\right)
\dN{\dvp2}(\kvi)+ \left(\frac{k}{aH}\right)^2\dvp2(\kvi)
+\left(\frac{\Upp}{H^2}-{24 \pi G}(\dN{\vp_{0}})^2\right)
\dvp2(\kvi) \,.
\end{equation}
Figure~\ref{fig:src-vs-others} then shows the absolute magnitude of
both $S$ and $T$.  It is clear that for the scale $\kwmap$ the contributions to the
source term are only of comparable
magnitude during the early stages of the simulation.  
Figure~\ref{fig:s-over-t-3ks}
shows a comparison of $|S|/|T|$ for three different $k$ values. 
The magnitude of $S$ is closer to that of the rest of the terms for the
larger $k$ mode.
A priori, the range of $k$ modes where $S$ will be large for a particular chosen
potential is not known. However, once the relevant values of $k$ have been
determined, it may be possible to significantly reduce the time
required for the simulation by restricting the calculation of $S$ to those regions
where it is important. Figure~\ref{fig:s-over-t-3ks} shows that it is not possible
to arbitrarily ignore the contribution of $S$ either inside or outside the horizon.
The full calculation on sub- and super-horizon scales is important for the evolution
of the second order perturbation at different scales.

%
%

\begin{figure}[htbp]
\centering
\includegraphics[width=0.8\textwidth]{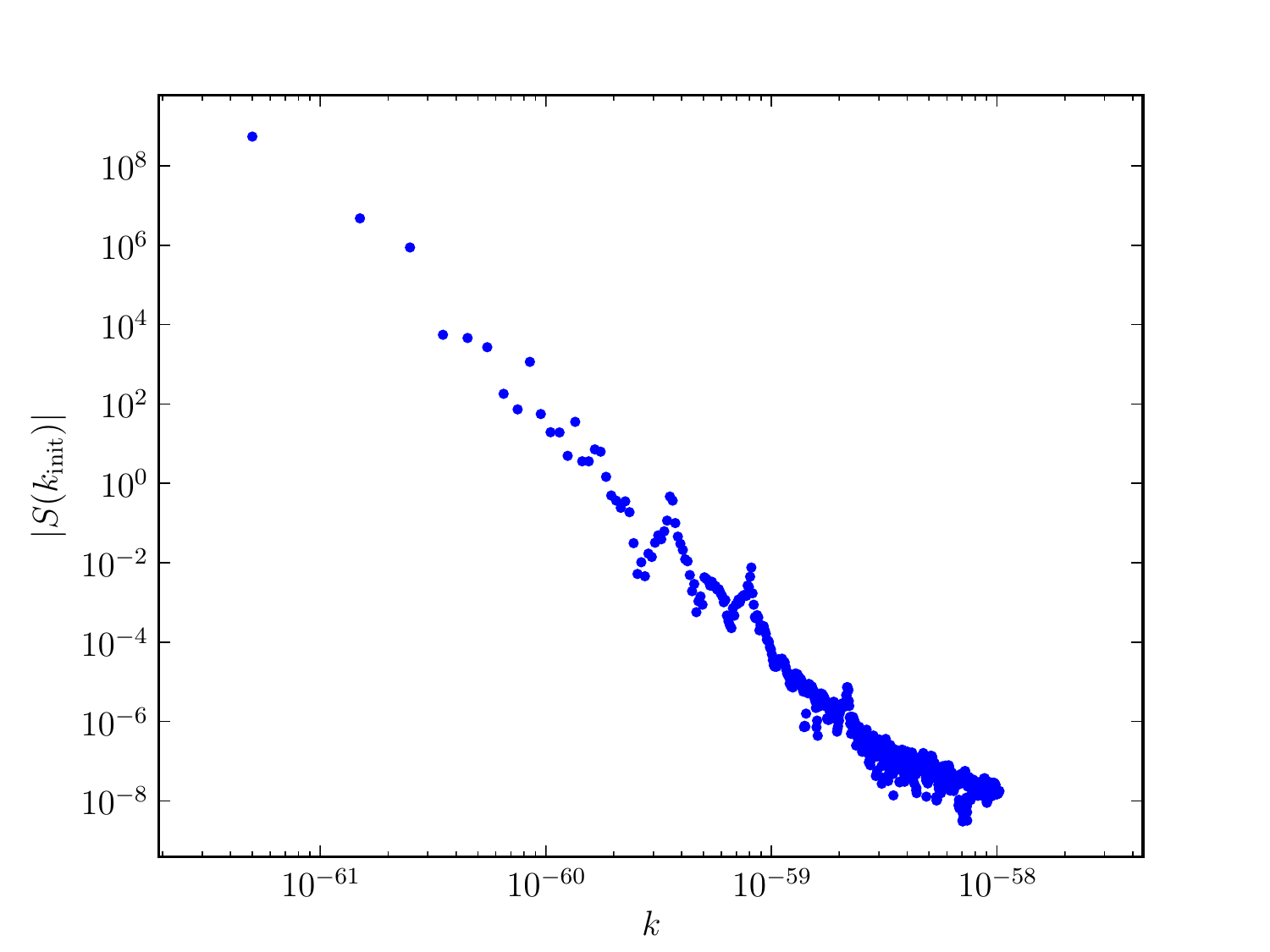}
 \caption[Source Term at Initialisation]{The absolute magnitude of the source 
term at the initial start time for each $k$ value when $k/aH = 50$ deep inside
the
horizon. The results are for the range $K_1$.}
\label{fig:src-kinit}
\end{figure}

In Figure~\ref{fig:src-kinit} the value of $|S|$ at the initialisation time
for each $k$ mode is shown. The initial magnitude of the source term is much
smaller for larger values of $k$ (smaller scales). 
Because the smaller $k$ modes begin their evolution earlier, the relative difference
in $|S|$ is not as pronounced when measured at a single time step (see for example
Figure~\ref{fig:src-3ks}).
It should also be remembered that the magnitude of the other terms in the second
order ODE is small for larger $k$ modes as shown by the ratio $|S|/|T|$ in
Figure~\ref{fig:s-over-t-3ks}.

\begin{figure}[htbp]
\centering
\includegraphics[width=0.8\textwidth]{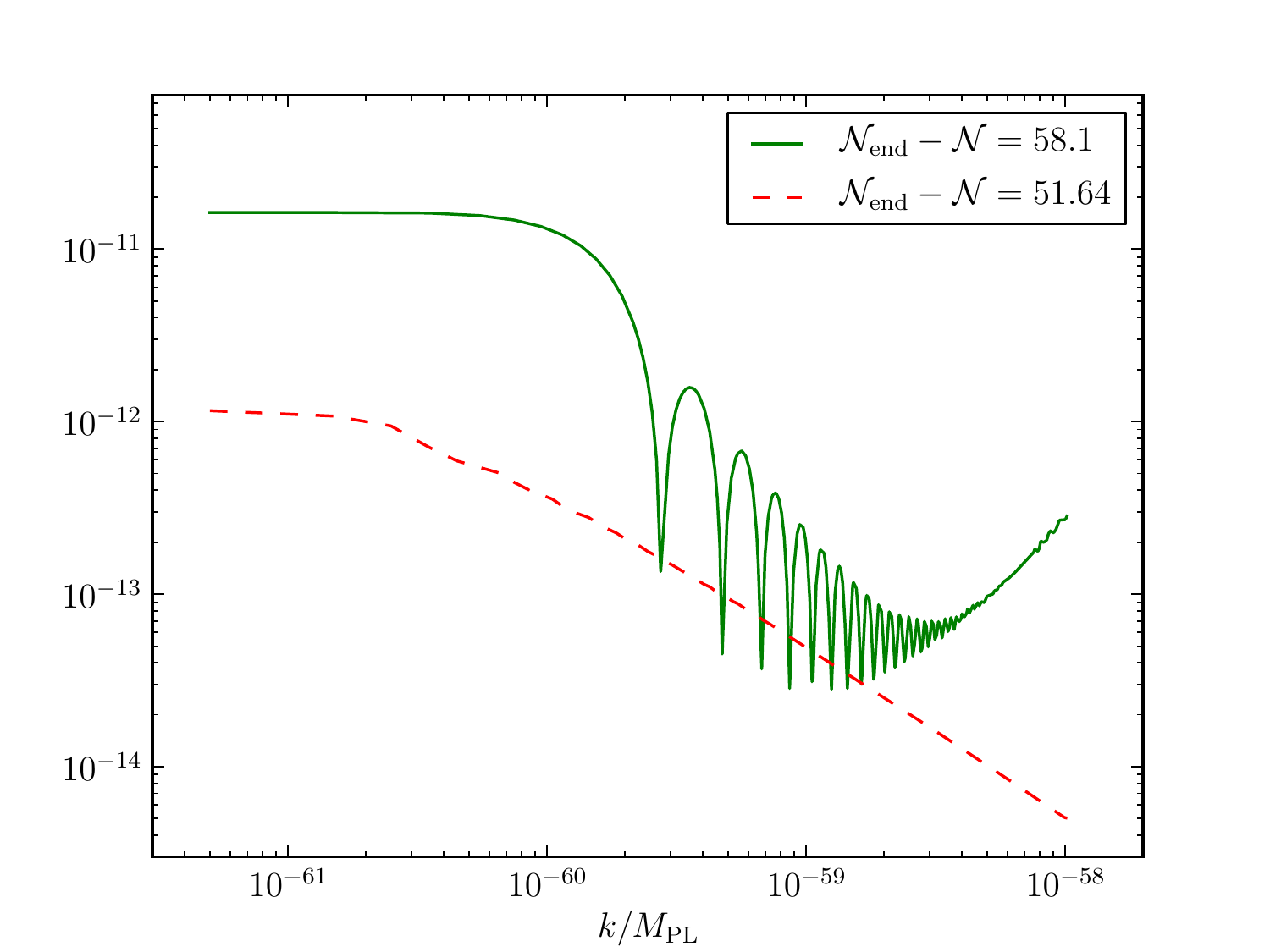}
\caption[Source Term at Two Different Times]{The absolute magnitude of the source 
term for all $k$ values in the range $K_1$ at two different time steps. The green line shows
$|S|$ when all modes have been initialised. The lower red dashed line shows $|S|$ approximately 
52 e-foldings before the end of inflation, when all modes have
exited the horizon.}
\label{fig:src-3ns}
\end{figure}

The source term can also be compared at different time steps over all the $k$ values.
In Figure~\ref{fig:src-3ns} the green
line shows $|S|$ when all $\dvp1$ modes have started to evolve. The lower
red line illustrates $|S|$
after all modes have exited the horizon, around 52 e-foldings before the end of
inflation.
%

\subsection{Comparison of \texorpdfstring{$V(\vp)=\msqphisq$}{m-squared Model} Results with Analytic
Solution}
\label{sec:analytic-res}
In this section results for the quadratic model will be compared with an analytic solution for this
model. However, an analytical result is difficult to obtain for the case of the full first
order solution in terms of Hankel functions with the phase information included. The analytical
solution we will use, therefore, is the non-interacting de Sitter
space solution with the phase information ignored. The first order perturbations are then given by
\begin{equation}
 \dvp1(\eta, \kvi) = \frac{1}{a\sqrt{2 k}} \left( 1 - \frac{i}{k \eta}\right) \,,
\end{equation}
and the derivative in terms of $\N$ is 
\begin{equation}
 \dN{\dvp1}(\eta, \kvi) = -  \frac{1}{a\sqrt{2 k}} \left( 1 -
                        \frac{i}{k \eta}\right) \left(1 + \frac{1}{aH\eta}\right)
                         - \frac{i}{a^2 H \sqrt{2}}\sqrt{k}\,.
\end{equation}

The source term is found using \eq{eq:KG2-source-ntime} and the values of the background
quantities. The analytical solution of \eq{eq:KG2-source-ntime} for this choice of first order
solution is given in Appendix~\ref{sec:analyticsrc-apx} in Eqs.
(\ref{eq:Jdefns-apx}-\ref{eq:JD-apx}).

\begin{figure}
 \centering
 \includegraphics[width=0.8\textwidth]{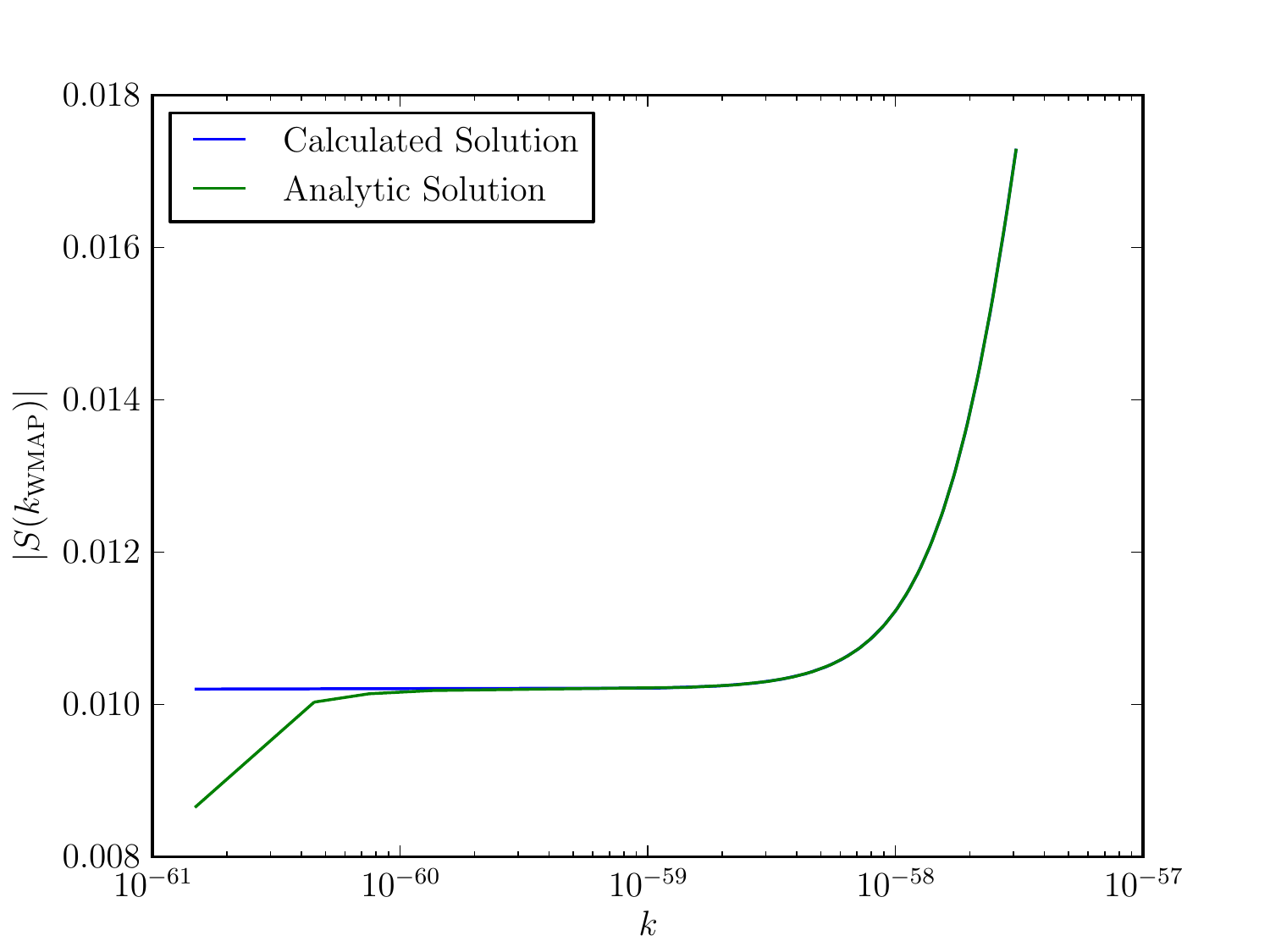}
 \caption[Comparison of Analytical and Calculated Source Terms]{A comparison of the
analytical and calculated solutions for the source term at one time step, approximately 60
e-foldings before the end of inflation.}
 \label{fig:analytic-61before-res}
\end{figure}

\begin{figure}
 \centering
 \includegraphics[width=0.8\textwidth]{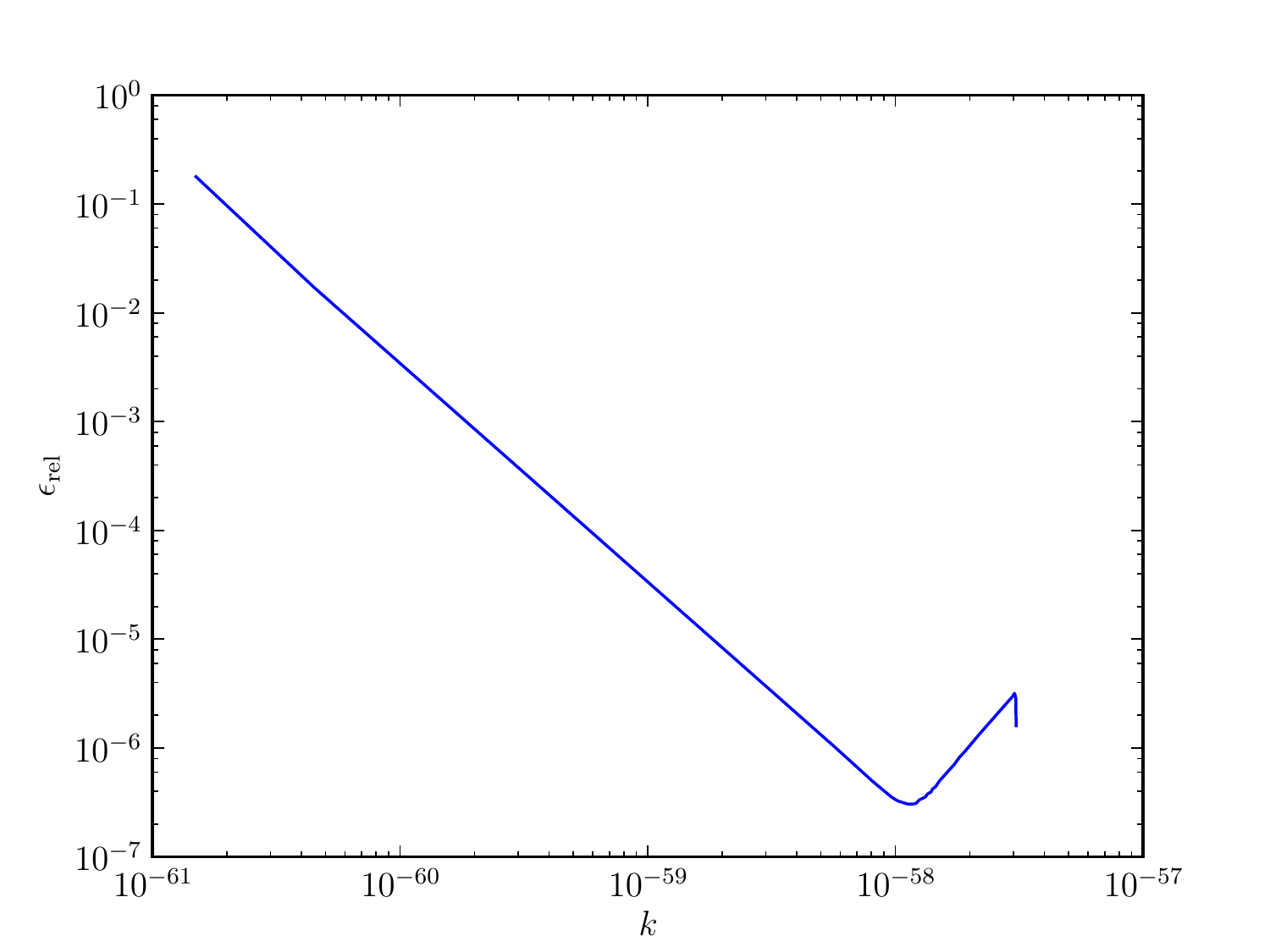}
 \caption[Relative Error Of Calculated Solution at One Time Step]{The relative error of the
calculated solution compared to the analytical solution for the source term. The error is shown
for all $k$ values at one time step, approximately 60 e-foldings before the end of inflation.}
 \label{fig:analytic-61before-errors-res}
\end{figure}

In Figure~\ref{fig:analytic-61before-res} the analytical and calculated solutions are plotted for
one timestep about 60 e-foldings before the end of inflation. At a single time step the correlation
between the two solutions for $S$ is very good. There is a deviation at small
values of $k$ due to the analytical solution getting rapidly smaller as $k$ approaches zero. This
 is not replicated in the calculated version. However, this only strongly affects the
result for the smallest values of $k$ and for $\kwmap$ for example the relative error of the
calculated solution compared to the analytical solution is about $10^{-4}$. The relative error of
the calculated solution is shown in Figure~\ref{fig:analytic-61before-errors-res} for a single time
step.

\begin{figure}
 \centering
 \includegraphics[width=0.8\textwidth]{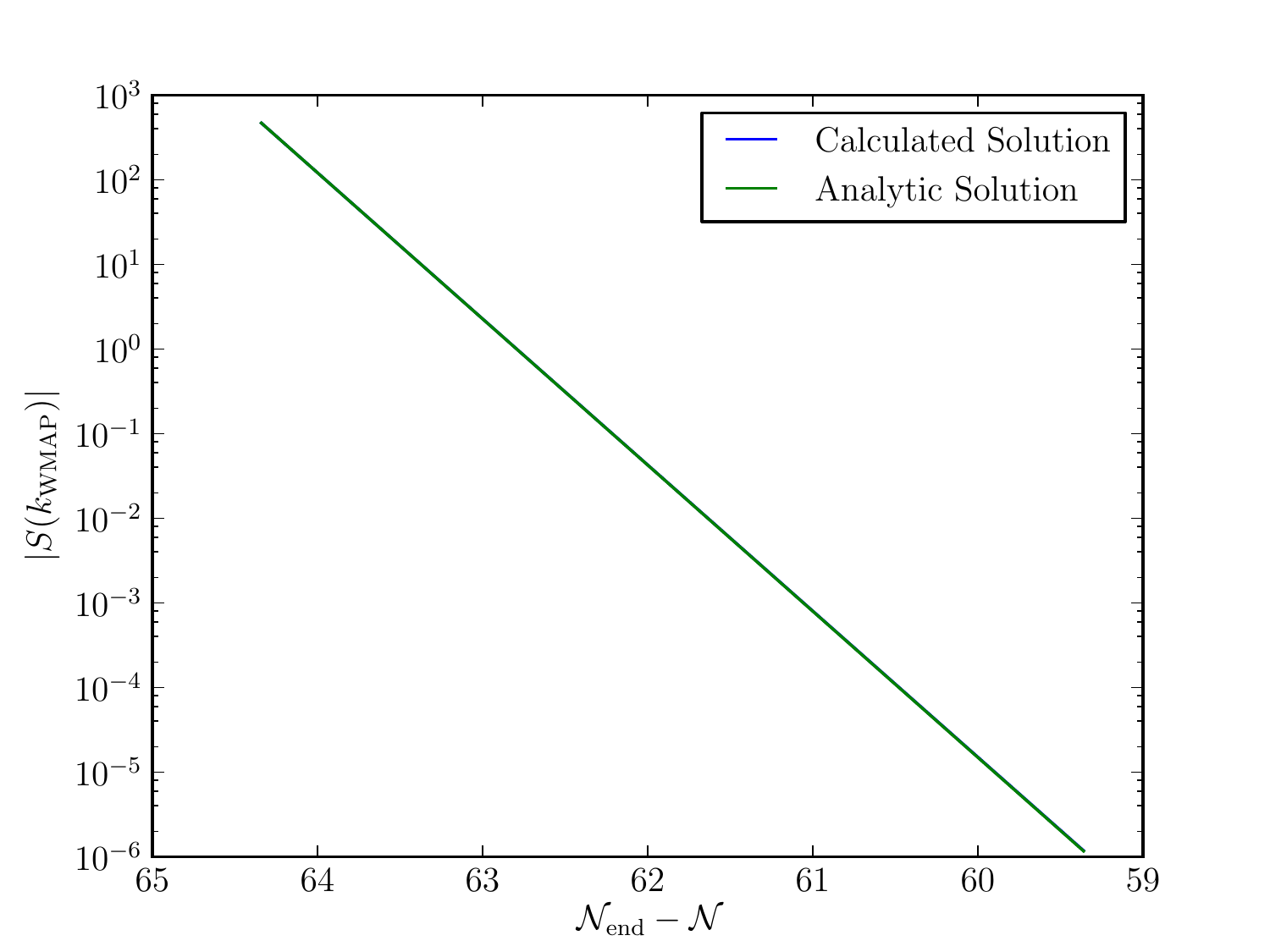}
 \caption[Comparison of Solutions Before Horizon Crossing]{A comparison of the analytical and
calculated results for the source term of the $\kwmap$ mode before horizon crossing.}
 \label{fig:analytic-prehorizon-res}
\end{figure}

The analytical and calculated values for the source term can also be compared for a single $k$
value across a range of time steps. In Figure~\ref{fig:analytic-prehorizon-res} the absolute
magnitude of the source term for $\kwmap$ is plotted for a range of a few e-foldings before horizon
crossing. The analytical and calculated results are extremely similar and not distinguishable in the
plot. The relative error of the calculated solution is around $10^{-4}$. Because the first order
perturbations do not include any phase information the result is much smoother than the result
generated using the full first order phase information. However, as the phase angle is a function of
$|\kvi-\qvi|$ it cannot be trivially ignored in the computation of the full convolution integral. 

\subsection{Comparison of Models}
\label{sec:compare-res}
All the results quoted so far have been for the quadratic potential. In this
section
the results for all four potentials will be compared using the $K_2$ range. 
\begin{figure}[htbp]
 \centering
\includegraphics[width=0.8\textwidth]{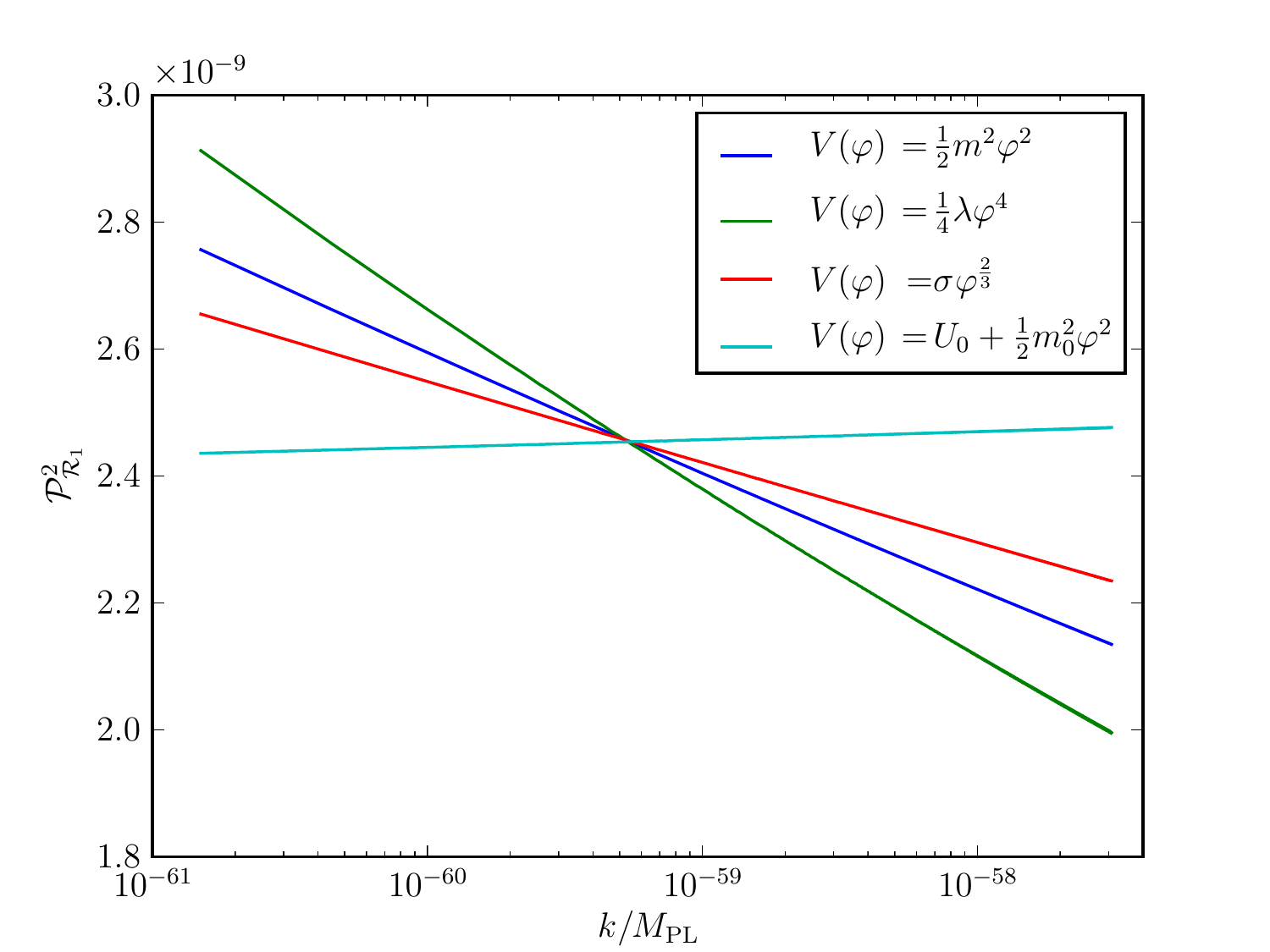}
\caption[Comparison of $\mathcal{P}^2_{\mathcal{R}_1}$ for the Four
Potentials]{Comparison of the power spectrum $\mathcal{P}^2_{\mathcal{R}_1}$ for
the four different models. The three models with potentials $\msqphisq$,
$\lambdaphifour$ and
$\phitwooverthree$ have red spectra ($n_s <1$) while the $\msqphisqwithV$
model has a blue spectrum ($n_s>1$).}
\label{fig:cmp-Pr}
\end{figure}
Figure~\ref{fig:cmp-Pr} shows the power spectrum of first order curvature
perturbations, $\mathcal{P}^2_{\mathcal{R}_1}$, for each
potential. The $\msqphisq$, $\lambdaphifour$ and $\phitwooverthree$ models all
clearly have
a red spectrum with $n_s <1$. On the other hand, the $\msqphisqwithV$ model has
a blue
spectrum ($n_s>1$) when $U_0$ to chosen to be $5\e{-10}\Mpl^4$, as specified in
Section~\ref{sec:pots-num}. 
The values of $n_s$ obtained for the four potentials are given in Table~\ref{table:ns-res}. 
\begin{table}
\begin{center}
\begin{tabular}{ccr}
\toprule
Potential & $n_s$ & $n_s - 1$ \\
\midrule
$\msqphisq$ & 0.965 & -0.035 \\ 
$\lambdaphifour$ & 0.949 & -0.051 \\
$\phitwooverthree$ & 0.977 & -0.023 \\
$\msqphisqwithV$ & 1.002 & 0.002 \\
\bottomrule
\end{tabular}
\caption[Spectral Index Values for the Four Potentials]
{The spectral index for scalar perturbations for each of the four
potentials used. These values are calculated for the $\kwmap$ scale, five e-foldings
after it crosses the horizon. The potential parameters are listed in
Table~\ref{tab:params-num}. The value $U_0=5\e{-10}\Mpl^{4}$ was chosen to ensure
a blue spectrum ($n_s>1$).
}
\label{table:ns-res}
\end{center}
\end{table}

\begin{figure}[htbp]
\centering%
\subfloat[$V(\vp)=\msqphisq$]{
 \includegraphics[width=0.43\textwidth]{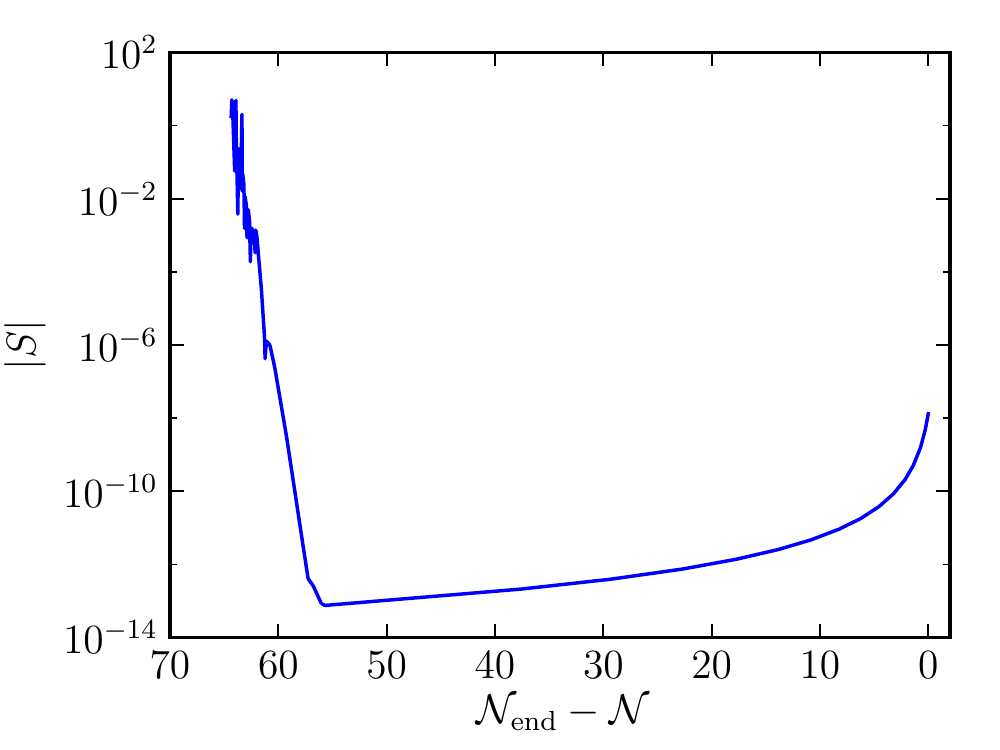}
}\qquad%
\subfloat[$V(\vp)=\lambdaphifour$]{
 \includegraphics[width=0.43\textwidth]{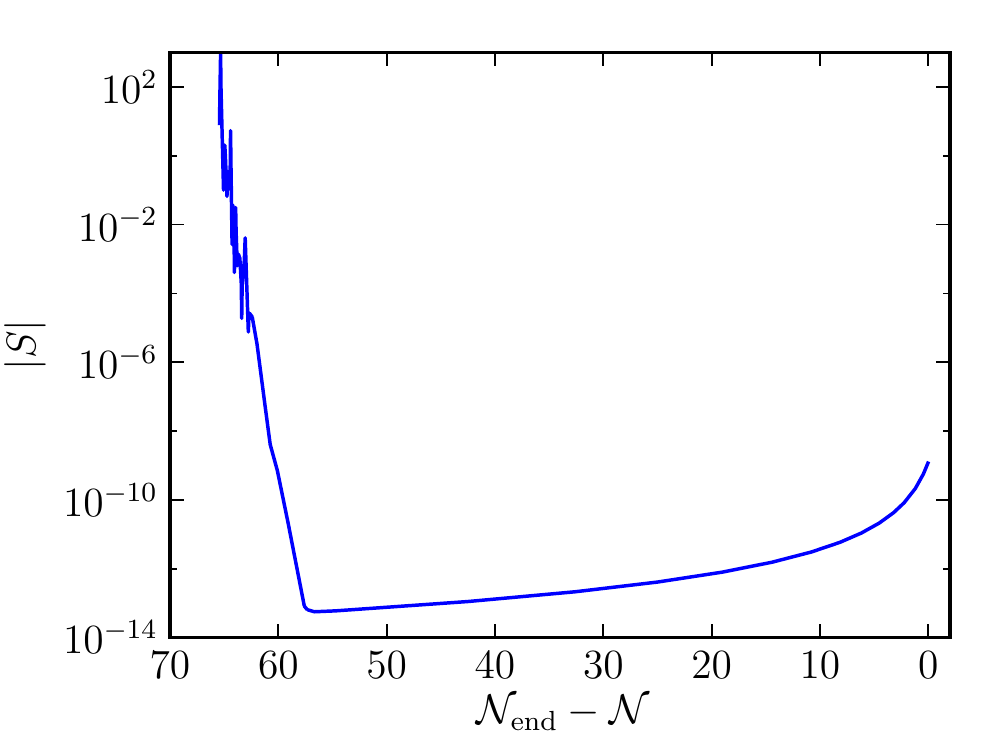}
}\\%
\subfloat[$V(\vp)=\phitwooverthree$]{
 \includegraphics[width=0.43\textwidth]{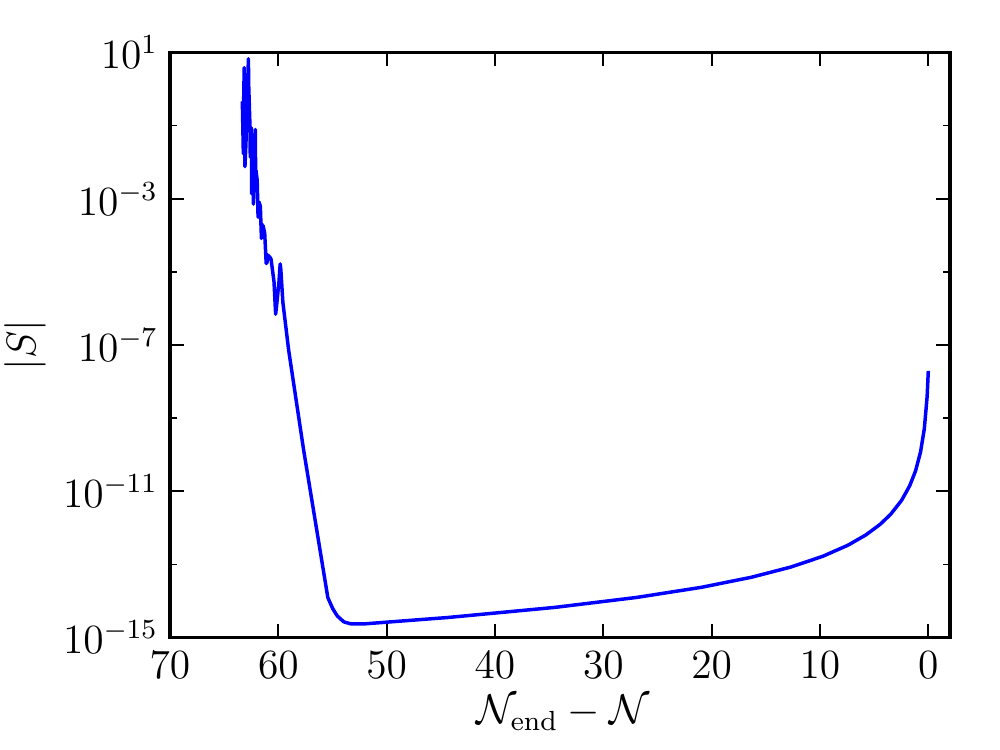}
}\qquad%
\subfloat[$V(\vp)=\msqphisqwithV$]{
 \includegraphics[width=0.43\textwidth]
  {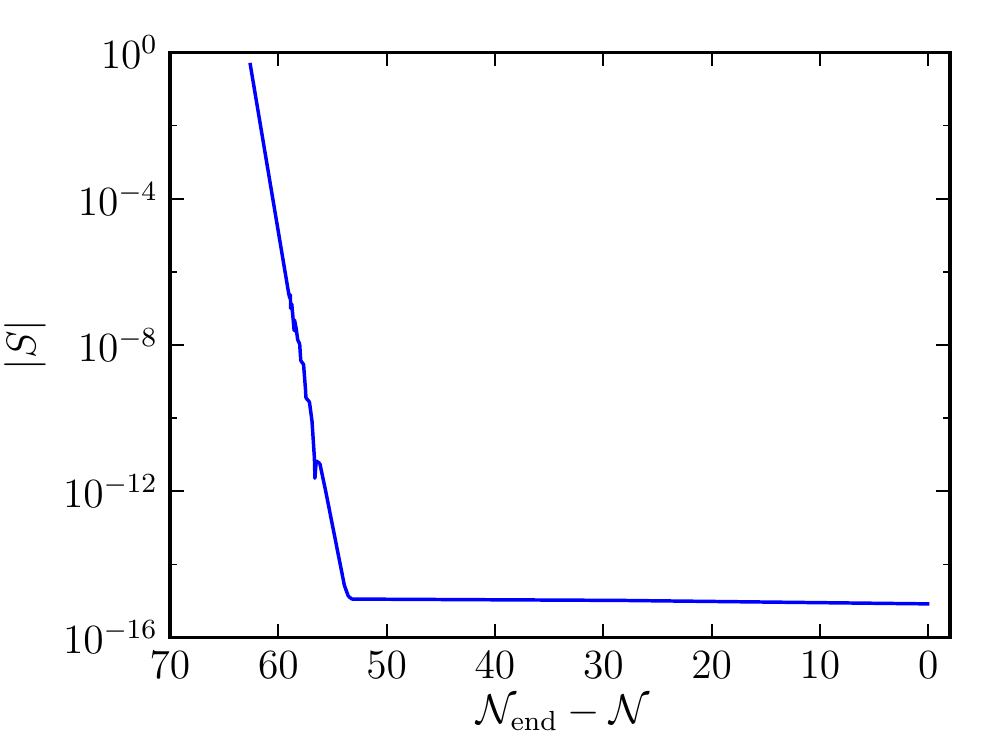}
}
\caption[Source Term for the Four Potentials]{Plots of the source
term for the
four different potentials studied.}
\label{fig:sourcecomparison-res}
\end{figure}
\begin{figure}[htbp]
 \centering
 \includegraphics[width=0.8\textwidth]{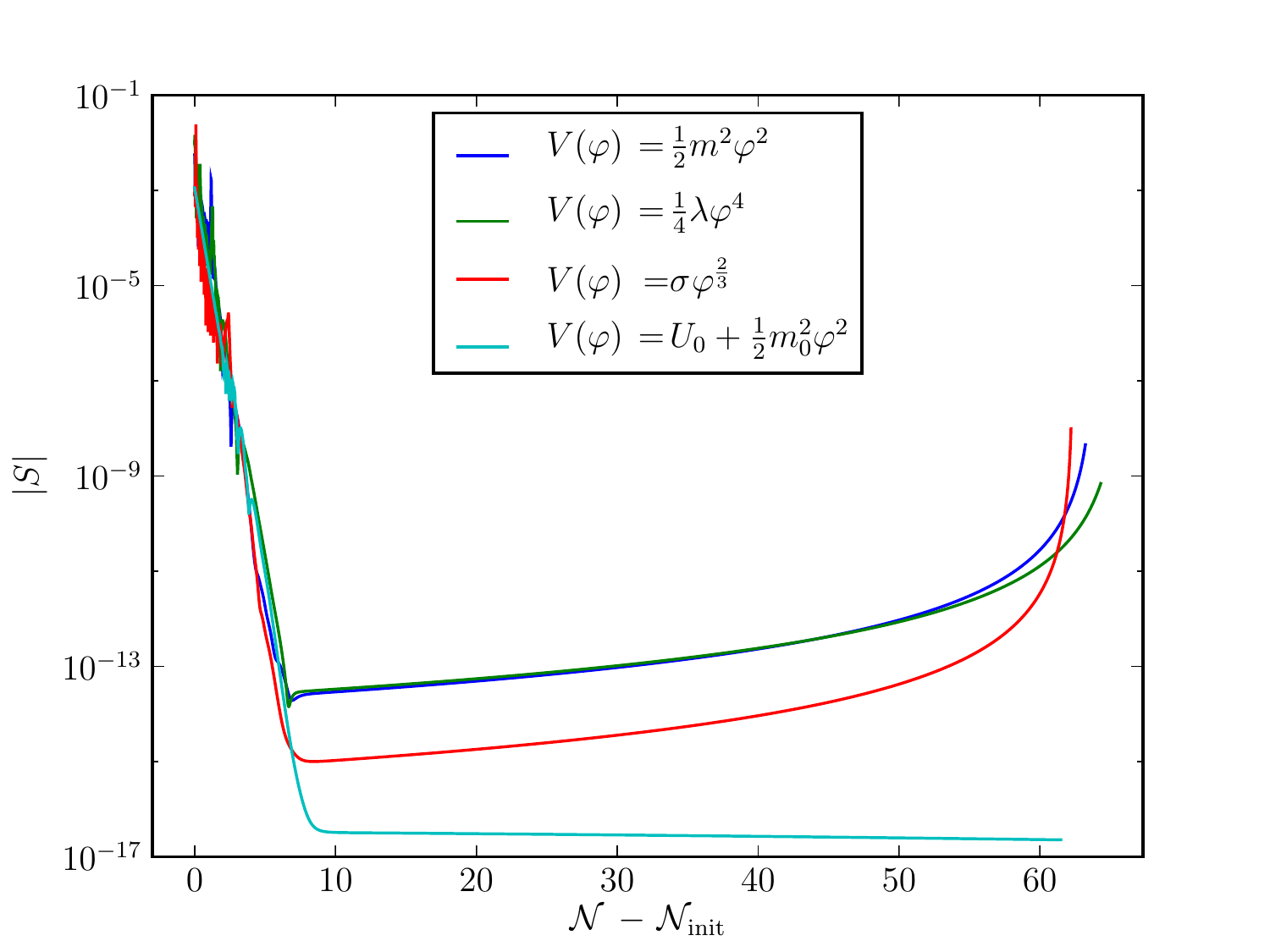}
 \caption[Comparison of Source Term for the Four Potentials]{Comparison of the
source term
evolution for the four different models. After horizon crossing the magnitude of the
source term is larger for the quadratic and quartic models than for the other two.
Towards the end of the numerical calculation there is a marked increase in $|S|$ for
three of the models as $\bar{\varepsilon}_H$ increases towards unity. The end time of
inflation is specified by hand for the contrived toy model, so this effect is
not seen.}
\label{fig:cmp-src-kwmap}
\end{figure}

The source term for each model is shown separately in Figure~\ref{fig:sourcecomparison-res} for
$\kwmap$ using the $K_2$ range\footnotemark. 
\footnotetext{These plots use a different $k$ range to the ones comparing $V(\vp)=\msqphisq$ and
$\lambdaphifour$ in \Rref{hustonmalik}.}
Although these terms are qualitatively similar,
differences between them are apparent. Figure~\ref{fig:cmp-src-kwmap} brings
together the source terms at $\kwmap$ to enable a direct comparison to be made. The
$\kwmap$ mode begins at different times for the
different models. Each result is therefore plotted in terms of the
initialisation time for that mode.  This change in duration is a consequence of
allowing $H$ to evolve during the calculation. 

\begin{figure}[htbp]
 \centering
 \includegraphics[width=0.8\textwidth]{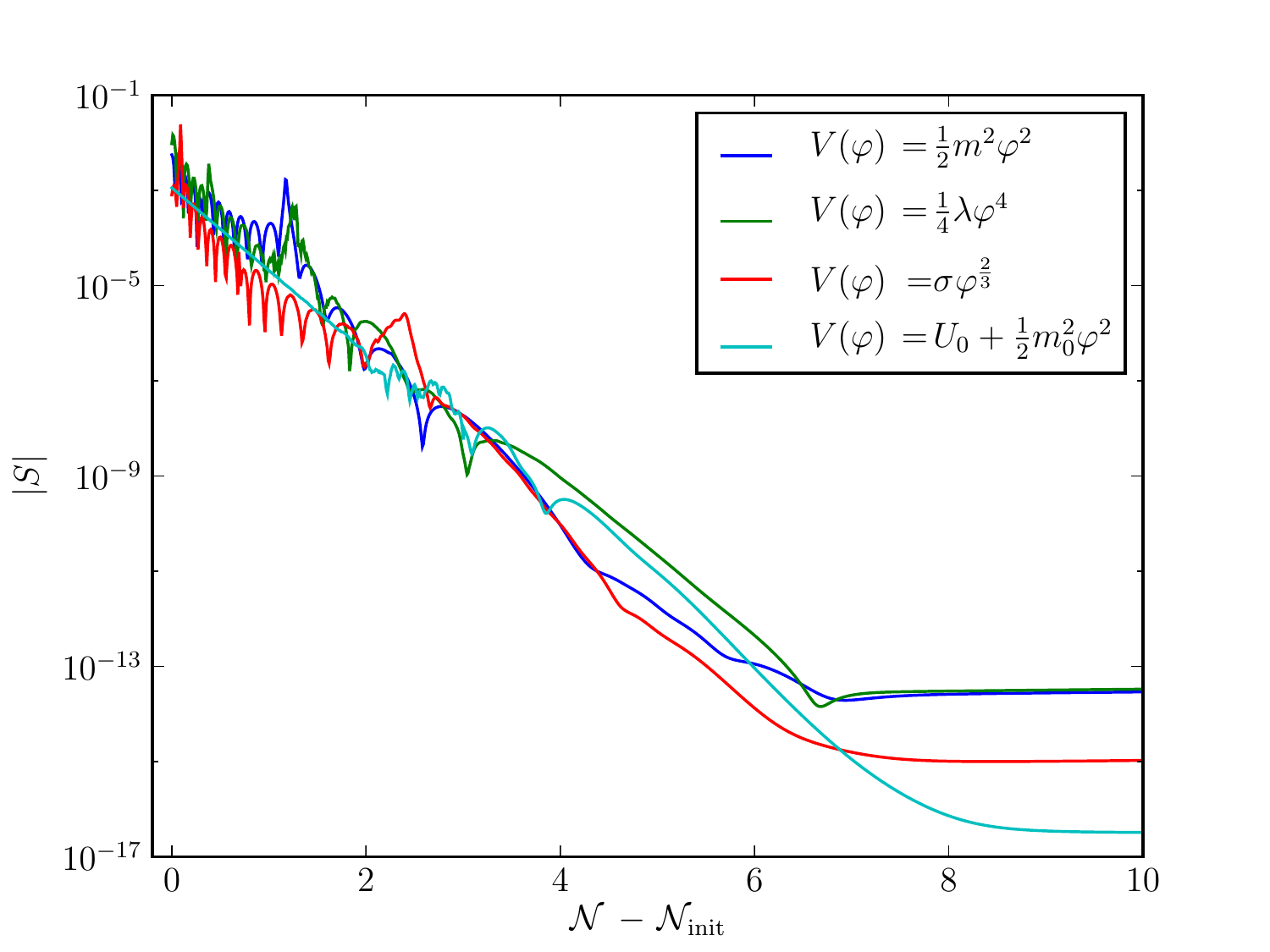}
 \caption[Comparison of Source Term at Early Times]{Comparison of the
source term evolution for the four different models at early times. This figure
highlights the early evolution of the four models shown in
Figure~\ref{fig:cmp-src-kwmap}. Before horizon crossing the magnitude of the source
term is comparable for each model. After horizon crossing differences between the
models become apparent.}
\label{fig:cmp-src-zoom-kwmap}
\end{figure}

The source term results for the quadratic and quartic potentials are very similar.
Indeed, from horizon crossing to near the end of inflation the results appear to
coincide.
The $\lambdaphifour$ mode has a slightly longer duration and at late times is reduced in comparison
with the $\msqphisq$ one. Figure~\ref{fig:cmp-src-zoom-kwmap} shows that at early times the
relationship is more complicated with the $\lambdaphifour$ mode being larger for a significant
period.

In the early stages the amplitude of the $V(\vp)=\phitwooverthree$ model is very
similar to the other two
results described above. After horizon crossing, however, there is a significant drop
in the
amplitude of $S$ in comparison with the $\msqphisq$ and $\lambdaphifour$ models. This
continues
until late in the evolution when $|S|$ increases swiftly to reach levels above the
others.
The duration of the mode in this model is shorter than in the other two models
described so far. 

The fourth model, with potential $V(\vp)=\msqphisqwithV$, is a contrived toy model.
As described in Section~\ref{sec:pots-num}, in order to perform the single field
calculation, the end time of inflation must be specified by hand. In this simulation
$\vp\simeq 8$
is taken as the end time. The potential is extremely flat in this region and the
effect of this can
be seen in the source term of the model. Before horizon crossing it is of comparable magnitude to
the other terms. However, a steep decrease in $|S|$ ensures that it is a few orders
of magnitude
smaller than
the other terms after horizon crossing. In contrast to the behaviour of the other models, the source
term does not increase close to the end of inflation. This is due to the enforced end time cut-off
which means that $\bar{\varepsilon}_H$ does not become large.

In this section we have described the results for four different single field potentials. As
expected for single field slow roll models they exhibit similar properties. In the next section
plans to extend the calculation to deal with more interesting models will be outlined.

\section{Future Directions}
\label{sec:next-res}

There are many possible ways to improve the program outlined in
Chapter \ref{ch:numericalsystem}. Chief amongst these is the implementation of the
full second order source term given in \eqs{eq:SOKG-real-num} and
\eqref{eq:Fdvk1-fourier-num}. 
As we have seen the slow roll approximation is very helpful in reducing the equations of motion to
a manageable size. However, many interesting models break the assumptions of a slowly rolling field
and to investigate these models it is necessary to use the full field equations.

Models in which the field potential is not smooth due to the presence of a feature are
particularly interesting examples of single field inflation for which slow roll is broken. 
As the
derivatives of the potential can be large around the feature, these models must necessarily be
handled without assumptions about the size of the slow roll parameters. 
In \Rref{Adams:2001vc} a model with a step potential was proposed which takes the form
\begin{equation}
 V(\vp) = \frac{1}{2}m^2\vp^2\left(1 + c\tanh\left(\frac{\vp -\vp_s}{d}\right)\right)\,,
\end{equation}
where $\vp_s$, $c$ and $d$ parametrise the location, height and width of the step feature. A bump
model has also been proposed \cite{Chen:2008wn}, with potential 
\begin{equation}
 V(\vp) = \frac{1}{2}m^2\vp^2\left(1 + c\,\mathrm{sech}\left(\frac{\vp-\vp_b}{d}\right)\right)\,,
\end{equation}
where again $\vp_b$, $c$ and $d$ parametrise the feature. At first order these models introduce
noticeable differences in the scalar power spectrum. They are also known to be able to produce
significant amounts of non-Gaussianity in shapes which are not similar to either the local or
equilateral types described in Section~\ref{sec:fnl-intro} \cite{Chen:2006xjb, Chen:2008wn}.

It will also be important to go beyond slow roll in the multiple field case.
To obtain analytic results, the study of multi-field models has often been restricted to those with
either sum or product separable potentials. Even very simple
models with two fields such as the double inflation model with the potential given by
\cite{Turner:1986te, Silk:1986vc}
\begin{equation}
 V(\vp,\chi) = \frac{1}{2}m_\vp^2\vp^2 + \frac{1}{2}m_\chi^2 \chi^2\,,
\end{equation}
can violate slow roll when the fields $\vp$ and $\chi$ are close to
equality. To go some way towards considering the full range of possible multi-field models with
arbitrary inflationary potentials then requires that the full non-slow roll evolution equations are
used.

As far as the implementation of the code is concerned, the extension to the non-slow roll single
field case is the next step.
Although clearly
more complicated than the slow roll case of \eq{eq:KG2-fourier-sr-ntime},
only three more $\theta$ dependent terms need to be added to the $\A$--$\D$ terms
listed in \eq{eq:AtoD-num}.  The four potentials considered above all result in 
slow roll inflation. Therefore, it is not expected that
using the full source equation will result in an appreciably
different outcome in these models until near the end of the inflationary phase. Once
the field has stopped rolling slowly, new observable features are expected to
arise, as is indeed the case at first order.

\eqs{eq:SOKG-real-num} and \eqref{eq:Fdvk1-fourier-num} must be written in
terms of $\N$, with the $\theta$ dependent terms grouped together, in order to set
up the numerical system completely at second order. 
The main equation
becomes
\begin{multline}
 \label{eq:fullso-res}
\ddN{\dvp2}(\kvi) + \left(3 + \frac{\dN{H}}{H}\right) \dN{\dvp2}(\kvi) +
\left(\frac{k}{aH}\right)^2 \dvp2(\kvi) \\
+ \frac{1}{H^2}\left[ \Upp + 8\pi G\left(2\dN{\vp_0}\Uphi + 8\pi G
\left(\dN{\vp_0}\right)^2 \U \right)\right] \dvp2(\kvi)
+ S_\mathrm{full}(\kvi) = 0\,,
\end{multline}
where the full source equation is given by
\begin{align}
\label{eq:fullsrc-res}
S_\mathrm{full}(\kvi) = \frac{1}{(2\pi)^2}\int \d q q^2 &\Bigg\{ 
\frac{1}{\left(H\right)^2} \left[ \Uppp + 3(8\pi G)\dN{\vp_0}\Upp\right]
 \dvp1(\qvi) \A(\kvi, \qvi) \nonumber \\
&+\frac{(8\pi G)^2 \dN{\vp_0}}{(aH)^2}\left[ 2a^2\dN{\vp_0}\Uphi +\dN{\vp_0}Q
-\frac{Q^2}{2(aH)^2}\right] \dvp1(\qvi) \A(\kvi, \qvi) \nonumber \\
&- \frac{(8\pi G)^2}{(aH)^2}\frac{(\dN{\vp_0})^2 Q}{2} \dN{\dvp1}(\qvi)
\A(\kvi, \qvi) \nonumber \\
&+ \frac{2(8\pi G)Q}{(aH)^2} \dvp1(\qvi)\wt{\C}(\kvi, \qvi) 
+ \frac{8\pi G \dN{\vp_0}}{2} \dN{\dvp1}(\qvi) \wt{\C}(\kvi, \qvi) \Bigg\} \nonumber
\\
&+ F_\mathrm{full}(\dvp1(\kvi), \dN{\dvp1}(\kvi))\,.
\end{align}
The $F_\mathrm{full}$ term in \eq{eq:fullsrc-res} requires the use of three further
$\theta$
integrals in addition to those presented in \eq{eq:AtoD-num}. These take the form
\begin{align}
\label{eq:efg-terms-res}
 \E(\kvi, \qvi) &= \int_0^\pi \cos^2(\theta) \sin(\theta) \dvp1(\kvi-\qvi)\d \theta
\,,\nonumber \\
\F(\kvi, \qvi) &= \int_0^\pi \frac{\sin^3(\theta)}{|\kvi-\qvi|^2} \dvp1(\kvi-\qvi)\d
\theta \,,\nonumber \\
\wt{\G}(\kvi, \qvi) &= \int_0^\pi \frac{\sin^3(\theta)}{|\kvi-\qvi|^2}
\dN{\dvp1}(\kvi-\qvi)\d \theta \,.
\end{align}
It is worth noting that the term $\sin^3(\theta)/|\kvi-\qvi|^2$ tends to zero in the
limit where $k=q$ and
$\theta\rightarrow 0$.
The $F_\mathrm{full}$ term can now be written in terms of $\E$, $\F$ and $\wt{\G}$:
\begin{align}
 \label{eq:fullfterm-res}
F_\mathrm{full} &= \frac{8\pi G}{(2\pi)^2}\frac{1}{(aH)^2} \int \d q\, q^2 \Bigg\{
\nonumber \\
&  \dN{\vp_0}\Bigg[ \left(2k^2 + \left(\frac{7}{2} - \frac{8\pi
G}{4}(\dN{\vp_0})^2 \right)q^2 + \frac{3}{4}\frac{8\pi G}{(aH)^2} X^2 \right)
\dvp1(\qvi) \nonumber \\
& \qquad + (8\pi G) Q \dN{\vp_0} \left(\frac{3}{4} + \frac{q^2}{k^2}\right)
 \dN{\dvp1}(\qvi) \Bigg] \A(\kvi, \qvi) \nonumber \\
&+ \Bigg[ \left( 2Q \frac{q}{k} \left(1- \frac{8\pi G}{(aH)^2} Q \dN{\vp_0}\right)
-\frac{9}{2} \dN{\vp_0} k q - \dN{\vp_0} \frac{q^3}{k}\right) \dvp1(\qvi) \nonumber
\\
&\qquad - 2Q (8\pi G) (\dN{\vp_0})^2\frac{q}{k} \dN{\dvp1}(\qvi)\Bigg] \B(\kvi, \qvi)
\nonumber \displaybreak[0]\\
&+ \Bigg[ \left(-2 + (8\pi G)(\dN{\vp_0})^2 \left(\frac{1}{4} + \frac{1}{2aH}\right)
\right) Q \dvp1(\qvi) \nonumber \\
&\qquad + \left(\frac{8\pi G}{4}(\dN{\vp_0})^2 -2\right) \dN{\vp_0} (aH)^2
\dN{\dvp1}(\qvi)\Bigg] \wt{\C}(\kvi, \qvi)\nonumber \\
&+ \Bigg[ 2Q \frac{k}{q}\dvp1(\qvi) + \left(2\frac{k}{q}-\frac{q}{k}\right)
\dN{\vp_0} (aH)^2 \dN{\dvp1}(\qvi) \Bigg]\wt{\D}(\kvi, \qvi) \nonumber \\
&+ (8\pi G) \dN{\vp_0} \Bigg[ \left(\frac{1}{4}(\dN{\vp_0})^2q^2 +
\frac{Q^2}{2(aH)^2}\right) \dvp1(\qvi) + \frac{Q}{2}\dN{\vp_0} \dN{\dvp1}(\qvi)
\Bigg] \E(\kvi, \qvi) \nonumber \displaybreak[0]\\
&+ (8\pi G)^2 \dN{\vp_0} Q \Bigg[ -\frac{Q}{2(aH)^2}\left(\frac{k^2}{2}+q^2\right)
\dvp1(\qvi) - \frac{1}{4} \dN{\vp_0} k^2 \dN{\dvp1}(\qvi) \Bigg] \F(\kvi, \qvi)
\nonumber \\
&+ (8\pi G)^2 (\dN{\vp_0})^2 \Bigg[ -\frac{Q}{2}\left(\frac{k^2}{2} +
\frac{q^2}{aH}\right)\dvp1(\qvi) -\frac{(aH)^2}{4}\dN{\vp_0}k^2 \dN{\dvp1}(\qvi)
\Bigg] \wt{\G}(\kvi, \qvi) \Bigg\} \,.
\end{align}
\eqs{eq:fullsrc-res} and \eqref{eq:fullfterm-res} are clearly more complicated than
the slow roll versions used in
Chapter~\ref{ch:numericalsystem}. The numerical complexity is not significantly
greater, however, once the three terms in \eq{eq:efg-terms-res} have been calculated.
The running time of the full calculation will clearly be a significant constraint.

With this in mind, the performance of the numerical simulation could be improved by
analysing the most time consuming processes and investigating what
optimisations might be implemented. The current, perhaps
inelegant, procedure will allow any performance improvements to be benchmarked for
accuracy as well as for speed.
As discussed above, $N_k$ was set to $1025$ for the test runs. This provides good
coverage of the
WMAP $k$ range, but it is not clear whether it sufficiently
approximates the integral to infinity for the source term.  Logistical factors,
including the running time and memory usage of the code,
restrict the choice of $N_k$. By optimising the routines
for reduced memory and increased speed it is hoped that the range of scales can be
extended and the resolution enhanced.

Beyond these considerations, the next significant step is to implement a multi-field
version of the system. This would allow the investigation of models that inherently
produce large second order perturbations. In \Rref{Malik:2006ir} the
second order Klein-Gordon equation for multiple fields was presented and upgrading
the
simulation to use these equations should be a straight-forward (if lengthy) process.
Extending the current data-structures and routines to a fixed number of extra fields
will increase the numerical complexity and the run-time of the code. 

For example, let us suppose that the second order perturbations of two scalar fields,
$\vp$ and $\chi$, are to
be calculated. Let $V$ denote the potential and $V_0$ its background value. As the
coding environment we have used can easily handle arrays of variables, it is useful
to write the equations in vector form. The following definitions will be used:
\begin{align}
\label{eq:vector-defns-res}
 \bm{\vp_0} &= \begin{pmatrix}
                \vp_0 \\
		\chi_0
               \end{pmatrix} \,, &
 &\bm{\dvp1} = \begin{pmatrix}
                \dvp1 \\
		\delta\chi_1
               \end{pmatrix} \,, &
 &\bm{\dvp2} = \begin{pmatrix}
                \dvp2 \\
		\delta\chi_2
               \end{pmatrix} \,, \\
 \bm{V_1} &= \begin{pmatrix}
             \Uphi \\
	     V_{,\chi} 
            \end{pmatrix} \,, &
 &\bm{V_2} = \begin{pmatrix}
             \Upp & V_{,\vp\chi} \\
	     V_{,\chi\vp} & V_{,\chi\chi}
            \end{pmatrix} \,, &
&\bm{V_3} = \begin{pmatrix}
            \Uppp & V_{,\vp\vp\chi} \\
	    V_{,\vp\vp\chi} & V_{,\vp\chi\chi} \\
	    V_{,\vp\chi\chi} & V_{,\chi\chi\chi}
           \end{pmatrix} \,.
\end{align}
In conformal time the Friedmann equation becomes
\begin{equation}
 \H^2 = \frac{8\pi G}{3} \left( \frac{1}{2}\left(\vp_0'\right)^2 
        + \frac{1}{2}\left(\chi_0'\right)^2 + a^2 \U \right) 
= \frac{8\pi G}{3}\left( \frac{1}{2} \left( \bm{\vp_0}' \right)^{T} \bm{\vp_0}' + a^2 \U \right) \,,
\end{equation}
where $\bm{\vp}^{T}$ denotes the transpose of $\bm{\vp}$.
The background vector equation of motion is given by
\begin{equation}
 \bm{\vp_0}'' + 2\H \bm{\vp_0}' + a^2 \bm{V_1} = \bm{0}\,,
\end{equation}
where $\bm{0}$ is the zero vector.
The first order vector equation takes the form
\begin{multline}
 \bm{\dvp1}''(\kvi) + 2\H \bm{\dvp1}'(\kvi) \\
 + \left( k^2 \bm{1}  + \bm{V_2} 
 + \frac{8\pi G}{\H}\left\{ 
  \bm{\vp_0}' \bm{V_1}^{T} + \bm{V_1}\left(\bm{\vp_0}'\right)^T
  + \frac{8\pi G}{\H}\U \bm{\vp_0}' \left(\bm{\vp_0}'\right)^T
\right\}
\right)\bm{\dvp1}(\kvi) = \bm{0}\,,
\end{multline}
where $\bm{1}$ is the identity matrix.

We will outline the second order vector equation using the slow roll approximation.
In the multi-field case there are many more slow roll parameters than in the
single field scenario. Extending the definition of $\bar{\varepsilon}_H$ in
\eq{eq:bareps-defn-perts} to two fields gives
\begin{align}
 \bar{\varepsilon}_\vp &= \sqrt{4\pi G} \left( \frac{\vp_0'}{\H} \right) \,,\\
 \bar{\varepsilon}_\chi &= \sqrt{4\pi G} \left( \frac{\chi_0'}{\H} \right)\,.
\end{align}
There are now four $\eta_H$-type parameters corresponding to the different
combinations of second derivatives of $V$. These can be written together in matrix form
as
\begin{equation}
 (\eta_{IJ}) = \bm{\eta}_H = \frac{a^2}{3\H^2} \bm{V_2} \,,
\end{equation}
where $I,J = \vp,\chi$.
The magnitude of $\eta_{IJ}$ is only small in the adiabatic direction, so terms
including $\eta_{IJ}$ are included when making the slow roll approximation
\cite{Malik:2006ir}.

The second order, slow roll, vector equation for the perturbations is given by 
\begin{align}
 \bm{\dvp2}''(\kvi) &+ 2\H\bm{\dvp2}'(\kvi) 
 + \left( k^2\bm{1} + a^2\bm{V_2} - 24\pi G \bm{\vp_0}' \left(\bm{\vp_0}'\right)^T \right)
       \bm{\dvp2}(\kvi) \nonumber\\ 
 &+ \bm{S}(\kvi) = \bm{0}\,,
\end{align}
where the slow roll source term equation is
\begin{align}
\label{eq:src-vector-res}
\bm{S}(\kvi) =\,  
&\frac{1}{(2\pi)^3}\int \d^3p\ \d^3q\ \delta^3(\kvi-\pvi-\qvi) \Bigg\{ \\
&a^2 \Bigg[ 
\begin{pmatrix}
\dvp1(\pvi) & \delta \chi_1(\pvi) & 0  \nonumber\\
0           & \dvp1(\pvi)         & \delta \chi_1(\pvi) 
\end{pmatrix}
\bm{V_3}\bm{\dvp1}(\qvi) \nonumber \\
&+ \frac{8\pi G}{\H}\left( 
    \bm{\dvp1}^T(\pvi) \bm{V_2} \bm{\dvp1}(\qvi)
\right) \bm{\vp_0}' \Bigg] \nonumber\\
&+ \frac{16\pi G a^2}{\H}\left(\left(\bm{\vp_0}'\right)^T \bm{\dvp1}(\pvi) \right)
    \bm{V_2} \bm{\dvp1}(\qvi) \nonumber \\
&+ \frac{8\pi G}{\H} \Bigg[ 2\frac{p_l q^l}{q^2} 
      \left( \left(\bm{\vp_0}'\right)^T \bm{\dvp1}'(\qvi) \right) \bm{\dvp1}'(\pvi)
 + 2p^2 \left( \left(\bm{\vp_0}'\right)^T \bm{\dvp1}(\qvi) \right) \bm{\dvp1}(\pvi) \nonumber \\
&+\left( \frac{p_l q^l + p^2}{k^2}q^2 - \frac{p_l q^l}{2} \right) 
      \left( \bm{\dvp1}^T(\pvi) \bm{\dvp1}(\qvi) \right) \bm{\vp_0}' \nonumber \\
&+\left(\frac{1}{2} -\frac{p_l q^l + q^2}{k^2}\right) 
      \left( \left(\bm{\dvp1}'\right)^T(\pvi) \bm{\dvp1}'(\qvi) \right) \bm{\vp_0}' \Bigg] \Bigg\}
\,.
\end{align}

Following the method of Section~\ref{sec:eqs-num}, the $\d^3p$ integral is evaluated
and the $\d^3 q$ integral is written in spherical polar coordinates. The $\theta$
dependent terms, which are equivalent to \eq{eq:AtoD-num} in the single field case,
are given by
\begin{align}
\label{eq:AtoD-vectors-res}
 \A_\vp(\kvi,\qvi) &= \int_0^\pi \sin(\theta) \dvp1(\kvi-\qvi) \d\theta \,,
\nonumber \\ \displaybreak[0]
 \A_\chi(\kvi,\qvi) &= \int_0^\pi \sin(\theta) \delta \chi_1(\kvi-\qvi) \d\theta \,,
\nonumber\\ \displaybreak[0]
 \bm{\A}(\kvi,\qvi) &= \int_0^\pi \sin(\theta) \bm{\dvp1}(\kvi-\qvi) \d\theta \,,
\nonumber\\ \displaybreak[0]
 \bm{\B}(\kvi,\qvi) &= \int_0^\pi \cos(\theta)\sin(\theta) \bm{\dvp1}(\kvi-\qvi)
\d\theta \,,\nonumber\\ \displaybreak[0]
 \bm{\C}(\kvi,\qvi) &= \int_0^\pi \sin(\theta) \bm{\dvp1}'(\kvi-\qvi) \d\theta \,,
\nonumber\\
 \bm{\D}(\kvi,\qvi) &= \int_0^\pi \cos(\theta) \sin(\theta) \bm{\dvp1}'(\kvi-\qvi)
\d\theta \,.
\end{align}
The first two equations are not vector equations but are needed for the explicit
matrix term in \eq{eq:src-vector-res}. We rewrite that equation with these
definitions to obtain:
\begin{align}
 \bm{S}(\kvi) =\, &\frac{1}{(2\pi)^2}\int\d q \Bigg\{ \nonumber \\
&a^2 q^2\Biggl[ \begin{pmatrix}
                 \A_\vp(\kvi,\qvi) & \A_\chi(\kvi,\qvi) & 0 \\
		 0                 & \A_\vp(\kvi, \qvi) & \A_\chi(\kvi, \qvi)
                \end{pmatrix}
		\bm{V_3} \bm{\dvp1}(\qvi) \nonumber \\
& \qquad +\frac{8\pi G}{\H} \left( \bm{\A}^T(\kvi,\qvi) \bm{V_2}
\bm{\dvp1}(\qvi)\right) \bm{\vp_0}'
	\Biggr] \nonumber \\
&+ \frac{16\pi G}{\H}a^2 q^2 \left( \left(\bm{\vp_0}'\right)^T \bm{\A}(\kvi,\qvi) \right)
	\bm{V_2}\bm{\dvp1}(\qvi) \nonumber \\
&+ \frac{8\pi G}{\H} \Bigg[ 2\left( \left(\bm{\vp_0}'\right)^T \bm{\dvp1}'(\qvi) \right)
	\left( kq \bm{\D}(\kvi,\qvi) - q^2 \bm{\C}(\kvi, \qvi) \right) \nonumber \\
&\qquad +2q^2 \left( \left(\bm{\vp_0}'\right)^T \bm{\dvp1}(\qvi) \right)
	\left( (k^2 + q^2)\bm{\A}(\kvi,\qvi) - 2kq\bm{\B}(\kvi, \qvi) \right)\notag\\
&\qquad + \left(\left[ \frac{3}{2}q^4 \bm{\A}^T(\kvi,\qvi) 
	- \left( \frac{1}{2}kq^3 + \frac{q^5}{k}\right) \bm{\B}^T(\kvi,\qvi)
	\right] \bm{\dvp1}(\qvi) \right) \bm{\vp_0}' \nonumber \\
&\qquad + \left(\left[ \frac{1}{2}q^2 \bm{\C}^T(\kvi,\qvi) 
	- \frac{q^3}{k} \bm{\D}^T(\kvi,\qvi)
	\right] \bm{\dvp1}'(\qvi) \right) \bm{\vp_0}' \Bigg] \Bigg\} \,.
\end{align}
This expression for $\bm{S}$ reduces to \eq{eq:KG2-src-sr-aterms} when only one
field is considered. At least in the slow roll case, the multi-field source term
equation is not considerably more complex than the single field one. The extra
numerical complexity arises from the calculation of the new $\theta$
dependent terms in \eq{eq:AtoD-vectors-res}.

%
%
%
\section{Discussion}
\label{sec:disc-num}

Part~\ref{part:numerical} of this thesis has described the numerical
solution of the evolution equations for second order scalar perturbations. The
closed form of the Klein-Gordon equation \eqref{eq:KG2-fourier-sr-num} has been
employed for the first time. We have shown that direct
calculation of the field perturbations beyond
first order using perturbation theory is readily achievable, although
it is non-trivial.

This first demonstration has been limited to considering
the slow roll approximation of the source term in \eq{eq:KG2-fourier-sr-num} which
is quadratic in first order perturbations. Slow roll has not been imposed on the
background or first order equations. Four different
potentials were
used to demonstrate the capabilities of the
system. The singularity at $k=0$, which arises as larger and larger
scales are considered, is avoided by implementing a cutoff at small
wavenumbers below $\kmin$. 
This is a pragmatic choice necessary for
the calculation.
It is also necessary to
specify a maximum value of $k$. This choice is dictated by computational
resources and with reference to observationally relevant scales. In
this demonstration, $k$ ranges have been used which are comparable with the
scales observed by the WMAP satellite. By comparing the analytical results of the
convolution integral with the numerical calculation, values of the parameters
$N_\theta, N_k$ and $\Delta k$ were chosen in such a way that the numerical error
was minimised. The convolution scheme that has been implemented
works best when $\Delta k>\kmin$.

We have seen explicitly that the second order calculations for the chosen
potentials can be completed once the cut-off for $\kmin$ is
imposed. As expected for these potentials the magnitude of second order perturbations
is extremely
suppressed in the slowly rolling regime, in comparison with the first order
amplitude.
We have also shown that the
evolution of the source term during the inflationary regime can be
readily calculated.

By computing the perturbations to second order, we have direct access
to the non-Gaussianity of $\dvp{}$. When used to investigate models
that predict a large non-linearity parameter, $\fnl$, this technique
could yield greater insight into the formation and development of the
non-Gaussian contributions by studying the effects of the different
terms in the source equation \eqref{eq:KG2-source-ntime}.
It was shown recently that $\fnl$ can be calculated directly from the field
equations \cite{Musso:2006pt,Seery:2008qj}, instead of
using the standard method based on a Lagrangian formalism
\cite{Maldacena:2002vr}. The method presented here
will
therefore eventually allow a full numerical calculation of $\fnl$ to be made.

Our numerical code evolves the second order perturbation itself and gives an insight into how this
field behaves through the full course of the inflationary era. This is in contrast to other
approaches which only consider the result for the three point function of the field, or
alternatively of the curvature perturbation.
The computational system handles perturbations with scales both inside and outside the horizon. Any
effects of horizon crossing are visible and no assumptions need to be made about the form of the
solution inside the horizon.

The numerical code, when developed with the full equation, will not require any simplifying
assumptions about the form of the potential used. This allows models which are not amenable to
analytic analysis to be examined. Examples of models which require consideration beyond the slow
roll approximation include single field models with a step or other feature in the potential, and
multi-field double inflation models where the field values are roughly equal.

The code we have developed is also applicable in other physical circumstances. Beyond scalar
perturbations the form of the source term is similar in other interesting cosmological physics. 
The generation and evolution of non-Gaussian curvature perturbations is, of course, directly related
to the behaviour of the second order scalars as has been described in Section~\ref{sec:fnl-intro}
and Section~\ref{sec:observable-perts}. Investigating and classifying non-Gaussian signatures for
inflationary models is the main goal of our future work.

The generation of vorticity in a cosmological setting has physical parallels with the equations we
have studied. This second order effect arises through the vector perturbations which we have not
considered in this thesis. Vorticity in the early universe could also lead to the generation of
primordial magnetic fields, an area which is of increasing interest
\cite{Christopherson:2009bt,1950ZNatA...5...65B}.
The wave equations for tensor mode perturbations also exhibit the same form as the scalar equations
with a source term at second order. The code we have developed could be modified to examine the
behaviour of gravitational waves in the early universe at second order.

In summary, we have demonstrated that numerically solving the closed
Klein-Gordon equation for second order perturbations is possible. The slow roll
version of the source term was used in the calculation, but
as described in Section~\ref{sec:next-res}, the extension of the system to include
the full source term is achievable. 
The analytic and numerical solutions for the convolution terms were
compared directly and found to be in good agreement.
The models used have been shown to have negligible second order
perturbations in line with known analytic results. 


\part{Conclusion}
\label{part:conclusion}

%

\chapter{Conclusion and Discussion}
\label{ch:conclusions}

As the number of viable cosmological models increases, the need to constrain
them becomes more
important. At the same time, the quantity and quality of observational data continue
to improve.
There now exists the opportunity to go beyond linear statistical analyses
and confront the predictions of models with observational data from the non-linear
regime. In this thesis both analytic and numerical methods have been developed to
constrain inflationary models.

The framework used in this thesis is the Friedmann-Robertson-Walker universe,
reviewed in Chapter~\ref{ch:introduction}. During the accelerated expansion of the
inflationary period, quantum fluctuations seeded energy density variations, which in turn
gave rise to the diverse structure of the present universe. First order
cosmological perturbation theory
is necessary to describe the evolution of these fluctuations. The main observable
quantities can be calculated at horizon crossing by using the Bunch-Davies vacuum
initial conditions. The departure of the perturbations from a purely Gaussian random
field is parametrised by $\fnl$, which is described in two limits, local and
equilateral. 
In addition to canonical actions, we also introduced
non-canonical models, for which the speed of sound of the perturbations plays a
crucial role. When the sound speed is small, the amplitude of $\fnleq$ for these
models is large.

In Part~\ref{part:dbi}, analytic methods were developed to constrain
string theory inspired non-canonical inflationary models. The Dirac-Born-Infeld
scenario was
outlined in Chapter~\ref{ch:dbi-intro}. In this model, a D3-brane propagates in a
six-dimensional warped throat. The radial position of the brane from the tip of the
throat assumes the role of the inflaton field. The non-canonical nature of the DBI
action restricts the kinetic energy of this field no matter how steep the potential.
This allows an inflationary period of sufficient duration to occur.
This model has been widely regarded as a very promising realisation of an
inflationary
model in a string-theory context.

In \Rref{bmpaper}, Baumann \& McAllister used the Lyth bound \cite{lyth} to limit
the tensor-scalar ratio. Their analysis was based on the conservative assumption that the brane
could
not propagate further than the full length of the throat. In Chapter~\ref{ch:dbi},
we showed that this bound can be tightened by applying it over the
portion of the throat through which the brane passes during the directly observable
stage of inflation. Restricting the field variation to be over these
approximately four e-foldings constrains $r$ to be less than $10^{-7}$ for standard
parameter values.

The most optimistic estimates of advances in experimental techniques and data
analysis, including foreground reduction, indicate that observations of $r>10^{-4}$
might be achievable in the future \cite{Baumann:2008aq,vpj}. Therefore, the new bound in
\eq{eq:upperbound} immediately rules out the observation of a tensor mode signal
from this model.

In addition to this, we also derived a lower bound on $r$ in \eq{eq:lowerbound}.
This depends on observable quantities, namely the scalar spectral index and the
equilateral non-Gaussianity. Saturating the WMAP5 observational limit on $\fnleq$
and taking the best fit value for $n_s$, we found that the most conservative lower
limit is $r>0.005$. This is clearly incompatible with the previously derived upper
bound. Therefore, for standard parameter values, the D3-brane DBI scenario is not
viable. Numerical simulations by Peiris \etal in \Rref{Peiris:2007gz} have
demonstrated the lower bound, in the relativistic limit, using the
Hamiltonian flow approach.

In Section~\ref{sec:relaxing-dbi}, a phenomenological approach was taken to easing
the upper bound on the tensor-scalar ratio. By considering a DBI-type action with unspecified field
functions,
$f_i$, we showed that the generalised lower and upper bounds can be consistent
if the product of $f_A$ and $f_B$ is sufficiently large on observable scales. This provides a guide
to the types of models which could evade the inconsistency of the bounds on $r$. For more general
models with a non-canonical action, a bound on $r$ which relates the geometry of the throat, the
number of e-foldings of observable inflation, and the derivatives of the action has been derived in
\eq{eq:LHbound}. This bound, although it does not in general relate to observational quantities can
be used when the details of a particular physical model are known.

The discovery of the incompatible bounds on $r$ for DBI inflation has had a noticeable
impact on the research community, spurring interest in finding models which evade
these bounds. Many such models have been proposed with varying degrees of success. In
Section~\ref{sec:others-dbi} these were categorised according to whether they
featured single or multiple fields, and single or multiple branes. Some of these models are still
constrained by the bounds on $r$ but not to the same extent as the standard DBI scenario. For
example the parameter space of the models with wrapped brane configurations is still extremely
limited by the observational values from WMAP5 \cite{Alabidi:2008ej}. For other models an analysis
in terms of the bounds derived in this thesis has yet to be undertaken. As the observational limits
on $\fnleq$ and $r$ continue to improve, an important step in ensuring the validity of DBI based
models is to check whether equivalent bounds to those derived here exist, and whether they can be
met for any significant proportion of the parameter space.

General actions which relax the bounds on $r$, were derived in
Chapter~\ref{ch:multibrane}. 
We found a class of actions similar in form to that of the DBI model. 
However, instead of a square-root in the kinetic term, the index of the main term of these
actions depends on the constant of proportionality between $\Lambda$ and the sound
speed of inflaton fluctuations.
The upper bound on $r$ can be derived for these general actions and when the index
is below the critical value of $1/2$ (corresponding to the standard DBI scenario), this bound is
significantly relaxed.  
When new models are proposed in the future, our phenomenological derivation of this family of
actions will allow those models for which the bound on $r$ is less stringent to be easily
identified.

One such model is the single field, multi-coincident brane scenario of
Thomas \& Ward \cite{thomasward}. When $n$ D3-branes propagate in a throat, the
non-Abelian interactions between the branes result in major departures from the
single brane case. This is in contrast to the non-interacting branes model, in which
the total action is simply the sum of copies of the single brane action.

In the limit of a large number of branes being coincident, the effective action is
similar to $n$ times the single brane model, with the addition of a fuzzy potential
term. 
Indeed this model is of the type considered phenomenologically in Section~\ref{sec:relaxing-dbi} and
the bounds derived in that section can be applied. 
These bounds on $r$ can be somewhat relaxed when this potential is large,
but the model is still strongly constrained by current observations. For an $AdS_5$
throat, standard parameter choices limit the number of allowed branes to be
less than 150, at which point the assumption of arbitrarily large $n$ becomes
questionable.

More promising is the finite $n$ limit of the coincident brane model. The
non-Abelian nature of the interactions leads, in this case, to a recursive relation for
the $n$-brane action in terms of the $n=2$ one. In
Section~\ref{sec:finiten-multi}, we showed that the action for finite $n$ is one of
the class of bound-evading actions described above. This identification is possible
because the last term of the recursive sum dominates in the relativistic limit. This
approximation is valid at least when $n<10$ and the backreaction of the multiple
branes is kept well under control for this range of $n$.

Although the bounds on $r$ are eased
for this model, we showed that observations strongly constrain the possibility of
an observable tensor signal being generated. If an observable tensor-scalar ratio is
considered to be $r>10^{-4}$, then only the two or three brane cases are capable of
producing such a signal. This bound on $n$ depends on the WMAP5 limit on $\fnleq$.
If, as expected, the observational limits on $\fnleq$ tighten considerably in the
future, the possibility of an observable tensor signal from the multi-coincident
brane model could be ruled out.

On the other hand, the choice of $r>10^{-4}$ as the threshold of an observable signal is very
optimistic. If foreground removal techniques and the signal-to-noise ratios of future experiments
cannot reach this threshold, and instead reach $r>10^{-3}$, no number of branes will be able to
produce an observable tensor signal when combined with the current limits on the non-Gaussianity.
There will then be little possibility of a distinguishing
observational signature for these coincident brane models.

In Part~\ref{part:numerical} of this thesis, numerical methods were used to test inflationary
models up to second order in cosmological perturbation theory. 
The Klein-Gordon equation at second order was derived in \Rref{Malik:2006ir} for the
multi-field case. 
In Chapter~\ref{ch:perts}, second order gauge transformations were outlined
and the Klein-Gordon equation reproduced for a single scalar field model. In contrast
to the $\Delta \N$ approach, this equation is valid on all scales, both inside and
outside the horizon.

In Fourier space, the second order Klein-Gordon equation \eqref{eq:SOKG-real-num} contains a
convolution term of the first order scalar field perturbation. For this first
demonstration of the numerical system, a slow roll approximation of the full
equation was used.

Calculating the second order scalar field perturbations provides the possibility of
a unique insight into the generation and evolution of non-linear contributions to
the scalar curvature perturbation. 
One advantage of using the inflaton field equations is that we can directly
investigate how the perturbations are generated.
If, instead, we integrated the evolution
equation for a derived observable quantity, there would be a degree of
separation from the physical origins of this process. Indeed, there is, as
yet,
no known evolution equation for the main observable quantity, the comoving
curvature perturbation, at second order. 
Using cosmological perturbation theory also provides control over the calculation.
The domain of applicability of the perturbative expansion is well defined and the
resultant equations are certain to be valid in this domain.

The main observable quantity is not however the second order scalar perturbation, but rather the
departure from Gaussianity in the CMB temperature map, parametrised by the amplitude of the
bispectrum of the perturbations. In Section~\ref{sec:observable-perts} we outlined how $\fnl$ could
be calculated from the numerically found $\dvp2$ both for the local type and more generally using
the bispectrum of the uniform density curvature perturbation. As the observational limits on $\fnl$
are tightened over the course of the remaining WMAP releases and future Planck data, the importance
of comparing the predictions for $\fnl$ of inflationary models with the observed values will only
increase. In this thesis we have not computed $\fnl$ for the models we have considered, but this is
an important future step that will be undertaken.

The long term aim of this continuing project is
to analyse multi-field, non-slow roll models, in which non-Gaussian effects are
expected to play an important role.
As a step towards this goal, we described the implementation of a numerical
calculation of the single field, slow roll, second order equation in
Chapter~\ref{ch:numericalsystem}. The construction and evaluation of the convolved
source term in \eq{eq:KG2-src-sr-aterms} proved to be the most numerically complex
step required. 

To allow a numerical calculation, an energy scale cutoff must be implemented. We used
a sharp cutoff at small wavenumbers below which the perturbations were taken to be
identically zero. Another cutoff at small scales (or equivalently large wavenumbers) was dictated
by practical considerations of calculation size and computation time. 

We defined, in \eq{eq:AtoD-num}, four $\theta$ dependent integrals
into which the convolution term can be decomposed. By comparison with an analytic
solution for a particular smooth choice of the first order perturbation, an estimate
was made of the relative error present in the integration of each term. The
number of discrete wavenumber values, their spacing, and the number of discrete
$\theta$ values were chosen to minimise the error in one of the integrals. From
these parameter values, three
different finite ranges of discrete values of the wavenumber were defined.
These all contain the WMAP pivot scale at $\kwmap=0.002\Mpc^{-1}$ and cover the
WMAP observed scales to varying degrees.
Despite the $k$ ranges having being chosen to minimise the relative error in the
integral of only
one of the $\theta$ dependent terms, the integrals of the three other terms also
display small relative errors for these ranges. 
The analytic solutions which have been found will form an
important part of the verification of any future modifications to the
numerical code.

The execution of the code is in four stages, building on previous calculations of
first order perturbations in Refs.~\cite{Martin:2006rs, Ringeval:2007am,
Salopek:1988qh}. To begin, the background equations are solved and the end time of
inflation is fixed. The initial conditions for the first order perturbation can then
be set and solutions found for the evolution equations.
Despite the large volume of calculations required at each
time step, the easily parallelisable nature of the source term calculations allows
the run time of the third stage to be reduced significantly. 
The final stage of the calculation uses the source term results to solve the second
order perturbation equations.

The initial conditions for the second order perturbations are taken to be $\dvp2=0$ and
$\dN{\dvp2}=0$ as described in Section~\ref{sec:initconds-num}. For this choice of initial
conditions the homogeneous part of the solution of the second order equation is zero at all times.
As the perturbations are supposed to become more Gaussian the further back in time they are
considered, in the limit of the far past the second order perturbations should be zero. It remains
to be investigated whether the choice of initialisation time is sufficiently far in the past for
this assumption to be accurate. At first order it is known that the perturbations are well
 approximated by the Bunch-Davies vacuum initial conditions even just a few e-foldings before
horizon crossing. However, this choice of initialisation time may not be the most appropriate for
the second order perturbations. In future work it would be worth considering whether the analytic
Green's function solution for $\dvp2$ at very early times could be integrated until the numerical
initialisation time and used as the initial condition for the perturbation.

To test the code, four different, large field, monomial potentials were used. These were the
standard quadratic and quartic potentials, a fractional index potential derived from the monodromy
string inflation model and a toy model in which inflation is stopped by hand and a blue spectrum is
produced.
Each potential depends on a single parameter, which was fixed by comparing the resultant scalar
curvature power spectrum with the WMAP5 normalisation. The slow roll
approximation can be applied to all four potentials. These potentials are not meant to represent an
exhaustive survey of single field slow roll models but are sufficiently different to exhibit
different power spectra and second order source terms.

We presented the results for each potential in Chapter~\ref{ch:results}. The first
order results match those in Refs.~\cite{Martin:2006rs, Ringeval:2007am, Salopek:1988qh}. The
results of the source term calculation show that before
horizon crossing, the source term amplitude decays rapidly for all four potentials.
The amplitude changes less after horizon crossing, until later times when it increases
as the slow roll approximation breaks down.

The four different potentials have similar amplitudes before horizon crossing but reach different
values after horizon crossing. The differences in the slow roll parameters for each potential are
compared with the source term values in Appendix~\ref{sec:apx-srcdisc}. The slow roll parameters
do not appear to be directly related to the amplitudes of the source terms, at least in a linear
fashion.

The choice of wavenumber range affects the amplitude of the source term, as
expected, due to the implementation of a sharp cutoff at large scales. This
dependence is only apparent, however, before horizon crossing. For a particular
range,  the magnitude of the source term decreases as wavenumber increases. However,
the ratio of the source term to the other terms in the Klein-Gordon equation increases with
wavenumber. 

As expected, the amplitude of the second order scalar perturbations is much smaller
than that of the first order ones. After the generation of the second order
perturbations at early times, their evolution is that of a damped harmonic
oscillator similar to the first order evolution.

We have shown that the magnitude of the source term can be important throughout the
full evolution and that it is not sufficient to calculate this term only for
modes either entirely inside or outside the horizon, \iec taking a short or long
wavelength approximation respectively. We have been able to access both of these
regimes, by solving the evolution equations of the
inflaton field perturbation. This is in contrast to other approaches which could
have been taken, for example using the $\Delta\N$
formalism, which is only applicable in the large scale limit.

The construction of any numerical code involves a considerable commitment of time
and resources so it is important to understand why such an endeavour has been
undertaken. 
The numerical calculation of first order cosmological perturbations is an invaluable
part of the cosmologist's toolkit. It allows analytic predictions of inflationary
models to be confirmed where these exist, but also generates predictions where no
analytic solution is possible. Another important use is to test predictions based
on the slow roll approximation  against the full evolution equations. 
We have taken the first step towards upgrading this standard numerical calculation to
include second order scalar perturbations. 
The ability to check the predictions of inflationary models at second order will
be a powerful tool to constrain these models and check the consistency of any
analytic assumptions that have been made. Solving the inflaton field equations
provides the most direct access to the non-linear effects that the increase in
available statistics have made observationally important.

We have presented the
first numerical calculation of the Klein-Gordon equation for second order scalar
perturbations which was derived in \Rref{Malik:2006ir}. Although we have restricted
ourselves in this thesis to the single field, slow roll version of the second order
equation,
the expertise gained and the lessons learned in the development of the numerical
system will be of significant assistance when the next steps towards a full
multi-field calculation are taken. 

In the past, a numerical calculation on the scale we have achieved would have been
the preserve of dedicated super-computing facilities. We have demonstrated that a
calculation of this scope is now possible using relatively modest local resources.
If the computational power available increases, the practical limits on the
resolution and extent of the $k$ ranges will ease. Further improvements in the
efficiency of the code will also loosen these constraints.

Another consideration in the development of a numerical system is the possibility of
code re-use. One of our future goals is to develop our code into a numerical toolkit which
can be applied to a variety of physical situations.
The equations of motion of the inflaton scalar field are similar in form
to the governing equations of other important cosmological phenomena. Therefore, it
should be possible to adapt the numerical system we have constructed and apply it to
other areas of interest. The form of the second order equation and source term are similar to those
applicable in the evolution of tensor perturbations and the
generation of vorticity in the early universe. The flexibility of the numerical system we have
developed will be a positive factor in any attempt to apply our code to these physical systems.

The numerical calculation described is the first step towards a system capable of
handling the multi-field, non-slow roll models for which non-linear contributions
are important. In Section~\ref{sec:next-res}, the next steps towards this goal were
outlined. 
Continuing our work by calculating the second order perturbations for both the
non-slow roll, single field case and the slow
roll, multi-field case will pave the way for the eventual calculation of the
non-slow roll, multi-field equations.

The full single field, non-slow roll, second order equation can be treated using the
method already described. 
Three more $\theta$ dependent terms are necessary to
compute the full convolution integral in this case. It will be important to find
analytic solutions for these three extra terms, as already done for the terms in the
slow roll case, in
order to gauge the effectiveness of the extended code. When the extension to non slow-roll models
is complete, it will be possible to investigate models with a step or other feature in their
potential. These models can exhibit large amounts of non-Gaussianity produced around the feature
with a shape dependence that is more general than that of the local and equilateral forms. 

The Klein-Gordon equation for the multi-field case introduces further complexity. We plan to
expand the numerical system to encompass two or three scalar fields. The differences
between single and multi-field models are already apparent for these cases. We have
presented the slow roll source term equation for multiple fields in vector notation.
The definitions of the four $\theta$ dependent terms used in the single
field, slow roll model were also extended to the multi-field case. Beyond the slow
roll approximation, the full multi-field equation should be treatable in a similar
manner to the single field case, by introducing further $\theta$ dependent terms.

To conclude, it is worth reiterating our opening remarks. Cosmology has moved from
being a theorists' playground to a genuine scientific discipline. Inflationary models can
now be strongly tested by observations and the next generation of experiments will
place even tighter limits on the viable parameter space of such models. In this thesis, analytic
arguments have constrained string theory inspired inflationary models and numerical
methods have paved the way to calculating higher order cosmological perturbations.

\appendix
%

\chapter{Appendix}
\label{ch:appendix}

The following materials supplement the calculations and discussions in the main thesis.

\section{Analytic Solution of Generalised Sound Speed Relation}
\label{sec:apx-multi}
\eq{eq:defalpha} can be analytically solved in full 
generality without imposing the limits (\ref{eq:Plimits}) on the 
derivatives of the kinetic function. This allows us to determine the 
most general class of models where the non-linearity parameter 
satisfies the condition $\fnleq \propto 1/\cs^2$ at leading order. 

In general \eq{eq:defalpha} takes the form 
\begin{equation}
\label{eq:genPXeqn-multi}
(2-\alpha ) P_{,X}P_{,XX} + 4XP^2_{,XX} = \frac{2\alpha }{3}
X P_{,X}P_{,XXX}
\end{equation}
and this reduces to 
\begin{equation}
\label{eq:genreduce-multi}
\alpha \Upsilon_{,X} = (6-\alpha ) \Upsilon^2 + \frac{3(2-\alpha )}{2}
\frac{\Upsilon}{X} \, ,
\end{equation}
where $\Upsilon \equiv P_{,XX}/P_{,X}$. 
\eq{eq:genreduce-multi} can be transformed into the 
linear equation
\begin{equation}
\label{eq:lineargen-multi}
U_{,X}+ \frac{3(2-\alpha )}{2\alpha} \frac{U}{X} = \frac{\alpha -6}{\alpha}
\end{equation}
after the change of variables $U \equiv 1/\Upsilon$
and the general solution to \eq{eq:lineargen-multi} is given by
\begin{equation}
\label{eq:gensolnlinear-multi}
\frac{P_{,XX}}{P_{,X}} = \frac{1}{X\left[ f_2(\varphi) X^{(\alpha -6)/2\alpha}
-2 \right] } \, .
\end{equation}
Integrating a second time implies that
\begin{equation}
\label{eq:secondint-multi}
P_{,X} = -f_1 (\varphi ) \left( 1- f_2(\varphi ) X^{-s} \right)^{1/(2s)}  \, ,
\end{equation}
where $s \equiv (\alpha -6 )/(2 \alpha)$ and we have redefined 
the arbitrary integration functions $f_i(\varphi )$.  
Finally \eq{eq:secondint-multi} can be formally integrated 
in terms of a hypergeometric function
\begin{equation}
 \label{eq:thirdint-multi}
 P= -f_1X \,{_2}F_1 \left( -\frac{1}{s}, -\frac{1}{2s}; 1-\frac{1}{s}, f_2X^{-s}
\right)  \, ,
\end{equation}
which represents the most general solution for this class of models. 
Note that we have set the
remaining constant of integration to zero to ensure 
that the kinetic function vanishes in the limit of
zero velocity. In fact this expression admits many 
different classes of solution, arising as limits
of the expansion of the hypergeometric function.

\section{Generalised BM bound for Finite \texorpdfstring{$n$}{n} Models}
\label{sec:apx-genbmbound}

For completeness we should also consider the 
BM bound \eqref{eq:genBMbound} for the finite $n$ multi-coincident brane models. This is
given by 
\begin{equation}
\label{eq:BMAdS-multi}
r_* < -\frac{42}{N \Neff^2}\sqrt{1 +(n-1)^2Y}\fnleq \,,
\end{equation}
and in the case of an $AdS_5 \times X_5$ throat simplifies to
\begin{equation}
\label{eq:bmadsbound}
r_* < -\frac{5}{\Neff^2} 
\frac{\fnleq}{(n-1)\sqrt{N}} \,.
\end{equation}
Comparing the limits in Eqs.~\eqref{eq:AdSupper-multi} and
\eqref{eq:bmadsbound} 
implies that the bound \eqref{eq:LHbound} is stronger than the corresponding BM
bound \eqref{eq:genBMbound} if 
\begin{equation}
\label{eq:LHstrongerads}
n > 1 -5.5 \times 10^{-14} N^{3/2} \Neff^2 \fnleq \,,
\end{equation}
and this condition is always satisfied if 
\begin{equation}
\label{eq:allNbound-multi}
-5.5 \times 10^{-14} N^{3/2} \Neff^2 \fnleq  <1  \, .
\end{equation}
Moreover, the bound \eqref{eq:allNbound-multi} will itself be satisfied for 
all values of $\fnleq$ and $N$ if it is satisfied when the limits 
$\fnleq =-151$ and $N=75852$ are imposed. Hence, we conclude that the bound
\eqref{eq:LHbound} 
is stronger for $\Neff < 75$. 
In general, it is difficult to quantify 
the magnitude of $\Neff$ without 
imposing further restrictions on the parameters of the models 
and, in particular, on the functional form of the inflaton potential. 
However, if the ratio $\varepsilon_H/P_{,X}$ remains approximately 
constant during the final stages of inflation, one would anticipate that 
$\Neff \lesssim 60$. Nevertheless, if $N \ll 75852$, the bound 
\eqref{eq:LHstrongerads} will only be violated for $n \le 3$ if 
$\Neff \gg 60$.

\section{Discussion of Homogeneous Solution for Second Order Equation}
\label{sec:init-apx}

The homogeneous equation for the second order perturbations is  
\begin{equation}
\label{eq:homogen-apx}
 \dvp2''(\eta, \kvi) + 2\H \dvp2'(\eta, \kvi) 
+ \left[ k^2 + a^2\Upp - 24\pi G (\vp_0')^2 \right]\dvp2(\eta,\kvi) = 0\,.
\end{equation}
During slow roll, with the slow roll variables $\varepsilon_H$ and $\eta_H$ defined in
Chapter~\ref{ch:introduction}, this becomes
\begin{equation}
 \dvp2'' + 2\H \dvp2' + \left[k^2 + 3\H^2(\eta_H + \varepsilon_H)\right]\dvp2 = 0\,.
\end{equation}
If we let $u=a\dvp2$, this equation can be rewritten as
\begin{equation}
\label{eq:uH-apx}
 u'' + \left[k^2 +\H^2(3\eta_H - 2\varepsilon_H - 2)\right]u = 0\,.
\end{equation}
When $\varepsilon_H$ is small, the conformal time $\eta$ is given by 
\begin{equation}
 \eta \simeq -\frac{1}{\H(1-\varepsilon_H)}\,,
\end{equation}
so we can rewrite \eq{eq:uH-apx} as
\begin{equation}
 u'' + \left[k^2 + \frac{1}{(-\eta)^2}\frac{3\eta_H -2\varepsilon_H -2}{(1-\varepsilon_H)^2} 
       \right]u = 0\,.
\end{equation}
If the derivatives are taken in terms of $(-\eta)$ instead of $\eta$ this is in the form of a
Bessel equation with solutions in terms of Hankel functions given by
\begin{equation}
 u_{1,2} = \sqrt{-\eta} H_\nu^{(1,2)}(-k\eta)\,, 
\end{equation}
where $H_\nu^{(1,2)}$ are the Hankel functions (Bessel functions of the third kind), and $\nu$ is
given by
\begin{equation}
 \nu^2 = \frac{6\varepsilon_H -12\eta_H + 7}{4(1-2\varepsilon_H)}\,.
\end{equation}

The full solution for $u$ is then
\begin{equation}
 u_\mathrm{full} = C_1 \sqrt{-\eta} H_\nu^{(1)}(-k\eta) +
                   C_2 \sqrt{-\eta} H_\nu^{(2)}(-k\eta)\,,
\end{equation}
where $C_1, C_2\in \mathbb{C}$. When the (real) argument of the Hankel functions goes to $+\infty$
they have the following asymptotic form \cite{abramowitz+stegun}:
\begin{align}
 H_\nu^{(1)}(z) &\rightarrow \sqrt{\frac{2}{\pi z}} e^{i(z-\frac{\pi}{2}\nu - \frac{\pi}{4})} \,,\\
 H_\nu^{(2)}(z) &\rightarrow \sqrt{\frac{2}{\pi z}} e^{-i(z-\frac{\pi}{2}\nu - \frac{\pi}{4})} \,.
\end{align}

So at early times when $\eta\rightarrow -\infty$ and $-k\eta\rightarrow +\infty$ we have the
following expressions for u:
\begin{align}
 u_{i} &= \sqrt{\frac{2}{\pi k}}\left( C_1 e^{-i(k\eta +\frac{\pi}{2}\nu + \frac{\pi}{4})}
                + C_2 e^{+i(k\eta +\frac{\pi}{2}\nu + \frac{\pi}{4})} \right) \,,\\
 u_{i}' &= i k\sqrt{\frac{2}{\pi k}}\left( -C_1 e^{-i(k\eta +\frac{\pi}{2}\nu + \frac{\pi}{4})}
                + C_2 e^{+i(k\eta +\frac{\pi}{2}\nu + \frac{\pi}{4})} \right) \,,\\
\end{align}
where we have assumed that $\nu$ is slowly varying far in the past, \iec the derivatives of the slow
roll parameters are very small. 

As explained in Section~\ref{sec:initconds-num}, the results given for $\dvp2$ are for the full solution
including the homogeneous part. To remove the homogeneous part of the solution the initial conditions
for the full $\dvp2$ should be chosen such that $C_1=C_2=0$ at all times.

\section{Analytic Tests for \texorpdfstring{$\B,\wt{\C}$ and $\wt{\D}$}{B, C and D} Terms}
\label{sec:apx-codetests}

%
%
Analytic solutions can also be found for the $\B$, $\wt{\C}$ and $\wt{\D}$ terms.
The $\B$ term integral, $I_\B$, is given by
\begin{align}
 \label{eq:bintegral-num}
I_\B &= 2\pi \int_{\kmin}^{\kmax} \d q\, q^2\dvp1(\qvi)\B(\kvi,\qvi) \nonumber\\
     &= 2\pi\alpha^2 \int_{\kmin}^{\kmax} \d q\, q^{\frac{3}{2}}
\int_{0}^{\pi} \d\theta\, (k^2 + q^2 -2k q \cos{\theta})^{-1/4}
\cos{\theta}\sin{\theta}\,,
\end{align}
and has the following analytic solution when $\dvp1(q) = \alpha/\sqrt{q}$:
\begin{align}
\label{eq:intb-soln-num}
 I_\B = -\frac{\pi\alpha^2}{168 k^2}\Bigg\{ 
        &-63 k^4 \Bigg[ \log\Biggl(\frac{\sqrt{k}}{\sqrt{k+\kmin} + \sqrt{\kmin}}
                            \Biggr)
         + \log\Biggl( \frac{\sqrt{k+\kmax} +\sqrt{\kmax}}{\sqrt{\kmax-k} +
                      \sqrt{\kmax}}\Biggr) \nonumber \\
        &-\frac{\pi}{2} + \arctan\left( \frac{\sqrt{\kmin}}{\sqrt{k-\kmin}}\right)
        \Bigg] \nonumber\\
        &+\sqrt{\kmax}\Bigg[ \left(-65k^3 + 8k\kmax^2 \right)\left(\sqrt{k+\kmax} +
          \sqrt{\kmax-k}\right) \nonumber \\
        &\qquad +\left(22k^2\kmax -16\kmax^3\right) \left(\sqrt{k+\kmax} -
         \sqrt{\kmax -k} \right) \Bigg] \nonumber \\
        &+\sqrt{\kmin}\Bigg[ \left(65k^3 - 8k\kmin^2 \right)\left(\sqrt{k+\kmin} -
          \sqrt{k-\kmin}\right) \nonumber \\
        &\qquad +\left(-22k^2\kmin +16\kmin^3\right) \left(\sqrt{k+\kmin} +
         \sqrt{k-\kmin} \right) \Bigg] \Bigg\} \,.
\end{align}
If, in addition to $\dvp1(q) = \alpha/\sqrt{q}$, we also take
\begin{equation}
 \dN{\dvp1}(q) = -\frac{\alpha}{\sqrt{q}} -i\frac{\alpha\sqrt{q}}{\beta}\,
\end{equation}
then the $\wt{\C}$ and $\wt{\D}$ terms can be integrated analytically.
The integral of the $\wt{\C}$ term is 
\begin{align}
 \label{eq:cint-num}
I_{\wt{\C}} &= \int\d^3 q\, \dvp1(\qvi) \dN{\dvp1}(\kvi-\qvi) 
    = 2\pi \int \d q\, q^2 \dvp1(\qvi) \wt{\C}(\kvi, \qvi) \nonumber\\
 &= -2\pi\alpha^2 \int_{\kmin}^{\kmax} \d q\, q^{\frac{3}{2}} \int_0^\pi 
     \left( \left(k^2 + q^2 -2kq \cos\theta\right)^{-\frac{1}{4}} \right.\nonumber \\
            &\qquad \qquad \left.+\frac{i}{\beta}\left(k^2 + q^2 -2kq
\cos\theta\right)^{\frac{1}{4}}
        \right) \sin\theta \d \theta\,,
\end{align}
and the analytic solution is given by
\begin{align}
\label{eq:cint-soln-num}
I_{\wt{\C}} = -I_\A -i\frac{\pi\alpha^2}{240 \beta k}\Bigg\{ 
        &15 k^4 \Bigg[ \log\Biggl(\frac{\sqrt{k+\kmin} + \sqrt{\kmin}}{\sqrt{k}}
                            \Biggr)
         + \log\Biggl( \frac{\sqrt{\kmax-k} + \sqrt{\kmax}}{\sqrt{k+\kmax}
                        +\sqrt{\kmax}}\Biggr) \nonumber \\
        &-\frac{\pi}{2} + \arctan\left( \frac{\sqrt{\kmin}}{\sqrt{k-\kmin}}\right)
        \Bigg] \nonumber\\
        &+\sqrt{\kmax}\Bigg[ \left(15k^3 + 136k\kmax^2 \right)\left(\sqrt{k+\kmax} +
          \sqrt{\kmax-k}\right) \nonumber \\
        &\qquad +\left(118k^2\kmax -48\kmax^3\right) \left(\sqrt{k+\kmax} -
         \sqrt{\kmax -k} \right) \Bigg] \nonumber \\
        &-\sqrt{\kmin}\Bigg[ \left(15k^3 + 136k\kmin^2 \right)\left(\sqrt{k+\kmin} +
          \sqrt{k-\kmin}\right) \nonumber \\
        &\qquad +\left(118k^2\kmin +48\kmin^3\right) \left(\sqrt{k+\kmin} -
         \sqrt{k-\kmin} \right) \Bigg] \Bigg\} \,.
\end{align}
The integral of the $\wt{\D}$ term is 
\begin{align}
 \label{eq:dint-num}
I_{\wt{\D}} &= 2\pi \int \d q\, q^2 \dvp1(\qvi) \wt{\D}(\kvi, \qvi) \\
 &= -2\pi\alpha^2 \int_{\kmin}^{\kmax} \d q\, q^{\frac{3}{2}} \int_0^\pi 
     \left( \left(k^2 + q^2 -2kq \cos\theta\right)^{-\frac{1}{4}} \right.\nonumber\\
        &\qquad\qquad\left.+\frac{i}{\beta}\left(k^2 + q^2 -2kq
\cos\theta\right)^{\frac{1}{4}}
        \right) \cos\theta\sin\theta \d \theta\,,
\end{align}
and the analytic solution is
\begin{align}
\label{eq:dint-soln-num}
I_{\wt{\D}} = -I_\B &-i\frac{\pi\alpha^2}{900\beta k^2}\Bigg\{ \nonumber \\
        &135 k^5 \Bigg[ \log\Biggl(\frac{\sqrt{\kmax-k} + \sqrt{\kmax}}{\sqrt{k}}
                            \Biggr)
         + \log\Biggl( \frac{\sqrt{k+\kmax} +\sqrt{\kmax}}{\sqrt{k+\kmin} +
                          \sqrt{\kmin}}\Biggr) \nonumber \\
        &\qquad -\frac{\pi}{2} + \arctan\left(
\frac{\sqrt{\kmin}}{\sqrt{k-\kmin}}\right)
        \Bigg] \nonumber\\
        &-\sqrt{\kmax}\Bigg[ \left(-185k^4 + 168k^2\kmax^2-32\kmax^4
            \right)\left(\sqrt{k+\kmax} - \sqrt{\kmax-k}\right) \nonumber \\
        &\qquad +\left(70k^3\kmax +16k\kmax^3\right) \left(\sqrt{k+\kmax} +
         \sqrt{\kmax -k} \right) \Bigg] \nonumber \\
        &+\sqrt{\kmin}\Bigg[ \left(-185k^4 + 168k^2\kmin^2 -32\kmax^4
            \right)\left(\sqrt{k+\kmin} - \sqrt{k-\kmin}\right) \nonumber \\
        &\qquad +\left(70k^3\kmin +16k\kmin^3\right) \left(\sqrt{k+\kmin} +
         \sqrt{k-\kmin} \right) \Bigg] \Bigg\} \,.
\end{align}
%

\section{Analytic Solution for Source Term}
\label{sec:analyticsrc-apx}
Suppose the first order perturbations are given by the non-interacting de Sitter space solution
such that
 \begin{equation}
 \dvp1(\eta, \kvi) = \frac{1}{a\sqrt{2 k}} \left( 1 - \frac{i}{k \eta}\right) \,,
\end{equation}
and the derivative in terms of $\N$ is 
\begin{equation}
 \dN{\dvp1}(\eta, \kvi) = -  \frac{1}{a\sqrt{2 k}} \left( 1 -
                        \frac{i}{k \eta}\right) \left(1 + \frac{1}{aH\eta}\right)
                         - \frac{i}{a^2 H \sqrt{2}}\sqrt{k}\,.
\end{equation}
The analytic solution of \eq{eq:KG2-source-ntime} for this choice of first order
solution can be written in terms of four integrals of the $\A$-$\wt{\D}$ terms:
\begin{equation}
\label{eq:KG2-source-jterms}
S(\kvi) = \frac{1}{(2\pi)^2} \left\{J_\A + J_\B + J_{\wt{\C}} + J_{\wt{\D}} \right\}\,,
\end{equation}
where
\begin{align}
\label{eq:Jdefns-apx}
 J_\A(\kvi) &= \int_{\kmin}^{\kmax} \d q\, \left( 
                \frac{\Uppp}{H^2} q^2 + \frac{8\pi G}{(aH)^2}\dN{\vp_{0}}\left[ 
 3a^2\Upp q^2  + \frac{7}{2}q^4 + 2k^2q^2 \right] \right) \dvp1(\qvi) \A(\kvi, \qvi) \,, \\
 J_\B(\kvi) &= \int_{\kmin}^{\kmax} \d q\, \frac{8\pi G}{(aH)^2}\dN{\vp_{0}}\left(-\frac{9}{2} -
\frac{q^2}{k^2}\right)kq^3 \dvp1(\qvi) \B(\kvi, \qvi) \,, \\
J_{\wt{\C}}(\kvi) &= \int_{\kmin}^{\kmax} \d q\, \left(
                        -8\pi G \dN{\vp_{0}} \frac{3}{2}q^2 \right) \dN{\dvp1}(\qvi)
\wt{\C}(\kvi, \qvi) \,, \\
J_{\wt{\D}}(\kvi) &= \int_{\kmin}^{\kmax} \d q\, \left(
                       8\pi G \dN{\vp_{0}} \left[2-\frac{q^2}{k^2}\right] kq \right)
\dN{\dvp1}(\qvi) \wt{\D}(\kvi, \qvi) \,. 
\end{align}

The analytic solution for $J_\A$ is given by
\begin{multline}
\label{eq:JA-apx}
J_\A = \left( 
                \frac{\Uppp}{H^2} + \frac{8\pi G}{(aH)^2}\dN{\vp_{0}}\left[ 
 3a^2\Upp    + 2k^2 \right] \right)\frac{ \alpha^2}{2880\eta ^2 k}\Bigg\{\\
240 k \arctan\left(\sqrt{\frac{\kmin}{k-\kmin}}\right) \left(\eta ^2 k^2-12 i \eta  k-24\right) \\
-120 k \pi  \left(\eta ^2 k^2-12 i \eta  k-24\right)
-80\sqrt{\kmax}
   \Bigg(\Bigg[3 \left(\sqrt{\kmax-k}-\sqrt{k+\kmax}\right) k^2 \\
-14 \kmax
   \left(\sqrt{\kmax-k}+\sqrt{k+\kmax}\right) k+8 \kmax^2
   \left(\sqrt{\kmax-k}-\sqrt{k+\kmax}\right)\Bigg] \eta ^2 \\
+48 i \left(\kmax
   \left(\sqrt{k+\kmax}-\sqrt{\kmax-k}\right)+k
   \left(\sqrt{\kmax-k}+\sqrt{k+\kmax}\right)\right) \eta \\
+72
   \left(\sqrt{k+\kmax}-\sqrt{\kmax-k}\right)\Bigg) \displaybreak[0]\\
-80 \sqrt{\kmin} \Bigg(\Bigg[3
   \left(\sqrt{k-\kmin}+\sqrt{k+\kmin}\right) k^2
-14 \kmin\left(\sqrt{k-\kmin}-\sqrt{k+\kmin}\right) k \\
+8 \kmin^2
   \left(\sqrt{k-\kmin}+\sqrt{k+\kmin}\right)\Bigg] \eta ^2
+12 i \Bigg[k
   \left(\sqrt{k-\kmin}-4 \sqrt{k+\kmin}\right) \\
+2 \kmin
\left(\sqrt{k-\kmin}-2
   \sqrt{k+\kmin}\right)\Bigg] \eta +72
   \left(\sqrt{k-\kmin}-\sqrt{k+\kmin}\right)\Bigg) \displaybreak[0]\\
+240 k \left(\eta ^2 k^2+24\right)\log\left(2 \sqrt{k}\right)
-240 k \left(\eta ^2 k^2+24\right)\log \left(2 \left(\sqrt{\kmax}+\sqrt{\kmax-k}\right)\right)\\
-240 k \left(\eta ^2 k^2+24\right) \log \left(2\left(\sqrt{\kmax}+\sqrt{k+\kmax}\right)\right)\\
+240 k \left(\eta ^2 k^2+24\right) \log
\left(2 \left(\sqrt{\kmin}+\sqrt{k+\kmin}\right)\right)\Bigg\} \displaybreak[0]\\
+ \frac{8\pi G}{(aH)^2}\dN{\vp_{0}}\frac{7}{2}\frac{\alpha ^2}{2880 \eta
^2 k} \Bigg\{3 \sqrt{2} \left(-317 \eta ^2 k^2+1000 i \eta  k+560\right) k^3 \\
+3\sqrt{2} \left(317 \eta ^2 k^2-1000 i \eta  k-560\right) k^3 \\
+45 \left(\eta ^2 k^2-12 i \eta k-16\right) \arctan\left(\sqrt{\frac{\kmin}{k-\kmin}}\right) k^3
+45\left(\eta ^2 k^2+8 i \eta k+16\right)
   \log \left(2 \sqrt{k}\right) k^3 \\
-45 \left(\eta ^2 k^2+8 i \eta  k+16\right) \log \left(2
   \left(\sqrt{\kmax}+\sqrt{\kmax-k}\right)\right) k^3 \\
-45 \left(\eta ^2 k^2-8 i \eta 
   k+16\right) \log \left(2 \left(\sqrt{\kmax}+\sqrt{k+\kmax}\right)\right) k^3 \\
+45\left(\eta ^2 k^2-8 i \eta  k+16\right) \log \left(2
\left(\sqrt{\kmin}+\sqrt{k+\kmin}\right)\right) k^3 -\frac{45}{2} \left(\eta ^2 k^2-12 i \eta 
k-16\right) \pi  k^3 \displaybreak[0]\\
-3 \sqrt{\kmax} \Bigg(15 \eta ^2
   \left(\sqrt{\kmax-k}-\sqrt{k+\kmax}\right) k^4 \\
 +10 \eta  (\eta  \kmax+12 i)
   \left(\sqrt{\kmax-k}+\sqrt{k+\kmax}\right) k^3 \\ 
+8 \left(\eta ^2 \kmax^2+10 i \eta 
   \kmax+30\right) \left(\sqrt{\kmax-k}-\sqrt{k+\kmax}\right) k^2 \\
-16 \kmax \left(11\eta ^2 \kmax^2-20 i \eta  \kmax-10\right)
   \left(\sqrt{\kmax-k}+\sqrt{k+\kmax}\right) k \\
+128 \kmax^2 \left(\eta ^2 \kmax^2-5i \eta  \kmax-5\right)
\left(\sqrt{\kmax-k}-\sqrt{k+\kmax}\right)\Bigg) \displaybreak[0]\\
-3 \sqrt{\kmin} \Bigg(15 \eta ^2 \left(\sqrt{k-\kmin}+\sqrt{k+\kmin}\right) k^4 \\
+10 \eta\left(\eta  \kmin \left(\sqrt{k-\kmin}-\sqrt{k+\kmin}\right)-6 i \left(3
   \sqrt{k-\kmin}+2 \sqrt{k+\kmin}\right)\right) k^3 \\
+8 \Bigg(\eta ^2
   \left(\sqrt{k-\kmin}+\sqrt{k+\kmin}\right) \kmin^2-5 i \eta  \left(3
   \sqrt{k-\kmin}-2 \sqrt{k+\kmin}\right) \kmin \\
+30
   \left(\sqrt{k+\kmin}-\sqrt{k-\kmin}\right)\Bigg) k^2 \displaybreak[0]\\
-16 \kmin \Bigg(11 \eta ^2
   \left(\sqrt{k-\kmin}-\sqrt{k+\kmin}\right) \kmin^2-10 i \eta 
   \left(\sqrt{k-\kmin}-2 \sqrt{k+\kmin}\right) \kmin \\+10
   \left(\sqrt{k-\kmin}+\sqrt{k+\kmin}\right)\Bigg) k \\
+64 \kmin^2 \Bigg(2 \eta ^2
   \left(\sqrt{k-\kmin}+\sqrt{k+\kmin}\right) \kmin^2+5 i \eta 
   \left(\sqrt{k-\kmin}-2 \sqrt{k+\kmin}\right) \kmin \\
+10
   \left(\sqrt{k-\kmin}-\sqrt{k+\kmin}\right)\Bigg)\Bigg)\Bigg\}\,. 
\end{multline}
The analytic solution for $J_\B$ is 
\begin{multline}
\label{eq:JB-apx}
J_\B = -\frac{8\pi G}{(aH)^2}\dN{\vp_{0}}\frac{9}{2}\frac{\alpha ^2}{2822400 \eta ^2
k} \Bigg\{29400 \left(4 \eta ^2 k^2+15 i \eta k+120\right)
\arctan\left(\sqrt{\frac{\kmin}{k-\kmin}}\right) k^3 \\
+29400 \left(4 \eta ^2 k^2-51 i \eta 
   k-120\right) \log \left(2 \sqrt{k}\right) k^3 \\
-29400 \left(4 \eta ^2 k^2-51 i \eta k-120\right) \log
   \left(2 \left(\sqrt{\kmax}+\sqrt{\kmax-k}\right)\right) k^3 \\
-29400 \left(4 \eta ^2 k^2+51 i \eta  k-120\right) \log \left(2
\left(\sqrt{\kmax}+\sqrt{k+\kmax}\right)\right) k^3 \\
+29400 \left(4 \eta ^2 k^2+51 i \eta  k-120\right) \log \left(2
   \left(\sqrt{\kmin}+\sqrt{k+\kmin}\right)\right) k^3 \\
-14700 \left(4 \eta ^2 k^2+15 i
\eta 
   k+120\right) \pi  k^3 
+280 \sqrt{\kmax (k+\kmax)} \Bigg(420 \eta ^2 k^4 \\
+5 \eta  (200 \eta \kmax+303 i) k^3 + \left(-416 \eta ^2 \kmax^2+6414 i \eta  \kmax+11592\right)
k^2 \\ 
- 24\kmax \left(8 \eta ^2 \kmax^2-87 i \eta  \kmax-126\right) k 
+48 \kmax^2\left(8\eta ^2 \kmax^2+53 i \eta  \kmax+84\right)\Bigg) \displaybreak[0]\\
-280 \sqrt{\kmax (\kmax-k)}
   \Bigg(420 \eta ^2 k^4-5 \eta  (200 \eta  \kmax+303 i) k^3 \\
+\left(-416 \eta ^2 \kmax^2+6414 i
   \eta  \kmax+11592\right) k^2+24 \kmax \left(8 \eta ^2 \kmax^2-87 i \eta 
   \kmax-126\right) k \\
+48 \kmax^2 \left(8 \eta ^2 \kmax^2+53 i \eta 
   \kmax+84\right)\Bigg) \displaybreak[0] \\
-280 \sqrt{\kmin} \sqrt{k+\kmin} \Bigg(420 \eta ^2 k^4+5
\eta 
   (200 \eta  \kmin+303 i) k^3 \\
+\left(-416 \eta ^2 \kmin^2+6414 i \eta \kmin+11592\right)k^2 -24 \kmin \left(8 \eta ^2 \kmin^2-87
i \eta  \kmin-126\right) k \\
+48 \kmin^2 \left(8 \eta ^2 \kmin^2+53 i \eta  \kmin+84\right)\Bigg) \displaybreak[0]\\
-280 \sqrt{(k-\kmin)
   \kmin} \Bigg(420 \eta ^2 k^4+5 \eta  (1083 i-200 \eta  \kmin) k^3 \\
-2 \left(208 \eta ^2
   \kmin^2+2547 i \eta  \kmin+5796\right) k^2+24 \kmin \left(8 \eta ^2
\kmin^2+67 i
   \eta  \kmin+126\right) k \\
+48 \kmin^2 \left(8 \eta ^2 \kmin^2-73 i \eta 
   \kmin-84\right)\Bigg)\Bigg\} \displaybreak[1]\\
- \frac{8\pi G}{(aH)^2}\dN{\vp_{0}}\frac{\alpha ^2}{2822400 \eta ^2 k^3} \Bigg\{105
\left(270 \eta ^2 k^2+2765 i \eta  k+4956\right) \arctan \left(\sqrt{\frac{\kmin}{k-\kmin}}\right)
   k^5 \\
+105 \left(270 \eta ^2 k^2-3745 i \eta  k-4956\right) \log \left(2 \sqrt{k}\right) k^5\\
-105\left(270 \eta ^2 k^2-3745 i \eta  k-4956\right) \log \left(2
   \left(\sqrt{\kmax}+\sqrt{\kmax-k}\right)\right) k^5 \\
-105 \left(270 \eta ^2 k^2+3745 i \eta 
   k-4956\right) \log \left(2 \left(\sqrt{\kmax}+\sqrt{k+\kmax}\right)\right) k^5 \\
+105\left(270  \eta ^2 k^2+3745 i \eta  k-4956\right) \log \left(2
   \left(\sqrt{\kmin}+\sqrt{k+\kmin}\right)\right) k^5 \\
-\frac{105}{2} \left(270 \eta ^2 k^2+2765 i \eta  k+4956\right) \pi  k^5\displaybreak[0]\\
-\sqrt{\kmax (\kmax-k)} \Bigg(28350 \eta ^2 k^6+525 \eta  (36
   \eta  \kmax-749 i) k^5 \\
+70 \left(216 \eta ^2 \kmax^2-3745 i \eta\kmax-7434\right) k^4 
-40 \kmax \left(3516 \eta ^2 \kmax^2-133 i \eta  \kmax+8673\right) k^3 \\
-48 \kmax^2 \left(1360 \eta ^2 \kmax^2-18655 i \eta  \kmax-22442\right) k^2 \\
+64 \kmax^3 \left(600 \eta ^2 \kmax^2-5110 i \eta  \kmax-5733\right) k \\
+256\kmax^4 \left(300 \eta ^2 \kmax^2+1855 i \eta  \kmax+2646\right)\Bigg) \displaybreak[0]\\
+\sqrt{\kmax(k+\kmax)} \Bigg(28350 \eta ^2 k^6-525 \eta  (36 \eta  \kmax-749 i) k^5 \\
+70 \left(216 \eta ^2  \kmax^2-3745 i \eta  \kmax-7434\right) k^4
+40 \kmax \left(3516 \eta ^2 \kmax^2-133 i \eta  \kmax+8673\right) k^3 \\
-48 \kmax^2 \left(1360 \eta^2 \kmax^2-18655 i \eta  \kmax-22442\right) k^2
-64 \kmax^3 \left(600\eta ^2 \kmax^2-5110 i \eta  \kmax-5733\right) k \\
+256 \kmax^4 \left(300 \eta^2 \kmax^2+1855 i \eta  \kmax+2646\right)\Bigg) \displaybreak[0] \\
-\sqrt{\kmin} \sqrt{k+\kmin}
   \Bigg(28350 \eta ^2 k^6-525 \eta  (36 \eta  \kmin-749 i) k^5 \\
+70 \left(216 \eta ^2 \kmin^2-3745 i \eta  \kmin-7434\right) k^4
+40 \kmin \left(3516 \eta ^2 \kmin^2-133 i \eta  \kmin+8673\right) k^3 \\
-48 \kmin^2 \left(1360 \eta^2 \kmin^2-18655 i \eta  \kmin-22442\right) k^2
-64 \kmin^3 \left(600\eta ^2 \kmin^2-5110 i \eta  \kmin-5733\right) k \\
+256 \kmin^4 \left(300 \eta^2\kmin^2+1855 i \eta  \kmin+2646\right)\Bigg) \displaybreak[0]\\
-\sqrt{(k-\kmin) \kmin}\Bigg(28350 \eta ^2 k^6+525 \eta  (36 \eta  \kmin+553 i) k^5 \\
+70 \left(216\eta ^2 \kmin^2+2765 i \eta\kmin+7434\right) k^4 
-40 \kmin \left(3516 \eta ^2\kmin^2-9247 i \eta\kmin-8673\right) k^3 \\
-48 \kmin^2 \left(1360 \eta^2 \kmin^2+15155 i \eta\kmin+22442\right) k^2 
+64 \kmin^3 \left(600\eta ^2 \kmin^2+3710 i \eta\kmin+5733\right) k \\
+256 \kmin^4 \left(300 \eta ^2\kmin^2-2555 i \eta\kmin-2646\right)\Bigg)\Bigg\}\,.
\end{multline}
The analytic solution for $J_{\wt{\C}}$ is 
\begin{multline}
\label{eq:JC-apx}
 J_{\wt{\C}} = -8\pi G \dN{\vp_{0}} \frac{3}{2} \frac{\alpha ^2 }{14400 \beta ^2 \eta ^4 k}
\Bigg\{-15 k \arctan\left(\sqrt{\frac{\kmin}{k-\kmin}}\right)
   \Bigg(9 \eta ^4 k^4-60 i \eta ^3 k^3-560 \eta ^2 k^2 \\
+960 i \eta  k-80 \beta ^2 \eta ^2 \left(\eta ^2 k^2-12 i \eta  k-24\right) 
+20 \beta  \eta\left(-3 i \eta ^3 k^3-32 \eta ^2 k^2+96 i \eta k+192\right)+1920\Bigg)
\displaybreak[0]\\
+\frac{15}{2} k \pi  \Bigg(9 \eta ^4 k^4-60 i \eta ^3 k^3-560 \eta ^2
k^2+960 i
   \eta  k-80 \beta ^2 \eta ^2 \left(\eta ^2 k^2-12 i \eta  k-24\right) \\
+20 \beta  \eta  \left(-3 i\eta ^3 k^3-32 \eta ^2 k^2+96 i \eta  k+192\right)+1920\Bigg)
\displaybreak[0]\\
-\sqrt{\kmax (\kmax-k)} \Bigg(9 \left(15 k^4+10 \kmax k^3-248 \kmax^2 k^2+336 \kmax^3 k
-128\kmax^4\right) \eta ^4 \\
+3840 i (k-\kmax)^2 \kmax \eta ^3 +80 \left(69 k^2-178 \kmax k+184 \kmax^2\right) \eta ^2 \\ 
+400 \beta ^2 \left(\left(3 k^2-14 \kmax k+8 \kmax^2\right) \eta^2 
+48 i (k-\kmax) \eta -72\right)\eta ^2+19200 i (k-\kmax) \eta \\
+320 \beta  \Bigg[12 i   (k-\kmax)^2 \kmax \eta ^3+\left(21 k^2-62 \kmax k+56 \kmax^2\right) \eta
^2 \\
+120 i(k-\kmax) \eta  -180\Bigg] \eta -28800\Bigg) \displaybreak[0] \\
+\sqrt{\kmax (k+\kmax)} \Bigg(9\left(15 k^4-10 \kmax k^3-248 \kmax^2 k^2-336 \kmax^3 k-128
\kmax^4\right) \eta ^4 \\
+3840 i \kmax (k+\kmax)^2 \eta ^3+80 \left(69 k^2+178 \kmax k+184 \kmax^2\right)\eta^2 \\
+400 \beta^2 \left(\left(3 k^2+14 \kmax k+8 \kmax^2\right) \eta ^2 -48 i (k+\kmax)
   \eta -72\right) \eta ^2 \\
-19200 i (k+\kmax) \eta +320 \beta  \Bigg[12 i \kmax
(k+\kmax)^2 \eta ^3+\left(21 k^2+62 \kmax k+56 \kmax^2\right) \eta ^2 \\
-120 i (k+\kmax) \eta -180\Bigg] \eta -28800\Bigg) \displaybreak[0] \\
+\sqrt{\kmin} \sqrt{k+\kmin} \Bigg(9 \left(-15 k^4+10  \kmin k^3+248 \kmin^2 k^2+336 \kmin^3 k+128
\kmin^4\right) \eta ^4 \\
-3840 i \kmin (k+\kmin)^2 \eta ^3-80 \left(69 k^2+178 \kmin k+184 \kmin^2\right)\eta^2 \\
-400 \beta ^2 \left(\left(3 k^2+14 \kmin k+8 \kmin^2\right) \eta ^2-48 i(k+\kmin)
   \eta -72\right) \eta ^2 \\
+19200 i (k+\kmin) \eta +320 \beta  \Bigg[-12 i \kmin
   (k+\kmin)^2 \eta ^3-\left(21 k^2+62 \kmin k+56 \kmin^2\right) \eta ^2 \\
+120 i(k+\kmin) \eta +180\Bigg] \eta +28800\Bigg) \displaybreak[0]\\
-\sqrt{(k-\kmin) \kmin} \Bigg(-9\left(15
   k^4+10 \kmin k^3-248 \kmin^2 k^2+336 \kmin^3 k-128 \kmin^4\right) \eta^4 \\
+60 i \left(15 k^3-54 \kmin k^2+8 \kmin^2 k+16 \kmin^3\right) \eta ^3
-80 \left(39k^2-38  \kmin k+104 \kmin^2\right) \eta ^2 \\
+400 \beta ^2 \left(\left(3 k^2-14 \kmin k+8 \kmin^2\right) \eta ^2+12 i (k+2 \kmin) \eta
+72\right) \eta ^2 \\
+4800 i (k+2 \kmin) \eta +20 i \beta  \Bigg[3 \left(15 k^3-54 \kmin k^2+8 \kmin^2 k+16
\kmin^3\right) \eta ^3 \\
+32 i \left(3 k^2+4 \kmin k+8 \kmin^2\right) \eta ^2+480 (k+2 \kmin) \eta -2880
i\Bigg]
   \eta +28800\Bigg) \displaybreak[0] \\
+15 k \left(9 \eta ^4 k^4-400 \eta ^2 k^2-320 \beta  \eta 
\left(\eta ^2
   k^2-12\right)+80 \beta ^2 \eta ^2 \left(\eta ^2 k^2+24\right)+1920\right) \log \left(2
\sqrt{k}\right) \\
-15k \Bigg(9 \eta ^4 k^4-400 \eta ^2 k^2-320 \beta  \eta  \left(\eta ^2 k^2-12\right)\\
+80 \beta ^2\eta ^2 \left(\eta ^2 k^2+24\right)+1920\Bigg) \log \left(2
   \left(\sqrt{\kmax}+\sqrt{\kmax-k}\right)\right) \\
-15 k \Bigg(9 \eta ^4 k^4-400 \eta ^2 k^2-320
   \beta  \eta  \left(\eta ^2 k^2-12\right) \\
+80 \beta ^2 \eta ^2 \left(\eta ^2
k^2+24\right)+1920\Bigg) \log \left(2 \left(\sqrt{\kmax}+\sqrt{k+\kmax}\right)\right) \\
+15 k \Bigg(9 \eta ^4 k^4-400\eta ^2 k^2-320 \beta  \eta  \left(\eta ^2 k^2-12\right) \\
+80 \beta ^2 \eta ^2 \left(\eta ^2 k^2+24\right)+1920\Bigg) \log \left(2
   \left(\sqrt{\kmin}+\sqrt{k+\kmin}\right)\right)\Bigg\} \,.
\end{multline}
The analytic solution for $J_{\wt{\D}}$ is 
\begin{multline}
\label{eq:JD-apx}
J_{\wt{\D}} = 8\pi G \dN{\vp_{0}}\frac{\alpha ^2}{302400\beta ^2\eta ^4 k} 
\Bigg\{-5040 k \arctan\left(\sqrt{\frac{\kmin}{k-\kmin}}\right) \Bigg(9 \eta ^4 k^4 
+230 \eta ^2 k^2+260 i \eta k\\
+20 \beta ^2 \eta ^2 \left(2 \eta ^2 k^2+13 i \eta  k-12\right) 
+10\beta \eta \left(27 \eta^2 k^2+52 i \eta  k-48\right)-240\Bigg) \displaybreak[0]\\
+2520 k \pi\Bigg(9 \eta ^4 k^4+230 \eta ^2 k^2+260 i \eta  k+20
\beta ^2\eta ^2 \left(2 \eta ^2 k^2+13 i \eta  k-12\right)\\
+10 \beta  \eta  \left(27 \eta ^2 k^2+52 i\eta k-48\right)-240\Bigg)
+\frac{16\sqrt{k+\kmax}}{\kmax^{3/2}} \Bigg(-35 \eta ^2 \Bigg(111 \eta ^2
   \kmax^2 \displaybreak[0]\\
+192 i \eta  \kmax+64 \beta  \eta  (3 i \eta  \kmax+1)+64\Bigg) k^4
-5\eta\Bigg(-294 \eta ^3 \kmax^3-717 i \eta ^2 \kmax^2+2656 \eta  \kmax \\
+960 \beta ^2\eta ^2 (3 \eta  \kmax-i)+\beta  \eta  \left(-717 i \eta ^2 \kmax^2+5536 \eta 
\kmax-1920 i\right)-960 i\Bigg) k^3 \\
-6 \Bigg(-588 \eta ^4 \kmax^4+1305 i \eta ^3 \kmax^3+9965
\eta ^2 \kmax^2+15520 i \eta  \kmax\displaybreak[0]\\
+20 \beta ^2 \eta ^2 \left(45 \eta ^2\kmax^2+776 i\eta\kmax+252\right)+5 \beta\eta\Bigg(261 i
\eta ^3 \kmax^3+2173 \eta ^2\kmax^2 
+6208 i\eta  \kmax+2016\Bigg) \displaybreak[0]\\
+5040\Bigg) k^2 +4 \kmax \Bigg(84 \eta ^4 \kmax^4-330 i\eta ^3 \kmax^3-3275 \eta ^2 \kmax^2+7065 i
\eta  \kmax-15 \beta ^2 \eta ^2 \Bigg(20\eta ^2 \kmax^2 \\
-471 i \eta  \kmax+1008\Bigg)-5 \beta\eta  \left(66 i \eta ^3\kmax^3+715\eta ^2 \kmax^2-2826 i \eta 
\kmax+6048\right)-15120\Bigg) k\\
+8 \kmax^2\Bigg(-84 \eta^4 \kmax^4+330 i \eta ^3 \kmax^3-1450 \eta ^2 \kmax^2+2385 i \eta 
\kmax+15 \beta^2 \eta ^2 \Bigg(20 \eta ^2 \kmax^2\\
+159 i \eta  \kmax+378\Bigg)+10 \beta  \eta\left(33 i \eta ^3 \kmax^3-115 \eta ^2 \kmax^2+477 i
\eta \kmax+1134\right)+5670\Bigg)\Bigg)\displaybreak[0]\\
+16\kmax^{-3/2} \sqrt{\kmax-k}\Bigg(35\eta ^2 \left(111 \eta ^2 \kmax^2+192 i \eta 
\kmax+64 \beta \eta  (3 i \eta \kmax+1)+64\right) k^4\\
+5 \eta  \Bigg(294 \eta ^3 \kmax^3+717 i \eta^2 \kmax^2-2656
   \eta  \kmax-960 \beta ^2 \eta ^2 (3 \eta  \kmax-i) \\
+i \beta  \eta  \left(717 \eta ^2  \kmax^2+5536 i \eta  \kmax+1920\right)+960 i\Bigg) k^3+6
\Bigg(-588 \eta ^4\kmax^4+1305 i \eta ^3 \kmax^3 \\
+9965 \eta ^2 \kmax^2+15520 i \eta\kmax+20 \beta
   ^2 \eta ^2 \left(45 \eta ^2 \kmax^2+776 i \eta  \kmax+252\right) \\
+5 \beta  \eta\left(261 i \eta ^3 \kmax^3+2173 \eta ^2 \kmax^2+6208 i \eta 
\kmax+2016\right)+5040\Bigg)k^2
+4 \kmax \Bigg(84 \eta ^4 \kmax^4 \\
-330 i \eta ^3 \kmax^3-3275 \eta ^2\kmax^2+7065 i\eta  \kmax-15 \beta ^2 \eta ^2 \left(20 \eta ^2
\kmax^2-471 i \eta \kmax+1008\right)\\
-5 \beta  \eta  \left(66 i \eta ^3 \kmax^3+715 \eta ^2\kmax^2-2826 i
   \eta  \kmax+6048\right)-15120\Bigg) k \displaybreak[0] \\
-8 \kmax^2 \Bigg(-84 \eta ^4 \kmax^4+330i \eta^3 \kmax^3-1450 \eta ^2 \kmax^2+2385 i \eta 
\kmax+15 \beta ^2 \eta ^2 \bigg(20\eta ^2 \kmax^2\\
+159 i \eta  \kmax+378\bigg)+10 \beta  \eta\left(33 i \eta ^3\kmax^3-115 \eta ^2 \kmax^2+477 i \eta 
\kmax+1134\right)+5670\Bigg)\Bigg)\displaybreak[0]\\
+16 \kmin^{-3/2}
   \sqrt{k+\kmin} \Bigg(35 \eta ^2 \left(111 \eta ^2 \kmin^2+192 i \eta  \kmin+64
\beta \eta  (3 i \eta  \kmin+1)+64\right) k^4 \\
-5 \eta  \Bigg(294 \eta ^3 \kmin^3+717 i \eta^2\kmin^2-2656 \eta  \kmin-960 \beta ^2 \eta ^2 (3
\eta  \kmin-i) \\
+i \beta  \eta\left(717 \eta ^2 \kmin^2+5536 i \eta  \kmin+1920\right)+960 i\Bigg) k^3
+6\Bigg(-588 \eta ^4 \kmin^4 \\
+1305 i \eta ^3 \kmin^3+9965 \eta ^2 \kmin^2+15520 i \eta 
\kmin+20 \beta ^2 \eta ^2 \left(45 \eta ^2 \kmin^2+776 i \eta  \kmin+252\right) \\
+5 \beta  \eta \left(261 i \eta ^3 \kmin^3+2173 \eta ^2 \kmin^2+6208 i \eta 
   \kmin+2016\right)+5040\Bigg) k^2\displaybreak[0]\\
-4 \kmin \Bigg(84 \eta ^4 \kmin^4-330 i \eta^3\kmin^3-3275 \eta ^2 \kmin^2+7065 i \eta  \kmin-15
\beta ^2 \eta ^2 \bigg(20\eta ^2\kmin^2\\
-471 i \eta  \kmin+1008\bigg)
-5 \beta  \eta  \left(66 i\eta ^3\kmin^3+715 \eta ^2 \kmin^2-2826 i \eta 
\kmin+6048\right)-15120\Bigg) k\\
-8 \kmin^2\Bigg(-84 \eta^4 \kmin^4+330 i \eta ^3 \kmin^3-1450 \eta ^2 \kmin^2+2385 i \eta 
\kmin+15 \beta^2 \eta ^2 \bigg(20 \eta ^2 \kmin^2 \\
+159 i \eta  \kmin+378\bigg)+10 \beta  \eta\left(33 i
   \eta ^3 \kmin^3-115 \eta ^2 \kmin^2+477 i \eta 
   \kmin+1134\right)+5670\Bigg)\Bigg)\displaybreak[0]\\
-16\kmin^{-3/2}\sqrt{k-\kmin}\Bigg(35\eta ^2 \left(111 \eta ^2 \kmin^2+192 i \eta  \kmin+64 \beta
 \eta  (3 i \eta \kmin+1)+64\right) k^4\\
+10 \eta  \Bigg(147 \eta ^3 \kmin^3-1104 i \eta ^2\kmin^2+1552
   \eta  \kmin+480 \beta ^2 \eta ^2 (3 \eta  \kmin-i)\\
+16 \beta  \eta  \left(-69 i \eta^2  
\kmin^2+187 \eta  \kmin-60 i\right)-480 i\Bigg) k^3+6 \Bigg(-588 \eta ^4
\kmin^4+480 i\eta ^3 \kmin^3\\
+8165 \eta ^2 \kmin^2+14720 i \eta  \kmin-20 \beta ^2 \eta ^2\left(45
   \eta ^2 \kmin^2-736 i \eta  \kmin-252\right)\\
+5 \beta  \eta  \left(96 i \eta ^3 \kmin^3+1453 \eta ^2 \kmin^2+5888 i \eta 
\kmin+2016\right)+5040\Bigg) k^2 \\
+4 \kmin \Bigg(84 \eta ^4 \kmin^4+120 i \eta ^3 \kmin^3-2675 \eta ^2\kmin^2+6165 i\eta  \kmin+15
\beta ^2 \eta ^2 \bigg(20 \eta ^2 \kmin^2 \\
+411 i \eta\kmin-1008\bigg)+5 i \beta  \eta  \left(24 \eta ^3 \kmin^3+475 i \eta ^2
\kmin^2+2466 \eta  \kmin+6048 i\right)-15120\Bigg) k\displaybreak[0]\\
+8 \kmin^2 \bigg(84 \eta ^4\kmin^4+120i \eta^3 \kmin^3+2050 \eta ^2 \kmin^2-3285 i \eta  \kmin+15
\beta ^2 \eta ^2 \bigg(20\eta ^2\kmin^2 \\
-219 i \eta  \kmin-378\bigg)+10 \beta  \eta  \left(12 i
\eta ^3\kmin^3+235  \eta ^2 \kmin^2-657 i \eta 
\kmin-1134\right)-5670\bigg)\Bigg) \displaybreak[0]\\
-5040 k\bigg(-9 \eta ^4 k^4-45 i \eta ^3 k^3-150 \eta ^2 k^2-340 i \eta  k+20 \beta ^2 \eta ^2
\left(2\eta ^2 k^2-17 i \eta  k+12\right)\\
+5 \beta  \eta  \left(-9 i \eta ^3 k^3-22 \eta ^2 k^2-136
i \eta k+96\right)+240\bigg) \log \left(2 \sqrt{k}\right)\\
+5040 k \bigg(-9 \eta ^4 k^4-45 i \eta^3 k^3-150 \eta^2 k^2-340 i \eta  k+20 \beta ^2 \eta ^2
\left(2 \eta ^2 k^2-17 i \eta  k+12\right)\\
+5 \beta \eta \left(-9 i \eta ^3 k^3-22 \eta ^2 k^2-136 i \eta  k+96\right)+240\bigg) \log
\left(2 \left(\sqrt{\kmax}+\sqrt{\kmax-k}\right)\right) \\
+5040 k \bigg(-9 \eta ^4 k^4+45 i \eta^3 k^3-150 \eta ^2 k^2+340 i \eta  k+20 \beta ^2 \eta ^2
\left(2 \eta ^2 k^2+17 i \eta k+12\right)\\
+5 \beta\eta  \left(9 i \eta ^3 k^3-22 \eta ^2 k^2+136 i \eta  k+96\right)+240\bigg) \log \left(2
   \left(\sqrt{\kmax}+\sqrt{k+\kmax}\right)\right)\\
-5040 k \bigg(-9 \eta ^4 k^4+45 i \eta^3 k^3-150 \eta ^2 k^2+340 i \eta  k+20 \beta ^2 \eta ^2
\left(2 \eta ^2 k^2+17 i \eta k+12\right) \\
+5 \beta \eta  \left(9 i \eta ^3 k^3-22 \eta ^2 k^2+136 i \eta  k+96\right)+240\bigg) \log \left(2
   \left(\sqrt{\kmin}+\sqrt{k+\kmin}\right)\right)\Bigg\} \displaybreak[1]\\
 +8\pi G \dN{\vp_{0}} \frac{\alpha ^2}{604800\beta ^2 \eta^4 k^3} 
\Bigg\{315 \Bigg(-9 \eta ^4 k^4+60 i \eta^3 k^3 -130 \eta ^2 k^2+300 i \eta  k\\
+20\beta ^2 \eta ^2 \left(4 \eta ^2 k^2+15 i \eta  k+120\right)+10
\beta  \eta  \left(6 i \eta ^3 k^3-5 \eta^2 k^2+60 i \eta  k+480\right)\\
+2400\Bigg)\arctan\left(\sqrt{\frac{\kmin}{k-\kmin}}\right)k^3 
+315 \Bigg(9 \eta ^4 k^4+10 i \eta ^3 k^3 +290 \eta ^2 k^2-1020 i \eta  k\\
+20 \beta ^2 \eta^2\left(4 \eta ^2 k^2-51 i \eta  k-120\right)
+10\beta  \eta  \left(i \eta ^3 k^3+37 \eta ^2 k^2-204i\eta k-480\right)\\
-2400\Bigg) \log \left(2\sqrt{k}\right) k^3 -315 \Bigg(9 \eta ^4 k^4+10 i \eta ^3
k^3+290\eta ^2 k^2-1020 i \eta  k\\
+20 \beta ^2 \eta ^2 \left(4 \eta ^2 k^2-51 i \eta 
k-120\right)+10\beta  \eta \left(i \eta ^3 k^3+37 \eta ^2 k^2-204 i \eta  k-480\right)\\
-2400\Bigg)\log \left(2 \left(\sqrt{\kmax}+\sqrt{\kmax-k}\right)\right) k^3 
-315 \Bigg(9 \eta ^4 k^4-10 i \eta^3 k^3\\
+290 \eta ^2 k^2+1020 i \eta  k+20 \beta ^2 \eta ^2\left(4 \eta ^2 k^2+51 i \eta 
k-120\right)+10 \beta\eta \bigg(-i \eta ^3 k^3+37 \eta ^2 k^2\\
+204 i \eta k-480\bigg)
-2400\Bigg) \log\left(2\left(\sqrt{\kmax}+\sqrt{k+\kmax}\right)\right) k^3+315 \Bigg(9 \eta ^4
k^4-10 i \eta^3 k^3\\
+290 \eta ^2 k^2+1020 i \eta  k+20 \beta ^2 \eta ^2 \left(4 \eta ^2 k^2+51 i\eta 
k-120\right)\\
+10   \beta  \eta  \left(-i \eta ^3 k^3+37 \eta ^2 k^2+204 i \eta 
k-480\right)
-2400\Bigg) \log\left(2  \left(\sqrt{\kmin}+\sqrt{k+\kmin}\right)\right) k^3 \displaybreak[0]\\
-\frac{315}{2} \Bigg(-9 \eta ^4 k^4+60 i \eta ^3 k^3-130 \eta ^2 k^2+300 i \eta  k+20 \beta ^2
\eta ^2 \left(4 \eta ^2 k^2+15 i \eta k+120\right)\\
+10 \beta  \eta  \left(6 i \eta ^3 k^3-5 \eta ^2 k^2+60 i \eta 
k+480\right)+2400\Bigg) \pi k^3\displaybreak[0]\\
+\sqrt{\kmax} \sqrt{k+\kmax} \Bigg\{2835 \eta ^4 k^6-630 i \eta ^3(5 \beta  \eta-3 i
   \kmax \eta +5) k^5+14 \eta ^2 \Bigg(1800 \beta ^2 \eta ^2\\
-1428 \kmax^2 \eta ^2+2710 i\kmax \eta +5 \beta  (542 i \eta  \kmax+3201) \eta +14205\Bigg) k^4\\
+4\eta \Bigg(2364 \eta^3 \kmax^3+6620 i \eta ^2 \kmax^2-9465 \eta  \kmax+75 \beta ^2 \eta ^2 (200
\eta \kmax+303 i)\displaybreak[0]\\
+5 \beta  \eta  \left(1324 i \eta ^2 \kmax^2+1107 \eta  \kmax+9090
   i\right)+22725 i\Bigg) k^3\\
-24 \Bigg(-1392 \eta ^4 \kmax^4+2500 i \eta ^3\kmax^3+13010 \eta^2 \kmax^2-16035 i \eta  \kmax+5
\beta ^2 \eta ^2 \bigg(208 \eta ^2\kmax^2\\
-3207 i \eta \kmax-5796\bigg)+10 \beta  \eta  \left(250 i \eta ^3 \kmax^3+1405 \eta ^2
   \kmax^2-3207 i \eta  \kmax-5796\right)-28980\Bigg) k^2\displaybreak[0] \\
+160 \kmax \Bigg(24 \eta^4  \kmax^4-88 i \eta ^3 \kmax^3-597 \eta ^2 \kmax^2+783 i \eta  \kmax-9
\beta ^2 \eta ^2 \bigg(8 \eta ^2 \kmax^2\\
-87 i \eta  \kmax-126\bigg)+\beta  \eta  \left(-88 i\eta ^3
   \kmax^3-669 \eta ^2 \kmax^2+1566 i \eta  \kmax+2268\right)+1134\Bigg) k\\
+320 \kmax^2 \Bigg(-24 \eta ^4 \kmax^4+88 i \eta ^3 \kmax^3-348 \eta ^2
\kmax^2+477 i\eta  \kmax\\
+9 \beta ^2 \eta ^2 \left(8 \eta ^2 \kmax^2+53 i \eta\kmax+84\right)+2
   \beta  \eta  \bigg(44 i \eta ^3 \kmax^3-138 \eta ^2 \kmax^2\\
+477 i \eta\kmax+756\bigg)+756\Bigg)\Bigg\}\displaybreak[1]\\
+\sqrt{\kmax} \sqrt{\kmax-k} \Bigg\{-2835 \eta^4 k^6-630 \eta ^3 (5 i \beta  \eta +3 \kmax \eta +5
i) k^5-14 \eta ^2 \Bigg(1800 \beta ^2\eta^2\\
-1428 \kmax^2 \eta ^2+2710 i \kmax \eta +5 \beta (542i \eta  \kmax+3201)\eta +14205\Bigg) k^4\\
+4\eta  \Bigg(2364 \eta ^3 \kmax^3+6620 i \eta ^2 \kmax^2-9465\eta\kmax+75 \beta ^2 \eta ^2 (200
\eta \kmax+303 i)\\
+5 \beta  \eta  \left(1324 i \eta ^2 \kmax^2+1107 \eta  \kmax+9090 i\right)+22725
i\Bigg) k^3+24 \Bigg(-1392 \eta ^4\kmax^4+2500 i \eta ^3 \kmax^3 \\
+13010 \eta ^2 \kmax^2-16035 i\eta\kmax+5 \beta^2 \eta ^2 \left(208 \eta ^2 \kmax^2-3207 i \eta 
\kmax-5796\right) \displaybreak[0]\\
+10 \beta  \eta\left(250 i \eta ^3 \kmax^3+1405 \eta ^2 \kmax^2-3207 i \eta 
\kmax-5796\right)-28980\Bigg) k^2\\
+160 \kmax \Bigg(24 \eta ^4 \kmax^4-88 i \eta ^3 \kmax^3-597
\eta ^2 \kmax^2+783 i \eta  \kmax-9 \beta ^2 \eta ^2 \bigg(8 \eta ^2 \kmax^2-87 i \eta \kmax\\
-126\bigg)+\beta  \eta  \left(-88 i \eta ^3 \kmax^3-669 \eta ^2\kmax^2+1566 i
   \eta  \kmax+2268\right)+1134\Bigg) k \displaybreak[0]\\
-320 \kmax^2 \Bigg(-24 \eta ^4 \kmax^4+88i \eta^3 \kmax^3-348 \eta ^2 \kmax^2+477 i \eta  \kmax\\
+9\beta ^2 \eta ^2 \left(8\eta ^2\kmax^2+53 i \eta  \kmax+84\right)+2 \beta  \eta  \bigg(44 i
\eta^3\kmax^3-138 \eta^2 \kmax^2+477 i \eta  \kmax+756\bigg)+756\Bigg)\Bigg\}\displaybreak[0]\\
-\sqrt{k-\kmin}\sqrt{\kmin} \Bigg\{-2835 \eta ^4 k^6+1890 i \eta ^3 (10 \beta  \eta +i \kmin \eta
+10) k^5+14 \eta ^2 \Bigg(1800 \beta ^2 \eta ^2\\
+1428 \kmin^2 \eta ^2-1660 i \kmin \eta +5 \beta(-332 i\eta  \kmin-1761) \eta -10605\Bigg) k^4-4
\eta  \Bigg(-2364 \eta ^3 \kmin^3\\
+13480 i\eta ^2 \kmin^2
+39465 \eta  \kmin+75 \beta ^2 \eta ^2(200 \eta  \kmin-1083 i)+5 \beta\eta 
   \bigg(2696 i \eta ^2 \kmin^2+10893 \eta  \kmin \displaybreak[0]\\
-32490 i\bigg)-81225 i\Bigg) k^3
-24 \Bigg(1392 \eta ^4 \kmin^4-1000 i \eta ^3 \kmin^3-10930 \eta ^2 \kmin^2+12735 i
\eta \kmin\\
+5 \beta ^2 \eta ^2 \left(208 \eta ^2 \kmin^2+2547 i \eta\kmin+5796\right)+10
   \beta  \eta  \bigg(-100 i \eta ^3 \kmin^3\\
-989 \eta ^2 \kmin^2+2547 i \eta \kmin+5796\bigg)+28980\Bigg) k^2+160 \kmin \Bigg(24 \eta ^4
\kmin^4+32 i \eta^3\kmin^3 \displaybreak[0]\\
-453 \eta ^2 \kmin^2+603 i \eta  \kmin+9 \beta ^2 \eta ^2 \left(8 \eta^2
   \kmin^2+67 i \eta  \kmin+126\right)+\beta  \eta  \bigg(32 i \eta ^3 \kmin^3-381
\eta ^2 \kmin^2\\
+1206 i \eta  \kmin+2268\bigg)+1134\Bigg) k+320 \kmin^2 \Bigg(24 \eta^4 \kmin^4+32 i \eta ^3
\kmin^3+492 \eta ^2 \kmin^2-657 i \eta  \kmin\\
+9\beta ^2 \eta ^2 \left(8 \eta ^2 \kmin^2-73 i \eta  \kmin-84\right)+2 \beta  \eta  \bigg(16 i
\eta ^3 \kmin^3\\
+282 \eta ^2 \kmin^2-657 i \eta\kmin-756\bigg)-756\Bigg)\Bigg\}\displaybreak[1]\\
+\sqrt{\kmin} \sqrt{k+\kmin} \Bigg(-2835 \eta^4 k^6+630 \eta ^3 (5 i \beta  \eta +3 \kmin \eta +5
i) k^5-14 \eta ^2 \bigg(1800 \beta ^2\eta^2\\
-1428 \kmin^2 \eta ^2+2710 i \kmin \eta +5 \beta  (542i \eta  \kmin+3201)\eta
   +14205\bigg) k^4-4 \eta  \bigg(2364 \eta ^3 \kmin^3\\
+6620 i \eta ^2 \kmin^2-9465\eta\kmin+75 \beta ^2 \eta ^2 (200 \eta  \kmin+303 i)+5 \beta  \eta 
\bigg(1324 i \eta ^2\kmin^2+1107 \eta  \kmin \displaybreak[0]\\
+9090 i\bigg)+22725 i\bigg) k^3+24 \Bigg(-1392 \eta^4
   \kmin^4+2500 i \eta ^3 \kmin^3+13010 \eta ^2 \kmin^2-16035 i \eta \kmin\\
+5 \beta^2 \eta ^2 \left(208 \eta ^2 \kmin^2-3207 i \eta  \kmin-5796\right)+10 \beta  \eta 
\bigg(250 i \eta ^3 \kmin^3+1405 \eta ^2 \kmin^2 \displaybreak[0]\\
-3207 i \eta\kmin-5796\bigg)-28980\Bigg) k^2-160 \kmin \Bigg(24 \eta ^4 \kmin^4-88 i \eta ^3
\kmin^3-597 \eta ^2 \kmin^2+783 i \eta  \kmin\\
-9 \beta ^2 \eta ^2 \left(8 \eta ^2 \kmin^2-87 i\eta \kmin-126\right)+\beta  \eta  \bigg(-88 i
\eta ^3 \kmin^3-669 \eta ^2\kmin^2\\
+1566 i \eta  \kmin+2268\bigg)+1134\Bigg) k-320 \kmin^2 \bigg(-24 \eta ^4 \kmin^4+88
i \eta^3 \kmin^3-348 \eta ^2 \kmin^2+477 i \eta  \kmin\\
+9 \beta ^2 \eta ^2 \left(8\eta ^2 \kmin^2+53 i \eta  \kmin+84\right)\\
+2 \beta  \eta  \left(44 i\eta ^3\kmin^3-138 \eta
   ^2 \kmin^2+477 i \eta  \kmin+756\right)+756\bigg)\Bigg)\Bigg\} \,.
\end{multline}

\section{Discussion of properties of source term for different potentials}
\label{sec:apx-srcdisc}
The evolution of the source term for the four potentials has been discussed in
Section~\ref{sec:compare-res}, with particular emphasis on the evolution after horizon crossing as
shown in Figure~\ref{fig:cmp-src-kwmap}. Here the differences apparent at early times, shown in
Figure~\ref{fig:cmp-src-zoom-kwmap} are commented on.

\begin{figure}
 \centering
 \includegraphics[width=0.75\textwidth]{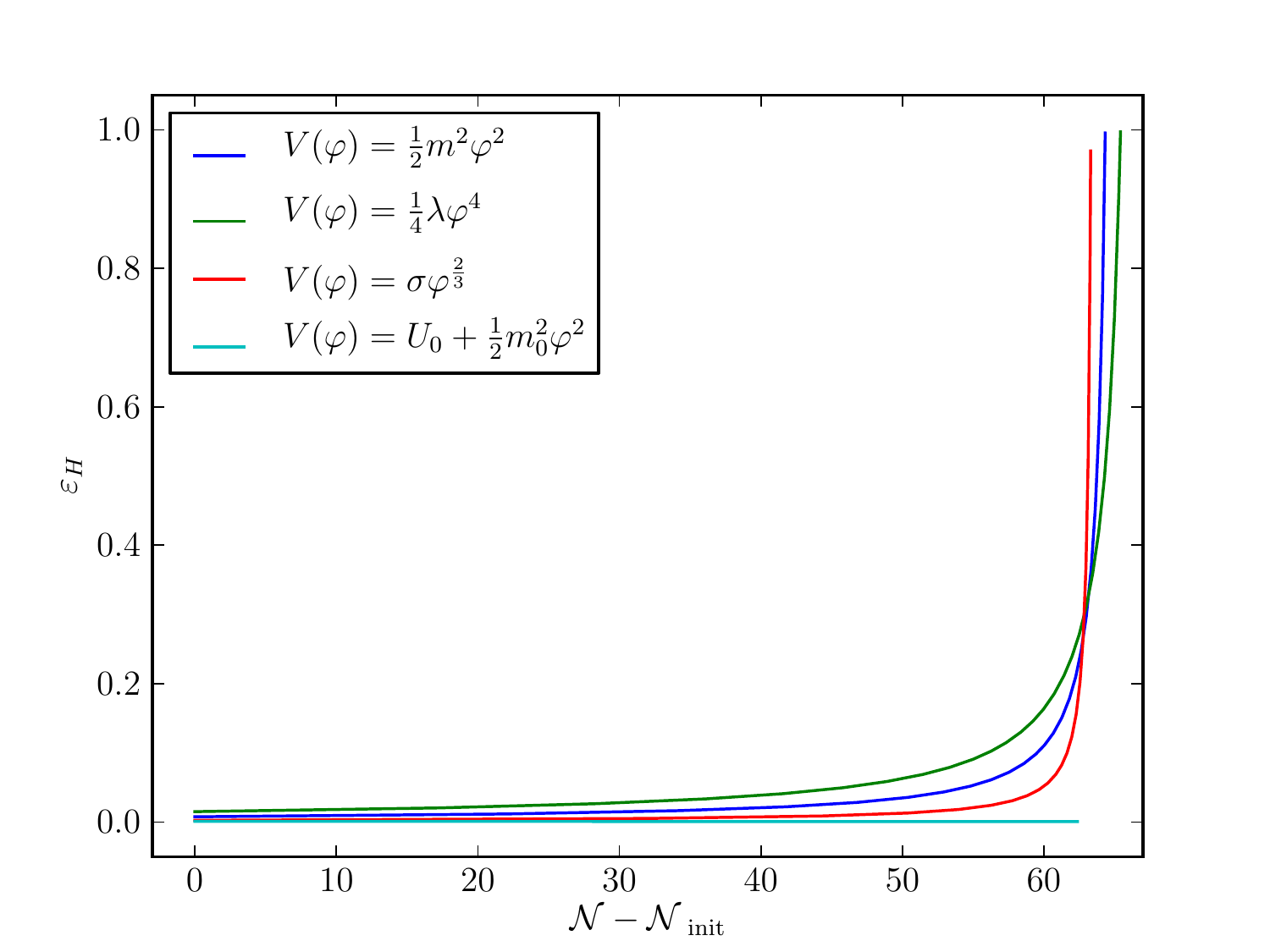}
 \caption[Value of $\varepsilon_H$ for the Four Potentials]{The value of $\varepsilon_H$ for
the four potentials.}
 \label{fig:eps-apx}
\end{figure}

\begin{figure}
 \centering
 \includegraphics[width=0.75\textwidth]{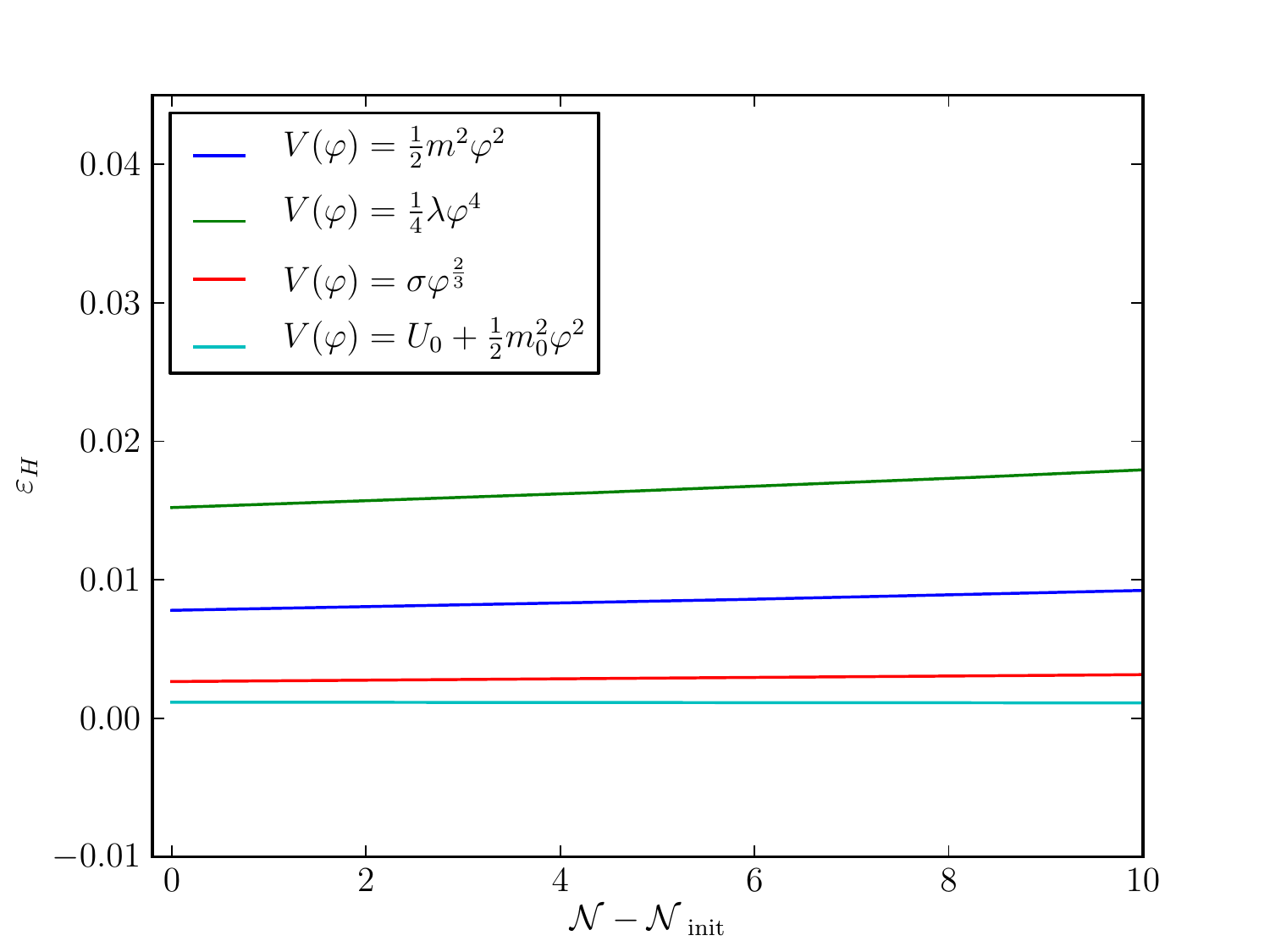}
 \caption[Value of $\varepsilon_H$ for the Four Potentials at Early Times]{The value of
$\varepsilon_H$ for the four potentials at early times.}
 \label{fig:eps-zoom-apx}
\end{figure}

\begin{figure}
 \centering
 \includegraphics[width=0.75\textwidth]{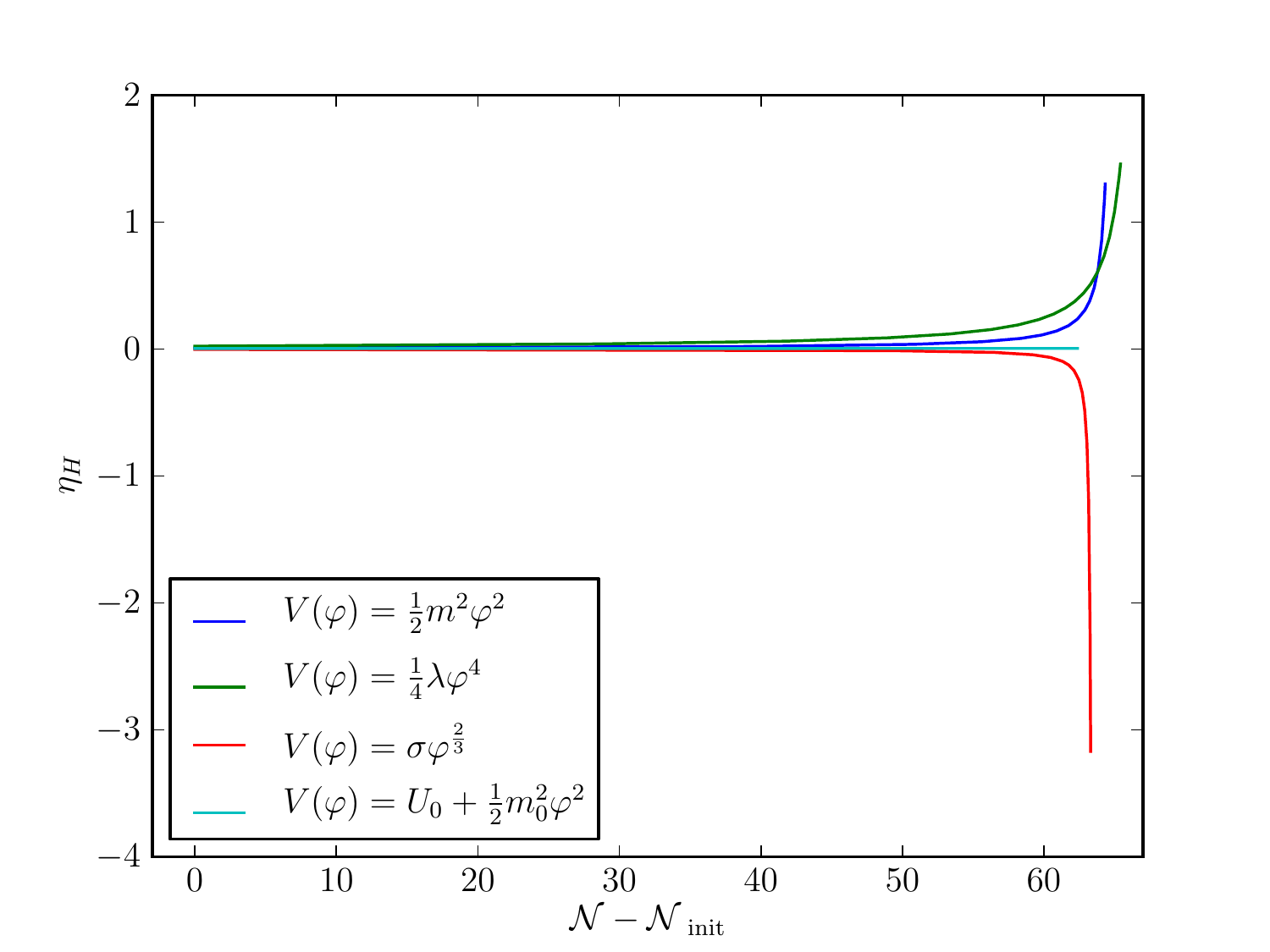}
 \caption[Value of $\eta_H$ for the Four Potentials]{The value of $\eta_H$ for the four
potentials.}
 \label{fig:eta-apx}
\end{figure}

\begin{figure}
 \centering
 \includegraphics[width=0.75\textwidth]{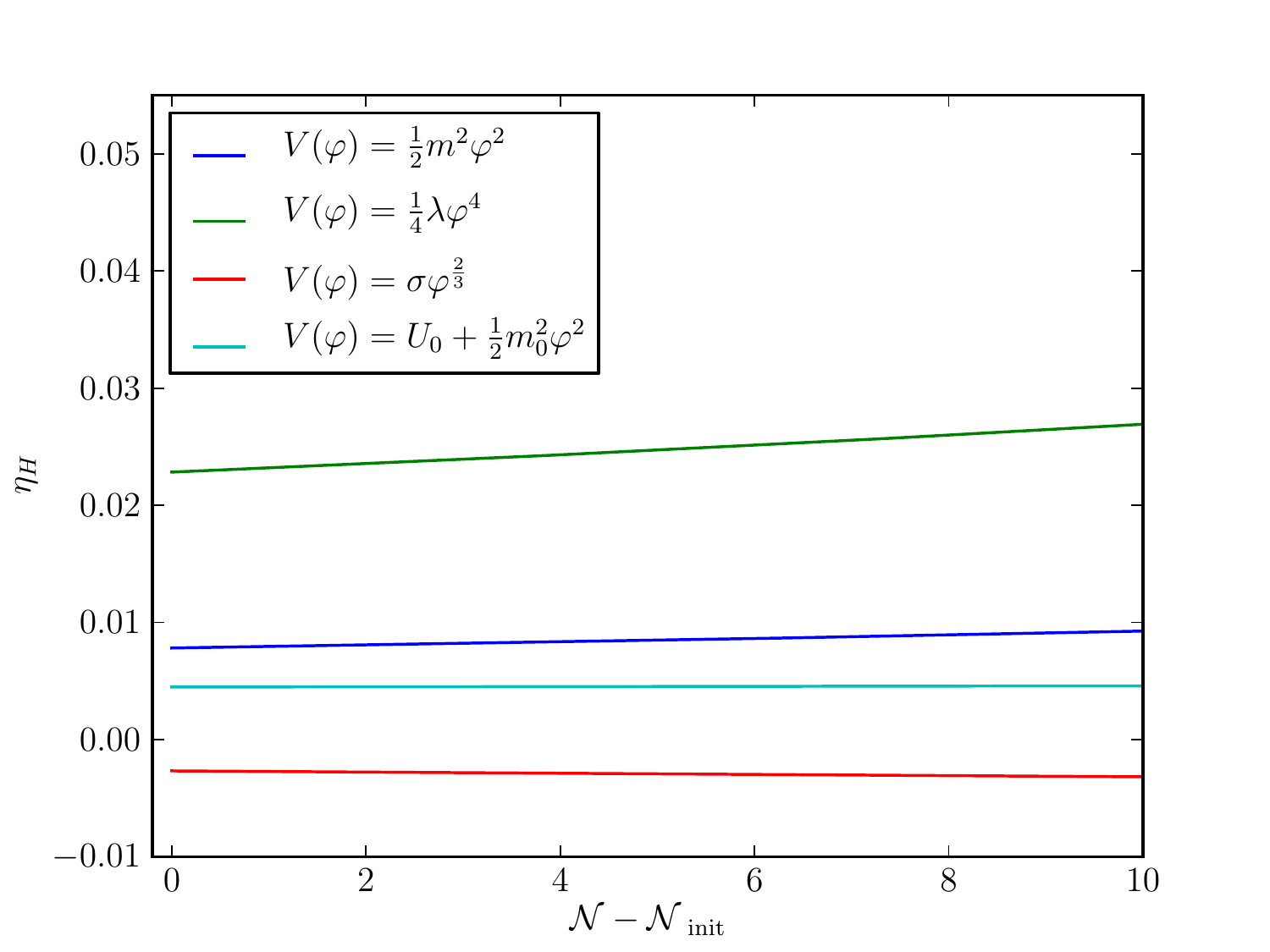}
 \caption[Value of $\eta_H$ for the Four Potentials at Early Times]{The value of $\eta_H$ for the
four potentials at early times.}
 \label{fig:eta-zoom-apx}
\end{figure}

At early times the first order perturbations are still very close to the Bunch-Davies initial
conditions as outlined in Section~\ref{sec:initconds-num}. In particular the perturbations are
highly oscillatory with phase $\exp(-k\eta)$, where $\eta$ is the conformal time. When
$\varepsilon_H$ is small this is given by 
\begin{equation}
 \eta = -\frac{1}{aH(1-\varepsilon_H)}\,.
\end{equation}
It is therefore instructive to plot the slow roll parameter $\varepsilon_H$ for the four potentials
at these early times, as has been done in Figures~\ref{fig:eps-apx} and \ref{fig:eps-zoom-apx}. For
completeness the other slow roll parameter $\eta_H$ defined in \eq{eq:etaHdefn-intro} has been
plotted in Figures~\ref{fig:eta-apx} and \ref{fig:eta-zoom-apx}. 

Figures~\ref{fig:eps-apx} and \ref{fig:eta-apx} show $\varepsilon_H$ and $\eta_H$ for the
four different models. Figures~\ref{fig:eps-zoom-apx} and
\ref{fig:eta-zoom-apx} show the early stages of the evolution as in
Fig~\ref{fig:cmp-src-zoom-kwmap}.

As is clear from these figures the change in the slow roll parameters is not easily related
to the differences in the profiles of the four potentials in Figure~\ref{fig:cmp-src-zoom-kwmap}. In
particular, although $\varepsilon_H$ and $\eta_H$ are quite different for the quadratic and quartic
models, the magnitude of $S$ after horizon crossing for these models is very similar.

\begin{figure}
 \centering
 \includegraphics[width=0.75\textwidth]{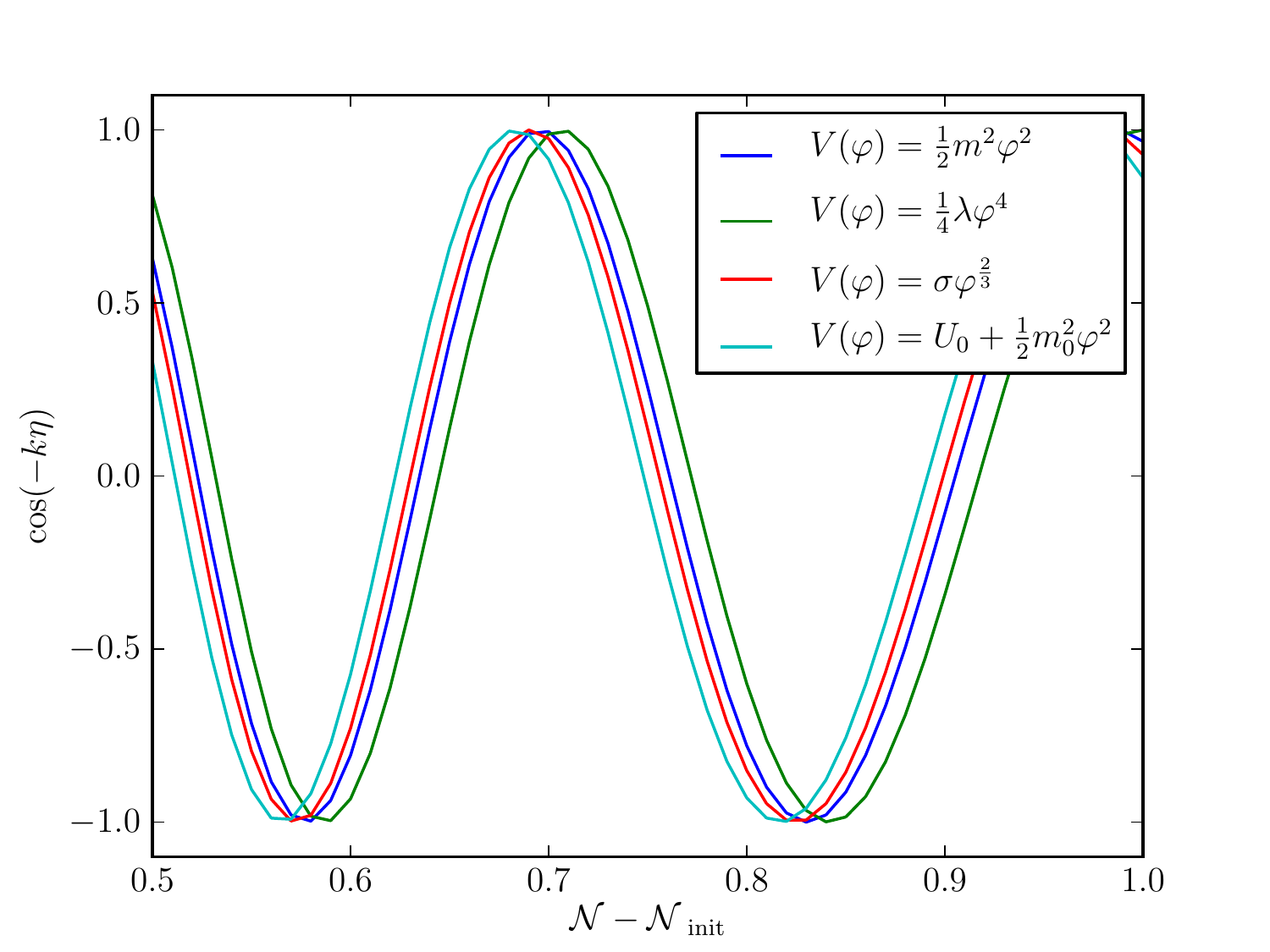}
 \caption[Real Part of the Phase at Early Times]{The real part of the phase in the Bunch Davies
initial conditions for the four different
potentials at early times.}
 \label{fig:cos-keta-apx}
\end{figure}

At the earliest stages of the calculation of $S$, one or two e-foldings after the initialisation of
the first order perturbation, there appear to be small oscillations which affect the models in
different ways. The highly oscillatory initial conditions, combined with the small but appreciable
differences in $\varepsilon_H$ and $\eta_H$ contribute to this effect. In
Figure~\ref{fig:cos-keta-apx} the real part of the phase of the initial condition for $\dvp1$ is
plotted just after initialisation for the four potentials. The small differences in phase for each
model combined with the sharp cutoff at large and small $k$ values could explain the variations
in $|S|$ at early times as seen in Figure~\ref{fig:cmp-src-zoom-kwmap}. 

\begin{singlespace}
\bibliography{thesis}
\bibliographystyle{utphys-ih}

\end{singlespace}
\end{document}